\documentclass[12pt]{article}
\pdfoutput=1

\usepackage{draft}
\usepackage{putex} 
\usepackage[all,cmtip]{xy}
\usepackage{fancybox}
  \usepackage{float} 
 \usepackage{extarrows}
 
\usepackage{graphicx,subfig}
\usepackage{cite}
\usepackage{mciteplus}
\usepackage{skak}
\usepackage{amsmath}
\DeclareFontFamily{OT1}{pzc}{}
\DeclareFontShape{OT1}{pzc}{m}{it}{<-> s * [1.10] pzcmi7t}{}
\DeclareMathAlphabet{\mathpzc}{OT1}{pzc}{m}{it}

\usepackage{multirow}

\newcommand{\TY}{{\rm TY}}

\def\({\left(}
\def\){\right)}

\newcommand{\pa}{\partial}

\newcommand{\ep}{\epsilon}
\newcommand{\Rep}{{\rm Rep}}

\renewcommand{\cH}{\mathcal{H}}
\newcommand{\eeq}{\end{equation}}
\newcommand{\ea}{\end{array}}

\def\eea{\end{eqnarray}}

\def\<{\langle}
\def\>{\rangle}

\def\bZ{\mathbb{Z}}

\def\cL{\mathcal{L}}
\def\cM{\mathcal{M}}
\def\cN{\mathcal{N}}
\def\cA{\mathcal{A}}
\def\cO{\mathcal{O}}
\def\cC{\mathcal{C}}
\def\cT{\mathcal{T}}
\usepackage{amsmath}
\usepackage{comment}
\usepackage{amssymb}
\usepackage{amsthm}
\usepackage{bbm}

\theoremstyle{definition}

\usepackage{color}
\usepackage{tikz-cd}
\usepackage{comment}

\usepackage[colorlinks=true]{hyperref}

\usepackage{adjustbox}

\def\[#1\]{%
  \begin{equation}#1\end{equation}%
}

\usepackage{fancybox}
\usepackage{cite}
\usepackage{framed}
\definecolor{shadecolor}{rgb}{0.9,0.9,0.95}
\definecolor{refkey}{rgb}{0.5,0.5,0}
\definecolor{labelkey}{rgb}{0.5,0.5,0}
\definecolor{citekey}{rgb}{0.5,0.5,0}
\definecolor{darkgreen}{rgb}{0,0.5,0}
\definecolor{darkblue}{cmyk}{0.9,0.9,0,0}
\definecolor{darkred}{rgb}{0.6,0,0.3}

\begin{document}
	
	\preprint{}

	\institution{Weizman}{Department of Condensed Matter Physics, Weizmann Institute of Science, Rehovot, Israel}
	\institution{PU}{Joseph Henry Laboratories, Princeton University, Princeton, NJ 08544, USA}
	\institution{CMSA}{Center of Mathematical Sciences and Applications, Harvard University, Cambridge, MA 02138, USA}
	\institution{HU}{Jefferson Physical Laboratory, Harvard University,
		Cambridge, MA 02138, USA}

	\title{
    \huge Fusion Category Symmetry II:\\ \Large Categoriosities at $c = 1$ and Beyond
	}

	\authors{Ryan Thorngren\worksat{\Weizman,\CMSA}    and Yifan Wang\worksat{\PU,\CMSA,\HU}}

	\abstract{We study generalized symmetries of quantum field theories in 1+1D generated by topological defect lines with no inverse. This paper follows our companion paper on gapped phases and anomalies associated with these symmetries. In the present work we focus on identifying fusion category symmetries, using both specialized 1+1D methods such as the modular bootstrap and (rational) conformal field theory (CFT), as well as general methods based on gauging finite symmetries, that extend to all dimensions. We apply these methods to $c = 1$ CFTs and uncover a rich structure. We find that even those $c = 1$ CFTs with only finite group-like symmetries can have  continuous fusion category symmetries, and prove a Noether theorem that relates such symmetries in general to non-local conserved currents. We also use these symmetries to derive new constraints on RG flows between 1+1D CFTs.
			 }
	\date{}

	\maketitle
	
	\tableofcontents
	
	\section{Introduction}
 
Symmetry is a guiding principle in the study of physical systems, especially quantum field theories (QFTs) where there are often very few other analytic methods. By exploring the constraints of symmetry, we have gained many insights into non-perturbative aspects of QFTs, such as renormalization group (RG) flows, phase diagrams, critical phenomena, and dualities.

Traditionally, symmetries have been formulated as transformations of the fields, or better yet as invertible operators on Hilbert space. Recently, it has been recognized that symmetries in QFT are associated with topological operators. For instance, symmetries which produce transformations of the fields define invertible topological defects by choosing boundary conditions across a hypersurface to that the fields on one side are glued to the transformed fields on the other side. This operator captures everything about the symmetry, so we might define a symmetry as an invertible codimension-one topological operator.

There is an obvious generalization, which is to relax the conditions of codimension-one and invertibility. By exploiting analogies with more familiar symmetries, general topological operators can be used to derive new symmetry principles in QFT. For example, topological surface operators in 3+1D that are generators of one-form symmetries have been used to elucidate confining and Higgs phases of gauge theories \cite{Gukov:2013zka,Kapustin:2013uxa,Gaiotto:2014kfa,Kapustin_2017,Gaiotto:2017yup}. More generally a $p$-form symmetry is given by a codimension-$(p + 1)$ invertible topological operator.

Relaxing invertibility is apparently more exotic, and the constraints from the associated symmetries are just beginning to be explored \cite{Kapustin:2009av,Kapustin:2010if,Fuchs:2012dt,Aasen:2016dop,Bhardwaj:2017xup,Bal:2018wbw,Chang:2018iay,Lin:2019hks,Thorngren:2019iar,Lichtman:2020nuw,Aasen:2020jwb,Komargodski:2020mxz,Chang:2020imq,Huang:2021ytb}.\footnote{We note there is a different notion of categorical symmetry presented in \cite{Ji:2019jhk,Kong:2020cie} for $2+1$D TQFTs in the presence of gapless boundaries. In their context, the symmetries themselves are invertible but the anomalies are not captured by a phase, thus non-invertible.} In this paper, which follows our previous work \cite{Thorngren:2019iar}, we show that these symmetries, which we dub ``fusion category symmetries", are actually rather ubiquitous, by giving methods for constructing them and applying those methods to $c = 1$ conformal field theories (CFTs) in 1+1D.
 
We use the term ``fusion category" not in the precise mathematical sense, but just to refer to the fact that topological defects form some kind of category (whose morphisms are given by junctions of defects \cite{Kapustin:2010ta}) with fusion data. We expect a general mathematical definition will eventually be found (see \cite{Johnson-Freyd:2020usu} for a recent proposal), but for our methods it will not be necessary.

For 1+1d CFTs, topological defect lines (TDLs) have been axiomatized in \cite{Bhardwaj:2017xup,Chang:2018iay}. TDLs come with a fusion algebra. Ordinary symmetries define invertible TDLs, but generally the algebra will also include \textit{non-invertible} TDLs, which are our primary interest here.

\begin{figure}[!htb]
	\centering
	\begin{minipage}{0.25\textwidth}
		\includegraphics[width=1\textwidth]{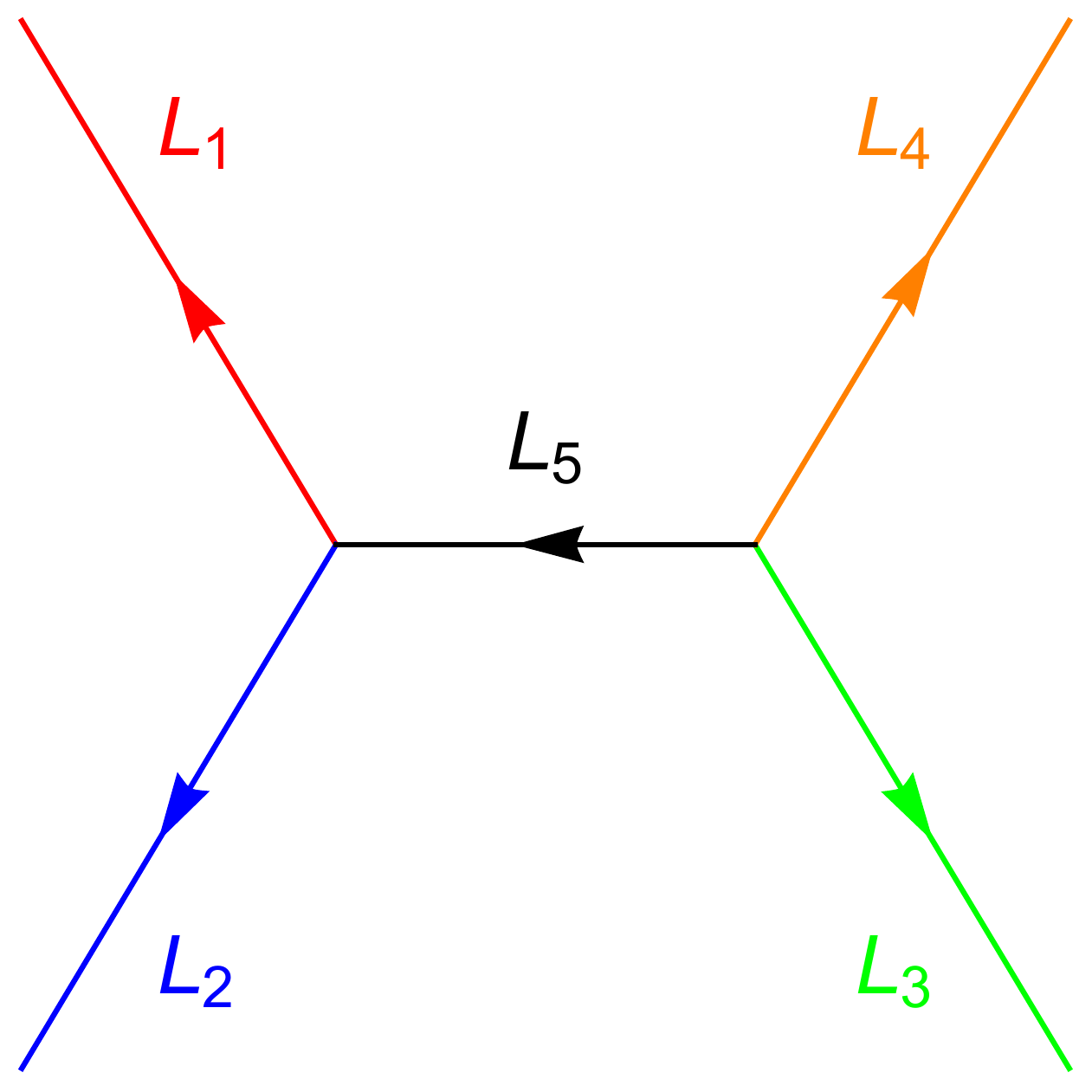}
	\end{minipage}%
	\begin{minipage}{0.20\textwidth}\begin{eqnarray*} =\sum_{{\rm simple}~\cL_6 } \cK^{\cL_1 \cL_4}_{\cL_2 \cL_3}(\cL_5,\cL_6) \\ \end{eqnarray*}
	\end{minipage}%
	\begin{minipage}{0.25\textwidth}
		\includegraphics[width=1\textwidth]{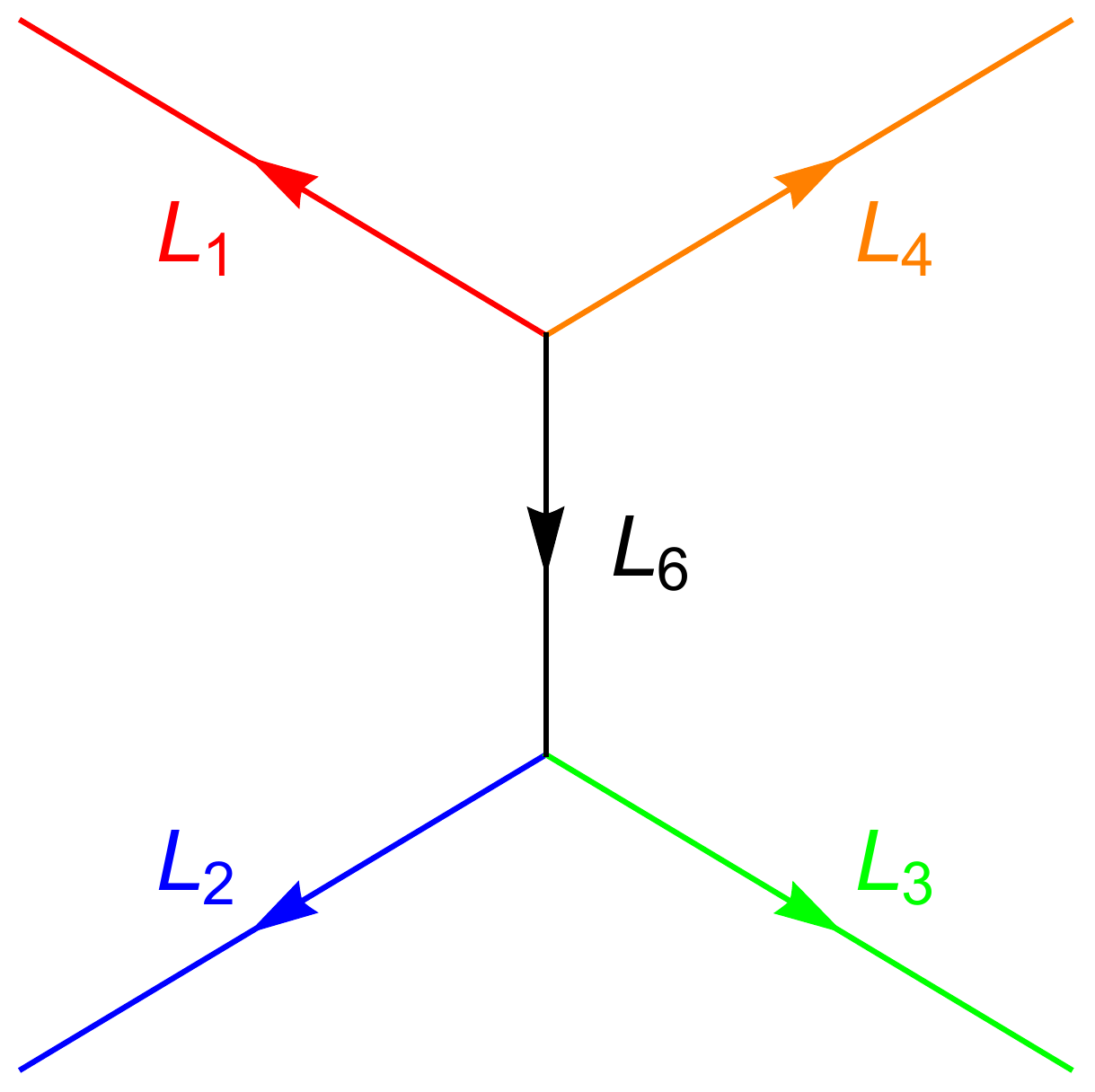}
	\end{minipage}%
	\caption{The F-move, that relates correlation functions containing two topologically distinct networks of TDLs. Here $\cK^{\cL_1 \cL_4}_{\cL_2 \cL_3}(\cL_5,\cL_6)$ are the F-symbols (also known as crossing kernels or $6j$ symbols).}
	\label{fig:Fsymbol}
\end{figure}

Furthermore we require the TDLs to obey the \textit{locality} condition \cite{Chang:2018iay}, meaning that the TDLs can fuse and split locally, creating branched networks. Topological invariance implies that the networks can be deformed by isotopy away from operator insertions without changing the value of any correlation function computed in the background of such a network. Meanwhile, recombinations of the network transform these correlation functions by certain universal factors. All such recombinations of generic (i.e. trivalent) networks are sequences of one basic recombination, shown in Figure~\ref{fig:Fsymbol}, referred to here as the F-move. The associated universal factors are known by many names, but we will refer to them as F-symbols. The collection of fusion rules and F-symbols defines an algebraic structure known as a fusion category \cite{etingof2015tensor}, hence the name ``fusion category symmetry".
 
In recent works \cite{Chang:2018iay,Thorngren:2019iar,Komargodski:2020mxz}, the authors (including us) have explored constraints of the fusion category symmetry on RG flows in 1+1D. In particular, the 't Hooft anomaly which plays an important role in understanding constraints of invertible symmetries on QFT dynamics has a natural generalization for  fusion category symmetries. As explained in \cite{Thorngren:2019iar}, given a fusion category $\cA$ realized by a CFT, the corresponding  fusion category symmetry is anomaly-free (meaning the theory admits an $\cA$-symmetric deformation to a gapped nondegenerate phase) if and only if $\cA$ admits a fiber functor (and such fiber functors describe the $\cA$-symmetric gapped nondegenerate phases, i.e. $\cA$-SPTs, themselves). By definition, RG flows protected by an \textit{anomalous} fusion category symmetry $\cA$ must either undergo $\cA$-symmetry breaking or end at a nontrivial $\cA$-symmetric CFT. For
 instance, any fusion category symmetry which involves a TDL of non-integer quantum dimension is anomalous \cite{Chang:2018iay}.

To utilize its power in constraining RG flows, it is thus very important to identify the fusion category symmetry of the (UV) CFT. Similar to trying to pinpoint ordinary global symmetries, this problem can be quite difficult in a generic QFT, as one needs to specify the action of the putative symmetries on general operators in the theory, some of which may be quite subtle, such as 1+1D vortex operators and 2+1D monopole operators, even when the theory has a Lagrangian description.

Sometimes TDLs can be identified directly using \textit{modular bootstrap} techniques. For example, in 1+1d rational CFTs (RCFTs), TDLs known as Verlinde lines \cite{Verlinde:1988sn} can be defined using the modular data of the chiral algebra (see \cite{Petkova:2000ip,Chang:2018iay} for explicit examples). However we will see that even this approach misses a huge portion of TDLs in familiar CFTs.

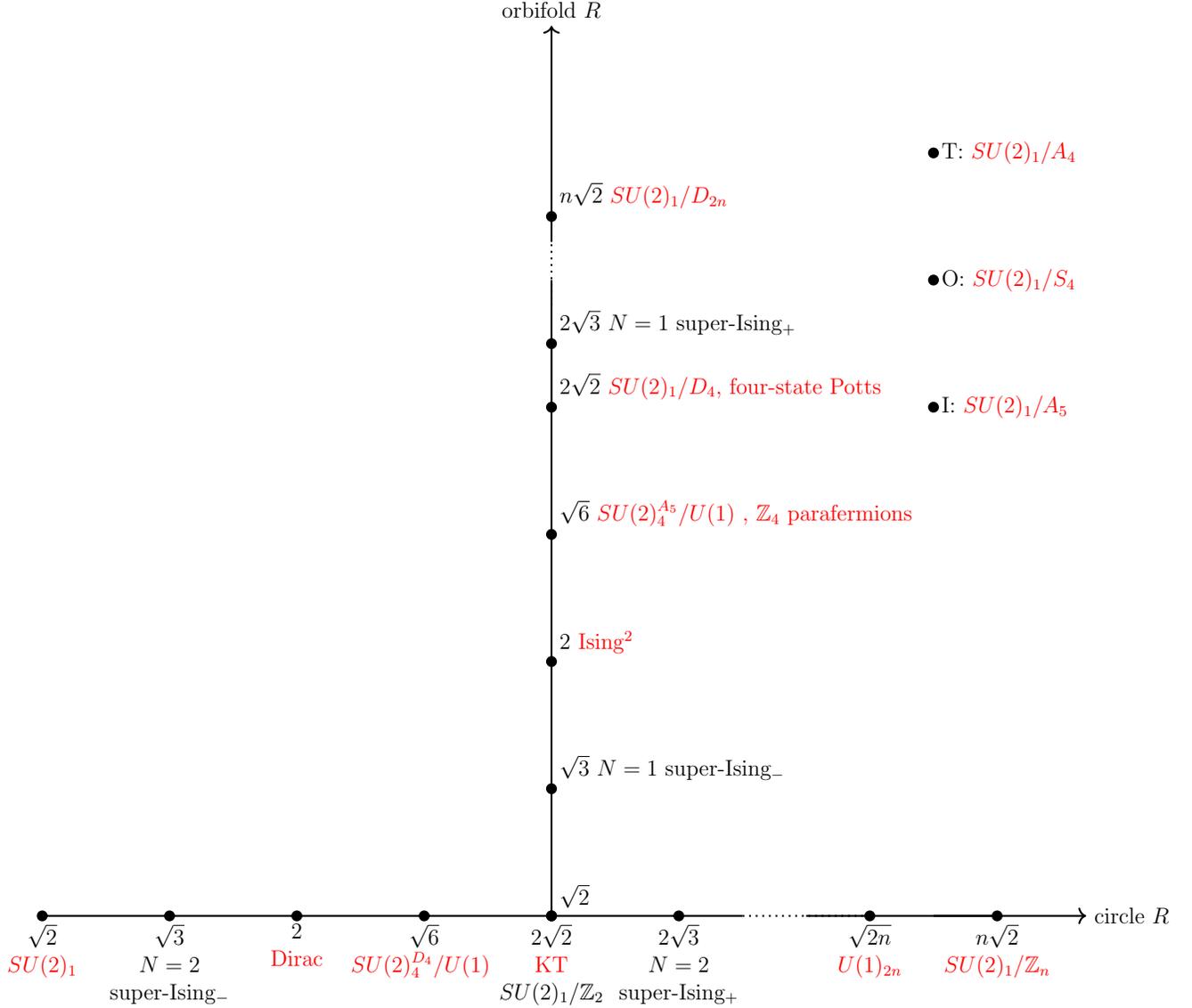
\begin{figure}[!htb]
  	\begin{tikzpicture}[thick,scale=1.9, every node/.style={scale=0.8}]
  	\draw[-] (2,0) -- (2.5,0) node[right] {};
  	\draw[-] (1,0) -- (1.5,0) node[right] {};
  	\draw[->] (3,0) -- (4.2,0) node[right] {circle~$R$};
  	\draw[-] (0,4) -- (0,5) node[right] {};
  	\draw[->] (0,5.3) -- (0,7) node[above] {orbifold~$R$};
  	
  	\filldraw 
  	(-4,0)   circle (1pt) node[align=center,   below] {$\sqrt{2}$\\ \hyperref[appSU2symmetry]{$SU(2)_1$}}--
  	(-3,0) circle (1pt) node[align=center,   below] {$\sqrt{3}$ \\ $N=2$ \\  super-Ising$_-$}--
  	(-2,0) circle (1pt) node[align=center,   below]{$2$ \\ \hyperref[subsecZkTYcompbos]{Dirac}}--
  	(-1.0,0) circle (1pt) node[align=center,   below]{$\sqrt{6}$ \\\hyperref[subsecZkTYcompbos]{$SU(2)^{D_4}_4/U(1)$ }}--
  	(0,0)circle (1pt) node[align=center,   below] {$2\sqrt{2}$\\ \hyperref[subsecKT]{KT} \\$SU(2)_1/\bZ_2$}--
  	(1,0)circle (1pt) node[align=center,   below] {$2\sqrt{3}$\\ $N=2$ \\ super-Ising$_+$ };

  	\draw [dotted] (1.5,0) -- (2.5,0) ; 
  	\filldraw   
  	(2.5,0) circle (1pt) node[align=center,   below]{$\sqrt{2n}$\\ \hyperref[subsecZkTYcompbos]{$U(1)_{2n}$} }
  	--
  	(3.5,0) circle (1pt) node[align=center,   below] {$n\sqrt{2}$\\\hyperref[subsecSU2continuum]{$SU(2)_1/\bZ_n$} }
  	;

  	\filldraw 
  	(0,0) circle (1pt) node[align=center, above right] {$\sqrt{2}$}--
  	(0,1)  circle (1pt) node[align=center, above right] {$\sqrt{3}$ $N=1$ super-Ising$_-$}--
  	(0,2)  circle (1pt) node[align=center, above right] {$2$  \hyperref[secRepH8Ising]{Ising$^2$}}--
  	(0,3)  circle (1pt) node[align=center, above right]  {$\sqrt{6}$ \hyperref[secZ4parafermion]{$SU(2)^{A_5}_4/U(1)$ , $\bZ_4$ parafermions}
  	}--
  	(0,4)  circle (1pt) node[align=center, above right] {$2\sqrt{2}$ \hyperref[secRepD84statepotts]{$SU(2)_1/D_4$, four-state Potts} }--
  	(0,4.5)  circle (1pt) node[align=center, above right] {$2\sqrt{3}$  $N=1$ super-Ising$_+$ };
  	\draw [dotted] (0,4.8) -- (0,5.3) ; 
  	\filldraw   
  	(0,5.5)   circle (1pt) node[align=center, above right]  {$n\sqrt{2}$ \hyperref[seccontinuumZ2orbifold]{$SU(2)_1/D_{2n}$}};

  	\filldraw 
  	(3,6) circle (1pt) node[align=center, right] {T:~\hyperref[sec:A4orb]{$SU(2)_1/A_4$}}
  	;
  	\filldraw 
  	(3,5) circle (1pt) node[align=center, right] {O:~\hyperref[sec:S4orb]{$SU(2)_1/S_4$}}
  	;
  	\filldraw 
  	(3,4) circle (1pt) node[align=center, right] {I:~\hyperref[sec:A5orb]{$SU(2)_1/A_5$}}
  	;

  	\end{tikzpicture} 
  	\caption{The  moduli space of $c=1$ CFTs and fusion category symmetries.}
  	\label{fig:c1moduli}
  \end{figure}

In particular, we focus on $c=1$ CFTs, which form a sort of edge case between rational and general {irrational} CFTs.\footnote{Here rational CFTs are CFTs that contain a finite number of conformal blocks with respect to a chiral algebra that extends the Virasoro algebra. CFTs that do not obey this property are irrational.} Familiar examples of $c=1$ CFTs include the free compact boson, its (reflection) $\mZ_2$ orbifold, and the $SU(2)_1$ WZW model. More general $c=1$ CFTs have been classified and form a moduli space consisting of two continuous branches and three isolated points \cite{Ginsparg:1987eb} (see Figure~\ref{fig:c1moduli}). The $c=1$ CFTs at generic points of the moduli space are \textit{irrational}, with no enhanced chiral algebra. Consequently it becomes much harder to identify the fusion category symmetries compared to the RCFT cases, and to date no broad study of fusion category symmetries in irrational theories have been attempted. However as we will show in this paper, by utilizing a combination of the generalized modular bootstrap method, 1+1D gauge theory arguments, and knowledge from the special rational points on the $c=1$ moduli space, we can capture a large zoo of fusion categories that inhabit different parts of the $c=1$ moduli space summarized in Figure~\ref{fig:c1moduli}.

The rest of the paper is organized as follows. In Section \ref{sectechniques}, we give a review of fusion category symmetry in 1+1D. We describe constraints imposed on the existence of these symmetries by the modular bootstrap equations. We provide general methods for constructing them including RCFT techniques, gauging finite groups (orbifolding), and the bosonization map from (para)fermions. We also prove a Noether theorem for continuous fusion category symmetry.

In Section \ref{seccircbranch} we apply these techniques to the circle branch of the $c = 1$ moduli space. We find no non-invertible symmetries at generic radius $R$. However, for $R = \sqrt{2k}$, the theory exhibits self-duality under $\bZ_k$ gauging corresponding to a pair of duality TDLs associated with two distinct $\bZ_k$ Tambara-Yamagami fusion categories. For $R \in \sqrt{2}\mathbb{Q}$, these fusion categories are enhanced to a continuum of TDLs parameterized by six parameters, coming from the $SO(4)$ symmetry enhancement at $R = \sqrt{2}$.

In Section \ref{secZ2genericorbifold} we study the $\bZ_2$ orbifold branch. We find self-dualities at arbitrary radius under gauging $\bZ_4$ and $\bZ_2 \times \bZ_2$ subgroups of the $D_8$ global symmetry. We identify the corresponding fusion categories as certain $\bZ_4$ Tambara-Yamagami categories in the former case and ${\rm  Rep}(D_8)$ and ${\rm Rep}(H_8)$ in the latter case by relating to more specialized constructions at the RCFT points described by the $\bZ_4$ parafermions, the Ising$^2$ CFT, and the four-state Potts model. We show at generic radius these dualities are enhanced to a four-parameter continuum of TDLs. At radius $R \in  \sqrt{2}\mathbb{Q}$, this continuum is enhanced to six parameters, similar to what happened on the circle branch. As an example, we describe an anomalous \textit{triality} of the Kosterlitz-Thouless (KT) theory which is associated with $\bZ_2 \times \bZ_2$ gauging. 

Finally, in Section \ref{secexceptional} we study the three isolated points of the moduli space, the exceptional orbifolds of the $SU(2)_1$ WZW model. Besides the Verlinde lines which we determine from the modular data, each of these orbifolds host a six-parameter continuum of TDLs arising from the parent $SU(2)_1$ CFT.

Some technical details and complementary discussions to those presented in the main text are given in the Appendix. A visual table of contents with links to relevant sections is shown in Figure~\ref{fig:c1moduli}.

\emph{Acknowledgements} We would like to thank Zohar Komargodski for collaborating on this work in its early stages. RT would also like to acknowledge Tsuf Lichtman, Erez Berg, Andy Stern, and Netanel Lindner for collaboration on a related project, as well as Dave Aasen and Dominic Williamson for many useful discussions. The work of YW is  supported in part by the Center for Mathematical Sciences and Applications and the Center for the Fundamental Laws of Nature at Harvard University. YW would like to thank Ofer Aharony and Xi Yin for useful discussions. YW is also grateful to the Weizmann Institute of Science for hospitality where the project was initiated during his visit.

  \section{Overview of Fusion Category Symmetries}\label{sectechniques}

  \subsection{Axiomatic Constraints and Generalized Modular Bootstrap}

  Like ordinary symmetries, fusion category symmetries act on local operators, but this is not the only data that defines them. For such an action to be associated with a topological defect line (TDL) $\cL$ and hence a fusion category, we would like to be able to define Euclidean correlation functions on arbitrary Riemann surfaces with local operator and TDL insertions. For this, it is necessary and sufficient to further specify the defect Hilbert space $\cH_\cL$, the OPE of defect operators, and the action of $\cL$ on them \cite{Chang:2018iay}.
  
\begin{figure}[!htb]
	\begin{minipage}{0.5\textwidth}
		\centering	\includegraphics[scale=.6]{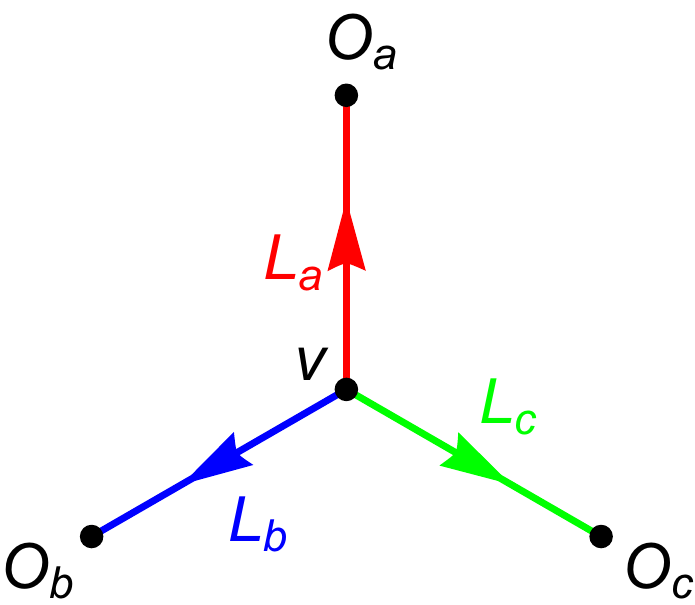}
	\end{minipage}%
	\begin{minipage}{0.5\textwidth}
		\centering
		\includegraphics[width=.6\textwidth]{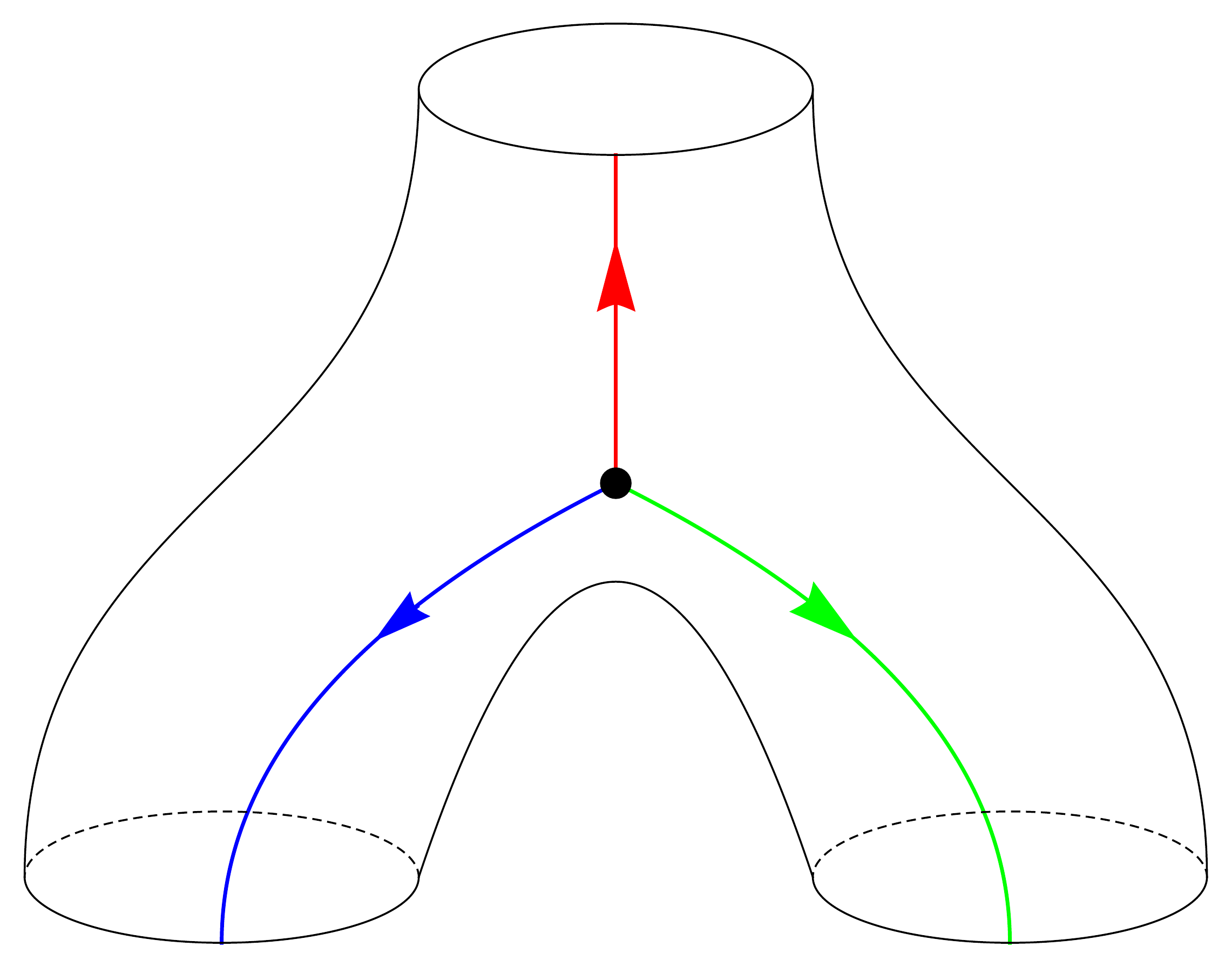}
	\end{minipage}%
	\caption{The three-point function $\la O_a O_b O_c\ra$ connected by TDLs $\cL_{a,b,c}$ and equivalently the CFT partition function on a pair of pants decorated by the TDLs. Here $v$ is a index denoting the fusion channel (topological junction), which we henceforth suppress from the notation and diagrams. The arrows denote orientations of the TDLs. Strictly speaking, the correlation functions depend on a co-orientation, i.e. a choice of normal direction for each TDL, but since we will only work on orientable surfaces, we can choose an orientation to make them equivalent.}
	\label{fig:3pf}
\end{figure}

  This is essentially a geometric  statement. The OPE of operators $\cO_a$ in the defect Hilbert spaces $\cH_{\cL_a}$ is governed by the three-point function $\la \cO_a \cO_b \cO_c\ra$, which is related to the partition function on a pair of pants by a conformal transformation as in Figure~\ref{fig:3pf}.

\begin{figure}[!htb]
	\centering
	\begin{minipage}{0.45\textwidth}
		\centering
		\includegraphics[scale=.3]{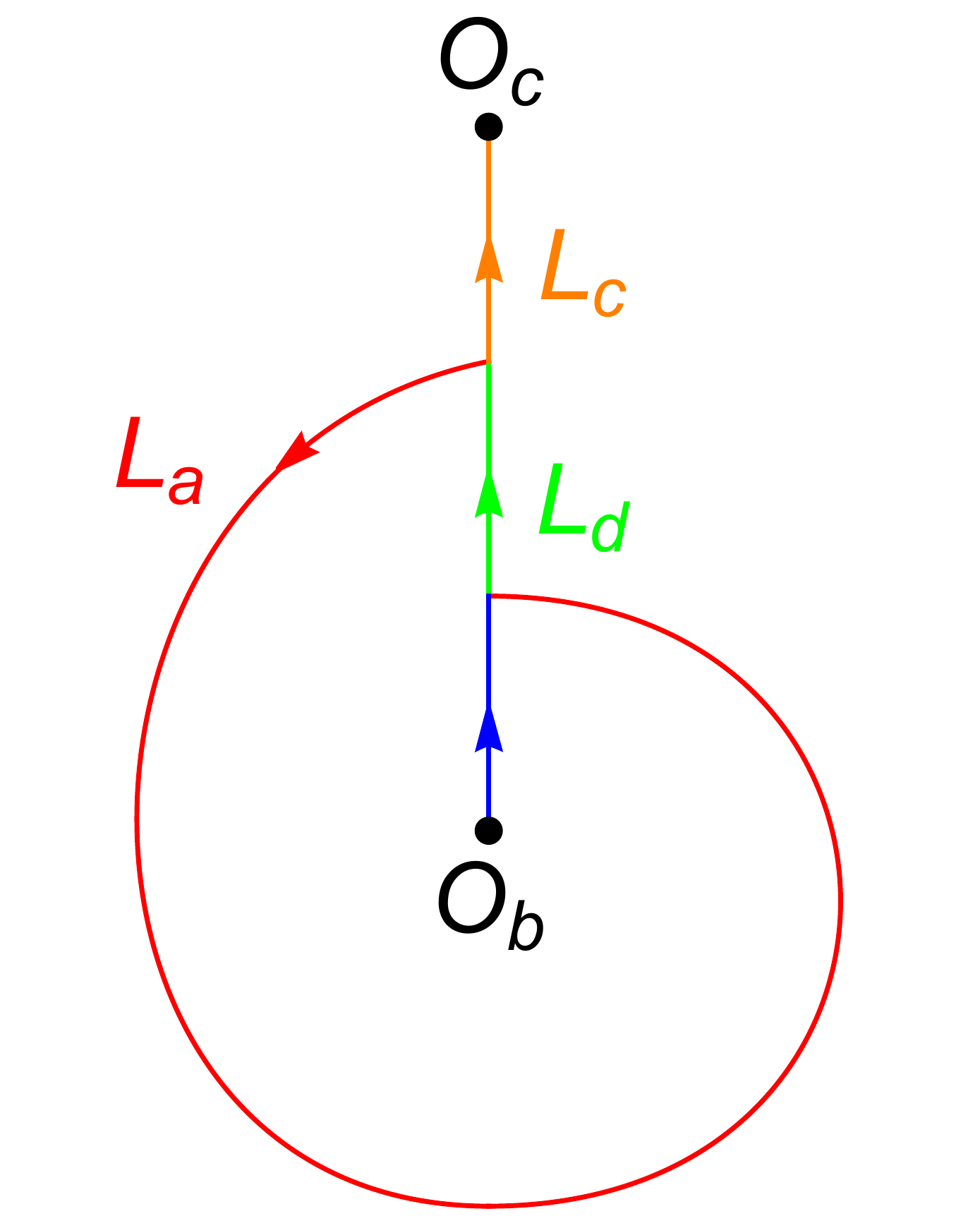}
	\end{minipage}%
	\begin{minipage}{0.45\textwidth}
		\centering
		\includegraphics[width=.6\textwidth]{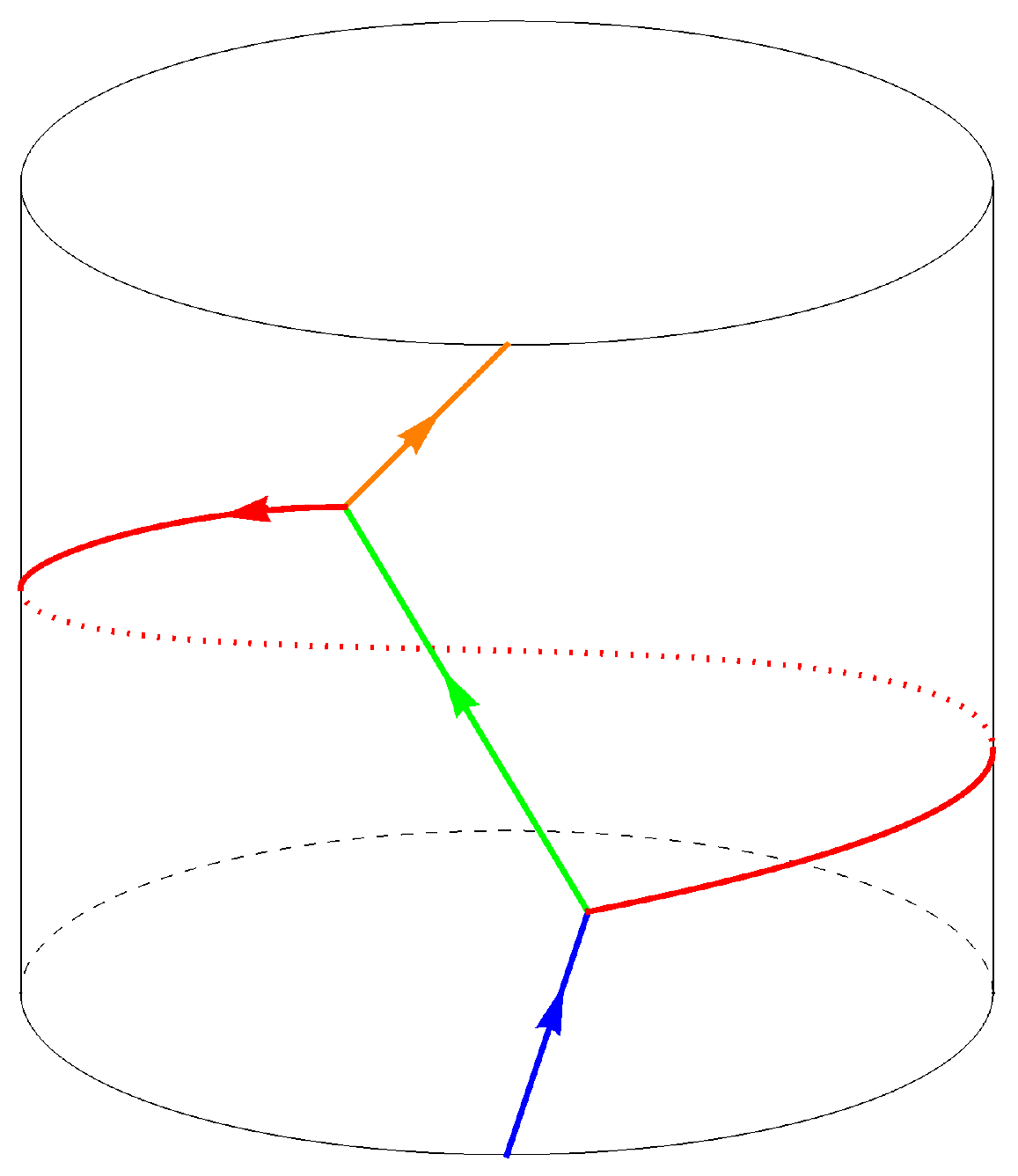}
	\end{minipage}%
	\caption{The \textit{lasso} diagram that depicts the two-point function of defect operators $O_b$ and $O_c$ joined by a web $X$ of TDLs $\cL_{a,b,c,d}$ and equivalently the partition function of the CFT on a cylinder decorated by the same TDL web $X$. They compute the matrix element of the TDL acting on the defect Hilbert spaces $(\widehat\cL_a)_X:\cH_b \to \cH_c$.}
	\label{fig:lasso}
\end{figure}

  The symmetry action of  a TDL $\cL_a$ on the defect Hilbert space $\cH_{\cL_b}$ is more subtle to define. It is determined by a certain two-point function of defect operators, which maps to a partition function on the cylinder with $\cL_a$ wrapping around and $\cL_b$ entering from the bottom. However, unlike ordinary symmetries, non-invertible TDLs generally do not preserve a given defect Hilbert space, for instance they often map ordinary local operators to defect operators, so a different TDL $\cL_c$ will exit at the top of this cylinder in general. Furthermore, we must resolve the crossing of $\cL_a$, $\cL_b$, and $\cL_c$ via a fourth TDL $\cL_d$ in the fusion product $\cL_c\cL_a$ and $\cL_b\cL_a$. We should consider all possible input and output spaces $\cH_b$ and $\cH_c$ and all possible ways of resolving these crossings as part of the symmetry action of $\cL_a$. We denote the corresponding operator as $(\widehat{\cL}_a)_X : \cH_b \to \cH_c$ where $X$ specifies the relevant TDL network on the cylinder as in Figure~\ref{fig:lasso}.

  Any Riemann surface $\Sigma$ with a network of TDLs and operator insertions can be obtained by gluing together these elementary pieces, known as a pants decomposition. Consistency demands the observable to be independent of the choice of pants decompositions, leading to nontrivial constraints on $\la \cO_a\cO_b\cO_c\ra$ and $(\widehat{\cL}_a)_X$. The different decompositions are related by the mapping class group $\Gamma(\Sigma)$. More explicitly, if $\Sigma$ is a Riemann surface with conformal moduli $\tau$ and $n$ marked points, and $X$ is a network of TDLs on $\Sigma$ that ends on the marked points, then the associated $n$-point function $\langle\cO_1,\ldots,\cO_n\rangle_{\Sigma,\tau,X}$ of appropriately-twisted operators sitting at the marked points transforms in the following way under a ``modular transformation", i.e. an element $\phi \in \Gamma(\Sigma)$ fixing the set of $n$ marked points,
    \ie
    \langle\cO_1,\ldots,\cO_n\rangle_{\Sigma,\tau,X} = \langle \phi(\cO_1),\ldots,\phi(\cO_n)\rangle_{\Sigma,\phi(\tau),\phi(X)}\,.
    \label{gminv}
    \fe
    Here crucially the modular transformation acts on the TDL network and operator insertions, hence this relation is also sometimes referred to as ``modular covariance".
    
    As argued in \cite{Chang:2018iay}, it is sufficient to verify invariance of CFT observables under two ``simple moves", namely the F-move on a four-punctured sphere and the S-transform on a one-punctured torus, which generate the mapping class group, and thus supply the complete set of constraints on the defect data.
    
    Let us outline one of the simplest nontrivial constraints among the infinite set, which arises from a torus $\Sigma = T^2$ with no puncture. In this case the mapping class group $\Gamma(\Sigma) = SL(2,\mZ)$ is generated by the familiar $S$ and $T$ transformations. It turns out that the constraints coming from \eqref{gminv} in this case are already quite stringent, and will allow us to constrain the spectrum of TDLs in a given CFT by a sort of generalized modular bootstrap.

\begin{figure}[!htb]
	\centering
	\begin{minipage}{0.20\textwidth}
		\includegraphics[width=1\textwidth]{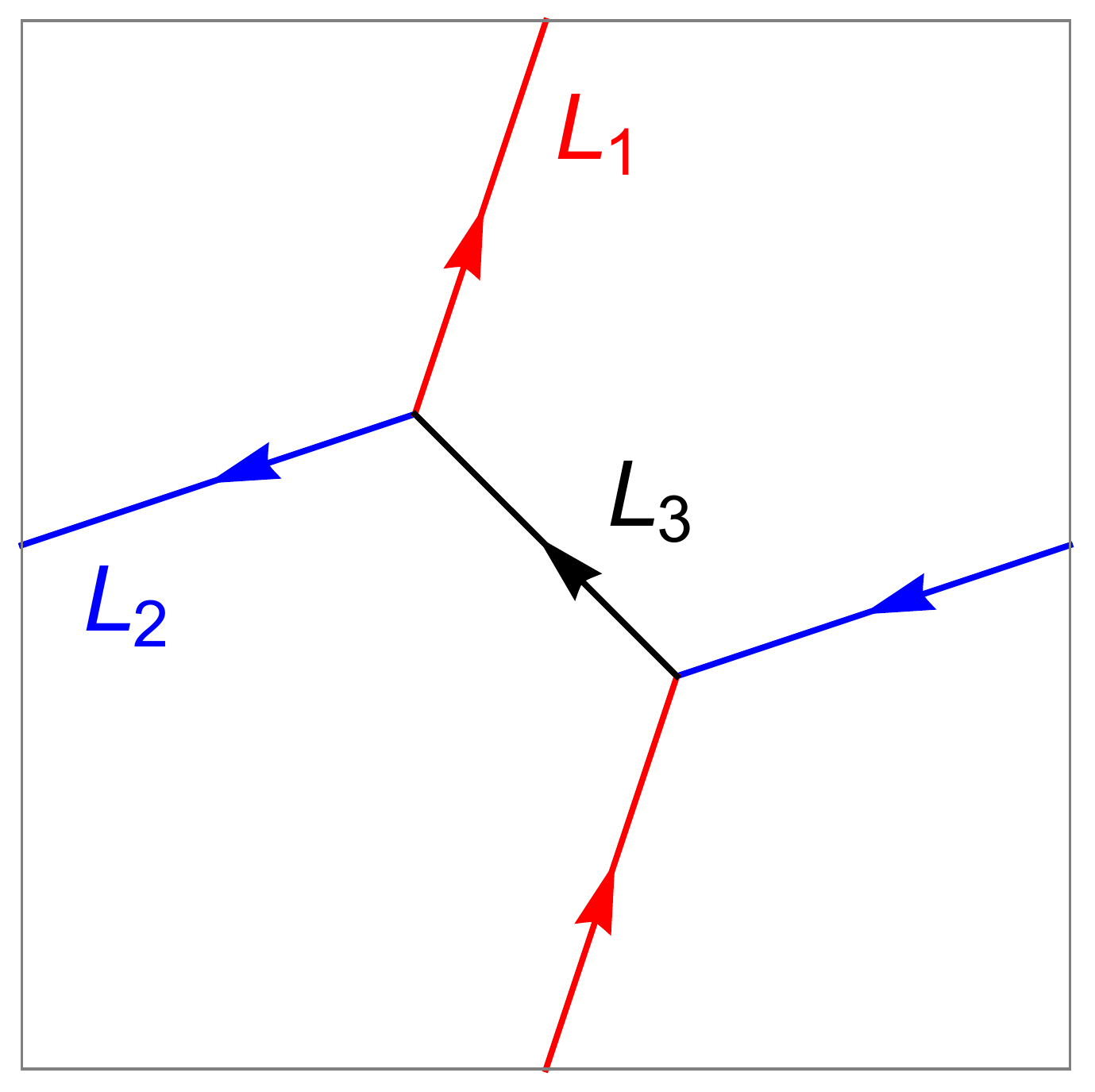}
	\end{minipage}%
	\begin{minipage}{0.08\textwidth}\begin{eqnarray*}~~\longrightarrow^{\!\!\!\!\!\!\!\!\! S}  \\ \end{eqnarray*}
	\end{minipage}%
	\begin{minipage}{0.20\textwidth}
		\includegraphics[width=1\textwidth]{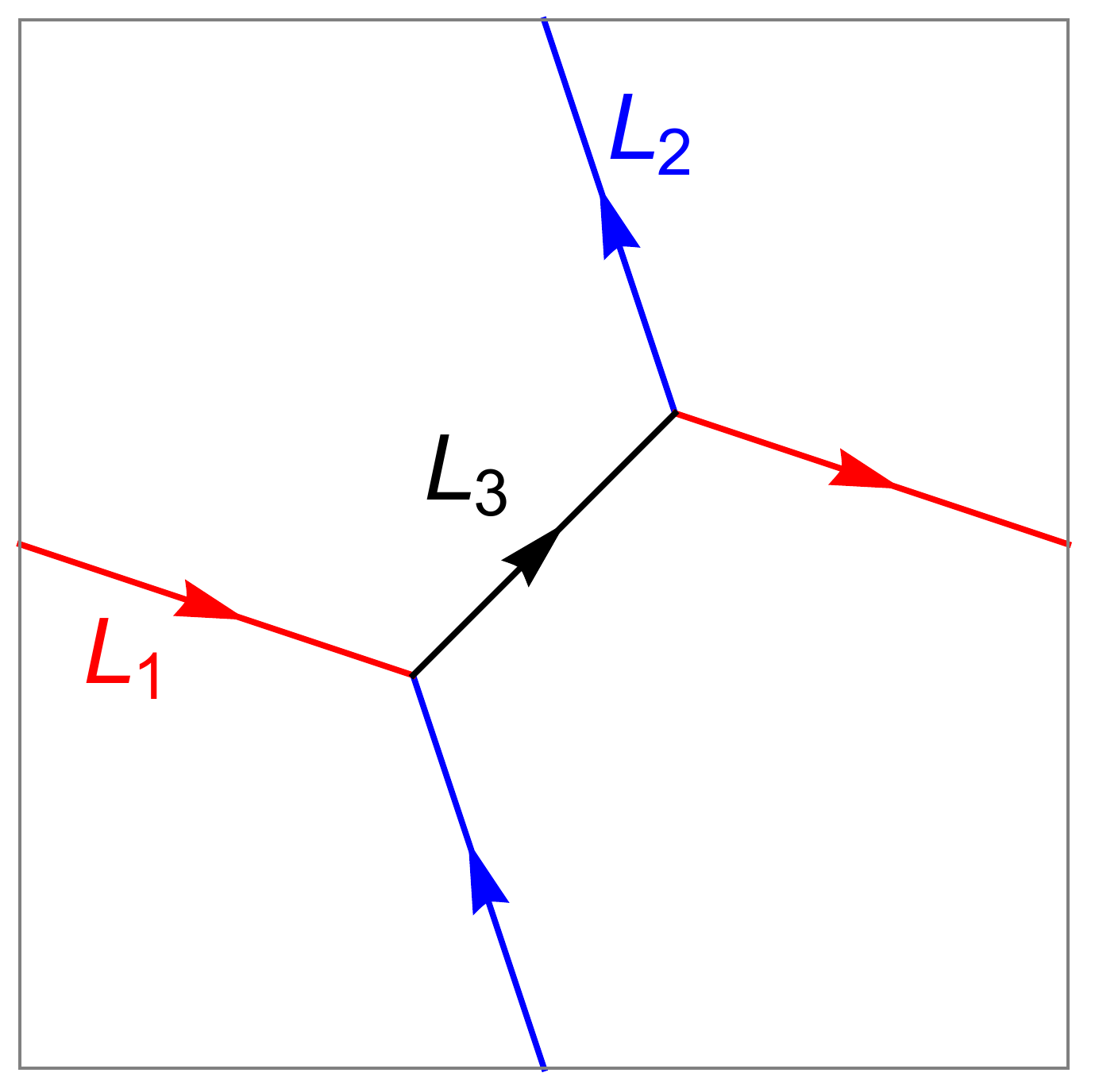}
	\end{minipage}%
	\begin{minipage}{0.25\textwidth}\begin{eqnarray*}~~ =\sum_{\cL_k } \cK^{\cL_1 \bar\cL_2}_{\cL_2 \bar\cL_1}(\cL_3,\cL_k) \\ \end{eqnarray*}
	\end{minipage}%
	\begin{minipage}{0.20\textwidth}
		\includegraphics[width=1\textwidth]{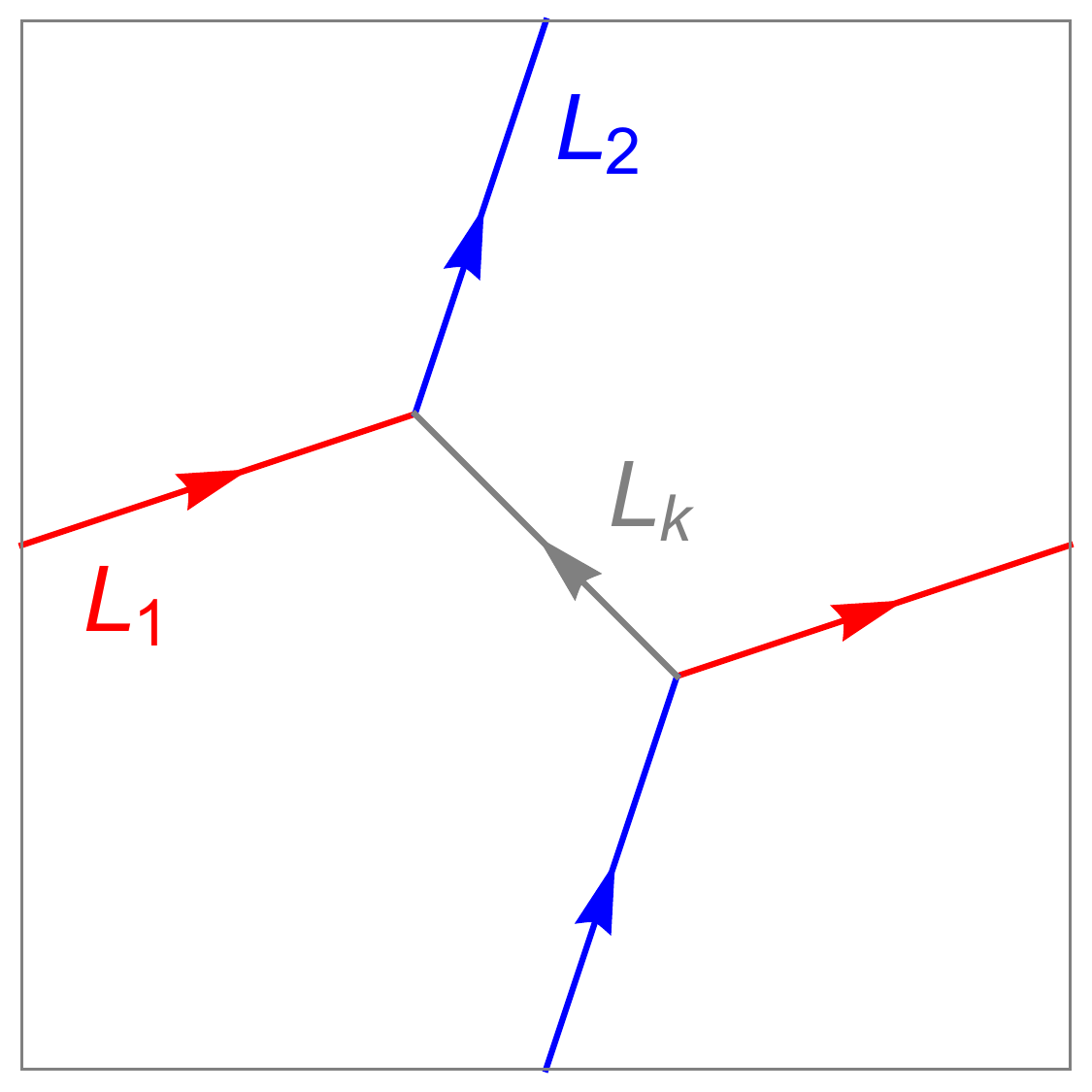}
	\end{minipage}%
	\caption{The first  diagram  on the left defines the twisted partition functions $Z^{\cL_3}_{\cL_1\cL_2}$. The second diagram is obtained from a modular $S$-transform. After a F-move, it is related to a combination of $Z^{\cL_k}_{\cL_2\bar\cL_1}$  for each simple $\cL_k$ that appear in the fusion product $\cL_2\bar\cL_1$.}
	\label{fig:L2L1PF}
\end{figure}

To be more specific,  we define the (twisted) torus partition function with simple TDLs $\cL_1,\cL_2$ inserted along the   time and spatial cycles of $T^2$  by 
\ie
Z_{\cL_1 \cL_2}^{\cL_3}(\tau,\bar\tau )=\tr_{\cH_{\cL_1}} (\widehat\cL_2)_{\cL_3} q^{L_0-{c\over 24}} \bar q^{\bar L_0-{c\over 24}} \,,
\label{tPFdef}
\fe
where $\cH_{\cL_1}$ is the defect Hilbert space for $\cL_1$ (the case with trivial $\cL_1=1$ corresponds to the usual CFT Hilbert space of point operators), $\cL_3$ is a simple TDL that specifies the resolution of the four-fold junctions between $\cL_1$ and $\cL_2$,
and $(\widehat \cL_{2})_{\cL_3}$ implements the corresponding action of the TDL $\cL_2$ on $\cH_{\cL_1}$ (as defined by the leftmost diagram in Figure~\ref{fig:L2L1PF}). We adopt the usual convention for the torus partition function with $q\equiv e^{2\pi i \tau}$ and $\bar q\equiv e^{-2\pi i \bar \tau}$.

Since the TDLs commute with the left and right Virasoro algebras, the twisted partition function $Z^{\cL_3}_{\cL_1 \cL_2}(\tau,\bar\tau)$ naturally decomposes into left and right moving Virasoro characters of the Virasoro primaries $\cH^{\rm prim}_{\cL_1}\subset \cH_{\cL_1}$,
\ie
Z^{\cL_3}_{\cL_1 \cL_2}(\tau,\bar\tau)=\sum_{(h,\bar h)\in \cH_{\cL_1}^{\rm prim}} (\widehat \cL_{2})_{\cL_3}^{h,\bar h}\chi_h \bar \chi_{\bar h}\,,
\label{tPFdec}
\fe
where $(\widehat \cL_{2})_{\cL_3}^{h,\bar h}$ specifies the action of $ \cL_{2}$ on the Virasoro primaries (hence on all states in $\cH_{\cL_1}$).

Likewise, $Z^{\cL_k}_{\cL_2 \bar\cL_1}(\tau)$ defines the twisted partition function where the roles of the two TDLs $\cL_1$ and $\cL_2$ are interchanged. Theses partition functions are related by the modular S-transform as (see Figure~\ref{fig:L2L1PF})
\ie
Z_{\cL_1 \cL_2}^{\cL_3}(-1/\tau) 
=&  \sum_{\cL_k}
\cK^{\cL_1 \bar\cL_2}_{\cL_2 \bar\cL_1}(\cL_3,\cL_k) 
Z_{\cL_2 \bar \cL_1}^{\cL_k}(\tau) \,,
\label{tPFst}
\fe
which translates into the constraint
\ie
\sum_{(h,\bar h)\in \cH_{\cL_1}^{\rm prim}}  (\widehat \cL_{2})_{\cL_3}^{h,\bar h} S_{h h'} S_{\bar h \bar h'}=
\begin{cases}
 \sum_{\cL_k }
 \cK^{\cL_1 \bar\cL_2}_{\cL_2 \bar\cL_1}(\cL_3,\cL_k) 
 (\widehat {\bar\cL_{1}})_{\cL_k}^{h',\bar h'}  & (h',\bar h')\in  \cH_{\cL_2}^{\rm prim} \,,\\
	0 & {\rm otherwise}\,.
\end{cases} 
\label{mbeqn}
\fe
The equation \eqref{mbeqn} applies to any triplet of TDLs $\cL_{1,2,3}$ with a nontrivial three-fold topological junction (i.e. $\bar\cL_3\in \cL_1\cL_2$) and can be generalized by considering more arbitrary TDL networks on $T^2$. We refer to them collectively as (generalized) modular bootstrap equations. The usual modular bootstrap equation  for 2d CFTs corresponds to the special case with trivial TDLs $\cL_1=\cL_2=1$.\footnote{The complete set of bootstrap equations for a 2d CFT enriched by TDLs includes  the F-crossing relations in Figure~\ref{fig:Fsymbol} and the modular covariance of torus one-point function involving a general point operator potentially connected to a nontrivial TDL network as in Figure~\ref{fig:L2L1PF} (which generalizes \eqref{mbeqn}).}

\subsection{RCFTs and Verlinde Lines}
\label{sec:verlinde}

A priori, the large set of bootstrap equations like \eqref{mbeqn} makes it a highly-constrained and difficult problem to identify non-trivial TDLs in a given CFT (especially when the set of primaries in $\cH_{\cL_1}^{\rm prim}$ is infinite).
An exception is when the CFT $\cT$ is rational. Suppose $\cT$ contains (higher spin) symmetry currents that generate a chiral algebra $\cV$ which extends the Virasoro algebra, and that the CFT is defined by a diagonal modular invariant torus partition function which contains a finite number of $\cV$ blocks corresponding to primary operators $\phi_i$. Then there is a special class of TDLs that commute with the large symmetry $\cV$ (both left and right) known as the Verlinde lines $\cL_i$, which are in one-to-one correspondence with the $\cV$ primaries $\phi_i$ and act diagonally on the $\cV$ blocks,
\ie
\widehat \cL_i |\phi_k\ra = {S_{ik}\over S_{0k}}   |\phi_k\ra\,,
\label{verlindeonops}
\fe
where $S_{ik}$ is the modular S-matrix. The fusion rules of these TDLs follow from the Verlinde formula
\ie
\cL_i \cL_j=\sum_k N_{ij}^k \cL_k ~~{\rm with}~~N_{ij}^k=\sum_{\ell}{S_{i\ell} S_{j\ell} S^*_{k\ell}\over S_{0\ell}}\,.
\label{verlindeformula}
\fe
The defect Hilbert space $\cH_{\cL_k}$ contains non-diagonal primaries $\phi_{i|j}$ of $\cV$ with degeneracy $N_{ij}^k$. 

The fusion category symmetry generated by the Verlinde lines $\cL_i$ turns out to be braided, and moreover a modular tensor category (MTC) equivalent to the representation category of the chiral algebra $\cC={\rm Rep}(\cV)$ \cite{Moore:1988qv,Moore:1989vd}.\footnote{We emphasize that the braiding structure is not necessary for general TDLs.} In this case, the bootstrap axioms reviewed in the last section boil down to a finite set of conditions on the defect structure constants which are satisfied in the RCFT. This was shown based on explicit constructions using the relation between the RCFT and the 2+1D topological field theory associated to the MTC $\cC$ together with a choice of certain algebra object in $\cC$
\cite{Fuchs:2002cm,Fuchs:2003id,Fuchs:2004dz,Fuchs:2004xi,Fjelstad:2005ua}.

For general TDLs in RCFTs and TDLs in general CFTs, we lose the luxury of rationality and have to face an infinite set of bootstrap equations.
However, if we can infer the existence of certain TDLs from alternative methods such as direct constructions,  the action of such TDLs on local (defect) operators and the corresponding fusion category can often be uniquely determined by analyzing a small subset of bootstrap equations of the form \eqref{mbeqn}. In the following subsections, we will analyze some ways of constructing TDLs directly in 2d CFTs.

  \subsection{Nonabelian Orbifolds}
  \label{sec:RepG}
  
  One common instance of fusion category symmetry in 2d CFT is $\cA=\Rep(G)$, the representation category of a finite group $G$. A CFT $\cT$ has such a symmetry if and only if it is a $G$-orbifold of another theory $\cT'$ \cite{Bhardwaj:2017xup}, 
\ie
\cT \cong {\cT'/ G} \,.
\label{discretegauging}
\fe 
We think about this orbifold by coupling $\cT'$ to a flat $G$ gauge field. These TDLs in $\cT$ correspond to $G$-Wilson lines $W_R$ labeled by $G$-representations $R$. Their fusion products follow from the usual tensor product of representations, and the corresponding F-symbols are identified with the $6j$ symbols. Moreover the quantum dimensions of the Wilson lines are simply the dimensions of the corresponding $G$-representations.
In particular, only the TDLs associated to one-dimensional representations are invertible, so if $G$ is nonabelian, $\cA$ contains non-invertible TDLs. These symmetries are present whether or not we include discrete torsion in \eqref{discretegauging}. 
Furthermore the action of the TDLs in ${\rm Rep}(G)$ on local operators in the theory $\cT$ is completely fixed. For an operator $\phi_{[g]}$ in the twisted sector labelled by the conjugacy class $[g]$, the Wilson line acts as
\ie
W_R |\phi_{[g]}\ra = \chi_R(g) |\phi_{[g]}\ra \,,
\fe
where $\chi_R(g)$ is the usual group character. This action obeys the expected fusion rules for the Wilson lines,
\ie
W_i W_j=\sum_k N_{ij}^k W_k~~{\rm with}~~N_{ij}^k={1\over |G|}\sum_{g\in G} \chi_i(g)\chi_j(g)\chi_k(g)^*\,,
\label{WLfusion}
\fe
as can be checked using the orthogonality property of the characters
\ie
{1\over |G|}\sum_k \chi_k(g)\chi_k(h)^*=\begin{cases}
{1\over |[g]|} & {\rm if}~h\in [g]\,,
 \\
 0 & {\rm otherwise}\,.
\end{cases}
\fe

This construction works in higher dimensions as well, again the Wilson lines give rise to TDLs, which can  probably be described as a kind of higher fusion category symmetry. It also works for higher form symmetries $G$ and higher group generalizations. While we will only be interested in gauging finite symmetries, with care one can sometimes find topological Wilson lines in gauge theories with continuous gauge group \cite{Bachas:2009mc,Chang:2018iay,Komargodski:2020mxz}.

  \subsection{Self-dual Orbifolds}\label{subsecoverviewselfdual}
  
  Another large class of fusion category symmetries that appears often is described by the Tambara-Yamagami (TY) fusion category, denoted by $\TY(G_{\rm ab},\chi,\ep)$, which is associated to an \textit{abelian} finite group $G_{\rm ab}$, a symmetric non-degenerate bilinear map $\chi:G_{\rm ab}\times G_{\rm ab} \to \mR/2\pi \mZ$ called the bicharacter, and a sign $\ep = \pm 1$ known as the Frobenius-Schur (FS) indicator \cite{tambarayam}. The TY category contains $G_{\rm ab}$ as the group of invertible lines $\cL_g $ and includes one extra simple TDL, namely the duality defect $\cN$ with quantum dimension $\sqrt{|G_{\rm ab}}|$, which acts as a Kramers-Wannier-like duality. The fusion rules are
  \ie
 & \cL_g \cL_{g'}=\cL_{gg'}\,,\quad 
   \mathcal{N} \mathcal{L}_{g} =\mathcal{L}_{g} \mathcal{N} = \mathcal{N}\,,\quad 
  {\cal N}^2 = \sum_{g \in G_{\rm ab}}{\cal L}_{g}\,.
  \label{TYfus}
  \fe
The F-symbols compatible with the above fusion rules and the pentagon equations are given in \cite{tambarayam} and the solutions are classified\footnote{We can redefine the correlation functions by local phase factors associated with the fusion vertices. We only classify F-symbols up to these redefinitions and we refer to this freedom as the F-gauge choice. } by $\chi$ and $\ep$. One such F-symbol is depicted below, where in solid blue we draw the duality defect, and in dashed red two group lines which end on it,
\begin{equation}\label{eqnmodcatF}
 	\adjincludegraphics[width=4cm,valign=c]{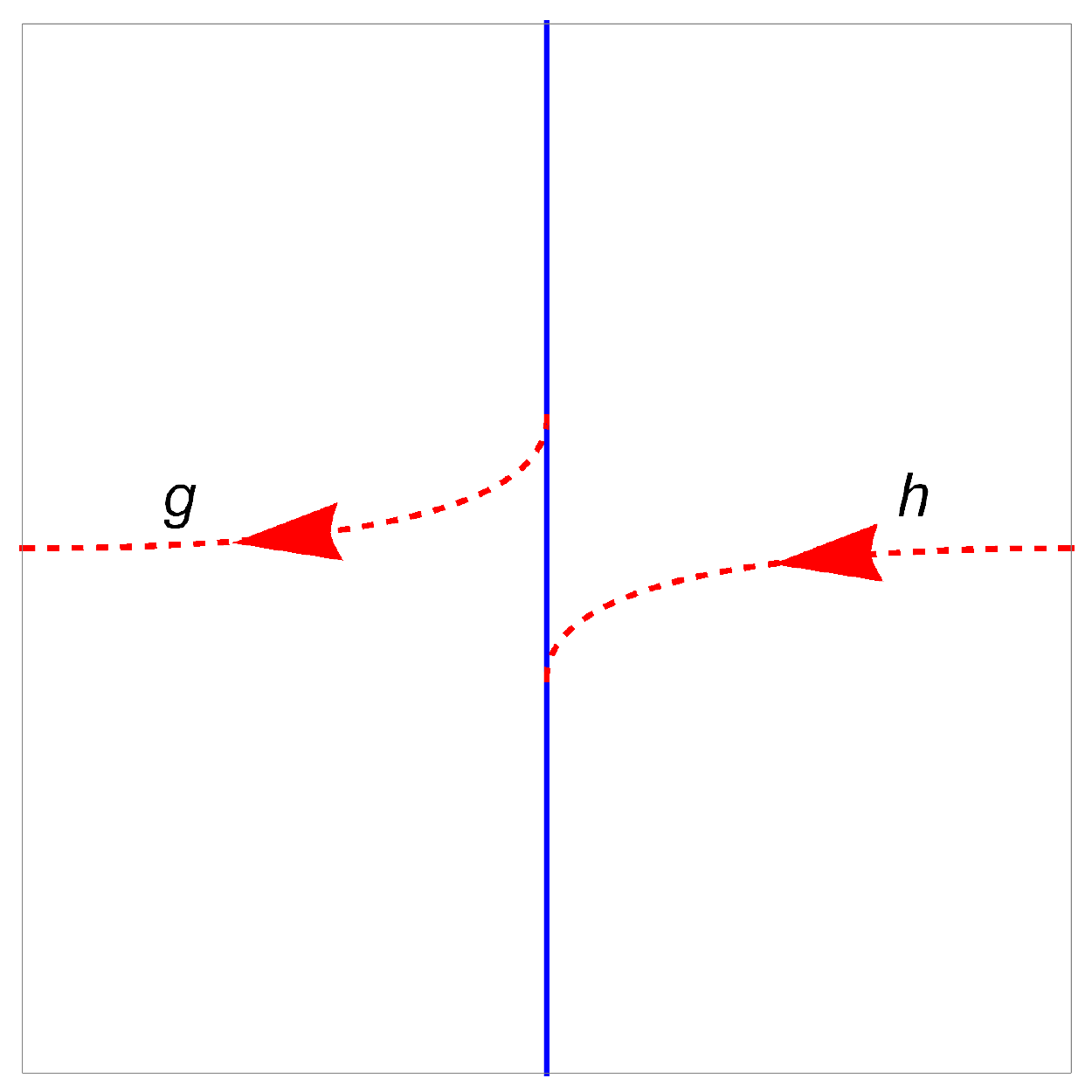} = e^{i \chi(g,h)}\adjincludegraphics[width=4cm,valign=c]{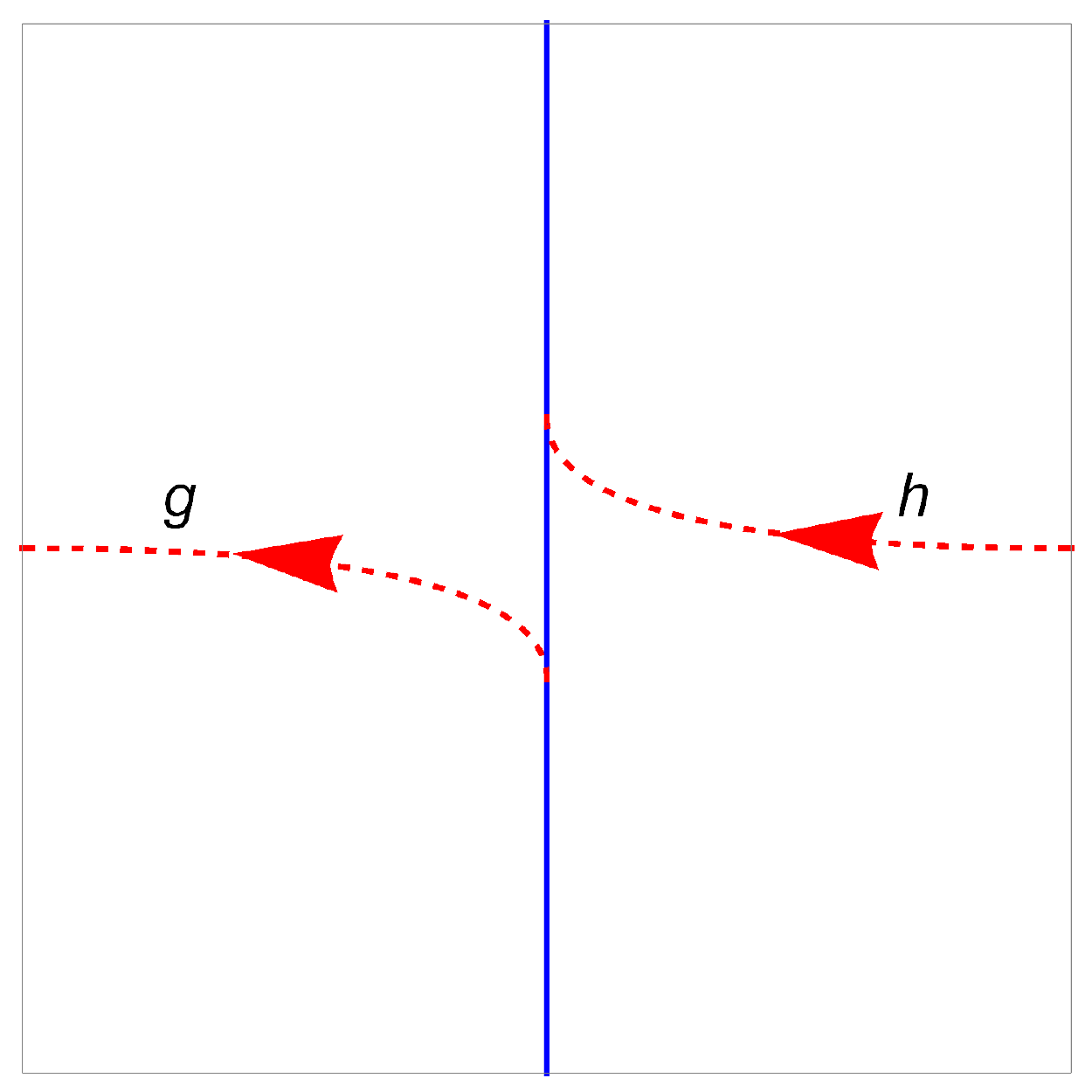}\,.
 	 	\end{equation}

Any  CFT $\cT$ that hosts a fusion category symmetry described by  $\TY(G_{\rm ab},\chi,\ep)$ has the curious feature that it is self-dual under $G_{\rm ab}$-gauging,
\ie
\cT\cong {\cT/ G_{\rm ab}} \,.
\label{selfgauging}
\fe
This is also a special case of \eqref{discretegauging}, and we have Wilson line TDLs from $\Rep(G_{\rm ab}) = \hom(G_{\rm ab},U(1))$. The self-duality maps $\cL_g$ with $g\in G_{\rm ab}$ to the Wilson line labelled by the abelian character $e^{i\chi(g,-)}$, and thus defines an isomorphism between the symmetry $G_{\rm ab}$ in $\cT$ and $\Rep(G_{\rm ab})$ in the orbifold $\cT/G_{\rm ab}$. Under this isomorphism, the bicharacter $e^{i\chi(g,h)}$ describes the $g$-charge of the $h$-twisted sector operators, as can be seen from the F-move above.

To prove the equivalence \eqref{selfgauging}, in any correlation function we nucleate a small loop of the duality defect and bring it around the system, using the fusion rule and F-moves when it circles around and meets itself \cite{Chang:2018iay,Aasen:2016dop}.

For instance, one consequence of \eqref{selfgauging} is an identity of twisted partition functions
\ie
Z(\Sigma,B) ={1\over \sqrt{| H^1(\Sigma,G_{\rm ab})|}} \sum_{A \in H^1(\Sigma,G_{\rm  ab})} Z(\Sigma,A) e^{i \int \chi(A, B) + \alpha(A) + \beta(B)}\,,
\label{eqnpartfnselfdualgauging}
\fe
where $A,B \in H^1(\Sigma ,G_{\rm ab})$ are flat $G_{\rm ab}$ gauge fields on the spacetime manifold $\Sigma$, (by abuse of notation) $\chi(A,B)$ is a pairing of such gauge fields derived from the bicharacter and cup product \cite{Thorngren:2019iar}, and $\alpha(A), \beta(B)$ captures the possible discrete torsion in $H^2(G_{\rm ab},U(1))$. If we apply the above technique to the torus partition function, we find
\begin{equation}
 	Z(T^2) = \frac{1}{|G_{\rm ab}|^{3/2}}\sum_{g,h \in G_{\rm ab}}\adjincludegraphics[width=4cm,valign=c]{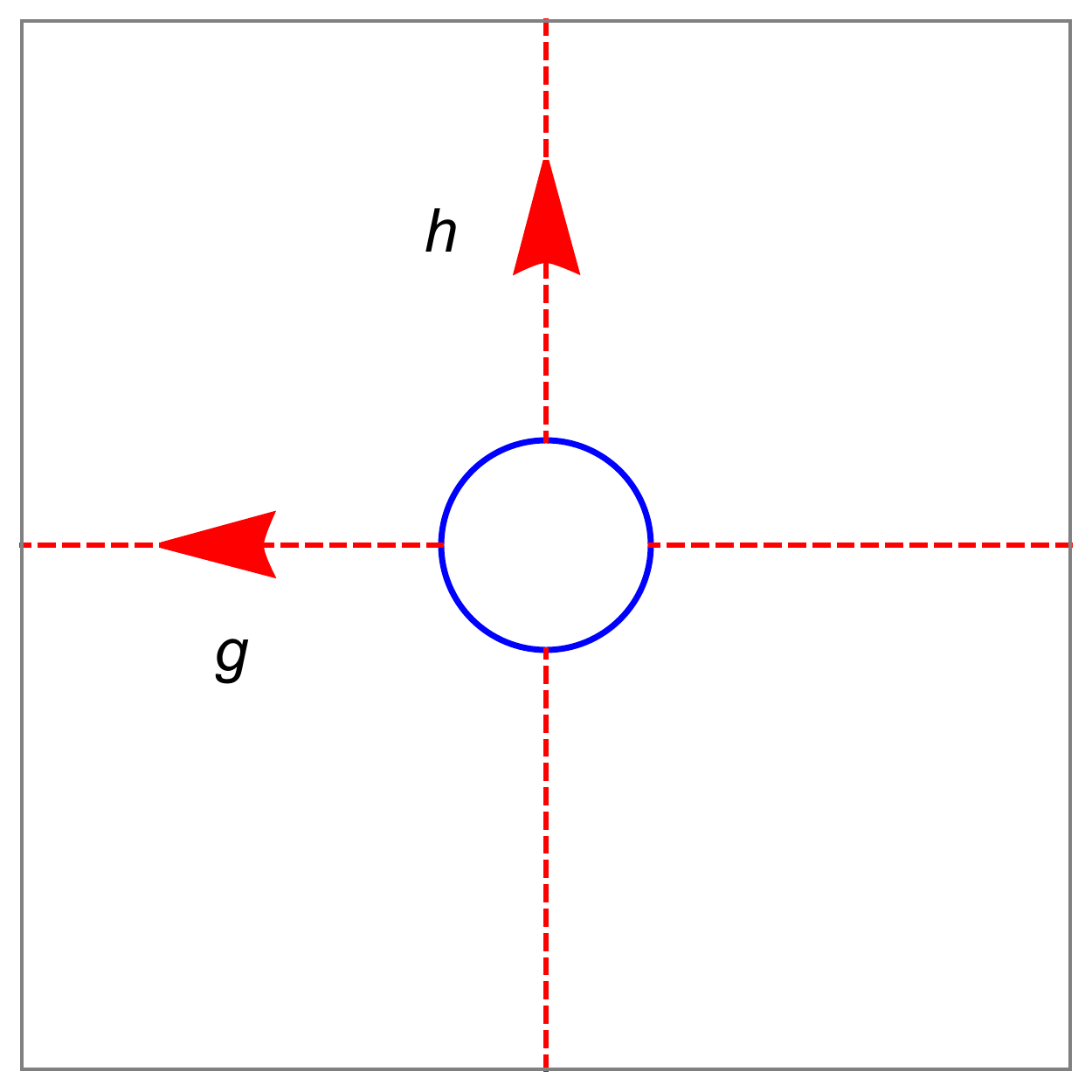}\,.
 	\label{Nloopfuse}
 	 	\end{equation}
The duality loop (in solid blue) may then be  pulled off of the group symmetry lines (in dashed red) and removed, introducing another factor of $|G_{\rm ab}|^{1/2}$ and a possible phase factor $\alpha(g,h)$ (coming from the F-symbols), thus obtaining an equality with the orbifold partition function with discrete torsion $\alpha$.

Conversely, if we have a theory $\cT$ which is self-dual under gauging a finite abelian symmetry $G_{\rm ab}$, it is guaranteed to host a duality defect $\cN$ which along with $G_{\rm ab}$ generates a fusion category symmetry. When $\alpha = -\beta$ (in which case they can be absorbed into the definition of $Z$), we can argue that this fusion category is a TY category. This duality defect may be defined as a topological interface between $\cT$ and $\cT/G$ by coupling to a flat $G$ gauge field which is constrained to live in a region $\mathcal{R}$ of the spacetime, with the  defect along $\partial \mathcal{R}$. To define this coupling properly, we have to choose appropriate boundary conditions for the gauge field, so that the gauge lines cannot end on $\partial \mathcal{R}$, or equivalently that Wilson lines \emph{can} end there, and modify the Hamiltonian so that it is gauge invariant. See Appendix \ref{appendixlattice} for lattice examples of this construction.

\begin{figure}[!htb]
    \centering
    \[
        \adjincludegraphics[width=7cm,valign=c]{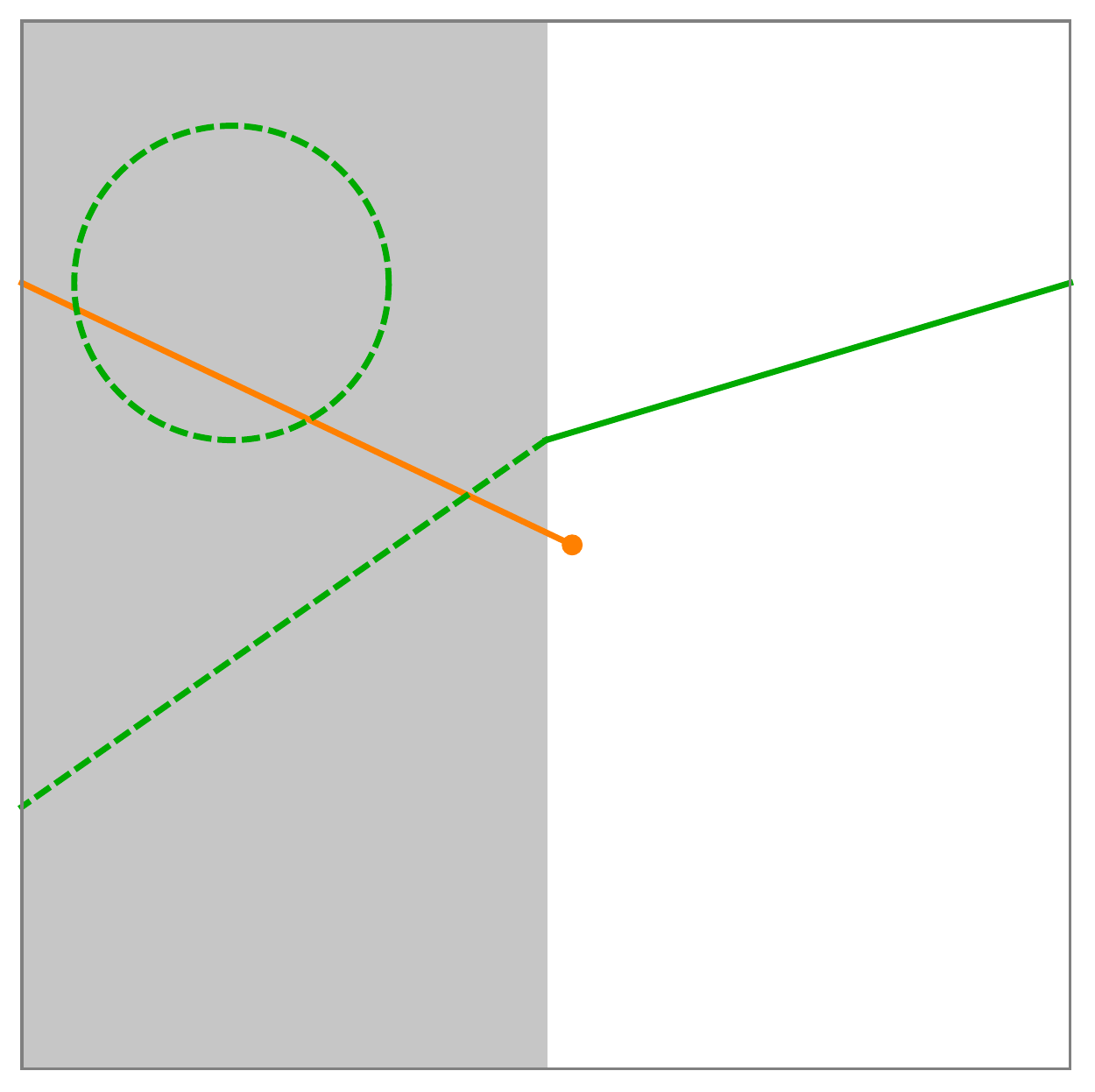} = \chi_R(g) \cdot
    \adjincludegraphics[width=7cm,valign=c]{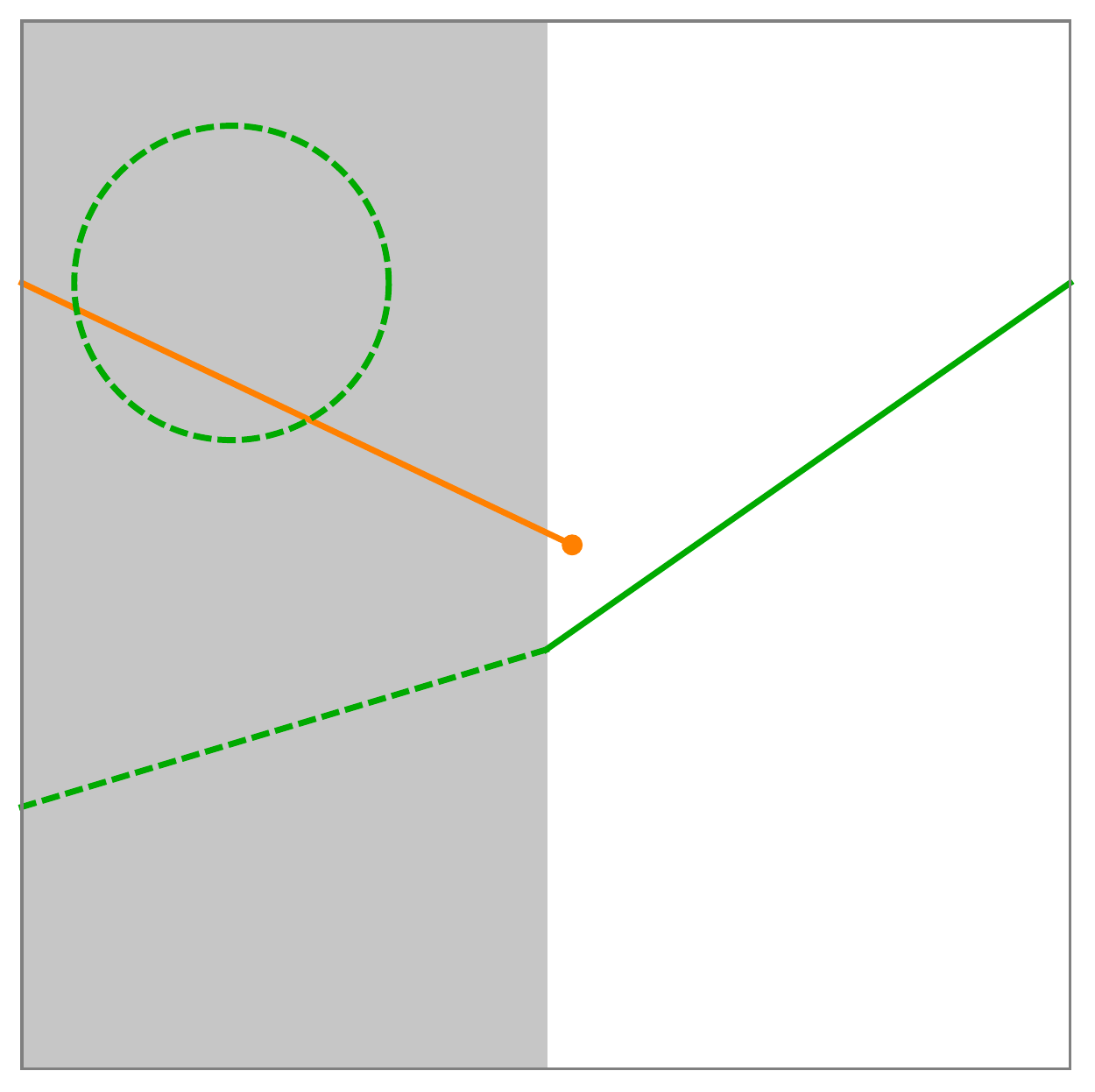}
\nonumber    \]
    
    \caption{A boundary between the region $\mathcal{R}$ (shaded gray), where $G$ is gauged (left) and a region where $G$ remains a global symmetry. In $\mathcal{R}$, we sum over all closed $G$ loops (dotted, green) which remain in the interior of $\mathcal{R}$, and the dual global symmetry acts via Wilson lines (orange), which simply count the number of crossings with the green lines. Such Wilson lines can end at the boundary in a gauge invariant way, so long as we place their endpoint (orange dot) just over the edge, so that the sum over closed loops in the interior is well-defined. Global $G$ lines (solid, green) can enter $\mathcal{R}$ from the ungauged side in a topological manner, joining the gauge lines (dotted, green). There is a non-trivial F symbol when Wilson lines and global $G$ lines meet from either side of the defect, which can be seen from the crossing of the solid green line and orange dot above, which changes the number of crossings between the Wilson line and the green $G$ lines, resulting in a phase. If the Wilson line is labelled by the representation $R$ and the $G$ line by $g \in G$, this phase is the character $\chi_R(g)$.}
    \label{fig:gaugedregion}
\end{figure}

These Wilson lines generate the $G_{\rm ab}$ \emph{global} symmetry inside $\mathcal{R}$ after gauging. In particular, $\cL_g = W_{\chi(g,-)}$, where $W_q$ represents a Wilson line of charge $q \in \hom(G_{\rm ab},U(1))$. As we have mentioned above, these Wilson lines are allowed to end on $\partial \mathcal{R}$. $G_{\rm ab}$ lines from outside $R$ can also ``end" on $\partial \mathcal{R}$---they pass through $\partial \mathcal{R}$ and join the gauge lines inside $R$. This allows us to derive the F-symbol above (see Figure~\ref{fig:gaugedregion}). We can also easily derive $\cN \cL_g = \cL_g \cN = \cN$.

Another perspective which is useful is to interpret the above construction in terms of boundary conditions of the folded theory $\cT\times \overline \cT$. If $|\cL_g\rangle$ is the boundary state corresponding to the group symmetry TDL $\cL_g$, the duality defect comes from the state
\ie
|\cN\rangle = \frac{1}{\sqrt{|G_{\rm ab}|}} \sum_{g\in G_{\rm ab}} |\cL_g\rangle\,. 
\label{Nfromgauging}
\fe
This state is not a Cardy state for $\cT\times \overline \cT$, but it enjoys a $G_{\rm ab}$ symmetry which acts only on one side of the defect (that is, on one layer of the folded theory). Gauging this one-sided $G_{\rm ab}$ symmetry results in an elementary Cardy boundary state for $\cT\times \overline \cT/G_{\rm ab}$, known as the \textit{regular brane} in the orbifold theory (see \cite{Recknagel:2013uja} for a review).
It has boundary entropy $g = \sqrt{|G_{\rm ab}|}$, which is identified with the quantum dimension of the topological interface between $\cT$ and $\cT/G_{\rm ab}$ after unfolding. Since the theory is self-dual under $G_{\rm ab}$ gauging, this defines a defect in the original theory $\cT$. It is easy to see that the cylinder amplitude $\langle \cN |e^{-\B H}| \cN \rangle$ in the folded theory $\cT\times \overline \cT$ is the torus partition function of $\cT$ where we sum over $g$-twisted sectors, capturing the fusion rule $\cN^2 = \sum_g \cL_g$ which is an immediate consequence of \eqref{Nfromgauging}.

The most familiar CFT with TY symmetry is the $c ={1\over 2}$ Ising CFT, with $G_{\rm ab} = \bZ_2$ acting as the spin-flip symmetry. The equivalence between Ising and its $\bZ_2$ orbifold is known as Kramers-Wannier duality. When expressed as a defect, by studying the modular bootstrap, it turns out to have $\epsilon =  1$ \cite{Thorngren:2019iar}, and meanwhile there is only one choice of bicharacter. The defect can be constructed using the methods above either on the lattice or in field theory, and it agrees with known constructions \cite{Oshikawa:1996dj,Aasen:2016dop}. We will explore an example of a self-duality (actually a triality) which does not have TY fusion rules in Section \ref{subsecKT}.

  \subsection{Shadows of Invertible Defects and Noether's Theorem}\label{subsecshadownoether}
 
For general orbifolds $\cT = \cT'/G$, we can ask what is the fate of topological defects in $\cT'$ after orbifolding? A TDL $\cL$ of $\cT'$ which is $G$-normalizing, meaning
\ie
\cL g= g' \cL
\label{Gnorm}
\fe
for $g,g'\in G$, will define a (possibly reducible) TDL of $\cT$. For invertible TDLs, this is the familiar fact that
global invertible symmetries of $\cT'$  in the normalizer of $G$ give rise to invertible symmetries of the orbifold $\cT$. Note that in the presence of nontrivial F-symbols between the TDL $\cL$ and $G$-symmetry TDLs, the fusion rule involving the resulting TDL in $\cT$ may change. For invertible $\cL$, this corresponds to a nontrivial extension of the symmetries due to a mixed 't Hooft anomaly with $G$ \cite{Tachikawa:2017gyf}.
 
 It turns out that general non-$G$-normalizing TDLs of $\cT'$ also define TDLs of $\cT$, after summing over their $G$-orbit. This sum can turn invertible TDLs into non-invertible ones. More precisely, if $\cL$ is a TDL of $\cT'$ with quantum dimension $\la \cL\ra$,
 \[\label{eqnsumoflines}\cL^G \equiv \sum_g g \cL g^{-1}\,,\]
 is a $G$-invariant TDL, and so defines a TDL of the gauge theory $\cT$ of quantum dimension $|G|\la \cL\ra $.\footnote{Note that $\cL^G$ can be reducible. In particular if $\cL$ is $G$-normalizing as in \eqref{Gnorm}, $\cL^G$ is equivalent to $|G| \cL$ in the orbifold theory.}
 
 This construction reveals a huge amount of TDLs hidden in familiar CFTs. For example, the compact boson at a generic radius has two continuous $U(1)$ symmetries which anti-commute with the $\bZ_2^C$ charge conjugation symmetry generated by $C$. When we gauge $C$ to go to the orbifold branch, there are no longer continuous ordinary symmetries, but there is an algebra of non-invertible TDLs of quantum dimension 2 labelled by continuous parameters! Such lines were recently discussed in \cite{Chang:2020imq} and we return to them in Section \ref{seccontinuumZ2orbifold}.
 
 Briefly, these defects can be defined as follows. We start with the compact boson before orbifolding. Let ${X}_1(x,t), {X}_2(x,t)$ be the compact scalar fields of radius $R$ to the left and right of the defect at $x = 0$. For the variational problem to be well defined, we require
 \ie
 (\D {X}_1 \pa_x {X}_1 - \D {X}_2 \pa_x {X}_2) |_{x=0}=0\,,
 \fe
A set of consistent conformal gluing conditions are 
\ie
({X}_1-  {X}_2) |_{x=0}=\A R,\quad(\pa_x{X}_1- \pa_x{X}_2) |_{x=0} = 0\,.
\fe
with $\A \in [0,2\pi]$. Furthermore, one can include a theta angle on the defect
\ie
  S_{\rm defect} = {\B\over R} \int_{x=0} d{X}_1\,,
\fe
with $\B \in [0,2\pi]$. They define a two-parameter $(\A,\B)$ family of topological defects in the compact boson theory which correspond to the $U(1)\times U(1)$ continuous symmetries. The $C$-invariant direct sum of the these defects given by $(\A,\B)\oplus (-\A,-\B)$ survive the orbifold identification ${X}_{1,2} \sim -{X}_{1,2}$ and give rise to the two-parameter family of TDLs in the orbifold theory which are now non-invertible. 
 
 TDLs such as these with continuous parameters satisfy a version of Noether's theorem. Indeed, the continuous parameter implies that there is a marginal operator $\cO$ defined on the defect.  If we fold the defect $\cL$ over the location of this operator, we see $\cO$ is defined at the end of (a fusion product in) $\cL \bar\cL$.\footnote{For a general line defect $\cL$ that is not necessarily topological, by state-operator correspondence via a conformal mapping, the operator $\cO$ on defect $\cL$ maps to a state in the defect Hilbert space $\cH_{\cL,\bar\cL}$ on $S^1$ that intersects twice with the ingoing and outgoing $\cL$ at antipodal points. If $\cL$ is a TDL, $\cH_{\cL,\bar\cL}$ is simply given by a direct sum of $\cH_{\cL_i}$ for each TDL $\cL_i$ in the fusion product $\cL \bar \cL$.} Because $\cO$ is marginal, its conformal dimension is fixed to be 1. Unitarity implies $\cO$ has spin 0 or 1 and topological invariance further requires the spin of $\cO$ to be 1. This captures the familiar Noether's theorem when $\cL$ is invertible, in which case $\cL \bar\cL$  is trivial, so $\cO$ is a local operator which we recognize as the conserved current.

 In the orbifold construction above, $\cO$ comes from the current in the theory $\cT'$ before gauging, and lives at the end of a Wilson line in $\cT = \cT'/G$ (in our example above the two currents are $d{X}_1$ and its dual $\star d{X}_1$ ). Conversely, if $\cO$ lives at the end of an anomaly-free line in $\cT$, that is, a TDL in an anomaly-free fusion subcategory of the full symmetry, then gauging this subcategory produces a theory $\cT'$ in which $\cO$ is a local operator. In fact if this line is invertible, it is guaranteed to be anomaly-free by the spin selection rule, which would otherwise forbid a spin 1 operator \cite{Chang:2018iay,Lin:2019kpn}.
 
 This construction works in all dimensions, on topological defects of any codimension, as does the Noether theorem above, producing a ``defect current" $\cO$ of dimension $D - k$, where $D$ is the spacetime dimension and $k$ is the codimension of the defect. For invertible defects, these are the conserved currents of continuous higher form symmetries \cite{Gaiotto:2014kfa}. In general, $\cO$ lives at the end of a line operator obtained by wrapping the defect on $S^{D-k-1}$. An example of a non-invertible 1-form symmetry in 3+1D was recently constructed by this method in \cite{Heidenreich:2021tna}. In particular, in $O(2)$ gauge theory in 3+1D the field strength is a conserved 2-form current which is not local but lives at the end of a Wilson line for the sign representation of $O(2)$.

\begin{figure}[!htb]
    \includegraphics[scale=.5]{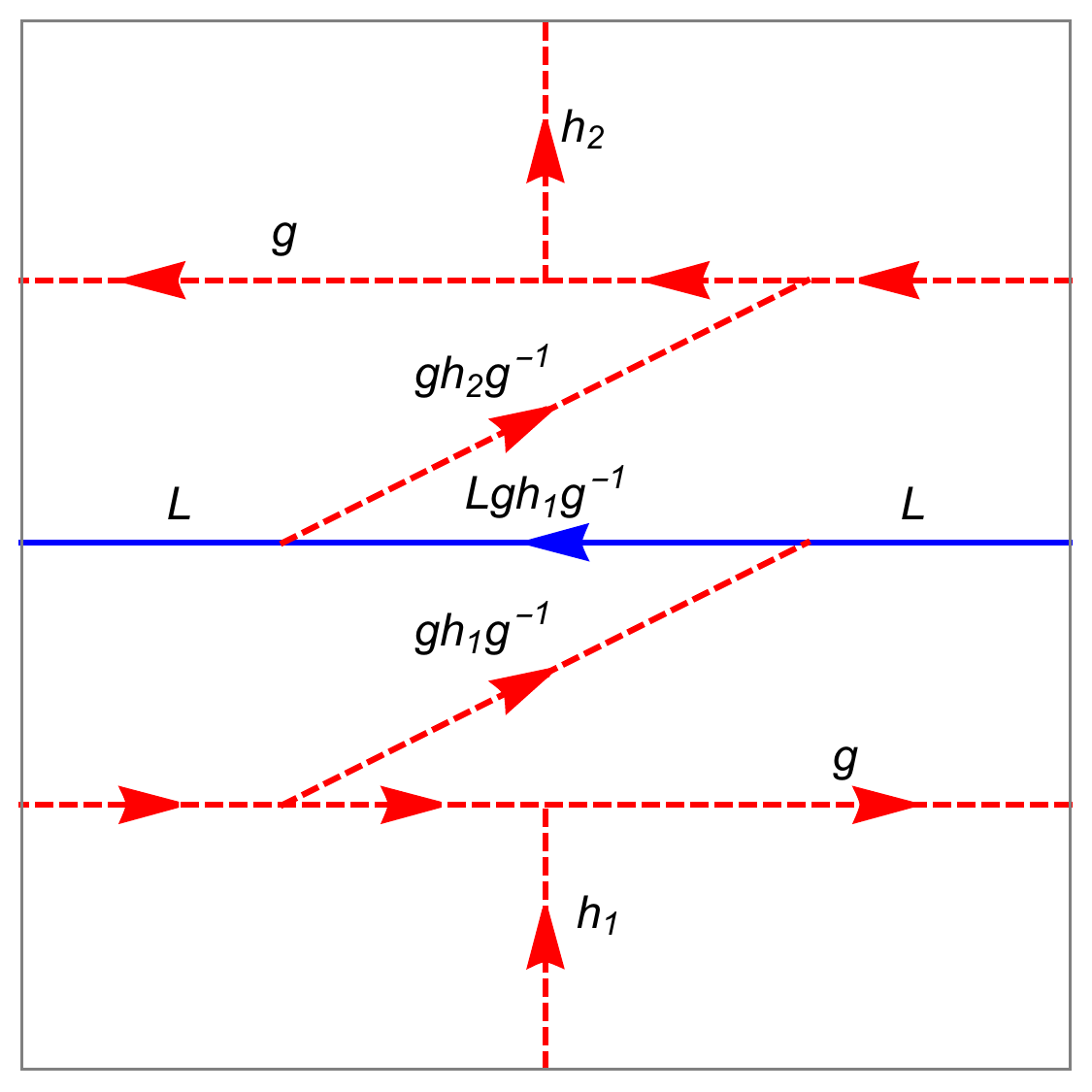}
    \centering
    \caption{To define the action of $\cL^G$ on all local operators of the $G$-orbifold, we need to consider the action of $\sum_g g \cL g^{-1}$ on operators in the $G$-twisted sector. This is equivalent to computing the cylinder partition function shown here, regarded as a map from the $h_1$-twisted to the $h_2$-twisted sectors, for $h_1,h_2 \in G$. The diagram is only nontrivial if the topological junctions exist.}
    \label{fig:actionontwistedsectors}
\end{figure}

 Above, \eqref{eqnsumoflines} defines the action of $\cL^G$ on local operators of the orbifold $\cT = \cT'/G$ in the untwisted sector (i.e. from local operators of $\cT'$). However, general local operators of $\cT$ come from the $G$-twisted sectors of $\cT'$ will be mapped to one another under $\cL^G$. To determine the action of $\cL^G$ in these cases, we must study TDL networks as in Figure \ref{fig:actionontwistedsectors} in the theory $\cT'$.

\subsection{Duality defects from (para)fermions}
\label{sec:TDLfromPF}
When the bosonic CFT is related by bosonization/fermionization duality to a fermionic CFT, anomalous discrete symmetries acting on the fermions are known to generate dualities of the bosonic theory after summing over spin structures (or GSO projection) \cite{Thorngren:2018bhj,Karch:2019lnn,Ji:2019ugf}. The most familiar case arises in the Ising CFT as the bosonization of one Majorana fermion, where the chiral fermionic number symmetry $(-1)^{F_L}$ corresponds to the Kramers-Wannier duality (which modifies the GSO projection by an SPT given by the Kitaev chain, also known as the Arf invariant). In this section, we will review the connection between symmetries in a fermionic theory and duality defects in its bosonized partner, and discuss the generalization to parafermionic theories.

In general, given a fermionic CFT $\cT_F$ with a non-anomalous fermion parity symmetry $(-1)^F$, one can obtain a bosonic CFT $\cT_B$ by summing over the spin structures $\rho$ of the underlying genus $g$ Riemann surface $\Sigma$, while preserving modular invariance on $\Sigma$. The corresponding partition functions on $\Sigma$ are related by 
\ie
Z_B(a)={1\over 2^g}\sum_{\rho} Z_F(\rho)  (-1)^{   {\rm Arf} (\rho+a)}\,,
\label{B2F}
\fe
which is known as the bosonization map. Here $a$ denotes the background gauge field for the dual $\mZ_2^B $ symmetry of the bosonic theory $\cT_B$. 
Conversely, the original fermionic CFT can be recovered from the bosonic theory via the fermionization procedure by gauging the dual $\mZ_2^B$ symmetry,
\ie
Z_F(\rho)={1\over 2^g} \sum_a Z_B(a) (-1)^{ {\rm Arf} (\rho+a)}\,.
\label{F2B}
\fe
The bosonization map (and its inverse fermionization map) defined with a given fermion parity is not unique. The  freedom comes from stacking the fermionic theory with a $1+1$D spin SPT given by $(-1)^{ {\rm Arf}(\rho)} $, 
\ie
Z_{F'}(\rho)=Z_F(\rho)(-1)^{  {\rm Arf} (\rho)}\,,
\fe
which modifies \eqref{B2F} and produces another bosonic theory $\cT_{B'}$ with partition function 
\ie
Z_{B'}(a)=&{1\over 2^g}\sum_{\rho} Z_F(\rho)  (-1)^{   {\rm Arf} (\rho+a)+{\rm Arf}(\rho)}\,.
\fe
Using the following property of the Arf invariant,
\ie
{1\over 2^g}\sum_a (-1)^{{\rm Arf}(\rho+a)+{\rm Arf}(\rho)+a\cup b)}=(-1)^{   {\rm Arf} (\rho+b)}\,,
\fe
it is clear that $\cT_{B'}$ is related to $\cT_B$ by gauging the  $\mZ_2^B$ symmetry. The relations between the bosonic and fermionic CFTs discussed above are summarized in the diagram below,
\ie 
\begin{tikzcd}[column sep=huge,row sep=huge]
\cT_B \arrow[leftrightarrow]{d}[swap,sloped]{{\rm gauging}} \arrow[leftrightarrow]{r}[sloped,above]{\text{bosonize}}[swap,sloped]{\text{fermionize}} & \cT_F \arrow[leftrightarrow]{d}[sloped]{\rm stacking} \\
\cT_{B'}=\cT_B/\mZ_2^B \arrow[leftrightarrow]{r}[sloped,above]{\text{bosonize}}[swap,sloped]{\text{fermionize}} & \cT_{F'}=\cT_F\otimes (-1)^{  {\rm Arf}(\rho)}  
\end{tikzcd}
\label{BFdiag}
\fe
If in addition the fermionic theory $\cT_F$ is invariant under the stacking operation with the spin SPT $(-1)^{ {\rm Arf}(\rho)}$,
\ie
Z_F(\rho)= Z_{F'}(\rho)\,,
\label{FFpeq}
\fe
the bosonic theory becomes self-dual under $\mZ_2^B$ gauging 
\ie
Z_B(a)=Z_{B'}(a)\,.
\fe
As explained in Section~\ref{subsecoverviewselfdual}, this implies a $\mZ_2$ duality TDL $\cN$ in $\cT_B$ that gives rise to a $\mZ_2$ TY categorical symmetry.
More generally, symmetries in the fermionic theory $\cT_F$ will give rise to TDLs in the bosonic CFT $\cT_B$ upon bosonization. The nature of the resulting TDLs depends crucially on the anomalies of the corresponding fermionic symmetries. Here  an anomalous chiral parity in $\cT_F$ explains the invariance \eqref{FFpeq} and is responsible for the corresponding duality TDL in $\cT_B$  \cite{Thorngren:2015gtw,Karch:2019lnn}. 

For example, let us consider $\cT_F$ described by $n$ Majorana fermions with total fermion parity $(-1)^F$. The theory has a diagonal chiral parity symmetry $(-1)^{F_L}$ with anomaly $n\in \mZ_8$ \cite{Fidkowski:2009dba,Ryu:2012he,Yao:2012dhg,Qi:2013dsa,Kapustin:2014dxa}. The bosonization map based on $(-1)^F$ gives the $Spin(n)_1$ WZW model, and the TDL corresponding to $(-1)^{F_L}$ is fixed by $n$ \cite{Numasawa:2017crf,Lin:2019kpn,Ji:2019ugf}, which we summarize in Table~\ref{table:majtospinWZW}.

\begin{table}[!htb]
    \centering
    \begin{tabular}{c|c|cc}
    $n~{\rm mod\,}8$    & $n$~{Majorana} & $Spin(n)_1$ \\
                 \hline
   $0$ &  \multirow{5}{*}{$(-1)^{F_L}$}&  anomaly-free $\mZ_2$
    \\
     $\pm 1$ & & Duality $\cN$ in $\TY(\mZ_2,+)$ \\
       $\pm 2$& & Anomalous $\mZ_4$\\
      $\pm 3$ & & Duality $\cN$ in $\TY(\mZ_2,-)$\\
        $4$ & & Anomalous $\mZ_2$\\
        \hline
    \end{tabular}
    \caption{The relation between the anomalous $(-1)^{F_L}$ parity symmetry of the Majorana fermions and TDLs in the $Spin(n)_1$ WZW model from bosonization. Here the anomalies for $n=\pm2,4{~\rm mod\,}8$ have order two.}
    \label{table:majtospinWZW}
\end{table}

The cases $n=1$ and $n=2$ are relevant for understanding TDLs in the $c=1$ CFT from bosonizing two Majorana fermions. There is a $\mZ_2\times\mZ_2$ anomaly-free fermion parity symmetry generated by $(-1)^{F^1}$ and $(-1)^{F^2}$ that act on the two fermions separately. Bosonizing with respect to the diagonal symmetry $(-1)^F=(-1)^{F^1+F^2}$, one obtains the $c=1$ CFT on the circle branch at the Dirac point $R=2$ (also known as the $U(1)_4$ CFT). The TDLs corresponding to the chiral fermion parities follow from Table~\ref{table:majtospinWZW} and are listed in Table~\ref{table:DiracTDLs}.

\begin{table}[!htb]
    \centering
    \begin{tabular}{c|c}
    Fermionic symmetry  & TDLs in the $U(1)_4$ CFT (Dirac point)\\
                 \hline
   $(-1)^{F^1_L},(-1)^{F^2_L}$ & Duality $\cN$ in $\TY(\mZ_2,+)$ 
     \\
    $(-1)^{F^1_L+F^2_L}$     & Anomalous $\mZ_4$\\
    \hline
    \end{tabular}
    \caption{TDLs in the $U(1)_4$ CFT from chiral parities in the fermionic theory.}
    \label{table:DiracTDLs}
\end{table}
Alternatively, one can bosonize the two Majorana fermions by gauging both $(-1)^{F^1}$ and $(-1)^{F^2}$, leading to the ${\rm Ising}^2$ CFT on the orbifold branch at $R=2$. The TDLs that are induced by the anomalous fermionic symmetries are summarized in Table~\ref{table:IsingsqTDLs}.
\begin{table}[!htb]
    \centering
    \begin{tabular}{c|c}
    Fermionic symmetry  & TDLs in ${\rm Ising}_1\otimes {\rm Ising}_2$\\
                 \hline
   $(-1)^{F^1_L}$ & Duality $\cN$ in $\TY(\mZ_2,+)_{{\rm Ising}_1}$
    \\
    $(-1)^{F^2_L}$ & Duality $\cN$ in $\TY(\mZ_2,+)_{{\rm Ising}_2}$\\
    $(-1)^{F^1_L+F^2_L}$     & Duality $\cN$ in $\Rep(H_8)$ \\
    \hline
    \end{tabular}
    \caption{TDLs in the ${\rm Ising}^2$ CFT from chiral parities in the fermionic theory.}
    \label{table:IsingsqTDLs}
\end{table}

There is a natural generalization of the fermionic CFTs where operators obey fractional statistics determined by a phase $\omega=e^{2\pi i\over k}$, known as $\mZ_k$ parafermionic CFTs $\cT_{PF}$. 
The case $k=2$ corresponds to the usual fermionic CFTs. The $\mZ_k$ parafermionic CFTs have a $\mZ_k$ symmetry that generalizes the $(-1)^F$ symmetry in the fermionic case. The spin structures for a fermionic CFT also have $\mZ_k$ counterparts, namely $\mZ_k$ paraspin structures. There is a bosonization procedure for the parafermionic theories $\cT_{PF}$ that produce bosonic CFTs $\cT_{B}$ by summing over the paraspin structures (at least on $T^2$). The resulting bosonic theories are invariant under a dual $\mZ_k^B$ symmetry. Familiar examples include the $SU(2)_k/U(1)$ coset CFTs.  
While the resulting bosonic CFTs are well-defined on an arbitrary Riemann surface $\Sigma$, the parafermionic theories have not been formulated for genus $g>1$ due to an incomplete understanding of the paraspin structure  \cite{Runkel:2018feb,Radicevic:2018okd,Yao:2020dqx}. Below we focus on $\Sigma=T^2$ and discuss relations between symmetries in the parafermionic theory $\cT_{{PF}}$ and its bosonic partner $\cT_{B}$. 

As in the fermionic case, the bosonization and parafermionization at the level of torus partition functions are given by \cite{Yao:2020dqx}\footnote{This is the continuum version of the Fradkin-Kadanoff transformation between the 1+1D $\mZ_k$ clock model and the discrete $\mZ_k$ parafermion \cite{Fradkin:1980th}. For $k=2$, it becomes the familiar Jordan-Wigner transformation between the 1+1D Ising model and the Majorana fermion.}
\ie
Z_B(a_1,a_2)=&\,{1\over k} \sum_{s_1,s_2\in \mZ_k} Z_{PF}(s_1,s_2) \bar\omega^{(s_1+a_1)(s_2+a_2)}\,,
\\
Z_{PF}(s_1,s_2)=&\,{1\over k} \sum_{a_1,a_2\in \mZ_k} Z_B(a_1,a_2) \omega^{(s_1+a_1)(s_2+a_2)}\,,
\label{PFandB}
\fe
where $a_1,a_2\in \mZ_k$ are holonomies around the two independent cycles of $T^2$ which specify the $\mZ_k$ background gauge field and $s_1,s_2\in \mZ_k$ label the $\mZ_k$ paraspin structure $\rho_k 
$. The phase $\omega^{s_1s_2}$ is a generalization of the $\mZ_2$ valued Arf invariant to Arf$_k$ on $T^2$.

Performing an $\mZ_k^B$ orbifold of the bosonic CFT $\cT_B$, we obtain the partition function for $\cT'_B$, 
\ie
Z_{B'}(a_1,a_2)=&\,{1\over k} \sum_{b_1,b_2\in \mZ_k}Z_B(b_1,b_2)\omega^{p(b_1 a_2-b_2 a_1)}\,,
\label{genZkorbPF}
\fe
where $1 \leq  p\leq k$ is coprime with $k$ and labels the bicharacter in the discrete gauging (see \eqref{eqnpartfnselfdualgauging}) corresponding to $e^{i\chi(a,b)}=\omega^{pab}$. Using \eqref{PFandB}, one have
\ie
Z_{B'}(a_1,a_2)={1\over k}\sum_{s_1,s_2\in \mZ_k}Z_{PF'}(s_1,s_2) \bar\omega^{(s_1+a_1)(s_2+a_2)}\,,
\fe
with the corresponding parafermion partition functions related by
\ie
Z_{PF'}(s_1,s_2)={1\over k}\sum_{t_i,a_i,b_i\in \mZ_k}Z_{PF}(t_1,t_2) \omega^{(s_1+a_1)(s_2+a_2)-(t_1+b_1)(t_2+b_2)+p(b_1a_2-b_2a_1)}\,.
\label{PForb}
\fe
We focus on the special cases with $p=\pm 1$ which will be relevant for the later discussion. In these cases, the above relation simplifies dramatically. For $p=1$, we have
\ie
Z_{PF'}(s_1,s_2) =Z_{PF^c}(s_1,s_2)\omega^{s_1 s_2}=Z_{PF}(s_1,-s_2)\omega^{s_1 s_2}
\label{PFrel1}
\fe
which involves a conjugation on the $\mZ_k$ charge together with stacking by the Arf$_k$ invariant, as summarized in
\ie 
\begin{tikzcd}[column sep=100pt,row sep=65pt]
\cT_{B} \arrow[leftrightarrow]{d}[swap,sloped]{\rm gauging} \arrow[leftrightarrow]{r}[sloped,above]{\text{bosonize}}[swap,sloped]{\text{parafermionize}} & \cT_{{PF}} \arrow[leftrightarrow]{d} [sloped]{\substack{\rm stacking\\+\,\rm conjugation}}\\
\cT_{B'}=\cT_{B}/\mZ^B_k \arrow[leftrightarrow]{r}[sloped,above]{\text{bosonize}}[swap,sloped]{\text{parafermionize}} & \cT_{{PF}'}=\cT_{PF^c}\otimes \omega^{ {\rm Arf}_k (\rho_k)}   
\end{tikzcd}
\label{BPFdiag}
\fe
which generalizes \eqref{BFdiag}. Similarly for $p=-1$, the relation between $\cT_{PF}$ and $\cT_{PF'}$ \eqref{PForb} becomes
\ie
Z_{PF'}(s_1,s_2) =Z_{PF}(-s_1,s_2)\omega^{s_1 s_2}\,.
\label{PFrel2}
\fe
From \eqref{BPFdiag}, if the parafermionic theory $\cT_{PF}$ is invariant under the combined stacking and conjugation,
we can conclude that the $\mZ_k^B$ invariant bosonic theory $\cT_B$ is self-dual under $\mZ_k^B$ gauging with $p=1$ in \eqref{genZkorbPF}. For more general $p$, this requires
\ie
Z_{PF'}(s_1,s_2)=Z_{PF}(s_1,s_2)\,.
\label{PFeq}
\fe

By the general argument in Section~\ref{subsecoverviewselfdual}, this property of $\cT_B$ implies a duality TDL that along with the $\mZ_k^B$ symmetry TDLs generates a $\mZ_k$ TY  symmetry. In analogy with the fermionic case, we naturally expect the duality to be associated to an anomalous symmetry in the parafermion theory. 

To be concrete, let us consider the $\mZ_k$ parafermion theory of \cite{Fateev:1985mm} and its bosonization which gives the $SU(2)_k/U(1)$ coset CFT with the diagonal modular invariant. The basic objects in the parafermion theory are the chiral parafermion fields $\psi_n(z)$ with $n=1,2,\dots k-1$ and their anti-chiral partner $\bar\psi_n(\bar z)$, with $\mZ_k$ charge $n$. They have fractional mutual statistics  as in Figure~\ref{fig:PFstat} 
and generalize the familiar Majorana fermion at $k=2$.
\begin{figure}[!htb]
\centering
\begin{tikzpicture}
\node (1) at (0,0) {};
\node (2) at (1,0) {};
\node (3) at (2.5,0) {};
\node (4) at (3.5,0) {};
\node (5) at (5,0) {};
\node (6) at (6,0) {};
\draw [fill] (1) circle (2pt) node  [left] {$\psi_n$};
\draw [fill] (2) circle (2pt) node [right] {$\psi_m$} ;
\draw[->] (2) to [out=110,in=70, looseness=1] (1);
\draw[->] (1) to [out=-70,in=-110, looseness=1] (2);
\node at (-1,-.8) {phase:};
\node at (0.5,-.8) {$\omega^{-mn}$};
\node at (3,-.8) {$\omega^{mn}$};
\node at (5.5,-.8) {$-1$};
\draw [fill] (3) circle (2pt) node  [ left] {$\bar\psi_n$};
\draw [fill] (4) circle (2pt) node [ right] {$\bar \psi_m$} ;
\draw[->] (4) to [out=110,in=70, looseness=1] (3);
\draw[->] (3) to [out=-70,in=-110, looseness=1] (4);
\draw [fill] (5) circle (2pt) node  [ left] {$\psi_n$};
\draw [fill] (6) circle (2pt) node [ right] {$\bar \psi_m$} ;
\draw[->] (6) to [out=110,in=70, looseness=1] (5);
\draw[->] (5) to [out=-70,in=-110, looseness=1] (6);
\end{tikzpicture}

\caption{The mutual statistics of $\mZ_k$ parafermions.}
\label{fig:PFstat}
\end{figure}
As we explain in Appendix~\ref{app:cosetduality}, in addition to the $\mZ_k$ symmetry, the theory has a non-chiral charge conjugations symmetry $\mZ_2^C$
\ie
C:~\psi_n(z) \to   \psi_{k-n}(z)\,,\quad \bar \psi_n(\bar z) \to  \bar\psi_{k-n}(\bar z)\,,
\fe
and a chiral parity symmetry $\mZ_2^S$ 
\ie
S:~\psi_n(z) \to  (-1)^n \psi_n(z)\,,\quad \bar \psi_n(\bar z) \to  \bar\psi_{k-n}(\bar z)\,.
\fe
The chiral parities $S$ and $CS$ lead to the equality \eqref{PFrel1} for $p=1$  and \eqref{PFrel2} for $p=- 1$ respectively and explain the two self-dualities of the $SU(2)_k/U(1)$ coset under $\mZ_k^B$ gauging. In particular the corresponding duality defects generate a pair of $\mZ_k$  TY symmetries in the coset CFT as summarized in Table~\ref{table:cosetTDLs} (see Appendix~\ref{app:cosetduality} for details). For $k=4$, the coset CFT describes the $R=\sqrt{6}$ point on the $c=1$ orbifold branch. See Section~\ref{secZ4parafermion} for further discussions on the corresponding duality TDLs.

\begin{table}[!htb]
    \centering
    \begin{tabular}{c|c}
    Parafermionic symmetry  & TDLs in $SU(2)_k/U(1)$ diagonal coset CFT\\
                 \hline
   $S$ & Duality $\cN$ in $\TY(\mZ_k,\omega,+)$
   \\    \hline
   $CS$ & Duality $\cN$ in $\TY(\mZ_k,\bar\omega,+)$
   \\
    \hline
    \end{tabular}
    \caption{TDLs in the $SU(2)_k/U(1)$ CFT from symmetries in the parafermionic theory.}
    \label{table:cosetTDLs}
\end{table}

Finally we emphasize that in general fermionic and parafermionic theories, the chiral parity symmetries discussed above are often absent (e.g. broken by interactions). In these cases, the general relations between bosonic and (para)fermionic theories in \eqref{BFdiag} and \eqref{BPFdiag} still apply, but the bosonic theory $\cT_B$ and its orbifold $\cT_{B'}$ are now distinct. On the $c=1$ moduli space, this happens for example on the circle branch at $R=\sqrt{3}\,,2\sqrt{3}$ where the theories are denoted by $\cN=2$ super-Ising$_\pm$ in Figure~\ref{fig:c1moduli}, and also on the orbifold branch at the same radii described by $\cN=1$ super-Ising$_\pm$.\footnote{Here the orbifold by $\mZ_2^C$ of the $\cN=2$ super-Ising$_\pm$ CFT projects out half of the supercurrents, leaving behind an $\cN=1$ super-Virasoro symmetry.}
The underlying fermionic theories correspond to $\cN=2$ and $\cN=1$ super-Virasoro minimal models at $c=1$ respectively. The $\pm$ subscripts are interchanged by stacking with the $(-1)^{\rm Arf(\rho)}$ SPT which leads to different fermionic and bosonic theories in these cases.

	\section{Fusion Categories of the Compact Free Boson}\label{seccircbranch}
	
	In this section we explore the fusion category symmetries of the $c=1$ compact boson along the circle branch. At radius $R = \sqrt{2k}$, these theories are rational with an enhanced chiral algebra  $U(1)_{2k}$. We discuss the associated Verlinde lines in Section \ref{subseccircularverlinde}, which for this chiral algebra are all invertible. Also at this radius, the theory is self-dual under gauging a $\bZ_k$ symmetry, and there is an associated Tambara-Yamagami symmetry (which is beyond the Verlinde lines), which we discuss in Section \ref{subsecZkTYcompbos}. Finally, at the radii $k\sqrt{2}$, which are obtained by gauging a $\bZ_k$ symmetry of the $SU(2)_1$ theory, the enhanced continuous symmetry of $SU(2)_1$ gives rise to a continuum of TDLs we discuss in Section \ref{subsecSU2continuum}.
	
	\subsection{Operator Content of the Compact Boson}
	
	There are several ways of describing the compact free boson in 1+1D, which we call the compact boson for short. The most direct description is in terms of a $2\pi R$-periodic field $X$, where $R$ is the radius of the target space circle. We define the $2\pi$-periodic field $\theta = X/R$ and its $2\pi$-periodic conjugate momentum $\phi$. They may be expressed in terms of the left and right moving fields $X_{L,R}$ as\footnote{We note there is an important subtlety which is that while it is possible to define $\theta$ and $\phi$ as $2\pi$-periodic, there is no way to consistently assign periodicities to $X_L$ and $X_R$, because the above change of variables is not invertible over $\bZ$ for any $R$. This may be accounted for by introducing certain cocycles into the definition of the vertex operators (see Section 5 of \cite{Harvey:2017rko}). One can avoid this subtlety by working with $\theta$ and $\phi$ but to make contact with the CFT literature we include transformation rules for $X_L$ and $X_R$ as well, with the caveat that these rules only unambiguously define the symmetry actions on the oscillator part of the spectrum.}
	\ie\theta = R^{-1}(X_L + X_R)\,,\quad 
\phi = R(X_L - X_R)/2\,.
\label{sfandX}
	\fe
		The compact boson at a generic radius $R$ has symmetry
\ie G_{\rm bos} = (U(1)^\theta \times U(1)^\phi) \rtimes \bZ_2^C\,,
	\label{Gbos}
	\fe
	where $U(1)^\theta$ and $U(1)^\phi$ act as shift symmetries of $\theta$ and $\phi$, respectively, and ``charge conjugation" $C$ acts as
\ie
C:\begin{cases}
	\theta \mapsto -\theta \,, \\ \phi \mapsto -\phi\,.
	\end{cases}
\fe
		The primary local operators in this theory consist of vertex operators
	\ie\label{eqncompbossvertop}
	\begin{gathered}
	V_{n,w} = :e^{i \left({n\over R}+{wR\over 2}\right) X_L}e^{i \left({n\over R}-{wR\over 2}\right) X_R}: = :e^{in\theta} e^{iw\phi}:  \qquad n,w \in \bZ \,,\\
	(h,\bar h) = \left(\frac{1}{2}\left(\frac{n}{R}+\frac{wR}{2}\right)^2, \frac{1}{2}\left(\frac{n}{R}-\frac{wR}{2}\right)^2\right),
	\end{gathered}
	\fe
 where $(h,\bar h)$ label the conformal weights, 
	as well as normal ordered Schur (symmetric) polynomials in the $U(1)$ currents $j_1 = \partial X_L$ and $\bar j_1 = \bar \partial X_R$ and their derivatives, which we denote by
	\ie
	j_{n^2} \bar j_{m^2} \qquad (h,\bar h) = (n^2,m^2)\,.
	\fe
	There is a duality known as T-duality which takes
	\ie\label{eqnTdualitybos}
	\begin{gathered}
	R \mapsto 2/R\,, \\ X_R \mapsto - X_R
	\,,\\ \theta \leftrightarrow \phi\,,
	\end{gathered}
	\fe
	and will play a crucial role in our story. As written, this duality commutes with charge conjugation $C$ and exchanges $U(1)^\theta$ and $U(1)^\phi$.
	
	There are several important subtleties of this duality, emphasized in \cite{Harvey:2017rko}, which can be seen at the self-dual radius $R = \sqrt{2}$. At this radius, there is an enhanced $SO(4) = (SU(2)_L \times SU(2)_R)/\bZ_2$ symmetry in which we expect to find T-duality. To define it unambiguously we need to specify the $SO(4)$ matrix by which it acts. The following specifies the vector representation of the $SO(4)$,
	\[(\cos \theta, \sin \theta, \cos \phi,\sin \phi)\,,\label{SO4basis}\]
	and in this basis we choose the swapping matrix
	\[\label{eqnTdualitychoice}T=\begin{bmatrix} 0 && I_{2 \times 2} \\ I_{2 \times 2} && 0 \end{bmatrix}\,,\]
	which agrees with the usual prescription \eqref{eqnTdualitybos}. This differs from the choice in \cite{Harvey:2017rko}, which is order 4 and anomalous. The above is anomaly-free, and generalize to self-dualities at the other points on the circle branch, as we will see.

	\subsection{RCFTs on the circle branch}\label{subseccircularverlinde}
	
	Special points on the $c=1$ moduli space (see Figure~\ref{fig:c1moduli}) are described by rational CFTs, which have a finite number of conformal blocks with respect to an enhanced chiral algebra. These arise when the radius satisfies $R^2 \in \mathbb{Q}$. For these theories we can study the Verlinde lines, which turn out to be invertible, and generate a finite subgroup of the ordinary global symmetry.
	
	We start with the case $R=\sqrt{2k}$ for $k\in \mZ$. The enhanced chiral algebra is generated by
	\ie
	T\,,~j_L\,,~e^{\pm i \sqrt{2k} X_L}
	\fe
	of spin $2,1,k$ respectively, and likewise for the right-movers. We will refer to this as the $U(1)_{2k}$ chiral algebra (the associated 3d Chern-Simons theory is $U(1)_{2k}$).  There are $2k$ chiral primaries as listed in Table~\ref{table:circleRCFT}.

	\begin{table}[htb!]
		\begin{center}
			\renewcommand{\arraystretch}{1.5}
			\begin{tabular}{ |c|c|  }
				\hline
				Primary & $e^{i n \theta}$ $(n=-k+1,-k+2,\dots,k)$
				\\\hline
				$h$ &  ${n^2\over 4k}$  
				\\\hline
				Characters &  $K^k_n(\tau)\equiv {1\over \eta(\tau)}\sum_{r \in \mZ}q^{k\left( r+{n\over 2k}\right)^2}$
				\\\hline
			\end{tabular}
		\end{center}
		\caption{The $2k$ chiral primaries in $U(1)_{2k}$.}
		\label{table:circleRCFT}
	\end{table}
The $U(1)_{2k}$ characters satisfy the following identities
	\ie
	K_m^k=K_{-m}^k=K^k_{2k+m}\,.
	\fe 
Under the modular $S$ transformation, we have
\ie
K_n^k(-1/\tau) ={1\over \sqrt{2k}}\sum_{m \in \mZ_{2k}} e^{- {\pi i mn\over k}} K_{m}^k(\tau)\,,
\fe
while under the $T$ transformation, they become
 \ie
K_n^k(\tau+1) = e^{\pi i n^2\over 2k } e^{-{\pi i \over 12}} K_{n}^k(\tau)\,.
\fe
The partition function of the theory is given by the diagonal modular invariant
	\ie
	Z_{\rm bos}(R=\sqrt{2k})=\sum_{m=-k+1}^k |K^k_m|^2
	\fe
	and we will refer to this theory as the  $U(1)_{2k}$ CFT.
	
	The Verlinde lines in this theory are labelled by the primaries $e^{i n \theta}$, $n \in \bZ_{2k}$. Let us denote the corresponding line by $\cL_n$. As for any Verlinde line, the action of $\cL_n$ on local operators is determined by the modular $S$ matrix, such that an operator in the $|K_m^k|^2$ block gets a phase $e^{- i \pi m n/k}$. We recognize this action as the $\bZ_{2k}$ shift symmetry $\theta \mapsto \theta - \pi m/k$.

    These lines are all invertible. Indeed, the Chern-Simons theory $U(1)_{2k}$ which governs the properties of these lines is abelian. In the following, we will find there are also non-invertible lines at these and other special radii.
	
	The more general RCFTs on the circle branch are associated to non-diagonal modular invariants of the chiral algebra for $U(1)_{2k}$. The RCFT at $R=\sqrt{2 p\over q}$ with positive coprime integers $p,q$ has torus partition function  
	\ie
	Z_{\rm bos}\left (R=\sqrt{2 p\over q} \right )
	=
	\sum_{r=0}^{2q-1}\sum_{s=0}^{2p-1} K^{pq}_{pr+qs}(\tau) K^{pq}_{pr-qs}(\bar \tau)\,,
	\fe
	which can be easily derived using the fact that this CFT is the $\mZ_q$ (shift symmetry) orbifold of the $U(1)_{2pq}$ CFT. A $\mZ_{2p}$ subset of the $\mZ_{2pq}$ Verlinde lines in the $U(1)_{2pq}$ CFT survives the orbifold and becomes a part of the global symmetry in the non-diagonal RCFT at $R=\sqrt{2 p\over q}$.\footnote{General TDLs that are transparent to the chiral algebra in a non-diagonal RCFT can be identified using the corresponding 3d TQFT on a slab with a surface operator parallel to the boundaries that defines the modular invariant \cite{Kapustin:2010if,Fuchs:2012dt,Carqueville:2017ono}. Here the TDLs come from bulk anyons and their fusion products in the presence of the surface operator \cite{Komargodski:2020mxz}. The corresponding map between the fusion category symmetries underlying the diagonal and non-diagonal RCFT is known as  
 the  $\A$-induction \cite{Longo:1994xe,Bockenhauer:1998ca,Bockenhauer:1998in,Bockenhauer:1998ef,Bockenhauer:1999wt,2001math.....11139O} .}

	\subsection{$\bZ_k$ Tambara-Yamagami symmetry}\label{subsecZkTYcompbos}

    The Verlinde lines discussed above on the circle branch are all invertible. There are also non-invertible TDLs at these radii, indicated by the self-duality under $\bZ_k$ gauging, which as we have discussed in Section \ref{subsecoverviewselfdual}, gives rise to a $\bZ_k$ Tambara-Yamagami (TY) symmetry. Recall the $\bZ_k$ TY category is characterized by a nondegenerate bicharacter $\bZ_k \times \bZ_k \to U(1)$, which may be written as
	\ie
	e^{i\chi(a,b)}=\omega^{ab}\,,
	\label{Zkchi}
	\fe
	with $\omega$ a primitive $k$-th root of unity, as well as the Frobenius-Schur indicator $ \epsilon = \pm 1$ of the duality defect $\cN$ \cite{tambarayam}. In this subsection,  we will constrain this data and identify the TY symmetry action from knowledge of the operator spectrum on the $c=1$ circle branch. We find an continuum of such symmetry actions parametrized by a sign $\pm$ and a number $\alpha \in [0,1)$,
		\ie\label{circleTYonop}
	\widehat \cN_{\A,\pm} :\begin{cases}
		V_{n,w} \to  
		\sqrt{k} e^{\pi i n w + 2\pi i \alpha(n \mp  kw)}
		V_{\pm wk,\pm {n\over k}} \quad {~\rm for~} n \in k\mZ\,,
		\\
		(\pa X_L,\bar\pa X_R) \to ( \pm \sqrt{k} \pa X_L,\mp \sqrt{k} \bar\pa X_R)\,,
		\\
		{\rm other\ primaries} \to {\rm non-local\ operators}\,,
	\end{cases}
	\fe
	where $V_{n,w}$ are the vertex operators defined in \eqref{eqncompbossvertop}. The corresponding  fusion category is
	\ie
	\cN_{\A,\pm } \to \TY(\mZ_k,\chi_\pm,1)\,,
	\label{circleTYf}
	\fe 
	where $\chi_\pm(a,b)\equiv  \pm {2\pi a b\over k}$, as we will derive in the following sections.

	\subsubsection{Self-duality under gauging}\label{subsubsecZkselfdualgauging}
	
	Let us consider the (anomaly-free) cyclic $\bZ_k^\theta$ subgroup of $U(1)^\theta$, which acts as
	\[\bZ_k^\theta: \theta \mapsto \theta + \frac{2\pi}{k}\,.\]
	We can think of this as a $k$-fold rotation symmetry of the circle target space. It is known that when one gauges a discrete geometric symmetry of a sigma model, if that symmetry acts freely on the target space, the result is again a sigma model but now the target space is replaced by its quotient \cite{Ginsparg:1987eb}, in this case a circle of radius $R\over k$. Indeed, when we gauge this subgroup, we restrict the $\theta$ vertex operators to have momentum numbers $n = k \tilde n$, while the twisted sectors contribute $\phi$ vertex operators of fractional winding $w = \tilde w/k$. Comparing with \eqref{eqncompbossvertop}, we see the spectrum of the $V_{k \tilde n, \tilde w/k}$ at radius $R$ matches that of the $V_{\tilde n, \tilde w}$ at radius $R\over k$.
	
	At the special radius $R = \sqrt{2k}$, ${R\over k} = \sqrt{2\over k}$ is the T-dual radius, so this theory is self-dual under gauging $\bZ_k^\theta$. As in \eqref{eqnpartfnselfdualgauging}, one consequence of this self-duality is an identity of twisted partition functions
	\ie
	Z_{\rm bos}(R = \sqrt{2k},\Sigma ,B) ={1\over \sqrt{| H^1(\Sigma,\bZ_k)|}} \sum_{A \in H^1(\Sigma,\bZ_k)} Z_{\rm bos}(R = \sqrt{2k},\Sigma,A) \omega^{\int_\Sigma A \cup B}\,,
	\label{PFsd}
	\fe
	\noindent
	where $A,B \in H^1(\Sigma,\bZ_k)$ are background $\bZ_k$ gauge fields on the spacetime manifold $\Sigma$, $\cup$ denotes the cup product of gauge fields, and $\omega$ is a primitive $k$-th root of unity which defines the bicharacter for the TY category.
	
	Let us now determine the possible $\omega$ which can occur in this formula for the theory at hand. Indeed an immediate consequence of the self-duality condition \eqref{PFsd} applied to the $U(1)_{2k}$ CFT on $T^2$ is that $\omega=e^{\pm {2\pi i\over k}}$. This follows by studying the coefficients of the term $q^{{1\over 4k}-{1\over 24}}\bar q^{{1\over 4k}-{1\over 24}}$ on the two sides of the equality, with $B$ chosen to have holonomy equal to the $\xi:\theta \to \theta +{2\pi \over k}$ generator of $\mZ_k^\theta$ around the temporal cycle, and trivial holonomy around the spatial cycle. On the LHS, the two momentum operators $V_{\pm 1,0}$ contribute $e^{\pm {2\pi i\over k}}$ for a total  of $2\cos {2\pi \over k}$. On the RHS, this term comes from the $\mZ_k^\theta$ symmetry generator $\xi$ acting on twisted sector operators. Now the $\xi^n$-twisted sector ground states for $0\leq n \leq k-1$ correspond to operators with fractional winding $e^{{i n  \over k}\phi}$ or $e^{{i (n-k)  \over k}\phi}$ depending on which one has the lower scaling dimension. Clearly $h=\bar h={1\over 4k}$ is only attainable for $n=1,k-1$. Furthermore since these operators are scalars, whereby the $T$-transformation acts trivially, they carry vanishing $\mZ_k^\theta$ charge. Meanwhile the bicharacter contributes $\omega$ and $\omega^{k-1} = \bar \omega$, respectively. Consequently, we arrive at the promised equality
\ie
2\cos {2\pi i\over k}= \omega +\bar \omega\,,
\label{Zksdconstraint}
\fe
 which determines $\omega$ up to a conjugation ambiguity.
 
  Generally speaking, there is a certain ambiguity in the expression \eqref{PFsd}, whereby the Galois action on $\omega$ can be equated with the action of ${\rm Aut}(\bZ_k^\theta)$  on the background gauge field $B$. The partition function of the compact boson is invariant under $B \mapsto -B$ because of the charge conjugation symmetry $C$, which acts to negate the generator of $\bZ_k^\theta$. One can check that it has no further ${\rm Aut}(\bZ_k^\theta)$ ambiguities by studying the operators $V_{\pm 1, 0}$  with momentum charge $\pm 1$ and zero winding charge. Thus, the identity above cannot distinguish between $\omega$ and $\bar \omega$. In fact, by fusing a $\bZ_k$ TY TDL with a certain bicharacter $e^{i\chi}$ by the $C$ TDL we obtain a $\bZ_k$ TY TDL with the conjugate bicharacter $e^{-i\chi}$, so they always come in pairs (see Figure~\ref{fig:stackf}).
	
    When we study the TDL action on local operators in the CFT, there are new constraints on $\omega$. We will analyze these constraints using the duality twisted partition functions.

	\subsubsection{Duality acting on local operators}

The duality is obtained by gauging and then applying T-duality. Using the definition of T-duality in \eqref{eqnTdualitybos}, we obtain the following action of the TDL $\cN$ on the primaries
	\ie
	\widehat \cN :\begin{cases}
		V_{n,w} \to  
		\sqrt{k} e^{\pi i n w}
		V_{wk,{n\over k}} \quad {~\rm for~} n \in k\mZ\,,
		\\
		(\pa X_L,\bar\pa X_R) \to ( \sqrt{k} \pa X_L,- \sqrt{k} \bar\pa X_R)\,,
		\\
		{\rm other\ primaries} \to \text{non-local primaries}\,,
	\end{cases}
	\label{TYZk}
	\fe
	corresponding to the special case $\widehat \cN_{\A=0,+}$ in \eqref{circleTYonop}.
	The extra phase $e^{\pi i n w}$ comes from the mutual non-locality of $\theta$ and $\phi$ variables. It is easy to check that $\widehat \cN$ satisfies the $\bZ_k$ TY fusion rule
	\[\widehat \cN^2 = \sum_{m \in \bZ_k} U^\theta(2\pi m/k)\,,\]
	where $U^\theta(2\pi m/k)$ generates a $2\pi m/k$ shift of $\theta$.
	
	This solution is not unique---there are actually many $\bZ_k$ duality defects in the $U(1)_{2k}$ CFT. For instance we can compose $\cN$ with the charge conjugation symmetry $C$ (which as we mentioned above yields a $\bZ_k$ duality TDL $C\cN$ with the conjugate bicharacter). Generally besides this $C$ ambiguity we can also modify the action of $\cN$ on the operators $V_{n,w}$ \eqref{TYZk} by a phase $\m(n,w)$. Consistency with the OPE and the TY fusion rule requires that $\m(n,w)=e^{2\pi i \A (n-kw)}$ for $\A \in [0,1)$. This phase corresponds to an invertible defect 
	\ie
	\cL^{U(1)_R}_{\A}\equiv e^{i\A\sqrt{2k}\oint d\bar z \bar \pa X_R}\,,
	\fe
	associated to the right-moving $U(1)$ symmetry, and we can write the modified duality defect as $\cN_{\A,+}\equiv \cN \cL^{U(1)_R}_{\A} $. It is easy to check that it satisfies the $\mZ_k$ TY fusion rule as a consequence of \eqref{TYZk}. Similarly, we have the modified TY duality defect  $\cN_{\A,-}\equiv C\cN  \cL^{U(1)_L}_{\A}$
	using the left-moving $U(1)$ symmetry. 
 Therefore the $U(1)_{2k}$ RCFT contains a family of $\mZ_k$ TY categories indexed by $\A$ (and a complementary family related by $C$).\footnote{$\bar \cN_{\alpha,\pm} \cN_{\alpha,\pm} = \cN_{\alpha,\pm} \cN_{\alpha,\pm}$ contains the trivial line, and the  Noether current obtained from this family by the method of Section \ref{subsecshadownoether} is just the local $U(1)_R$ current $\bar \partial X_R$.}  Since they are related by invertible defects which are already known, it is enough to show that just  one of these symmetry actions in \eqref{circleTYonop} is associated with a TDL, so we will focus on the case $\cN=\cN_{\A=0,+}$ below.\footnote{As we will see, the compositions with the chiral symmetries $\cL^{U(1)_R}_{\A}$ and $\cL^{U(1)_L}_{\A}$ do not lead to further changes to the TY category.}

	Let us study the torus partition function of the $U(1)_{2k}$ CFT with the $\cN$  TDL inserted along a spatial cycle. For even $k$, $\cN$ commutes with the left-moving chiral algebra, and  its action on the right-moving side projects to the sector with $n=wk$, namely the vacuum block with respect to the chiral algebra.\footnote{As a consequence, the duality defect $\cN \cL_\A^{U(1)_R}$ has the identical twisted partition function as for $\cN$.} Therefore we have 
	\ie
	Z_{1\cN}(\tau,\bar\tau)
	=&\sqrt{k} K_0^k(\tau) \left (\sum_{m \in \mZ_{\geq 0}} (-1)^{m} \chi_{m^2}(\bar\tau)\right)
	=
	{\sqrt{k}\over |\eta(\tau)|^2}
	\sum_{n,m \in \mZ}   
	(-1)^{m}q^{ kn^2}  \bar q^{m^2} 
	={ \sqrt{k}\theta_3(2k\tau) \theta_4(2\bar \tau)\over  |\eta|^2}
	  \,,
	\label{ZkN1}
	\fe 
	where $K_0^k$ is the identity character for the $U(1)_{2k}$ chiral algebra (see Table~\ref{table:circleRCFT}).\footnote{In contrast to $\cN$, the duality defect $C\cN$ commutes with the right-moving chiral algebra. The corresponding twisted partition function is the parity-flip $\tau \leftrightarrow \bar\tau$ of \eqref{ZkN1}. The same relation holds for other twisted partition functions discussed here and in the following sections.\label{footnote:genTY}} 
We have used the fact that the right-moving $U(1)_{2k}$ vacuum character twisted by $\cN$ gives,
	\ie 
	K_0^k (\bar\tau) \xlongrightarrow[]{\cN}  \sum_{ n \in \mZ} (-1)^m   \chi_{m^2}(\bar \tau)\,,
	\fe	
	and also the following relation for the degenerate Virasoro characters
	\ie
	\sum_{m \in \mZ_{\geq 0}}(-1)^m\chi_{m^2} =\sum_{m\in \mZ} {(-1)^mq^{m^2}\over \eta}\,.
	\fe
One can immediately check that the twisted partition function is compatible with modular invariance and integrality.
	Indeed the modular S-transform gives
	\ie
	Z_{\cN1}(\tau,\bar\tau)
	=&{1\over 2|\eta(\tau)|^2}
	\sum_{n,m \in \mZ}   
	q^{n^2\over 4k}  
	\bar q^{(m+1/2)^2\over 4}
	= 
 \sum_{n=-k+1}^k 	\sum_{ m \in \mZ_{\geq 0}}    K_n^k(\tau) 
	 \chi_{(m+1/2)^2\over 4}(\bar \tau)\,,
	\label{ZkNm2}
	\fe 
	which decomposes into  $U(1)_{2k}$ and Virasoro characters with integral degeneracies, as required.

	For odd $k$, a similar analysis gives the $\cN$ twisted torus partition function
	\ie
	Z_{1\cN}(\tau,\bar\tau)
	=&{\sqrt{k}\over |\eta(\tau)|^2}
	\left( \sum_{ n \in \mZ} (-1)^n   q^{kn ^2}\right) \left( \sum_{  m \in \mZ} (-1)^m \bar q^{m^2}\right) 
	={ \sqrt{k}\theta_4(2k\tau) \theta_4(2\bar \tau)\over  |\eta(\tau)|^2} \,,
	\label{ZkN1oddk}
	\fe 
	where we have used the fact that the $U(1)_{2k}$ vacuum character twisted by $\cN$ gives,
	\ie 
	K_0^k (\tau) \xlongrightarrow[]{\cN} {1\over \eta(\tau)}\sum_{ n \in \mZ} (-1)^n   q^{kn ^2}\,.
	\label{NtwisteK0}
	\fe
 Performing an S-transform, we obtain
	\ie
	Z_{\cN1}(\tau,\bar\tau)
	=&{1\over 2|\eta(\tau)|^2}
	\sum_{n,m \in \mZ}   
	q^{(n+1/2)^2\over 4k}  
	\bar q^{(m+1/2)^2\over 4}
		= 
	2\sum_{n,m \in \mZ_{\geq 0}}   
	\chi_{(n+1/2)^2\over 4k}  (\tau)
		\chi_{(m+1/2)^2\over 4}(\bar \tau)\,.
	\label{ZkNm1}
	\fe  
Note that for $k=1$, the duality defect becomes invertible and generates  a  non-anomalous $\mZ_2$ T-duality symmetry \eqref{eqnTdualitychoice}. The twisted partition function and its S-transform  for this case are
	\ie
	Z_{1\cN}(\tau,\bar\tau)
	={ \theta_4(2\tau) \theta_4(2\bar \tau)\over  |\eta(\tau)|^2} 
	\,,\quad
	Z_{\cN1}(\tau,\bar\tau)
	={ \theta_2( \tau/2) \theta_2( \bar \tau/2)\over  2|\eta(\tau)|^2}  \,.
	\label{HNodd}
	\fe
	
	\subsubsection{Spin selection rules}\label{subsecspinselectionrulesZkTY}
	Here we derive relations between the spectrum of spins in the defect Hilbert space $\cH_\cN$ for the duality TDL and the F-symbols of the TY category. As we will see, they lead to further constraints on the F-symbol data of $\mZ_k$ TY symmetries, namely the primitive $k$-th root of unity $\omega$ and FS indicator $\ep$.

\begin{figure}
    \includegraphics[width=4cm]{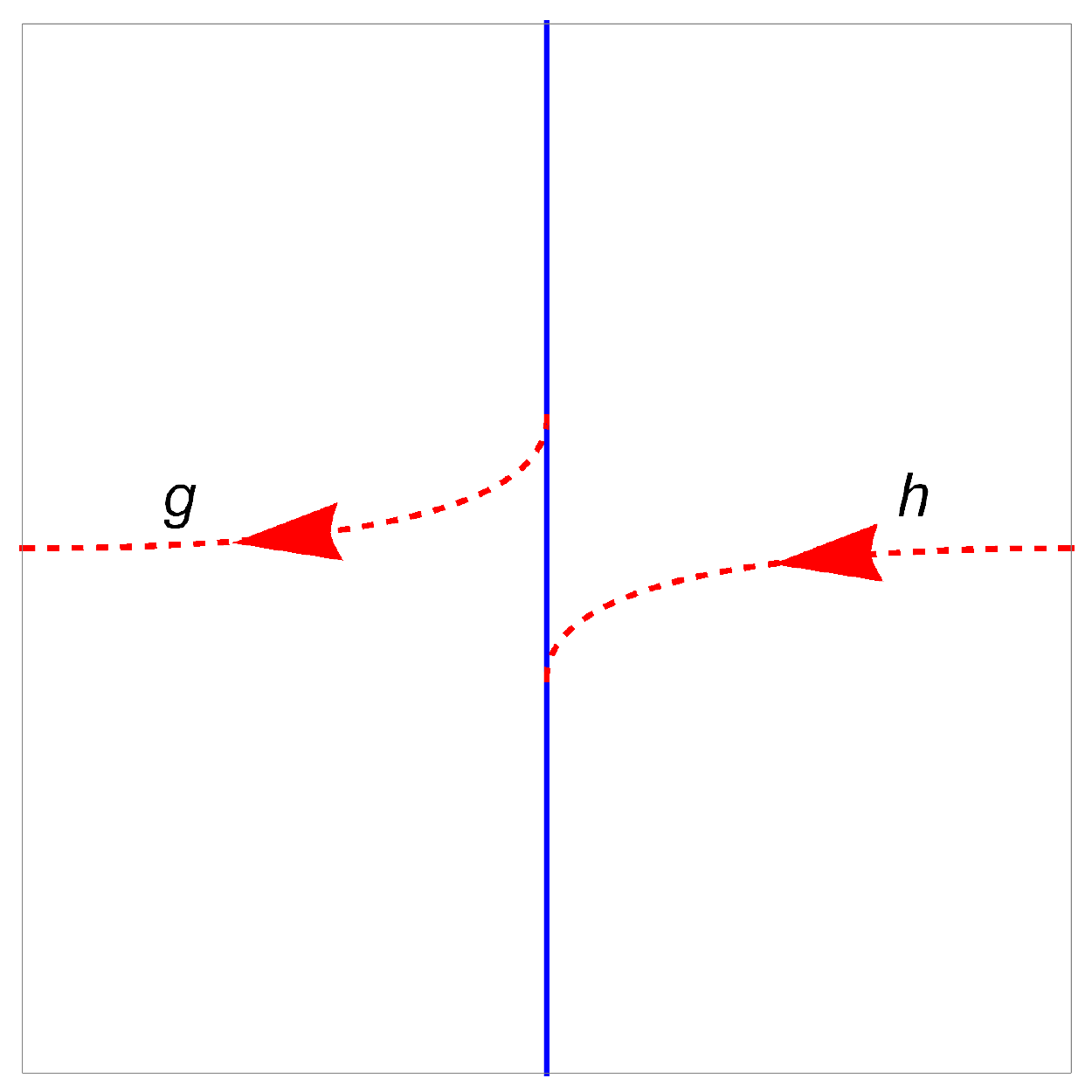}
    \centering
    \caption{To define the $\bZ_k$ symmetry action in the $\cN$-twisted sector, we must choose a resolution of the crossings into 3-fold junctions. We refer to the resolution depicted in the figure as the ``minus" resolution, and the operators associated with group elements $g = h = r^a \in \bZ_k$ as $\widehat{(r^a)}_-$. This is related to the other resolution by a phase $\widehat{(r^a)}_+ = \omega^{a^2} \widehat{(r^a)}_-$.}
    \label{figleftaction}
\end{figure}

	 We define the action of the $\mZ_k$ symmetry TDL $\cL$ on the defect Hilbert space $\cH_\cN$ as $\hat\cL_-$ as in  Figure~\ref{figleftaction}.  	
The bicharacter $e^{i\chi(a,b)}=\omega^{ab}$  enters in the crossing relations involving both the duality TDL $\cN$ and  the $\mZ_k$ generators $r^a$ and $r^b$. In particular,
	\ie
	\widehat{(r^a)}_- \cdot \widehat{(r^b)}_-=\omega^{ab}\widehat{(r^{a+b})}_-\,,
	\fe
	which implies 
	\ie
	\widehat{(r^a)}_-=\omega^{-{a(a-1)\over 2}} (\widehat{r}_-)^a\,,
	\fe
	and
	\[\label{eqnrminusorder}
	(\widehat{r}_-)^k=\omega^{k(k-1)\over 2}=(-1)^{k-1}\,.\]
	
	We would like to derive a relation between the spin spectrum in $\cH_\cN$ and the choice of bicharacter \eqref{Zkchi}. We proceed as in \cite{Chang:2018iay} to  consider the cylinder amplitude between a pair of states $|\psi\ra,|\psi'\ra \in \cH_\cN$ of equal conformal weights  
	\ie
	\la \psi' |q^{L_0-{c\over 24}}\bar q^{\bar L_0-{\bar c\over 24}} |\psi\ra \,.
	\fe
\begin{figure}[H]
  	\centering	\begin{minipage}{0.20\textwidth}
  		\includegraphics[width=1\textwidth]{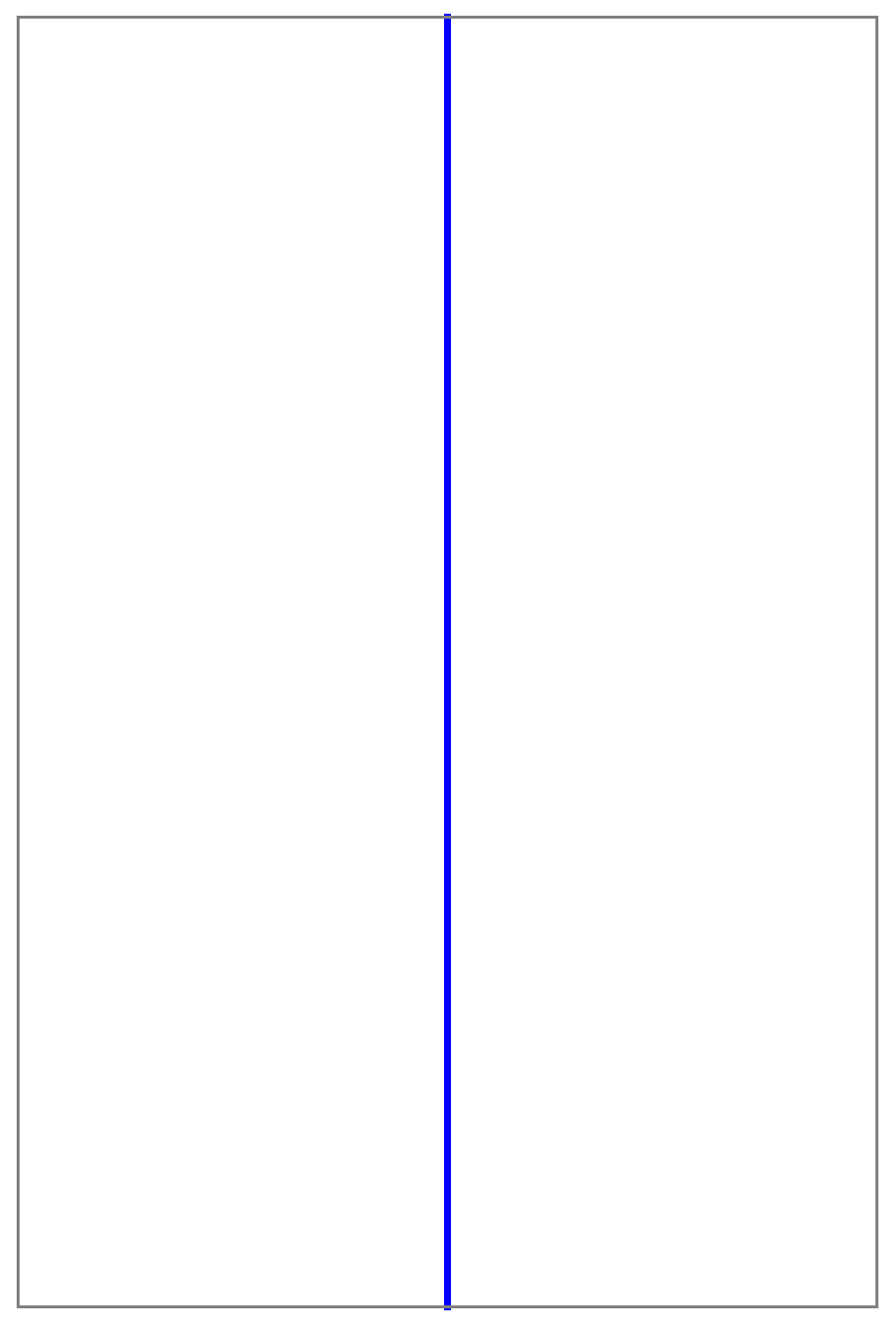}
  	\end{minipage}%
  		\begin{minipage}{0.08\textwidth}\begin{eqnarray*}~~\longrightarrow^{\!\!\!\!\!\!\!\!\! T^2}  \\ \end{eqnarray*}
	\end{minipage}%
	\begin{minipage}{0.20\textwidth}
  		\includegraphics[width=1\textwidth]{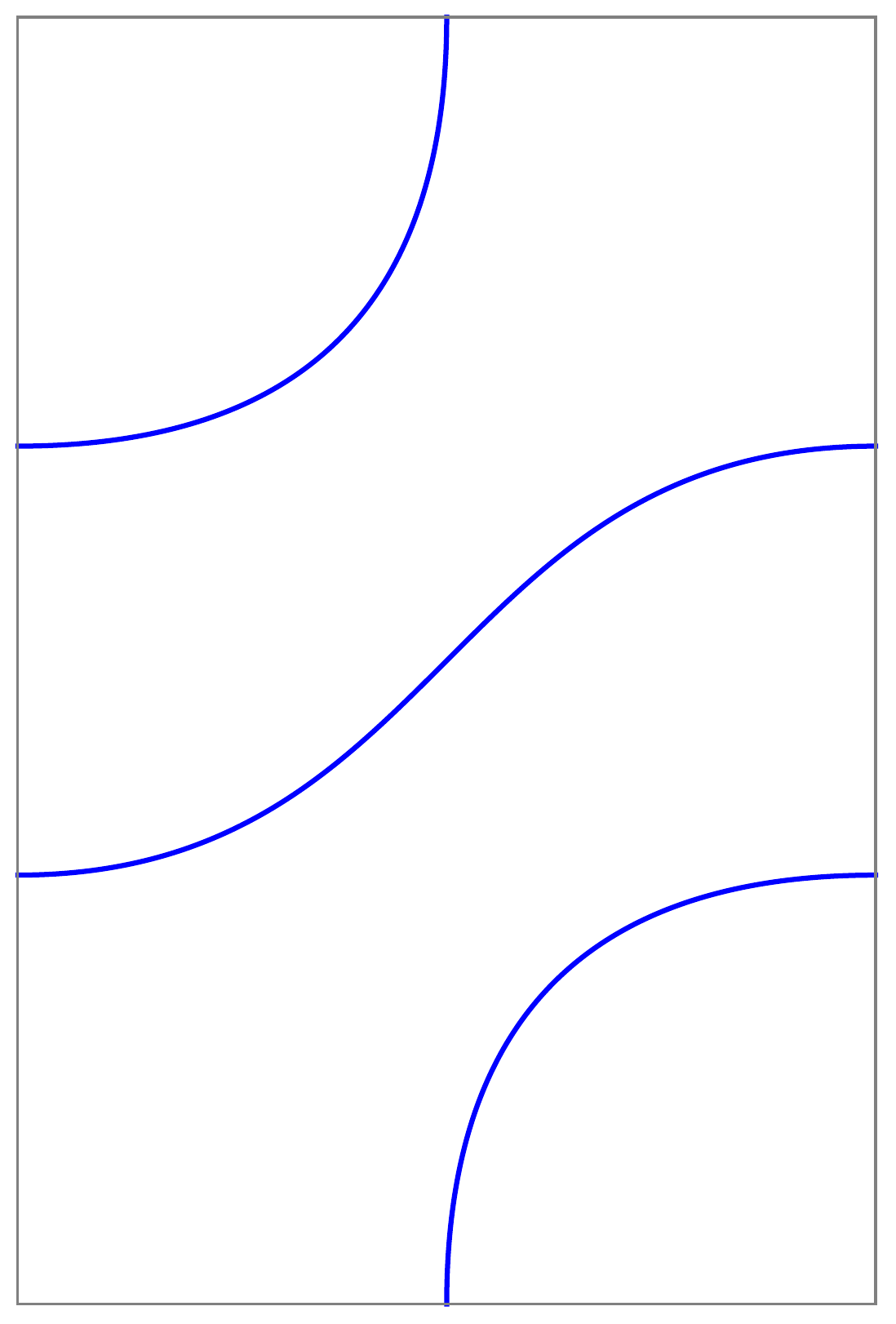}
  	\end{minipage}%
  	\begin{minipage}{0.14\textwidth}\begin{eqnarray*}~= {\epsilon\over \sqrt{k}}\sum_{a=0}^{k-1} \\ \end{eqnarray*}
  	\end{minipage}%
  	  \begin{minipage}{0.20\textwidth}
  	\includegraphics[width=1\textwidth]{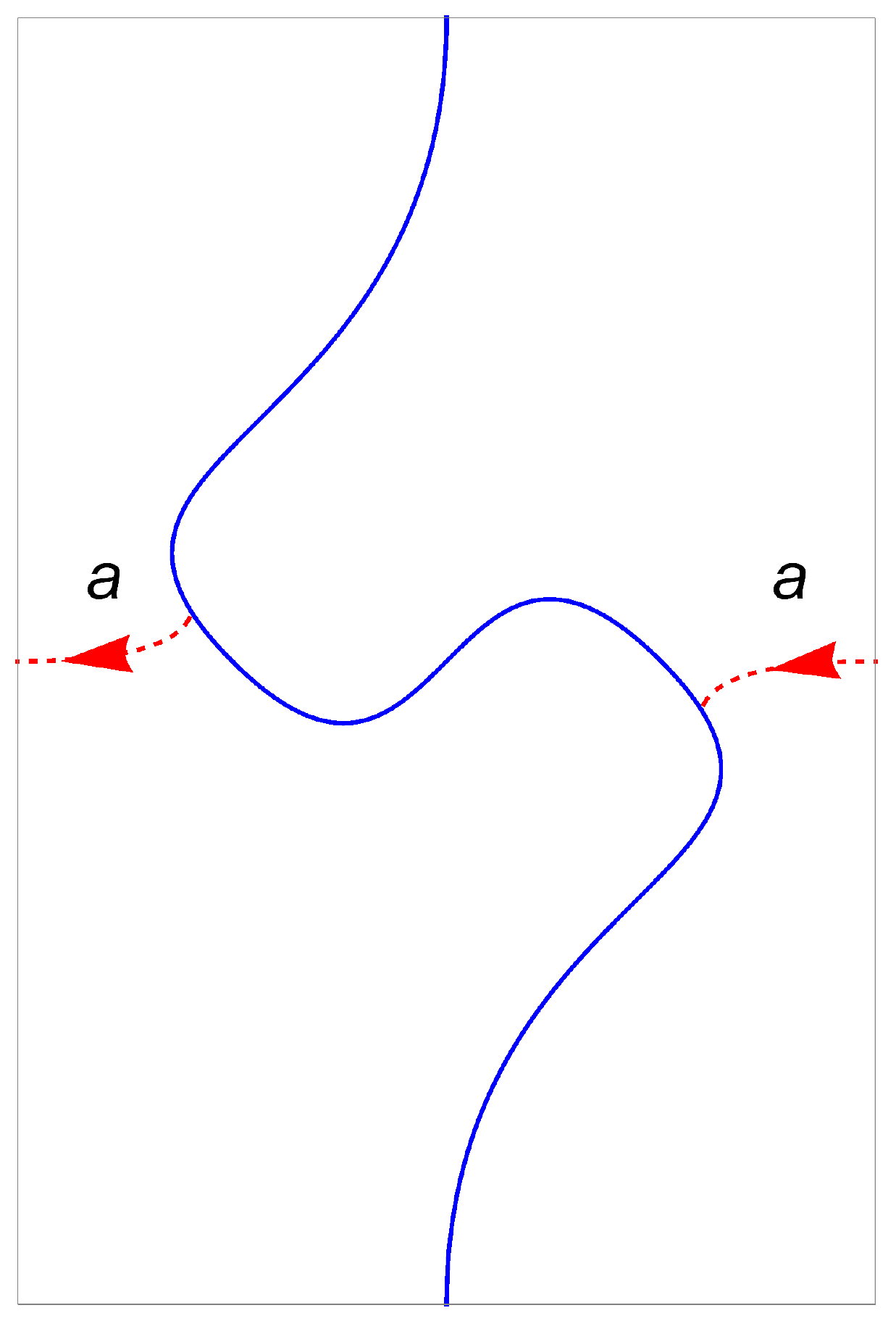}
  \end{minipage}%
  	\caption{Applying partial fusion and crossing to the $T^2$ transformation of ${\cal N}$ line in time direction.}
  	\label{fig:ssrT2}
  \end{figure}
  By performing a $T^2$ transformation $\tau \to \tau+2$ on this configuration (which involves the duality $\cN$ in the time direction), and applying  partial fusion and crossing relations, as in Figure~\ref{fig:ssrT2}, we reach the equation
	\ie
	e^{4\pi i s}\la \psi' | \psi \ra
	=
	{\ep \over \sqrt{k}}
	\la \psi' | 
	\sum_{a=0}^ {k-1}\widehat{(r^a)}_-
	|\psi\ra \,.
	\label{ssr}
	\fe
 Let us restrict our attention to $\mZ_k$ eigenstates in $\cH_\cN$. Since $\omega$ is a primitive $k$-th root of unity, these states satisfy
	\ie
	\hat r_- |\psi\ra =\omega^q |\psi \ra \,,
	\label{HNZkcharge}
	\fe
	with $q\in \mZ$ for $k$ odd and $q\in \mZ+{1\over 2}$ for $k$ even (cf. \eqref{eqnrminusorder}). Then \eqref{ssr} can be rewritten as
	\ie
	e^{4\pi i s}=
	{\ep \over \sqrt{k}}
	\sum_{a=0}^ {k-1} \bar\omega^{a(a-1)\over 2} \omega^{qa}
	=
	{\ep \over \sqrt{k}}
	\sum_{a=0}^ {k-1}  \omega^{2qa-a(a-1)\over 2} 
	=
	{\ep \over \sqrt{k}}
	\sum_{a=0}^ {k-1}  \omega^{-{1\over 2}{(a-{2q+1\over 2})^2}} \omega^{ (2q+1)^2\over 8} \,.
		\fe
 For $k$ odd, using the periodicity in $a \to a+k$, we have 
	\ie
	e^{4\pi i s}=
	{\ep \over \sqrt{k}}
	\omega^{ (2q+1)^2\over 8} 
	\sum_{a=0}^ {k-1}  \omega^{-{(a-1/2)^2\over 2}}  \,.
	\label{ssrZkodd}
	\fe
		Similarly for $k$ even, using the periodicity in $a \to a+k$, redefining $p=q+{1\over 2}\in \mZ_k$\footnote{Here the fractional powers of $\omega$ are taken in the standard branch. Consequently for fixed $p\in \mZ_k$, the RHS of \eqref{ssrZkeven} in general changes as $\omega \to  w e^{2\pi i}$. However the set of $e^{4\pi i s}$ for $p\in \mZ_k$ does not depend on such a change of branch (it can be canceled by shifting $p \to p+{k\over 2}$). }
	\ie
	e^{4\pi i s}=
	{\ep \over \sqrt{k}}
	\sum_{a=0}^ {k-1}  \omega^{-{(a-p)^2\over 2}} \omega^{ p^2\over 2} 
	=
	{\ep \over \sqrt{k}} \omega^{ p^2\over 2} 
	\sum_{a=0}^ {k-1}  \omega^{-{a^2\over 2}} \,.
		\label{ssrZkeven}
		\fe
	 These Gauss sums constrain  the allowed spins in the defect Hilbert space $\cH_\cN$ in terms of the F-symbol data of the TY category ($\ep$ and $\omega$ here) and vice versa.

    The second Gauss sum \eqref{ssrZkeven} may be found in Appendix 4 of \cite{milnor1973symmetric}, where it is shown that
    \[\frac{1}{\sqrt{k}}\sum_{a = 0}^{k-1} \omega^{-{a^2\over 2}} = e^{-{2\pi i \sigma\over 8}},\]
    and $\sigma$ is the signature of the associated quadratic form on $\mZ_k$
    \[q(a) = \omega^{a^2\over 2}.\]
	Namely, we must find an (even) integer bilinear form $\langle -,-\rangle$ on some $\bZ^n$ lattice with discriminant group $\bZ_k$ and associated quadratic form $q$, and then $\sigma$ is the signature of $\langle -,-\rangle$. There is an algorithm for producing such a lift by Wall \cite{WALL1963281}. This lift is not unique and can only define a signature modulo 8 because of the existence of even bilinear forms with nonzero signature and trivial discriminant group, such as the $E_8$ root lattice. See \cite{wang2020abelian} for a discussion of these facts in another physics context. In particular, if $\omega = e^{\pm {2\pi i\over k}}$, $e^{2\pi i\sigma\over 8} = e^{\pm {\pi i\over 4}}$ which comes from a rank one bilinear form $\la x,y\ra \equiv \pm kxy$ for $x,y\in \mZ$.
	
	It turns out the first Gauss sum \eqref{ssrZkodd} is also related to a signature \cite{turaev_1998}. Namely, it is\footnote{Note that unlike the $k$ even case, here the summand $q(a)\equiv \omega^{-{(a-1/2)^2\over 2}}$ defines a quadratic function on $\mZ_k\to U(1)$ but not a quadratic form on $\mZ_k$ (i.e. $q(na)\neq q(a)^{n^2}$). }
	\[\frac{1}{\sqrt{k}}\sum_{a=0}^ {k-1}  \omega^{-{(a-1/2)^2\over 2}} = e^{-{2\pi i\sigma\over 8}},\]
	where now we must find an integer bilinear form $\langle -, - \rangle$ with discriminant group $\mZ_k$ associated to the bilinear form $\omega^{ab}$, and a Wu (characteristic) element $w \in \mR^n$ such that
	\[\langle x,w\rangle = \langle x, x \rangle \mod 2 \qquad \forall x \in \bZ^n\,,\]
	and such that $w$ lifts the generator of the discriminant group $1 \in \bZ_k$. 
	In particular, if $\omega = e^{\pm {2\pi i\over k}}$, we can take a rank one lattice with bilinear form $\la x,y\ra=\pm kxy$ and Wu  element $w=\pm{1\over k}$. Consequently we find $e^{2\pi i\sigma\over 8} = e^{\pm {\pi i\over 4}}$ and again the signature $\sigma$ has a mod 8 ambiguity due to the $E_8$ lattice.
	
	To summarize, the spin selection rules we have derived for $\cH_\cN$ with bicharacter $\chi_\pm$ and general FS indicator $\ep$ are,
	\ie
e^{4\pi i s}=\begin{cases}
 \ep e^{\pm{ \pi i \over 4k }(2n+1)^2} e^{\mp \pi i /4} & k \in 2\mZ \,,
 \\
 \ep e^{\pm{ \pi i \over k }n^2} e^{\mp \pi i /4} & k \in 2\mZ+1 \,.
\end{cases}
\label{Zkssr}
\fe
	
		Let us now apply the above spin selection rules to infer about the F-symbols and determine which TY category symmetry is present in the $U(1)_{2k}$ CFT. From \eqref{ZkNm1} and \eqref{ZkNm2}, one can read off the spins in $\cH_\cN$,
	\ie
	s+{1\over 16}=\begin{cases}
		    {(2m+1)^2\over 16 k} +{1\over 2}\mZ  &  k\in 2\mZ+1\,,
		\\
		  {n^2\over 4 k}+{1\over 2}\mZ&  k\in 2\mZ\,,
	\end{cases} 
	\label{circspinspec}
	\fe
	where $m,n\in \mZ$.
	
	For $k$ odd, it is immediate to check that out of the four possibilities given by $\omega=e^{\pm {2\pi i \over k}}$ and $\ep=\pm 1$ for the TY symmetry, only the case $\omega=e^{{2\pi i \over k}}$ and $\ep=1$ produces the spin selection rule \eqref{ssrZkodd}  which is consistent with the above spin spectrum. This completes the identification of the TY category symmetry on the circle branch \eqref{circleTYf} for the $k$ odd case.
	
	For $k$ even, while  the spin selection rule  \eqref{ssrZkeven} with $\omega=e^{{2\pi i \over k}}$ and $\ep=1$ is  satisfied by the above spin spectrum \eqref{circspinspec}, other solutions to this constraint are possible at special values of $k$. To eliminate these spurious possibilities, we carry out a bootstrap analysis that utilizes the left-moving $U(1)_{2k}$ chiral algebra which commutes with the duality for even $k$.
	
\subsubsection{Bootstrap analysis for $k$ even}\label{sec:bootstrapZk}
The idea here is to analyze the modular bootstrap equation \eqref{mbeqn} for the torus partition function of the $U(1)_{2k}$ CFT decorated by the $\mZ_k$ symmetry TDL $r^a$ and the putative duality TDL $\cN$, to derive constraints on the $\mZ_k$ charge of the states in the twisted Hilbert space $\cH_{\cN}$ (e.g. $q$ in \eqref{HNZkcharge}). As we will see, this gives a refinement of the spin selection rule derived in the last section and determines the F-symbols for $\cN$ completely to be those in \eqref{circleTYf}. 

\begin{figure}[!htb]
	\centering
	\begin{minipage}{0.12\textwidth}\centering\begin{eqnarray*} \left. \widehat \cN^2 \right|_{\cH_{r^m}}= \\ \end{eqnarray*}
	\end{minipage}%
	\begin{minipage}{0.13\textwidth}
		\includegraphics[width=1\textwidth]{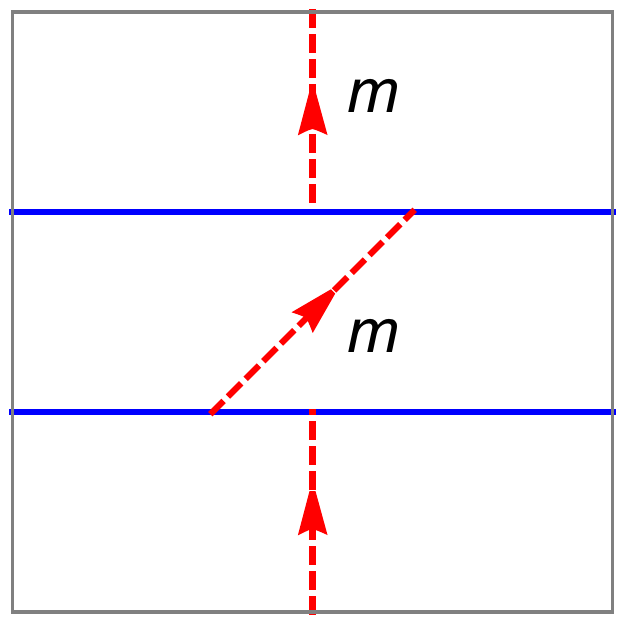}
	\end{minipage}%
	\begin{minipage}{0.12\textwidth}\centering\begin{eqnarray*}={\ep\over \sqrt{k}}\sum_{n=0}^{k-1}  \\ \end{eqnarray*}
	\end{minipage}%
	\begin{minipage}{0.13\textwidth}
		\includegraphics[width=1\textwidth]{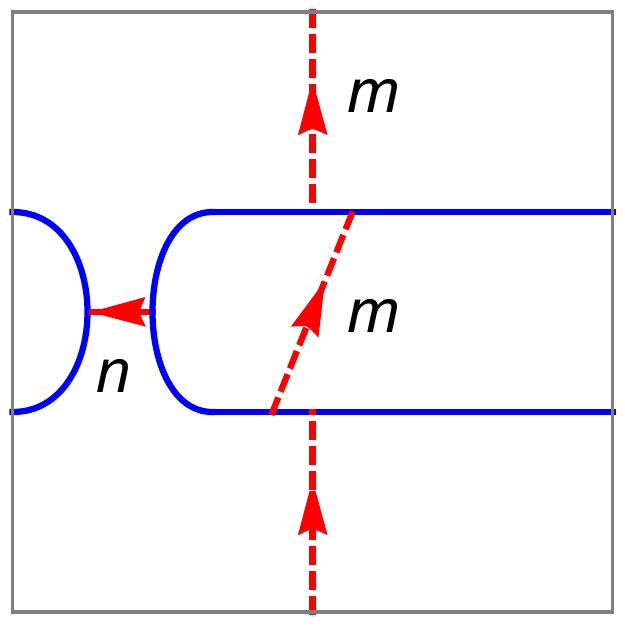}
	\end{minipage}%
		\begin{minipage}{0.21\textwidth}\centering\begin{eqnarray*}={\ep\over \sqrt{k}}\sum_{n=0}^{k-1}  \bar \omega^{m(m+n)} \\ \end{eqnarray*}
	\end{minipage}%
	\begin{minipage}{0.13\textwidth}
		\includegraphics[width=1\textwidth]{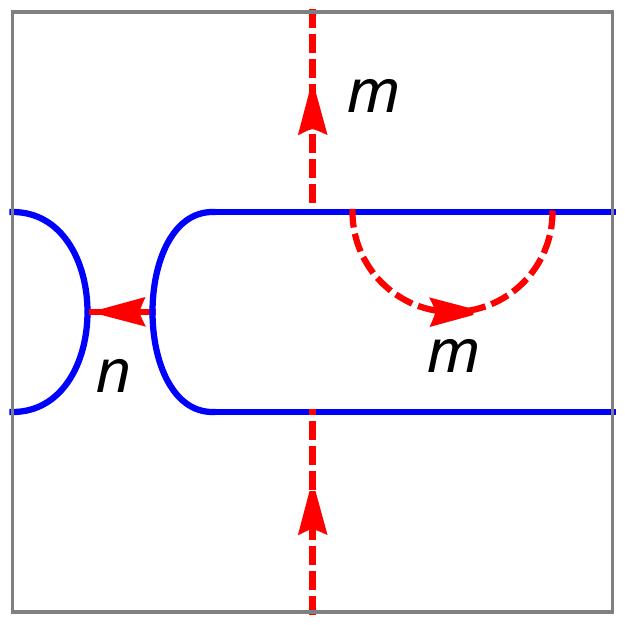}
	\end{minipage}%
	\\
	\begin{minipage}{0.21\textwidth}\centering\begin{eqnarray*}={\ep\over \sqrt{k}}\sum_{n=0}^{k-1} \bar \omega^{m(m+n)}  \\ \end{eqnarray*}
	\end{minipage}%
	\begin{minipage}{0.13\textwidth}
		\includegraphics[width=1\textwidth]{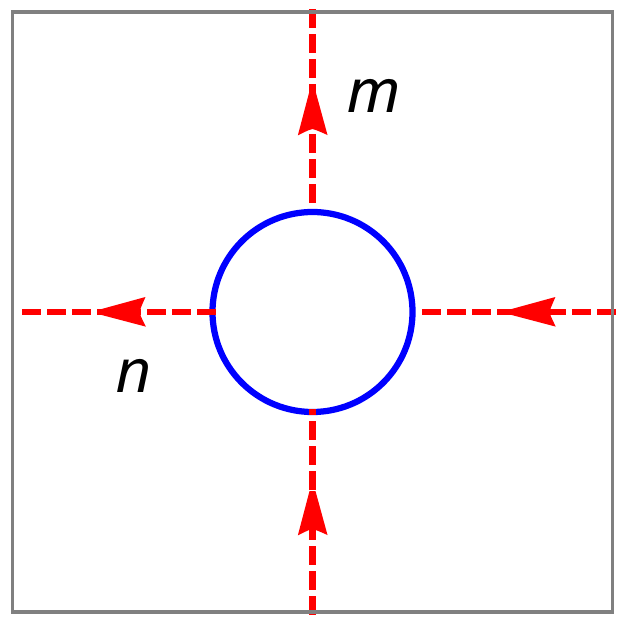}
	\end{minipage}%
		\begin{minipage}{0.16\textwidth}\centering\begin{eqnarray*}=\sum_{n=0}^{k-1}  \bar \omega^{m(m+n)} \\ \end{eqnarray*}
	\end{minipage}%
	\begin{minipage}{0.13\textwidth}
		\includegraphics[width=1\textwidth]{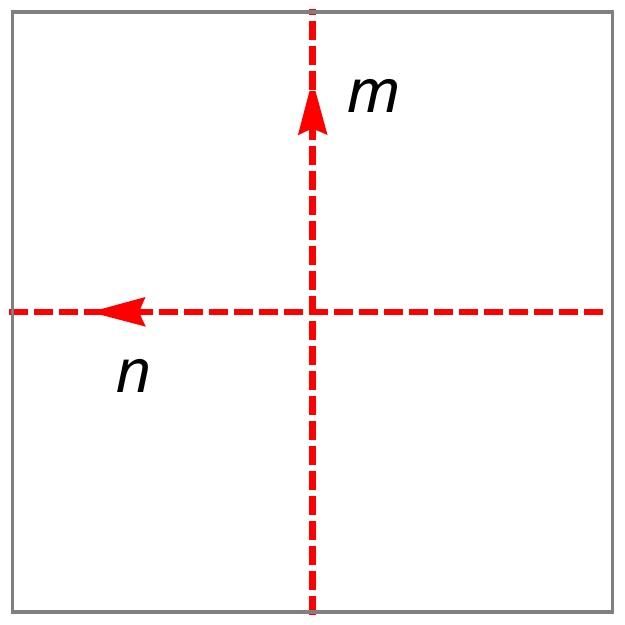}
	\end{minipage}%
	\caption{The sequence of F-moves that simplifies the squared action of the duality TDL $\widehat \cN^2$ on the defect Hilbert space $\cH_{r^m}$.  }
	\label{fig:Nsq}
\end{figure}

We start with the torus partition function with a $\mZ_k$ TDL along the spatial cycle,
\ie
Z_{1r}(\tau,\bar\tau)=\sum_{m=-k+1}^k e^{2\pi i m\over k} |K_m^k(\tau)|^2\,.
\fe
Its modular S-transform captures the states in the defect Hilbert space $\cH_r$,
\ie
Z_{r1}(\tau,\bar\tau)=\sum_{n=0}^{2k-1} K_n^k(\tau) K_{n-2}^k (\bar \tau)\,.
\label{ZktwistedPF}
\fe
If we insert a duality TDL along the spatial cycle, the corresponding partition function is the trace of $\widehat \cN$ in $\cH_r$. Using F-moves with the F-symbols parametrized by the undetermined parameters $\omega=e^{\pm {2\pi i \over k}}$ and $\epsilon=\pm 1$ as in Figure~\ref{fig:Nsq}, we obtain
\ie
\widehat \cN^2 =\bar\omega \sum_{n=0}^{k-1} \bar\omega^n  (\widehat r)^n\,.
\fe

This means that acting on $\cH_r$, $\widehat \cN$ annihilates a state unless it has $\mZ_k$ charge 1 if $\omega=e^{2\pi i\over k}$ and charge $-1$ if $\omega=e^{-{2\pi i\over k}}$. Note that each term $K_n^k(\tau)K^k_{n-2}(\bar \tau)$ in \eqref{ZktwistedPF} carries $\mZ_k$ charge $n-1$. Moreover since the duality acts as an automorphism of the right-moving $U(1)_{2k}$ chiral algebra, consistency with OPE requires that (non-local) right-moving primary operators $e^{i{n\over \sqrt{2k}}X_R}$ (from the block $K_n^k(\bar\tau)$) are mapped to $e^{-i{n\over \sqrt{2k}}X_R}$ under the duality action $\widehat \cN$. Consequently, the trace of $\widehat \cN$ can only receive contributions from the right moving vacuum block $K_0^k(\bar\tau )$, which transforms according to \eqref{NtwisteK0}. Now   $\cH_r$ does not contain states with $\mZ_k$ charge $-1$ if the right-moving part resides in $K_0^k(\bar\tau )$ which requires $n=2$ or $n=k-2$ in \eqref{ZktwistedPF}. As a result, we find 
\ie
Z_{r\cN}(\tau,\bar\tau)=\begin{cases}
\C e^{-{\pi i \over k}}   \sum_{m \in \mZ_{\geq 0}}K_2^k(\tau)(-1)^m\chi_{m^2}(\bar \tau) & \omega=e^{2\pi i \over k}\,,
 \\
0 & \omega=e^{-{2\pi i \over k}}\,,
\end{cases}
\label{etaN2c}
\fe
with $\C=\pm 1$.

The modular S-dual transform gives the partition function $Z_{\cN r}$ which computes the trace of the TDL $\hat r_-$ acting on the defect Hilbert space $\cH_\cN$. The defect ground state with weight $(h,\bar h)=(0,{1\over 16})$ is non-degenerate \eqref{ZkNm2}. Since $\hat r_-$ is invertible, this implies that $Z_{\cN r}$ cannot vanish identically. Therefore we can conclude that only the first solution in \eqref{etaN2c} is possible and $\omega=e^{2\pi i \over k}$ as desired. 

Next we want to derive the Frobenius-Schur indicator $\ep=1$. Along the way we will see that $\gamma=1$ as required by the bootstrap equation and the spin selection rule. Upon S-transform, we have
\ie
Z_{\cN r_-}(\tau,\bar\tau)=&\C e^{-{\pi i \over k}}\sum_{n=-k+1}^k 	\sum_{ m \in \mZ_{\geq 0}}  e^{2\pi  i n\over k}K_n^k(\tau) \chi^{(m+1/2)^2\over 4}(\bar \tau)\,.
\fe
From the above partition function, we see the defect ground state $|\psi_0\ra$ with spin $s=-{1\over 16}$ carries charge
\ie
\widehat r_- |\psi_0\ra =\C e^{-{\pi i \over k}} |\psi_0\ra\,.
\fe
The doubly-degenerate states $|\psi_{\pm 1}\ra$ of weight $(h,\bar h)=({1\over 4k},{1\over 16})$ have charges
\ie
\widehat r_- |\psi_{\pm 1}\ra =\C e^{-{\pi i \over k}} e^{\pm {2\pi i \over k}} |\psi_{\pm 1}\ra\,.
\fe
The spin selection rule \eqref{ssrZkeven} applied to these three states reads
\ie
\ep=e^{{\pi i  p^2\over k}}\,,\quad e^{\pi i \over k}\ep=e^{{\pi i  (p + 1)^2\over k}}=e^{{\pi i  (p- 1)^2\over k}}\,,
\fe
with $e^{{2\pi i  p\over k}}=\C$. It then immediately follows that $\C=1$ and $\ep=1$. This completes the determination of the TY fusion category symmetries in \eqref{circleTYf}.\footnote{Note that the parameter $\A$ for the duality defect does not change the analysis here for both $k$ even and odd. This is because the duality twisted partition function $Z_{1\cN}$ is independent of $\A$ (see around \eqref{ZkN1}). The analysis for the duality defect $C\cN$ (and its $\A$-family) is also completely parallel which involves a parity-flip of the twisted partition functions for $\cN$ (see Footnote~\ref{footnote:genTY}) and leads to a conjugate bicharacter $\chi(a,b)=-{2\pi i ab\over k}$ but the same FS indicator $\ep=1$.
}

	\subsubsection{Special cases}
	
	As a very simple example, taking $k = 2$ we find that the special radius where $2R  = {2\over R}$ is the Dirac radius $R = 1$. As explained in Section~\ref{sec:TDLfromPF}, the self-duality of this (bosonic) theory under $\bZ_2$ gauging arises from the chiral fermion parity of a single Majorana factor of the Dirac fermion (which can be thought of as two Majorana fermions with the diagonal GSO projection), just as the Kramers-Wannier duality does for the critical Ising model \cite{Thorngren:2018bhj,Ji:2019ugf} (see Table~\ref{table:DiracTDLs}). See also \cite{Karch:2019lnn}.

	Another important case is $k = 4$, as this radius $R =  \sqrt{8}$ describes the Kosterlitz-Thouless  (KT) point where the circle branch meets the $\bZ_2$ orbifold branch. We will see in Section \ref{subsecZ4TYorbifold} that that entire orbifold branch has $\bZ_4$ TY symmetries. This can be related to a chiral parity symmetry of $\bZ_4$ parafermions, which we explain in Section \ref{secZ4parafermion}.
	
\subsection{Symmetric RG flows from the circle branch}\label{sec:RGflowcirc}
The $\mZ_k$ TY symmetries are all anomalous \cite{Thorngren:2019iar}  (the corresponding TY categories do not admit fiber functors \cite{Tambara2000,Thorngren:2019iar}). Consequently any RG flow respecting such symmetries cannot end at a trivially gapped phase in the IR. Below we discuss the implications for RG flows from CFTs on the circle branch preserving such symmetries. 

In the case of the $U(1)_{2k}$ CFT on the circle branch, the theory admits relevant scalar operators commuting with the $\TY(\mZ_k,\chi_\pm,1)$ symmetry (with $\A=0$ and the $+$ sign in \eqref{circleTYf}) for $k=2,3$, given by linear combinations of 
\ie
k=2:\quad V_{2,0}+V_{0,1}\,,\quad V_{-2,0}+V_{0,-1}\,,\quad h=\bar  h={1\over 2}\,,
\\
k=3:\quad V_{3,0}+V_{0,1}\,,\quad V_{-3,0}+V_{0,-1}\,,\quad h=\bar  h={3\over 4}\,,
\label{z2z3relop}
\fe
and  marginal deformations for $k=4$\,
\ie
k=4:\quad V_{4,0}+V_{0,1}\,,\quad V_{-4,0}+V_{0,-1}\,,\quad h=\bar  h=1\,.
\label{z4mop}
\fe
Such deformations preserving the TY symmetry become irrelevant for $k\geq 5$. 

For simplicity let us consider the linear combination of the above deformations that preserve the $\mZ_2^C$ symmetry of the circle CFT as well. The anomalous fusion category symmetry requires the deformed theory to be a nontrivial CFT (or TQFT) in the IR which realizes the same symmetry or the symmetry must be spontaneously broken. Below we will see explicitly these deformed $c=1$ CFTs indeed retain the TY symmetries by flowing to a gapless phase.

For $k=4$, the CFT is at the Kosterlitz-Thouless point, which is the intersection of the circle and orbifold branches of the $c=1$ moduli space. The $\mZ_2^C$ symmetric combination of \eqref{z4mop} leads to precisely the exactly marginal deformation that moves onto the orbifold branch \cite{Dijkgraaf:1987vp}. Indeed we will show in Section~\ref{subsecZ4TYorbifold} that the $\TY(\mZ_4,\chi_\pm,1)$ symmetries persist on the entire orbifold branch. 

The $c=1$ CFT at $k=2$ is the diagonal bosonization of the Dirac fermion. From the bosonization map, it is easy to see that the $\mZ_2^C$ symmetric relevant deformation $V_{2,0}+V_{0,1}+V_{-2,0}+V_{0,-1}$ corresponds to the mass term for one of the two Majorana fermions. Consequently the theory flows to the Ising CFT in the IR, with the $\mZ_2$ TY symmetry acting as Kramers-Wannier duality. 

The RCFT at $k=3$ is described by the $SU(2)^{D_4}_4/U(1)$ coset with a non-diagonal modular invariant of the $D_4$ type \cite{Gepner:1986hr}. This is related to the $SU(2)^{A_5}_4/U(1)$  coset CFT with diagonal modular invariant of the $A_5$ type by gauging the $\mZ_2$ charge conjugation symmetry in \eqref{cosetsym}. Both CFTs come from two distinct bosonizations of the $\mZ_4$ parafermions in \cite{Fateev:1985mm} (see also Appendix~\ref{app:cosetduality} for this and a review of parafermions). As shown in \cite{Fateev:1991bv}, the $SU(2)^{A_{5}}_4/U(1)$ coset CFT admits an integrable RG flow to the tetracritical Ising model triggered by the parafermion bilinear deformation $e^{\pi i \over 4} \psi_1 \tilde \psi_1+e^{-{\pi i \over 4}} \psi_{3} \tilde \psi_{3}$ of weight $h=\bar h={3\over 4}$, which is $\mZ_2$ symmetric.\footnote{The general statement in \cite{Fateev:1991bv} involves an integrable RG flow from the $SU(2)^{A_{k+1}}_k/U(1)$ coset CFT deformed by $e^{\pi i \over k} \psi_1 \tilde \psi_1+(c.c)$ of weight $h=\bar h={k-1\over k}$ to the $(A_k,A_{k+1})$ minimal model for $k\geq 3$.} Moreover, this operator corresponds to the charge conjugation invariant combination of the operators in \eqref{z2z3relop} for $k=3$ (since it's the unique Virasoro primary with these weights that is $\mZ_2$ symmetric), as summarized in the diagram below,
\ie 
\begin{tikzcd}[column sep=huge,row sep=huge]
SU(2)^{A_5}_4/U(1) \arrow[rightarrow]{d}[swap]{\mZ_2\text{-symmetric RG}} \arrow[rightarrow]{r}[sloped,above]{\text{gauging}~\mZ_2} & SU(2)^{D_4}_4/U(1) \arrow[rightarrow]{d}[]{\TY\text{-symmetric RG}} \\
\text{tetracritical Ising} \arrow[rightarrow]{r}[sloped,above]{\text{gauging}~\mZ_2}&  {\text{three-state Potts}}  
\end{tikzcd}
\label{}
\fe
Gauging the $\mZ_2$ symmetry along the entire RG flow, this implies that the $SU(2)^{D_4}_4/U(1)$ coset CFT flows, under the symmetric relevant deformation in \eqref{z2z3relop}, to the three-state Potts model which is the $\mZ_2$ orbifold of the tetracritical Ising model. The three-state Potts is equivalently described by the $SU(2)_3^{A_4}/U(1)$ coset CFT, which evidently hosts the $\TY(\mZ_3,\chi_\pm,1)$ fusion category symmetries following the general discussions in Section~\ref{sec:TDLfromPF}.

	\subsection{Continuum of Topological Defects from $SU(2)_1$}\label{subsecSU2continuum}
 
    At the self-dual radius $R = \sqrt{2}$, the $(U(1)^\theta \times U(1)^\phi) \rtimes \bZ_2^C$ symmetry of the compact boson is enhanced to $SO(4)$. When gauging the $\bZ_n^\phi$ subgroup of $U(1)^\phi$, which takes us to the radius $R = n \sqrt{2}$, those $SO(4)$ rotations which do not commute with $\bZ_n^\phi$ define TDLs of quantum dimension $n$ according to the method of Section \ref{subsecshadownoether}.

    There are four defect currents associated with these TDLs, given by vertex operators with momentum and winding $(p,w)=(\pm n, \pm 1/n)$. Because of the fractional winding, these currents are not local, but sit at the end of $\bZ_n^\theta$ TDLs, which generate the magnetic symmetry of the orbifold. By then gauging a $\bZ_m^\theta$ symmetry, with $m$ coprime to $n$, we obtain a continuum family at radius $R = n \sqrt{2}/m$, with defect currents given by $(p,w)=(\pm n/m, \pm m/n)$.
    
    This family of TDLs was constructed previously in \cite{Fuchs:2007tx}, whose authors argued that they form a complete set of TDLs for the compact boson at these radii. For example, this family contains the $\bZ_k$ TY TDLs discussed in Section \ref{subsecZkTYcompbos} for $k = n^2$. To see this consider the $SO(4)$ element
    \[\begin{bmatrix} 0 && I_{2 \times 2} \\ I_{2 \times 2} && 0 \end{bmatrix}\,,\]
    which describes the action of T-duality in \eqref{eqnTdualitychoice}. At the self-dual point, this acts as a symmetry $V_{p,w} \mapsto (-1)^{pw} V_{w,p}$, with the sign coming from the relation $V_{p,w} ={:}V_{p,0} V_{0,w}{:} = (-1)^{pw}{:} V_{0,w} V_{p,0}{:}$. We will use $V_{p,w}'$ for the moment to denote the vertex operators at $R = n \sqrt{2}$. These operators come from $V_{p/n,nw}$ at the self-dual point when gauging the $\bZ_n^\phi$ symmetry therein. For $p \in n \bZ$ we thus obtain  the action on these operators using
    \[V'_{p,w} \sim V_{p/n,nw} \mapsto n (-1)^{pw} V_{nw,p/n} \sim n (-1)^{pw} V'_{n^2w,p/n^2}.\]
    The factor of $n$ comes from the sum over $\bZ_n^\phi$ orbit in \eqref{eqnsumoflines}. Thus we recover the action on vertex operators in \eqref{TYZk}.

	\section{Fusion Categories of the $\bZ_2$ Orbifold}\label{secZ2genericorbifold}

In this section, we study fusion category symmetries on the orbifold branch of the $c=1$ CFT which persist to all values of the marginal parameter. We will follow a similar strategy as in the previous section. In particular, the $\bZ_2$ orbifold has a global symmetry group $D_8$ and is self-dual under gauging various subgroups. This leads to two $\bZ_4$ TY categories TY$(\bZ_4,\chi_\pm,+)$ (Section \ref{subsecZ4TYorbifold}) and two $D_4=\bZ_2^2$ TY categories, $\Rep(D_8)$ and $\Rep(H_8)$ (Section \ref{subsecZ2Z2gauging}). We then discuss these symmetries at rational points on the orbifold branch where they can be understood using Verlinde lines and ordinary symmetries in (para)fermion theories. Furthermore we argue that these fusion categories are part of a continuum of TDLs which exists at all $R$ and are enhanced at special values $R = n \sqrt{2}$.

     	\subsection{Operator Content of the $\bZ_2$ Orbifold}
     	
     	The $c=1$ orbifold at radius $R$ (``the $\bZ_2$ orbifold") is obtained by gauging the $\mZ_2^C$ charge conjugation symmetry of the compact boson of the same radius. In terms of the (normalized) non-chiral compact boson $\theta\equiv {X_L+X_R\over R}$ and its T-dual $\phi\equiv{R(X_L -X_R)\over 2}$ (normalized such that both $\theta$ and $\phi$ have unit radii)
     	\ie
     	\mZ_2^C: (\theta, \phi)  \to  (-\theta, -\phi)\,.
     	\fe
     	The spectrum of Virasoro primaries on the $c=1$ orbifold branch consists of two sectors, the untwisted sector and the $\mZ_2$ twisted sector. The latter is charged under the magnetic symmetry $\mZ_2^{M}$.
     	
     	The twisted sector consists of two primary operators 
     	\ie\label{eqnsigmaops}
     	\sigma_1{\rm ~and~}\sigma_2,\quad (h,\bar h)=\left( {1\over 16}, {1\over 16}\right)
     	\fe
     	that correspond to the ground states at the two fixed points of $S^1/\mZ_2^C$, as well the first excited states
     	\ie
     	\tau_1{\rm ~and~}\tau_2,\quad (h,\bar h)=\left( {9\over 16}, {9\over 16}\right).
     	\fe
     	The untwisted sector includes reflection-invariant momentum-winding operators  $V^+_{n,w}$  defined by 
     	\ie
     	V^+_{n,w}={V_{n,w}+V_{-n,-w}\over \sqrt{2}}\,,
     	\fe
     	where $V_{n,w}$ denotes the usual momentum-winding operators in the unorbifolded theory \eqref{eqncompbossvertop}. 
     	     	The rest of the  Virasoro primaries in the untwisted sector are built from $\mZ_2^C$ invariant normal-ordered Schur polynomials in  the $U(1)$ currents $d\theta, d\phi$ and their derivatives in the unorbifolded theory \cite{Ginsparg:1987eb},
     	\ie
     	j_{n^2}j_{m^2}~{\rm with}~{m-n\in 2\mZ}\,,~(h,\bar h )=(n^2,m^2)\,.
     	\fe
     	For example,
     	\ie
     	j_{1}=\pa X_L\,, \quad j_4={:}j_1^4{:}-2{:}j_1\pa^2 j_1{:}+{3\over 2} {:}(\pa j_1)^2{:}\,.
     	\fe
     	The exactly marginal operator that gives rise to the orbifold branch is $j_1\bar j_1$. The spin-one $U(1)$ currents themselves are projected out in the orbifold, but there is a spin $4n^2$ left-moving (and right-moving) current for every $n \in \mZ_+$ at general $R$ and they generate a W-algebra of type $W(2,4)$ that extends the Virasoro algebra. At special radii, this chiral algebra can be further enhanced to an even larger W-algebra such that the CFT becomes rational, with a finite number of primaries and conformal blocks for this chiral algebra.

     	The orbifold branch has global symmetry 
     	\ie
     	G_{\rm orb}=D_8\,,
     	\fe
     	at generic $R$, which does not harbor 't Hooft anomalies. However we will see the orbifold branch hosts a rich zoo of non-invertible TDLs that include  TY categories associated to $\mZ_4$ and $D_4=\mZ_2\times \mZ_2$ subgroups of $D_8$.  We adopt the following convention for $D_8$
     	\ie\label{eqnD8conv}
     	D_8=\la s,r | s^2=r^4=(rs)^2=1\ra \,.
     	\fe
     	In the 2$\times$2 Pauli-matix representation of $D_8$, we have $s = \sigma^x$, $sr = \sigma^z$, $r = \sigma^x \sigma^z$, $r^2 = -1$, etc.
     	
     	The appearance of $D_8$ symmetry can be understood from the $\mZ_2^C$ discrete gauging as follows. Recall the unorbifolded boson at generic $R$ has global symmetry  
     	\ie
     	G=(U(1)^\theta\times U(1)^\phi)\rtimes \mZ_2^C\,,
     	\fe 
     	where $\mZ_2^C$ negates the parameters of both $U(1)$'s. The commutant of $\mZ_2^C$ in $G/\mZ_2^C$ is $\mZ_2^\theta\times \mZ_2^\phi$ generated by $\pi$-translations in $\theta$ or $\phi$. These symmetries continue to act as global symmetries in the orbifolded theory in both the twisted and untwisted sectors. In addition, we have the magnetic symmetry $\mZ_2^{M}$ which only acts non-trivially in the twisted sector, where it acts as a scalar $-1$. Due to a mixed anomaly between $\bZ_2^C$ and its commutant, we arrive at the group extension
     	\ie
     	1 \to  \mZ_2^{M} \to  D_8 \to \mZ_2^\theta\times \mZ_2^\phi \to 1\,.
     	\fe
     	See Appendix \ref{apporbifoldsymmetry} for more details. In particular, we find $\mZ_2^{M}$ is identified with the center of $D_8$, generated by $r^2$, and $D_8$ acts on the untwisted sector through its quotient $\mZ_2^\theta\times \mZ_2^\phi$,
     	\ie\label{eqnD8raction}
     	r: (\theta, \phi)\to (\theta+\pi, \phi+\pi)\,,~s:(\theta, \phi)\to (\theta, \phi+\pi)\,.
     	\fe
     	For the operators in the twisted sector, $D_8$ acts as 
     	\ie
     	&r: (\sigma_{1},\sigma_2) \to (i\sigma_{1},-i\sigma_2)\,,~(\tau_{1},\tau_2) \to (i\tau_{1},-i\tau_2)\,,
     	\\
     	&s: (\sigma_{1},\sigma_2) \to (\sigma_{2},\sigma_1)\,,~(\tau_{1},\tau_2) \to (\tau_{2},\tau_1)\,.
     	\fe
     	In Appendix \ref{apporbifoldsymmetry}, we show that this $D_8$ symmetry is anomaly-free, despite its chiral-looking action \eqref{eqnD8raction}.
     	
     	The $D_8$ symmetry has the following important subgroups
     	\[\langle r \rangle = \bZ_4^r\,, \quad \langle sr^3 \rangle = \bZ_2^\eta\,,\quad \langle sr \rangle = \bZ_2^\sigma\,,\]
     	where the superscript indicates the name of the TDL generating the symmetry, and
     	\ie
     	\langle r^2,s \rangle = D_4^A\,, \quad \langle r^2,sr \rangle = D_4^B\,,
     	\label{2D4s}
     	\fe
     	where $D_4 = \bZ_2 \times \bZ_2$.

     	Along the orbifold branch, the RCFTs locate at radius $R=\sqrt{2k}$ for $k\in \mZ$ \cite{Dijkgraaf:1989hb}. The enhanced chiral algebra  is generated by
     	\ie
     	T\,,~j_4\,,~\cos \sqrt{2k} X_L\,,
     	\fe
     	of spin $2,4,k$ respectively. There are $k+7$ chiral primaries listed in Table~\ref{table:orbifoldcharacters}.
     	\begin{table}[htb]
     		\begin{center}
     			\renewcommand{\arraystretch}{2}
     			\begin{tabular}{ |c|c|c|c|c|c|c| }
     				\hline
     				Primary & 1 & $j$ & $\phi^i_k$ & $\phi_m$ & $\sigma_i$ & $\tau_i$
     				\\\hline
     				$h=\bar h$ &  0 & 1 & ${k\over 4}$ & ${m^2\over 4k}$ & ${1\over 16}$ & ${9\over 16}$
     				\\\hline
     				Characters &  ${1\over 2}(K_0^k-W)$ & ${1\over 2}(K_0^k+W)$ & ${1\over 2} K^k_k $ & ${K_m^k}$ & ${1\over 2}(W_++W_-)$ & ${1\over 2}(W_+-W_-)$
     				\\\hline
     			\end{tabular}
     		\end{center}
     		\caption{The $k+7$ chiral primaries in the orbifold CFT $U(1)_{2k}/\mZ_2$. Here $i=1,2$ and $m=1,2,\dots, k-1$.}
     		\label{table:orbifoldcharacters}
     	\end{table}
     	\newline
     	Here in writing down the characters in Table~\ref{table:orbifoldcharacters} we have introduced \cite{Yang:1987wk}
     	\ie
     	W(\tau)\equiv {1\over \eta} \sum_{r\in \mZ}(-1)^r q^{r^2}\,,\quad W(\tau)_\pm \equiv {1\over \eta} \sum_{r\in \mZ}(\pm 1)^r q^{\left(r+{1\over 4}\right)^2}\,.
     	\fe
     	The partition function of the theory is given by the diagonal modular invariant
     	\ie
     	Z_{\rm orb}(R=\sqrt{2k})={1\over 2}Z_{\rm bos}(R=\sqrt{2k})+{1\over 2}|W|^2+|W_+|^2+|W_-|^2\,.
     	\fe
     	We will refer to such theories as the $U(1)_{2k}/\mZ_2$ CFTs.

	\subsection{$\bZ_4$ Tambara-Yamagami Symmetry}\label{subsecZ4TYorbifold}
	
	Let us denote the orbifold partition function at radius $R$ coupled to a $\bZ_4$ gauge field $A$ on a closed spacetime $\Sigma$ as $Z_{\rm orb}(R,\Sigma,A)$, where $\bZ_4$ is generated by the element $r \in D_8$ (cf. \eqref{eqnD8conv}). This theory is self-dual under gauging $\bZ_4$ at any radius, meaning\footnote{Recall there is a similar identity for the partition function on the circle branch \eqref{PFsd}. As we will see this is not a coincidence. The $\mZ_4$ TY symmetry at the KT point on the circle branch  can be continued to the orbifold branch since it preserves the corresponding marginal operator.}
	\[\label{orbZ4selfduality}
	Z_{\rm orb}(R,\Sigma,A) = \frac{1}{\sqrt{|H^1(\Sigma,\bZ_4)|}} \sum_{B \in H^1(\Sigma,\bZ_4)} e^{\frac{ \pm i\pi}{2} \int A \cup B} Z_{\rm orb}(R,\Sigma,B).
	\]
	For an explicit check with $\Sigma=T^2$ see Appendix \ref{appD8gauging}.
	
	As in Section \ref{subsecZkTYcompbos}, we expect that there is an associated $\bZ_4$ Tambara-Yamagami fusion category which acts on the theory. In general, there are four $\mZ_4$ TY categories labeled by the FS indicator $\epsilon=\pm$ of the duality defect and a choice for the bicharacter
\ie
\chi_\pm(r^a,r^b)\equiv (\pm i)^{ab}\,,
\fe
both of which are consistent with \eqref{orbZ4selfduality}.\footnote{None of the four TY categories admit fiber functors, and are thus anomalous according to \cite{Thorngren:2019iar}. Thus any CFT with such symmetries (including the orbifold CFTs here) cannot flow to trivially gapped phases under symmetric RG flows.}
The $c=1$ orbifold branch realizes the two $\mZ_4$ TY categories with $\epsilon=1$ at general $R$ and the two duality TDLs $\cN_{\mZ_4^\pm}$ are related by stacking with the $\mZ_2$ symmetry defect associated to $s\in D_8$:
\ie
\TY(\mZ_4, \chi_{+},+)
\xleftrightarrow{~~\cN_{\mZ_4^\pm}\ \mapsto\ s \cN_{\mZ_4^\pm}~~}
\TY(\mZ_4, \chi_-,+)\,.
\fe
We propose the following action on local primary operators
\ie
\widehat\cN_{\mZ_4^+}: &\begin{cases}
	V^+_{n,w}\to 2 i^{n-w} V_{n,w}^+ \quad {~\rm for~} n-w \in 2\mZ\,,
	\\
	j_{n^2} \bar j_{m^2} \to 2 j_{n^2} \bar j_{m^2}\,,
	\\
	\text{all other primaries} \to \text{non-local operators}\,,
\end{cases}
\\
\widehat\cN_{\mZ_4^-}: &\begin{cases}
	V^+_{n,w}\to 2 i^{n+w} V_{n,w}^+ \quad {~\rm for~} n-w\in 2\mZ\,,
	\\
	j_{n^2} \bar j_{m^2} \to 2 j_{n^2} \bar j_{m^2}\,,
	\\
	\text{all other primaries} \to \text{non-local operators}\,.
\end{cases}
\label{Z4TYonop}
\fe

These actions have the following desired features. First, the duality defects annihilate  operators that are charged under $\mZ_4$ and act on the $\mZ_4$ invariant operators with eigenvalue $\pm 2$ as required by the TY fusion rule $\cN^2=1+r+r^2 +r^3$. They also solve the modular bootstrap equation \eqref{mbeqn} for torus with a single duality twist around the space or time cycle.

More explicitly, 
the corresponding duality defects $\cN_{\mZ_4^+}$ and $\cN_{\mZ_4^-}=s\cN_{\mZ_4^+} $ define the following twisted partition functions at general $R\geq \sqrt{2}$,
\ie
Z_{1\cN_{\mZ_4^+}}(\tau,\bar\tau)=&
{1\over  |\eta|^2} \left(
\sum_{m,n\in \mZ,m-n\in 2\mZ}  
i^{n-m}q^{{1\over 2}\left({n\over R}+{mR\over 2}\right)^2}\bar q^{{1\over 2}\left({n\over R}-{mR\over 2}\right)^2}
+
\sum_{m,n\in \mZ}  (-1)^{m+n}q^{ m^2}\bar q^{ n^2}
\right)\,,
\\
Z_{1\cN_{\mZ_4^-}}(\tau,\bar\tau)=&
{1\over  |\eta|^2} \left(
\sum_{m,n\in \mZ,m-n\in 2\mZ}  
i^{n+m}q^{{1\over 2}\left({n\over R}+{mR\over 2}\right)^2}\bar q^{{1\over 2}\left({n\over R}-{mR\over 2}\right)^2}
+
\sum_{m,n\in \mZ}  (-1)^{m+n}q^{ m^2}\bar q^{ n^2}
\right)\,.
\label{Z4duality}
\fe
Their modular $S$-transform (using Poisson resummation) yields the spectrum in the duality-twisted sectors,
\ie
Z_{\cN_{\mZ_4^+}1}(\tau,\bar\tau)=&
{1\over  2|\eta|^2} \left(
\sum_{m,n\in \mZ,m-n\in 2\mZ}    
q^{{1\over 8}\left({n-1/2 \over R}+{(m+1/2) R\over 2}\right)^2}\bar q^{{1\over 8}\left({n-1/2\over R}-{(m+ 1/2) R\over 2}\right)^2}
+
\sum_{m,n\in \mZ}
q^{{1\over 4}(m+1/2)^2}\bar q^{{1\over4 }(n+1/2)^2}
\right)\,,
\\
Z_{\cN_{\mZ_4^-}1}(\tau,\bar\tau)=&
{1\over  2|\eta|^2} \left(
\sum_{m,n\in \mZ,m-n\in 2\mZ}    
q^{{1\over 8}\left({n+1/2 \over R}+{(m+1/2) R\over 2}\right)^2}\bar q^{{1\over 8}\left({n+1/2\over R}-{(m+ 1/2) R\over 2}\right)^2}
+
\sum_{m,n\in \mZ}
q^{{1\over 4}(m+1/2)^2}\bar q^{{1\over4 }(n+1/2)^2}
\right)\,,
\label{Z4dualityHS}
\fe
which decompose into Virasoro characters with positive integer degeneracies, as required, thanks to the shifts in the exponents of $q$ and $\bar q$.

     We can also study the spin selection rules by specializing \eqref{ssrZkeven} to $k=4$. If we define the two inequivalent bicharacters as
     \ie
     \chi_\pm (a,b)=(\pm i)^{ab}\,,
     \fe
     then
     	\ie
     	\chi_+,\epsilon=1:~&e^{4\pi i s}=\{1, \pm e^{-{\pi i \over 4}}\}\,,
     	\\
     	\chi_+,\epsilon=-1:~&e^{4\pi i s}=\{-1,  \pm e^{-{\pi i \over 4}}\}\,,
     	\\
     	\chi_-,\epsilon=1:~&e^{4\pi i s}=\{1,  \pm e^{{\pi i \over 4}}\}\,,
     	\\
     	\chi_-,\epsilon=-1:~&e^{4\pi i s}=\{-  1,\pm e^{{\pi i \over 4}}\}\,.
     	\label{ssruleZ4}
     	\fe
     	Since the set of allowed spins $s$ for the four $\mZ_4$ TY categories are non-identical, it provides a quick way to diagnose which $\mZ_4$ duality defect $\cN$ corresponds to by looking at the spins in $\cH_\cN$ (equivalently the partition function $Z_{\cN 1}$). This allows us to determine the Frobenius-Schur indicator to be $\epsilon = 1$ for both $\bZ_4$ duality lines.
     	
      In Section \ref{secZ4parafermion}, we will return to these duality defects at the $\bZ_4$ parafermion point, where they come from  chiral parity symmetries in the parafermion theory (see Section~\ref{sec:TDLfromPF}), and explicitly verify the action on local operators given above. At the KT point, they become the $\bZ_k$ self-duality studied in Section \ref{subsecZkTYcompbos} for $k = 4$.

	\subsection{$D_4$ Tambara-Yamagami Symmetry}\label{subsecZ2Z2gauging}
	
	Let us denote the orbifold partition function at radius $R$ coupled to $\bZ_2 \times \bZ_2 = D_4$ gauge fields $A_g$, $A_h$ on a closed spacetime $\Sigma$ for $g, h \in D_8$ commuting elements of order 2 as
	\[Z_{\rm orb}(\Sigma,R,A_g,A_h)\,.\]
	There are two choices of $g$ and $h$ up to conjugacy in $D_8$ and automorphism of $D_4$, namely $g,h = s, sr^2$ or $g,h = sr, sr^3$. The corresponding groups are denoted by $D_4^A$ and the  $D_4^B$ respectively in \eqref{2D4s}.
	
	This partition function is not uniquely defined by the action of $D_4$ on local operators since there is a possible discrete torsion class in $H^2(D_4,U(1)) = \bZ_2$ which can modify the action of $g$ on the $h$-twisted sector by a sign, and vice versa, meaning that the two partition functions on the torus $\Sigma = T^2$ in the sector $A_g = (1,g)$ ($g$-twist in time), $A_h = (h,1)$ ($h$-twist in space) differ by an overall sign, corresponding to tensoring the theory with an SPT.
	
	To fix the ambiguity, we look at the $g$-charge of the lightest operator in the $h$-twisted sector and choose the overall sign of $Z_{\rm orb}$ by requiring this operator to be neutral, so that in the low-temperature expansion, restricting to $q=\bar q$, the leading term is positive,
	\[Z_{\rm orb}(T^2,R,(1,g),(h,1)) = q^\alpha + \text{subleading terms} \qquad R>\sqrt{2}\,.\]
	This makes sense for the $\bZ_2$ orbifold because as long as $R$ is greater than its value at the KT point $R = \sqrt{2}$, there is a unique lightest operator in the $h$-twisted sector for any $\bZ_2$ symmetry generator $h$. For instance, with $h = r^2$, this operator is $e^{i \theta} - e^{- i \theta}$, of weight $h=\bar h={1\over 2R^2}$. At $R = \sqrt{2}$ the above partition function vanishes because of the chiral anomaly.
	
	It makes sense to define $Z_{\rm orb}(T^2,R,A_g,A_h)$ also for $R < \sqrt{2}$, since T-duality does not respect the $D_4$ symmetries of the orbifold, as it acts by an outer automorphism of $D_8$ taking $s \leftrightarrow sr$. As we continuously tune $R$ through the KT point $R = \sqrt{2}$ we find there is a level crossing in the $h$-twisted sector where two operators of opposite $g$-charge exchange places as the lightest operator in the $h$-twisted sector (this level crossing happens across the whole spectrum, giving the vanishing of the partition function we mentioned above). Thus, we have (again with $q=\bar q\ll 1$)
	\[Z_{\rm orb}(T^2,R,(1,g),(h,1)) = -q^\alpha + \text{subleading terms}  \qquad R < \sqrt{2}\,.\]
	This type of phenomenon was dubbed a ``topological transition" in \cite{Ji:2019ugf}.
	
	This level crossing is related to an anomalous accidental symmetry at the KT point related to T-duality which exchanges $D_4^A$ and $D_4^B$ (see Appendix~\ref{appKTsymmetry}). This complicates the usual discussion of the duality. The correct statement is that if we take our theory so-defined enriched with its $D_4^B$ symmetry and tune the marginal parameter $R$ through the KT point to the dual value $2\over R$, then we find a theory which may be equivalently described as a $D_4^{A\star}$-enriched orbifold at radius $R$, where the star indicates discrete torsion.
	
	We find that with the definitions above, the theory is self-dual under gauging $D_4^B$ for all $R$,
	\[Z_{\rm orb}(\Sigma,R,A_{sr},A_{sr^3}) = \frac{1}{4}\sum_{B_{sr}, B_{sr^3} \in H^1(\Sigma,\bZ_2)} (-1)^{\int_\Sigma A_{sr} B_{sr} + A_{sr^3} B_{sr^3}} Z_{\rm orb}(\Sigma,R,B_{sr},B_{sr^3})\,.\]
	On the other hand, if we define
	\[Z_{\rm orb}^\star(\Sigma,R,A_g,A_h) = (-1)^{\int_\Sigma A_g \cup A_h} Z_{\rm orb}(\Sigma,R,A_g,A_h)\]
	to be the theory obtained by taking the opposite convention as what we have considered above, equivalent to tensoring $Z_{\rm orb}$ with the $\bZ_2 \times \bZ_2$ SPT, we find instead it is self-dual under gauging $D_4^A$,
	\[Z_{\rm orb}^\star(\Sigma,R,A_{s},A_{sr^2}) = \frac{1}{4}\sum_{B_{s}, B_{sr^2} \in H^1(\Sigma,\bZ_2)} (-1)^{\int_\Sigma A_{s} B_{s} + A_{sr^2} B_{sr^2}} Z_{\rm orb}^\star (\Sigma,R,B_{s},B_{sr^2})\,.\]
	These two equations are related by T-duality and everything is consistent.
	
	Therefore we expect these self-dualities to give rise to T-dual pairs of $D_4^A$ and $D_4^B$ TY symmetries.\footnote{The two $D_4$ TY symmetries differ in their F-gauge. See discussions in Section~\ref{subsecoverviewselfdual} and around \eqref{Nloopfuse}.} There are four such TY categories labeled by the FS indicator $\epsilon=\pm$ of the duality defect and two bicharacters, $\chi_s$ coming from the product of the unique bicharacters of each $\bZ_2$ factor and $\chi_a$ related to $\chi_s$ by swapping the two $\bZ_2$ factors,
	\ie
\chi_s(i_1,i_2;j_1,j_2)=(-1)^{ i_1  j_1+i_2 j_2}
\,,\quad 
\chi_a(i_1,i_2;j_1,j_2)=(-1)^{i_1 j_2+i_2 j_1 }\,,
\label{D4bichar}
\fe
 where $i_{1,2},j_{1,2}\in \mZ_2$. 
	
	The self-dualities above evidently correspond to the bicharacter $\chi_s$. We will see below that the FS indicator is $+$, so this gives rise to the TY category $\TY(\mZ_2\times\mZ_2, \chi_s,+) = {\rm Rep}(H_8)$. We denote this duality line $\cN_{H_8,B}$ and its T-dual $\cN_{H_8,A}$. Let us focus on $\cN_{H_8}=\cN_{H_8,B}  $. The element $s \in D_8$ acts as the swapping automorphism of $D_4^B$, so fusing $\cN_{H_8}$ with the line corresponding to $s$ yields  $\TY(\mZ_2\times\mZ_2, \chi_a,+) = {\rm Rep}(D_8)$,

\ie
\TY(D_4^B, \chi_s,+)
\xleftrightarrow{~~\cN_{H_8}\ \mapsto\  s \cN_{H_8} = \cN_{D_8}~~}
\TY(D_4^B, \chi_a,+)\,.
\fe

We propose the following action on local operators
\ie\label{eqnD8action}
\widehat\cN_{D_8}: \begin{cases}
	V^+_{n,w}\to 2 i^n V_{n,w}^+ \quad {~\rm for~} n \in 2\mZ\,,
	\\
	j_{n^2} \bar j_{m^2} \to 2 j_{n^2} \bar j_{m^2}\,,
	\\
	\text{other primaries} \to \text{non-local operators}\,,
\end{cases}\quad
\fe
\ie\label{eqnH8action}
\widehat\cN_{H_8}: \begin{cases}
	V^+_{n,w}\to 2 i^n (-1)^w V_{n,w}^+ \quad {~\rm for~} n \in 2\mZ\,,
	\\
	j_{n^2} \bar j_{m^2} \to 2 j_{n^2} \bar j_{m^2}\,,
	\\
	\text{other primaries} \to \text{non-local operators}\,.
\end{cases}
\fe
These satisfy the desired TY fusion rules. The T-dual lines for the $D_4^A$ self-duality act as above by switching $n$ and $w$.

The twisted torus partition functions associated with this action are
\ie
Z_{1\cN_{D_8}}(\tau,\bar\tau)=&{1\over  |\eta|^2} \left(
\sum_{m,n\in \mZ }  
(-1)^{n}q^{{1\over 2}\left({2n\over R}+{mR\over 2 }\right)^2}\bar q^{{1\over 2}\left({2n\over R}-{mR \over 2 }\right)^2}
+
\sum_{m,n\in \mZ}  (-1)^{m+n}q^{ m^2}\bar q^{ n^2}
\right)\,,
\\
Z_{1\cN_{H_8}}(\tau,\bar\tau)=&{1\over  |\eta|^2} \left(
\sum_{m,n\in \mZ }  
(-1)^{m+n}q^{{1\over 2}\left({2n\over R}+{mR \over 2}\right)^2}\bar q^{{1\over 2}\left({2n\over R}-{mR\over 2 }\right)^2}
+
\sum_{m,n\in \mZ}  (-1)^{m+n}q^{ m^2}\bar q^{ n^2}
\right)\,.
\label{D4duality}
\fe
These solve the modular bootstrap equation \eqref{mbeqn} for the torus partition functions with one duality line inserted. 
In particular, the modular $S$-transform (from Poisson resummation formula) yields the duality-twisted sector spectra
\ie
Z_{\cN_{D_8}1}(\tau,\bar\tau)=&{1\over  2|\eta|^2} \left(
\sum_{m,n\in \mZ }    
q^{{1\over 2}\left({n\over R}+{(m+1/2) R\over 4 }\right)^2}\bar q^{{1\over 2}\left({n\over R}-{(m+ 1/2) R \over 4}\right)^2}
+
\sum_{m,n\in \mZ}
q^{{1\over 4}(m+1/2)^2}\bar q^{{1\over4 }(n+1/2)^2}
\right)\,,
\\
Z_{\cN_{H_8}1}(\tau,\bar\tau)=&{1\over  2|\eta|^2} \left(
\sum_{m,n\in \mZ }    
q^{{1\over 2}\left({n+1/2 \over R}+{(m+1/2) R \over 4 }\right)^2}\bar q^{{1\over 2}\left({n+1/2\over R}-{(m+ 1/2) R\over 4 }\right)^2}
+
\sum_{m,n\in \mZ}
q^{{1\over 4}(m+1/2)^2}\bar q^{{1\over4 }(n+1/2)^2}
\right),
\label{D4dualityHS}
\fe
which thanks to the shifts in the exponents of $q$ and $\bar q$ decomposes into Virasoro characters with positive integer degeneracies, as required.

The TY data also shows up in the spin selection rules (similar to those derived in Section~\ref{subsecspinselectionrulesZkTY}). For a general $D_4$ TY defect $\cN$ satisfying $\cN^2 = 1 + \eta + \sigma + \eta \sigma$ we have
     	\ie
     	&\chi_s:~e^{4\pi i s} \la \psi' |\psi\ra =&{\epsilon\over 2}\la \psi' |(1+\widehat \eta_- +\widehat \sigma_- + \widehat \eta_-   \widehat \sigma_-|\psi\ra \,,
     	\\
     	&\chi_a:~e^{4\pi i s} \la \psi' |\psi\ra =&{\epsilon\over 2}\la \psi' |(1+\widehat \eta_- +\widehat \sigma_- -\widehat \eta_-   \widehat \sigma_-|\psi\ra \,,
     	\fe
     	where we have used that $\eta_-,\sigma_-$ commute (regardless of the choice of bicharacter). Moreover from crossing relations we have
     	\ie
     	&\chi_s:~\widehat \eta_-^2=\widehat \sigma_-^2=-1 \,,
     	\\
     	&\chi_a:~\widehat \eta_-^2=\widehat \sigma_-^2= 1 \,.
     	\fe 
 	Consequently, the spin selection rules are
     	\ie
     	\chi_s,\epsilon=1:~&e^{4\pi i s}=\{1,i,-i\}\,,
     	\\
     	\chi_s,\epsilon=-1:~&e^{4\pi i s}=\{-1,i,-i\}\,,
     	\\
     	\chi_a,\epsilon=1:~&e^{4\pi i s}=\{1,-1\}\,,
     	\\
     	\chi_a,\epsilon=-1:~&e^{4\pi i s}=\{1,-1\}\,.
     	\label{ssruleD4}
     	\fe
     	This provides a quick way to determine whether the crossing data is associated to $\chi_s$ or $\chi_a$. The spin selection rules differentiate the two categories $\TY(D_4,\chi_s,\pm)$ but a more refined argument is needed to determine the FS indicator in $\TY(D_4,\chi_a,\pm)$.
     	
     	We will verify the action of local operators in Section \ref{secRepH8Ising} by studying Verlinde lines at the Ising$^2$ point and in Section \ref{secRepD84statepotts} at the four-state Potts point. This will also determine the Frobenius-Schur indicators.

     	\subsection{$\mZ_4$ Tambara-Yamagami and $\mZ_4$ Parafermions}\label{secZ4parafermion}
     
     	Let us consider the $\bZ_2$ orbifold at the special radius $R=\sqrt{6}$. This theory is rational and its RCFT structure is manifest from its $SU(2)_4/U(1)$ coset description of the $A_5$ type (diagonal modular invariant),\footnote{In the following all $SU(2)_k/U(1)$ coset CFTs are defined with respect to the diagonal modular invariant unless otherwise noted.}
     	\ie
     	{U(1)_6\over \mZ_2^C}\cong {SU(2)_4^{A_5}\over U(1)}\,.
     	\fe
     	The same CFT is also realized by bosonization from the $\mZ_4$ parafermions \cite{Fateev:1985mm} (see also Appendix~\ref{app:cosetduality}). 
    We will see that the $\bZ_4$ TY TDLs outlined in Section \ref{subsecZ4TYorbifold} consist of Verlinde lines (see Section~\ref{sec:verlinde}) at this point together with the duality TDLs coming from chiral symmetries for the $\mZ_4$ parafermions (see Section~\ref{sec:TDLfromPF}).
This also provides a check on the proposed actions \eqref{Z4TYonop} of dualities on local operators on the orbifold branch.
     		
The relevant chiral algebra is given by the W-algebra $W(2,3,4)$  which includes spin 3 and 4 generators,
     	\ie
     	\cos ({\sqrt{6} X_L})=\cos (3\theta+\phi)\quad {\rm and} \quad j_4\,,
     	\fe
     	in addition to the stress tensor.
     	With respect to $W(2,3,4)$, the $SU(2)_k/U(1)$ coset CFT has 10 primaries $\phi^j_m$ as  listed in Table~\ref{table:z4ops} with $j=0,{1\over 2},1,{3\over 2}, 2$ and $m+j\in\mZ$ and subject to the identification
     	\ie
     	\phi^j_m=\phi^j_{m+4}=\phi^{2-j}_{2+m}\,.
     	\fe
     	Correspondingly there are 10 TDLs in the $SU(2)_4/U(1)$  CFT realized by Verlinde lines $\cL_{(j,m)}$. In particular, from the fusion rules, we find that the $\mZ_4$ subgroup of $D_8$ is realized by the Verlinde lines as $r^i=\cL_{(0,i)}$ for $i=0,1,2,3$.  
     	
     	\begin{table}[htb]
     		\renewcommand{\arraystretch}{2.2}
     		\begin{center}
     			\begin{tabular}{ |c|c|c|c| }
     				\hline
     				Coset $\phi^j_m$ & $U(1)_6/\mZ_2^C$   &  $h=\bar h$   &  $\mZ_4$ charge \\\hline
     				$\phi^0_0$ & 1 & 0 & 0\\\hline
     				$\phi^0_1$ & $\cos{\sqrt{6} (X_L+X_R)\over 2}=\cos (3\theta)$ & ${3\over 4}$ & 2 \\\hline
     				$\phi^0_2$ & $\pa X_L\bar \pa X_R=d\phi d\theta $ & ${1}$ & 0 \\\hline
     				$\phi^0_3$ & $\cos{\sqrt{6} (X_L-X_R)\over 2}=\cos (\phi)$ & ${3\over 4}$ & 2\\\hline
     				$\phi^{1\over 2}_{-{1\over 2}}$ & $\sigma_1$ & ${1\over 16}$ & $3$ \\\hline
     				$\phi^{1\over 2}_{1\over 2}$ &  $\sigma_2$ & ${1\over 16}$ & $1$ \\\hline
     				$\phi^{1\over 2}_{3\over 2}$ &  $\tau_1$ & ${9\over 16}$ & $3$ \\\hline
     				$\phi^{1\over 2}_{5\over 2}$ &  $\tau_2$  & ${9\over 16}$ & $1$ \\\hline
     				$\phi^1_{-1}$ & $\cos{X_L+X_R\over \sqrt{6}}=\cos (\theta)$ &   ${1\over 12}$ & 2 \\\hline
     				$\phi^1_0$ &  $\cos{2(X_L+X_R)\over \sqrt{6}}=\cos (2\theta)$ &   ${1\over 3}$ & 0\\\hline
     			\end{tabular}
     		\end{center}
     		\caption{$SU(2)_4/U(1)$ coset primaries $\phi^j_{m}$ and their descriptions in the $U(1)_6/\mZ_2^C$ orbifold.}
     		\label{table:z4ops}
     	\end{table}
     	
     	As we have described in Section \ref{sec:TDLfromPF}, the $SU(2)_k/U(1)$ coset CFT is self-dual under $\mZ_k$ gauging, and there are two $\mZ_k$ TY symmetries corresponding to the fusion categories $\TY(\mZ_k,\chi_\pm,1)$ with $\chi_\pm(a,b)\equiv \pm {2\pi a b\over k}$.
     	Here we explicitly verify this for the corresponding duality defect $\cN$ in the $\mZ_4$ case (the analysis for the other duality defect $s\cN$ is similar).  Unlike the $\mZ_4$ symmetry TDLs, the duality defect $\cN$ does not commute with the full chiral algebra. Instead, it preserves the left-moving $W(2,3,4)$ while acting on the right-moving generators by a $\mZ_2$ automorphism that flips the sign of the spin 3 generator. Furthermore, the $\mZ_4$ invariant $W(2,3,4)$ primaries, namely the thermal operators $\phi_0^0,\phi^1_0,\phi^2_0$, transform under the duality as\footnote{The overall factor of 2 is required by the fusion relation $
     	\cN^2=I+r+r^2+r^3
     	$.}
     	\ie
     	\cN\cdot \phi^j_0 =2(-1)^j  \phi^j_0\,.
     	\fe
     	The above two conditions suffice to completely determine how $\cN$ acts on local operators (see Appendix~\ref{app:cosetduality} for details).

     	The $SU(2)_4/U(1)$ coset CFT has the diagonal modular invariant partition function
     	\ie
     	Z(\tau,\bar\tau)=|\eta|^2(|c^0_0|^2+2|c^0_2|^2+|c^0_4|^2+2|c^1_1|^2+2|c^1_3|^2+|c^2_2|^2+|c^2_0|^2)\,.
     	\fe
     	Here $\eta(\tau)c^{2j}_{2m}(\tau)$ are $W(2,3,4)$ characters (see Appendix~\ref{app:coset}).  For the thermal operators $\phi_0^j$ with $j=0,1,2$, they are given by      	\ie
     	\eta c^0_0(\tau)=&{1\over 2\eta(\tau)}\sum_{k\in \mZ} ((-1)^kq^{k^2}+q^{3k^2})\,,
     	\\
     	\eta c^2_0(\tau)=&{1\over \eta(\tau)}\sum_{k\in \mZ} q^{3(k+1/3)^2}\,,
     	\\
     	\eta c^4_0(\tau)=&{1\over 2\eta(\tau)}\sum_{k\in \mZ} (-(-1)^kq^{k^2}+q^{3k^2}) \,.
     	\fe
     As reviewed above (also Appendix~\ref{app:cosetduality}), to write down the duality twisted partition function, we just need the duality twisted characters for the thermal operators,
     	        	\ie
     	I_1(\tau)=&{1\over 2\eta(\tau)}\sum_{k\in \mZ} (q^{k^2}+q^{3k^2})(-1)^k \,,
     	\\
     	I_2(\tau)=&{1\over \eta(\tau)}\sum_{k\in \mZ} q^{3(k+1/3)^2} (-1)^k \,,
     	\\
     	I_3(\tau)=&{1\over 2\eta(\tau)}\sum_{k\in \mZ} (-q^{k^2}+q^{3k^2})(-1)^k \,,
     	\fe
     	where the extra sign comes from the odd parity of the spin 3 generator under the duality. 
     	
The  twisted partition function reads
     	\ie
     	Z_{1\cN}(\tau,\bar\tau )=2\bar\eta(\bar c_0^0I_1-  \bar  c^2_0    I_2 +\bar  c^4_0   I_3)\,,
     	\label{ZN4}
     	\fe
and has the following explicit form
     	\ie
     	Z_{1\cN}(\tau,\bar\tau)={1\over  |\eta|^2} \sum_{m,n\in \mZ} \left(
     	(-1)^{m }q^{3m^2}\bar q^{3n^2}
     	+(-1)^{m+n}q^{ m^2}\bar q^{ n^2}
     	-2(-1)^{m }q^{3(m+1/3)^2}\bar q^{3(n+1/3)^2}
     	\right)\,.
     	\label{Z4dualitypara}
     	\fe
     	Performing the modular $S$-transform using the Poisson resummation formula,
     	\ie
     	Z_{\cN1}(\tau,\bar\tau)=&{1\over  |\eta|^2} \sum_{m,n\in \mZ} \left(
     	{1\over 6}q^{{1\over 12}(m+1/2)^2}\bar q^{{1\over 12}n^2}
     	+ {1\over 2}q^{{1\over 4}(m+1/2)^2}\bar q^{{1\over4 }(n+1/2)^2}
     	- {1\over 3}q^{{1\over 12}(m+1/2)^2}\bar q^{{1\over 12}n^2}e^{(2(m+n)+1)\pi i\over 3}
     	\right)\,,
     	\\
     	=&{1\over 2|\eta|^2} \sum_{m,n\in \mZ} \left(
     	{1\over 3}(1-2\cos{(2(m+n)+1)\pi\over 3})q^{{1\over 12}(m+1/2)^2}\bar q^{{1\over 12}n^2}
     	+ q^{{1\over 4}(m+1/2)^2}\bar q^{{1\over4 }(n+1/2)^2}
     	\right)\,,
     	\\
     	=&{1\over 2|\eta|^2}  \left(
     	\sum_{m,n\in \mZ,m-n\in 2\mZ}
     	q^{{(n+3m+1)^2\over 16}}\bar 
     	q^{{(n-3m-2)^2\over 16}}
     	+ 
     	\sum_{m,n\in \mZ}
     	q^{{1\over 4}(m+1/2)^2}\bar q^{{1\over4 }(n+1/2)^2}
     	\right)\,,
     	\label{Z4dualityHSpara}
     	\fe
     	which indeed decomposes into Virasoro characters with non-negative integer degeneracies thanks to the overall factor of 2 in \eqref{ZN4}.\footnote{This is also the minimal choice. Since in the defect Hilbert space partition function $Z_{\cN1}$, certain operators appear with unit degeneracy, thus e.g. ${1\over 2}\cN$ would not be a well-defined TDL.}
     	This determines the defect Hilbert space $\cH_\cN$ completely. In particular, the spectrum of spins satisfies
     	\ie
     	e^{4\pi i s}=\{\pm e^{-{\pi i\over 4}},1\}\,,
     	\fe
     	which suggests, using the spin selections rules \eqref{ssruleZ4}, that the duality defect together with the $\mZ_4$ symmetry TDLs furnishes the $\TY(\mZ_4,\chi_+,1)$ category. By stacking the duality defect $\cN$ with the $\mZ_2$ symmetry TDL $s$, we obtain $\TY(\mZ_4,\chi_-,1)$ as explained in Section~\ref{subsecZ4TYorbifold}.
     	
     The marginal operator that moves away from the $SU(2)_4/U(1)$ CFT on the orbifold branch is $\phi^2_0=\phi_2^0$ in Table~\ref{table:z4ops}, which commutes with both $\mZ_4$ TY symmetries. Consequently this verifies the presence of these fusion category symmetries on the entire orbifold branch. In particular, the twisted partition functions here \eqref{Z4dualitypara} and \eqref{Z4dualityHSpara} matches with the general expressions \eqref{Z4duality} and \eqref{Z4dualityHS} once we set $R=\sqrt{6}$ as promised.

     	\subsection{${\rm Rep}(H_8)$ and Ising$^2$}\label{secRepH8Ising}
     	
     	At radius $R=2$ the $\bZ_2$ orbifold is rational and equivalent to a tensor product of two Ising CFTs, which we denote by ${\rm Ising}^2$. The operators $\sigma_{1,2}$ of the orbifold \eqref{eqnsigmaops} are identified with the spin operators of the two Ising sectors, while $\sigma_1 \sigma_2 = \cos \theta$, the sum of thermal operators $\epsilon_1 + \epsilon_2 = \cos \phi$, etc.
     	
     	This rational point was discussed in detail in Section 4 of our previous work \cite{Thorngren:2019iar}. We review it here for completeness. We will see that the self-duality of the $\bZ_2$ orbifold under $D_4^B$ gauging associated with the bicharacter $\chi_s$ gives rise to a TDL with Frobenius-Schur indicator $\epsilon = 1$ (the TY category is therefore isomorphic to ${\rm Rep}(H_8)$), and verify our proposed action \eqref{eqnH8action} on local operators. 
     	
     	The $D_8$ global symmetry is generated by the individual $\mZ_2$ symmetries $\eta \sim sr^3$ and $\sigma \sim sr$ of the Ising factors, along with the $\mZ_2$ TDL $s$  that swaps them. Each Ising factor is self-dual under gauging its $\bZ_2$ symmetry, and realizes the $\mZ_2$ TY category $\TY(\mZ_2,+)$ with duality defects $\cN_{1,2}$. These TDLs satisfy the fusion rules
     	\ie
     	\cN_1^2=1+\eta\,,\quad 
     	\cN_2^2=1+\sigma\,.
     	\fe
     	The tensor product of these two duality defects thus gives a $D_4^B = \langle sr, sr^3\rangle$ duality defect $\cN=\cN_1 \cN_2$. The bicharacter of this defect respects the product structure, and we recognize it as $\chi_s$ in \eqref{D4bichar}. Furthermore, since $\cN_{1,2}$ both have $\epsilon = 1$, so does $\cN_1 \cN_2$, so we recognize the TY category $\TY(D_4,\chi_s,1)$, which is equivalent to ${\rm Rep}(H_8)$.

     	Recall the  twisted partition function in the Ising model,
     	\ie
     	Z^{\rm Ising}_{1\cN_1}(\tau,\bar\tau)=&
     	\sqrt{2}(|\chi_{0}|-|\chi_{1\over 2}|^2)\,,
     	\fe
     	where $\chi_h$ here denotes the Virasoro character at $c={1\over 2}$
     	\ie
     	\chi_0=&{1\over 2}\left(
     	\sqrt{\theta_3\over \eta}+\sqrt{\theta_4\over \eta}
     	\right)\,,
     	\\
     	\chi_{1\over 2}=&{1\over 2}\left(
     	\sqrt{\theta_3\over \eta}-\sqrt{\theta_4\over \eta}
     	\right)\,,
     	\\
     	\chi_{1\over 16}=&{1\over \sqrt{2}} 
     	\sqrt{\theta_2\over \eta} \,.
     	\label{IsingChars}
     	\fe
     	Then in the Ising$^2$ CFT, we have
     	\ie
     	Z_{1\cN}(\tau,\bar\tau)=&
     	2(|\chi_{0}|^2-|\chi_{1\over 2}|^2)^2\,.
     	\fe
     	Using \eqref{IsingChars}, this becomes
     	\ie
     	Z_{1\cN} =
     	{1\over   |\eta|^2} \left(
     	\sum_{m,n\in \mZ }  
     	(-1)^{m+n} q^{(m+n)^2/2}\bar q^{(m-n)^2/2}
     	+
     	\sum_{m,n\in \mZ}  (-1)^{m+n}q^{ m^2}\bar q^{ n^2}
     	\right)\,,
     	\fe
     	and the modular S-transform gives
     	\ie
     	Z_{1\cN}(\tau,\bar\tau)=&{1\over  2|\eta|^2} \left(
     	\sum_{m,n\in \mZ}    
     	q^{{1\over 2}\left({ n  +m+1\over 2}\right)^2}\bar q^{{1\over 2}\left({n -m\over 2}\right)^2}
     	+
     	\sum_{m,n\in \mZ}
     	q^{{1\over 4}(m+1/2)^2}\bar q^{{1\over4 }(n+1/2)^2}
     	\right)\,,
     	\fe
     	which is agreement with our general proposal \eqref{D4duality} at $R=2$. Moreover, we can check the spin spectrum in $\cH_\cN$ satisfies
     	\ie
     	e^{4\pi i s}=\{\pm i, 1\}
     	\fe
     	as expected for the spin selection for ${\rm Rep}(H_8)$ \eqref{ssruleD4}.
     	
     	The tensor product of the thermal operators $\epsilon_{1,2}$ in the two Ising models, $\epsilon_1\epsilon_2$ has weight $h=\bar h=1$ and is exactly marginal. It is uncharged under $D_4^B$. Since $\epsilon_1\epsilon_2$ does not commute with the individual Ising $\mZ_2$ dualities, the duality defects $\cN_1$ and $\cN_2$ are no longer topological when we move away from the Ising$^2$ point. However this deformation commutes with the  duality defect $\cN$, 
     	\ie
     	\cN \cdot  \epsilon_1\epsilon_2= \la \cN \ra \epsilon_1\epsilon_2
     	=2  \epsilon_1\epsilon_2\,,
     	\fe
     	thus the $D_4^B$ duality defect $\cN$ and the associated category ${\rm Rep}(H_8)$ will persist along the entire orbifold branch, in agreement with our observation of self-duality under $D_4^B$ gauging.
     	
     	\subsection{${\rm Rep}(D_8)$ and the four-state Potts model}\label{secRepD84statepotts}
     	
     	At $R = 2 \sqrt{2}$ the $\bZ_2$ orbifold is again rational and is equivalent to the four-state Potts model and a $D_4$ orbifold of the $SU(2)_1$ WZW model,
     	\ie
     	{U(1)_8\over \bZ_2^C} \cong {SU(2)_1 \over D_4} \,.
     	\fe 
     	The theory has $S_4$ global symmetry and an enhanced chiral algebra $W(2,4,4)$ containing two spin 4 generators in addition to the stress tensor. There are eleven chiral primaries $1,j_a,\phi,\sigma_a,\tau_a$
     	with scaling dimension $h=\bar h=\{0,1,{1\over 4},{1\over 16},{9\over 16}\}$ respectively \cite{Dijkgraaf:1989hb}. The index $a=1,2,3$ transforms in the 3-dimensional irreducible representation of $S_4$.
     	
     	This rational point gives us access to the other $D_4^B$ TY symmetry of the $\bZ_2$ orbifold, associated with the bicharacter $\chi_a$ in \eqref{D4bichar}, allowing us to  determine the Frobenius-Schur indicator to be $\epsilon = 1$ and verify the action \eqref{eqnD8action} on local operators. Indeed, the Verlinde lines associated with the chiral primaries $1,j_a,\phi$ have $D_4$ TY fusion rules: $j_a$ correspond to the order 2 elements of $D_4^A$, and $\phi$ corresponds to the duality defect.

     	The untwisted  torus partition function of the four-state Potts model is
     	\ie
     	Z(\tau,\bar\tau)= |\chi_0|^2+3|\chi_1|^2+ |\chi_{1\over4 }|^2+ 3|\chi_{1\over16 }|^2+ 3|\chi_{9\over 16 }|^2\,,
     	\fe 
     	where $\chi_h$ denotes the $W(2,4,4)$ character for a weight $h$ primary.
          	The torus partition function twisted by the duality defect $\cN$ wrapping the spatial circle is given by the $S$ matrix by the braiding monodromy of $\phi$ (which is natural from the 2+1D TQFT perspective),
     	\ie 
     	Z_{1\cN}(\tau,\bar\tau)=& 2|\chi_0|^2+6|\chi_1|^2-2|\chi_{1\over4 }|^2 \,.
     	\label{4sp1N}
     	\fe
     	Its modular S-transform (see Table~\ref{tab:SDeven}) is
     	\ie 
     	Z_{\cN 1}(\tau,\bar \tau)=    (\chi_0+3\chi_1) \bar \chi_{1\over 4}+3 \chi_{1\over 16}\bar  \chi_{9\over 16}
     	) +c.c.)+3(|\chi_{1\over 16}|^2 +| \chi_{9\over 16}|^2)\,  .
     	\fe
     	The spin spectrum in $\cH_\cN$ satisfies
     	\ie
     	e^{4\pi i s}=\{\pm 1\}\,.
     	\fe
     	Compared to the spin selection rules in \eqref{ssruleD4}, this
     	implies $\cN$ is a duality defect with associated F-symbols determined by the bicharacter $\chi_a$ and a FS indicator $\epsilon=\pm 1$. Below we will carry out a refined version of the analysis around \eqref{ssruleD4} to show that $\epsilon=+1$.

     	Performing a $T^2$ transformation on $Z_{\cN 1}$, we find
     	\ie
     	Z_{\cN 1}(\tau+2,\bar \tau+2)={\epsilon\over 2}
     	\left(Z_{\cN 1} +Z_{\cN \eta}+ Z_{  \cN\sigma}+ Z_{ \cN\eta\sigma}
     	\right )\,,
     	\label{D8T2transf}
     	\fe
     	which constrains the action of the $D_4^B$ TDLs on the defect Hilbert space $\cH_\cN$.  Moreover, from partial fusion and crossing, one can derive that, 
     	\ie
     	\widehat\eta_-^2=\widehat \sigma_-^2=\widehat{\eta\sigma}_-^2=1\,,\quad \widehat\eta_- \widehat \sigma_-= \widehat \sigma_-\widehat\eta_- =-\widehat{\eta\sigma}_-
     	\label{esops}
     	\fe  
     	as linear operators acting on $\cH_\cN$. Since $\cH_\cN$ has a single defect chiral primary of weight $(h,\bar h)=(0,{1\over 4})$, from \eqref{D8T2transf}, we conclude that $\widehat\eta_-,\widehat \sigma_-,\widehat{\eta\sigma}_-$ acting on this operator has to be $-.-,-$ if $\epsilon=1$, and $+,+,-$ or its permutations if $\epsilon=-1$.
     	
     	We obtain additional constraints on $\widehat\eta_-,\widehat \sigma_-,\widehat{\eta\sigma}_-$ from the modular bootstrap equation \eqref{mbeqn} for the pair of TDLs $\eta$ and $\cN$, with two  solutions
     	\ie 
     	Z_{\cN\eta}=  &  \C  \left[
     	(\chi_0+3\chi_1) \bar \chi_{1\over 4}- \chi_{1\over 16}\bar  \chi_{9\over 16}
     	) +c.c.)-(|\chi_{1\over 16}|^2 +| \chi_{9\over 16}|^2)  
     	\right] \,,
     	\\
     	Z_{\eta\cN}  =&   2\C (\chi_0\bar \chi_1+\chi_1\bar{\chi}_0+2|\chi_1|^2-|\chi_{1\over 4}|^2)\,,
     	\fe
      parametrized by $\C = \pm 1$. From the discussion around \eqref{esops},  for $\epsilon=1$ we must have 
     	\ie
     	Z_{\cN\eta}= &     -\left[
     	(\chi_0+3\chi_1) \bar \chi_{1\over 4}- \chi_{1\over 16}\bar  \chi_{9\over 16}
     	) +c.c.)-(|\chi_{1\over 16}|^2 +| \chi_{9\over 16}|^2)  
     	\right] \,,
     	\\
     	Z_{\cN\sigma}= & -\left[
     	(\chi_0+3\chi_1) \bar \chi_{1\over 4}- \chi_{1\over 16}\bar  \chi_{9\over 16}
     	) +c.c.)-(|\chi_{1\over 16}|^2 +| \chi_{9\over 16}|^2)  
     	\right] \,,
     	\\
     	Z_{\cN\eta\sigma}= & -\left[
     	(\chi_0+3\chi_1) \bar \chi_{1\over 4}- \chi_{1\over 16}\bar  \chi_{9\over 16}
     	) +c.c.)-(|\chi_{1\over 16}|^2 +| \chi_{9\over 16}|^2)  
     	\right] \,,
     	\fe
     	while for $\epsilon=-1$ we end up with
     	\ie
     	Z_{\cN\eta}= &     \left[
     	(\chi_0+3\chi_1) \bar \chi_{1\over 4}- \chi_{1\over 16}\bar  \chi_{9\over 16}
     	) +c.c.)-(|\chi_{1\over 16}|^2 +| \chi_{9\over 16}|^2)  
     	\right] \,,
     	\\
     	Z_{\cN\sigma}= & \left[
     	(\chi_0+3\chi_1) \bar \chi_{1\over 4}- \chi_{1\over 16}\bar  \chi_{9\over 16}
     	) +c.c.)-(|\chi_{1\over 16}|^2 +| \chi_{9\over 16}|^2)  
     	\right] \,,
     	\\
     	Z_{\cN\eta\sigma}= &  -\left[
     	(\chi_0+3\chi_1) \bar \chi_{1\over 4}- \chi_{1\over 16}\bar  \chi_{9\over 16}
     	) +c.c.)-(|\chi_{1\over 16}|^2 +| \chi_{9\over 16}|^2)  
     	\right] \,,
     	\fe
     	or any permutations of the overall signs. However looking at the piece $\chi_{1\over 16}\bar \chi_{9\over 16}$, we see $\epsilon=-1$ cannot be compatible with \eqref{D8T2transf}. Thus we conclude $\ep=1$ for the $D_4$ TY symmetry in the four-state Potts model.
     	
\begin{figure}[H]
     		\centering
     		\includegraphics[width=.4\textwidth]{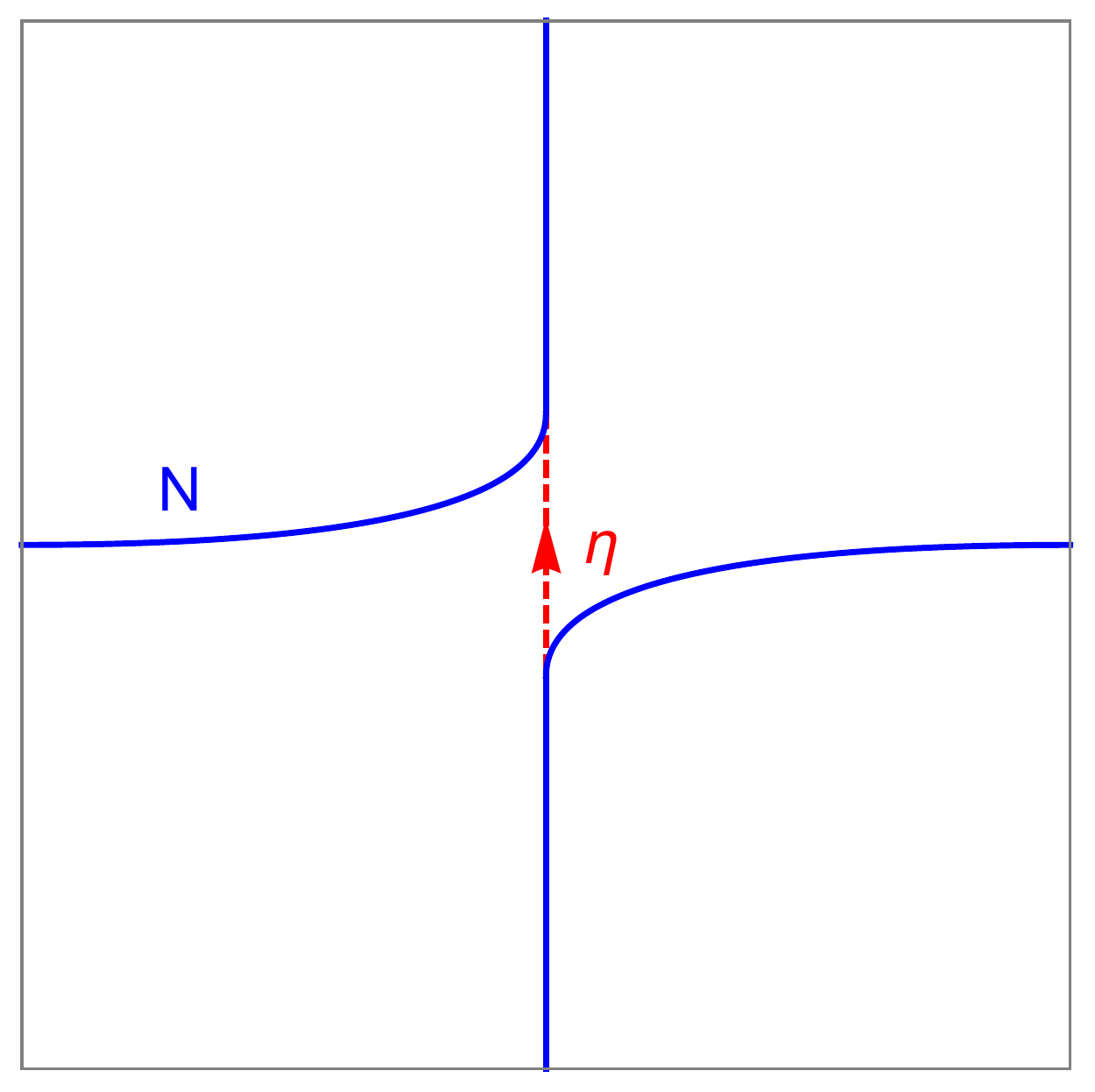}
     		\caption{Torus partition function $Z^X_{\cN \eta}$ decorated by a TDL network formed by duality TDL $\cN$ and symmetry TDL $\eta$ (similarly when $\eta$ is replaced by other symmetry TDLs in $D_4^B$.}
     		\label{fig:Xeta}
     	\end{figure}

     	We can also compute more general twisted partition functions that involve a network of  TDLs (see Figure~\ref{fig:Xeta}) which provide additional consistency checks on our identification of the TY symmetries.
     	By solving the modular bootstrap equations \eqref{mbeqn}, we find
     	\ie
     	Z^X_{\cN \eta}=Z^X_{\cN\sigma}=Z^X_{\cN \eta\sigma} =
     	\left[
     	i(\chi_0+3\chi_1) \bar \chi_{1\over 4}-\chi_{1\over 16}\bar  \chi_{9\over 16}
     	) +c.c.)+( |\chi_{1\over 16}|^2 +  | \chi_{9\over 16}|^2)  
     	\right] \,,
     	\fe
     	which satisfy the following relations
     	\ie
     	Z^{X}_{\cN\eta}(\tau+1)=&{1\over 2}(Z_{\cN}+Z_{\cN\eta}-Z_{\cN\sigma}- Z_{\cN\eta\sigma} )\,,
     	\\
     	Z^{X}_{\cN\sigma}(\tau+1)=&{1\over 2}(Z_{\cN}-Z_{\cN\eta}+Z_{\cN\sigma}- Z_{\cN\eta\sigma} )\,,
     	\\
     	Z^{X}_{\cN\eta\sigma}(\tau+1)=&{1\over 2}(Z_{\cN}-Z_{\cN\eta}-Z_{\cN\sigma}+ Z_{\cN\eta\sigma})\,,
     	\fe
     	that come from partial fusion and crossing.

     	Finally, we note that the four-state Potts model has three independent exactly-marginal operators $j_a$ which transform as the three-dimensional irreducible representation of $S_4$. Consequently, deformation from either of the three moves the CFT along the orbifold branch.
     	Now the duality defect $\cN$ commutes with all three marginal operators
     	\ie
     	\cN\cdot j_a \bar j_a = \la \cN \ra  j_a \bar j_a\,.
     	\fe
     	Thus the ${\rm Rep}(D_8)$ category of TDLs is preserved along the entire orbifold branch. 
     	
     	In particular we can compare to our general proposal \eqref{D4duality}, at $R=2\sqrt{2}$,
     	\ie
     	Z_{1\cN_{D_8}} =&{1\over  |\eta|^2} \left(
     	\sum_{m,n\in \mZ }  
     	(-1)^{m}q^{{1\over 4}\left({2n}+{ m }\right)^2}\bar q^{{1\over 4}\left({2n}-{ m }\right)^2}
     	+
     	\sum_{m,n\in \mZ}  (-1)^{m+n}q^{ m^2}\bar q^{ n^2}
     	\right)\,,
     	\fe
     	using the explicit forms for the $W(2,4,4)$ characters,
     	\ie
     	\chi_0(\tau)=&{1\over 2\eta}\sum_m (q^{4m^2}+(-1)^m q^{m^2}) \,,
     	\\
     	\chi_1(\tau)=&{1\over 2\eta}\sum_m (q^{4m^2}-(-1)^m q^{m^2})\,,
     	\\
     	\chi_{1\over 4}(\tau)=&{1\over  \eta}\sum_m  q^{(1+4m)^2\over 4}\,, 
     	\fe
     	we find that it agrees with $Z_{1\cN}$ in \eqref{4sp1N} as desired.

\subsection{Continuum of Topological Defects}\label{seccontinuumZ2orbifold}

As we have mentioned in Section \ref{subsecshadownoether}, and using the general method therein, we can construct a continuum of TDLs for every $R$ from the continuous $U(1)^\theta \times U(1)^\phi$ symmetry of the circle branch. In particular, if $\cL_\alpha^\theta$ is the TDL of the compact boson corresponding to the symmetry $\theta \mapsto \theta + \alpha$, then $\cL_\alpha^\theta + \cL_{-\alpha}^\theta$ is $C$ invariant and will define a noninvertible TDL $\cN^\theta_\alpha$ of the $\bZ_2^C$ orbifold with quantum dimension 2. It acts on local operators (for generic $\A$) by
\ie\label{eqncontaction}
\widehat\cN^{\theta}_\alpha: \begin{cases}
	V^+_{n,w}\to 2\cos(n\alpha)	V^+_{n,w}  \,,
	\\
	j_{n^2} \bar j_{m^2} \to 2 j_{n^2} \bar j_{m^2}\,,
	\\
	\text{other primaries} \to \text{non-local operators}\,,
\end{cases}\quad
\fe
and there is likewise a continuum family for the $\phi$ shift symmetry $\cL_\phi^\alpha$. See \cite{Chang:2020imq} for a recent discussion of these lines.

The Noether currents associated with these families are $d\theta$ and $d\phi$ respectively, which live at the end of the Wilson line for the $C$ gauge field. The TDL $\cN_{D_8}$ of \eqref{eqnD8action} appears in this family at $\alpha = {\pi\over 2}$. In fact the subalgebra of lines with $\alpha = {2\pi m\over k}$, $m = 1,\ldots,\lfloor {k-1\over 2} \rfloor$ generates a ${\rm Rep}(D_{2k})$ symmetry.\footnote{These non-invertible TDLs are in one-to-one correspondence with the $\lfloor {k-1\over 2} \rfloor$ two-dimensional irreducible representations of $D_{2k}$.} This can be seen by writing the $\bZ_2^C$ orbifold at radius $R$ as a $D_{2k}$ orbifold of the compact boson on the circle branch at radius $ R\over k$ by gauging $\bZ_k^\phi \rtimes \bZ_2^C = D_{2k}$. By multiplying with the $D_8$ generator $s$, we see that the ${\rm Rep}(H_8)$ symmetry similarly extends to an infinite family, as well as their T-dual TY symmetries, which arise from $\cL_\phi^\alpha$.

At the special radii $R = n \sqrt{2}$ this continuum is enhanced to an ``$SO(4)$" family, obtained from the $SO(4)$ invertible TDLs of $SU(2)_1$ upon gauging the $D_{2n}$ subgroup $\bZ_n^\phi \rtimes \bZ_2^C$. Besides $d \theta$ and $d\phi$ which live at the end of the Wilson line for the sign representation, there are four Noether currents $V_{\pm n, \pm 1/n}$, which live at the end of a Wilson line for two 2-dimensional representations of $D_{2n}$. If $m$ is coprime to $n$, by first gauging $\bZ_m^\theta$ and then $\bZ_n^\phi \rtimes \bZ_2^C$, we obtain a related family at radius $R = \frac{n}{m} \sqrt{2}$, with Noether currents $V_{\pm n/m, \pm m/n}$.

   	\subsection{The Kosterlitz-Thouless point and triality}\label{subsecKT}
   	
   	If we consider the $\bZ_2^C$ orbifold of the $SU(2)_1$ point, we land at the Kosterlitz-Thouless (KT) point, where the orbifold and circle branches meet. This can either be described in terms of the $C$-gauged fields $\theta$, $\phi$ and $C$-twist operators, or in terms of the (gauge invariant) circle branch fields $\tilde \theta$, $\tilde \phi$ (see Appendix \ref{appKTsymmetry}).
   	
   	Along with the continuous families of TDLs associated with the non-local currents $d\theta$ and $d\phi$, which persist along the entire orbifold branch, at the KT point we also have the four currents $V_{\pm 1, \pm 1}$ mentioned in the previous subsection (take $n = 1$). Two of these are local: $V_{1,\pm 1} + V_{-1,\mp 1}$, which we can identify with the currents $d \tilde \theta$, $d \tilde \phi$. These generate the familiar $U(1)_{\tilde  \theta} \times U(1)_{\tilde \phi}$ symmetry of the KT point, expressed as a circle branch theory. However, the other two currents $V_{1,\pm 1} - V_{-1,\mp 1}$ are non-local, and correspond to a two-parameter enhancement of the TDLs at this special point. Note this is the same four-parameter family we constructed in Section \ref{subsecSU2continuum} by considering the KT point as an orbifold of $SU(2)_1$ by $\bZ_2^\phi$, since $\bZ_2^\phi$ and $\bZ_2^C$ are conjugate in the full $SO(4)$ symmetry.

     	Some of the new TDLs at the KT point also have interpretations as  self-dualities under gauging. In part I \cite{Thorngren:2019iar}, we noted that by gauging $D_4$ using the bicharacter $\chi_a$, $\beta = 0$, and $\alpha$ the generator of $H^2(D_4,U(1)) = \bZ_2$ in \eqref{eqnpartfnselfdualgauging} we obtain an order three permutation of the $D_4$-symmetric gapped phases.
     	
     	It turns out this \textit{triality} is a category symmetry of the KT point. We can see this as follows. We identify $D_4^B$ with $\bZ_2^{\tilde \theta} \times \bZ_2^{\tilde C}$, where $\tilde C$ is the charge conjugation symmetry of the circle branch variables $\tilde \theta$, $\tilde \phi$. The theory is self-dual under gauging this symmetry: gauging $\bZ_2^{\tilde \theta}$ halves the radius, putting us at the $SU(2)_1$ point, then gauging $\bZ_2^{\tilde C}$ puts us back at the KT point. This self-duality defines a $D_4^B$ TY symmetry. A novelty for the KT point however is the chiral anomaly, one symptom of which is that partition functions which might detect the $D_4^B$ discrete torsion vanish identically (see Section \ref{subsecZ2Z2gauging}). We can attribute this to a cubic anomaly in $\bZ_2^{\tilde \theta} \times \bZ_2^{\tilde \phi} \times \bZ_2^{\tilde C}$, such that $\bZ_2^{\tilde \phi}$ ``toggles" the discrete torsion, mapping $\bZ_2^{\tilde C}$-twisted operators to ones of opposite $\bZ_2^{\tilde \theta}$ charge, etc (see Appendix \ref{apporbifoldsymmetry}). We expect that the triality is therefore realized as a fusion product of a $D_4^B$ TY TDL with the $\bZ_2^{\tilde \theta}$ TDL.
     	
     	To explore this in detail, and to see its three-fold nature, we can examine these symmetries as they arise from the $SO(4)$ symmetry of the $SU(2)_1$ point upon gauging $\bZ_2^C$. In terms of the symmetries and the fields $\theta,\phi$ before gauging, we have already identified the ${\rm Rep}(D_8)$ duality symmetry with a $\pi\over 2$ $U(1)^\theta$ rotation $R^\theta({\pi\over 2})$ (see Section~\ref{seccontinuumZ2orbifold}). Meanwhile $\bZ_2^{\tilde \phi}$ is generated by a conjugate ${\pi\over 2}$ rotation $R^{\phi'}({\pi \over 2})$ (this can be considered a $\pi \over 2$ rotation of an $SO(4)$-conjugate field $\phi'$). This squares to $C$ and becomes a $\bZ_2$ symmetry after gauging. As $SO(4)$ matrices in the basis \eqref{SO4basis}, these are two $2\times 2$ blocks overlapping in one position, and we find their composition is a $2\pi\over 3$ rotation in the combined $3 \times 3$ block,
     	\ie R^\theta\left({\pi\over 2}\right) =& \begin{bmatrix} 0 && 1 && 0 && 0 \\ -1 && 0 && 0 && 0 \\ 0 && 0 && 1 && 0 \\ 0 && 0 && 0 && 1  \end{bmatrix}\,,
     \\
     R^{\phi'}\left({\pi \over 2}\right) =& \begin{bmatrix} 0 && 0 && 1 && 0 \\ 0 && 1 && 0 && 0 \\ -1 && 0 && 0 && 0 \\ 0 && 0 && 0 && 1  \end{bmatrix}\,,
      \\
           	 Q = R^\theta\left({\pi\over 2}\right) R^{\phi'}\left({\pi \over 2}\right) =& \begin{bmatrix} 0 && 1 && 0 && 0 \\ 0 && 0 && -1 && 0 \\ -1 && 0 && 0 && 0 \\ 0 && 0 && 0 && 1  \end{bmatrix}\,,
           	 \\
           	 C = R^{\phi'}\left(\pi\right) =& \begin{bmatrix} -1 && 0 && 0 && 0 \\ 0 && 1 && 0 && 0 \\ 0 && 0 && -1 && 0 \\ 0 && 0 && 0 && 1  \end{bmatrix}\,.
           	 \fe
     	 See Appendix \ref{appKTsymmetry} for more details. We can thus as well consider the triality defect as coming from the invertible symmetry $Q$ of the $SU(2)_1$ point upon gauging $\bZ_2^C$, resulting in a triality TDL $\cL_Q$ of quantum dimension 2.
     	 
     	 The subgroup of the $SO(4)$ symmetry at the $SU(2)_1$ point generated by $C$ and $Q$ is the alternating group $A_4$.  We want to study what fusion category results from $A_4$ after gauging $\bZ_2^C$. This will be a fusion subcategory of the symmetries of the KT point containing the triality TDL we are interested in. Mathematically, $\bZ_2^C$ defines a module category $\cM$ for the grouplike category $\cA = {\rm Vec}_{A_4}$, such that the simple objects of $\cM$ are labelled by cosets in $A_4/\bZ_2^C$ \cite{Meir_2012,Thorngren:2019iar}. The fusion category symmetry we obtain after gauging is the ``dual" category $\cA^*_\cM = {\rm End}_{\cA}(\cM)$. Since $\bZ_2^C$ is not normal in $A_4$, $\cA^*_\cM$ is not grouplike (see Thm 3.4 in \cite{deepak}). We find there are six simple TDLs: four grouplike, corresponding to $\tilde C=sr$ (from $R^\theta(\pi)$) and the magnetic symmetry $r^2$, which together generate the $D_4^B \subset D_8$ symmetry of the KT point, as well as two triality defects of quantum dimension 2, associated with $Q$ and $Q^2 = Q^{-1}$. The fusion rules are
     	 \ie \cL_{Q} \cL_g = \cL_g \cL_{Q} = \cL_{Q}\,,\quad 
     	 \cL_{Q^2} \cL_g = \cL_g \cL_{Q^2} = \cL_{Q^2}\,,
     	 \fe
     	 for any $g \in D_4^B$. We also have
     	 \ie\cL_Q \cL_{Q^2} = \sum_{g \in D_4^B} \cL_g\,,\quad 
     	 \cL_Q^2 = 2 \cL_{Q^2}\,.
     	 \fe
     	 By analyzing the action of $\cL_Q$ on $D_4$-symmetric gapped phases (see Section 3.2.3 of \cite{Thorngren:2019iar}), we find that the fusion category is anomalous---there is no way to deform the KT point to a trivial gapped phase while preserving the triality. This also means that the category above is not ${\rm Rep}(H)$ for any Hopf algebra (or group) $H$.

  \subsection{Symmetric RG flows from the orbifold branch} \label{sec:RGfloworb}
Let us discuss deformations of the $c=1$ CFT on the orbifold branch preserving certain fusion category symmetries. We focus on the $D_4$ and $\mZ_4$ TY symmetries. Firstly, the $D_4$ TY symmetries presented on the orbifold branch are all non-anomalous as they admit fiber functors \cite{Tambara2000,Thorngren:2019iar}. On the contrary, the $\mZ_4$ TY symmetries are all anomalous (see also discussion in Section~\ref{sec:RGflowcirc}). Consequently, while the $D_4$ TY symmetries do not lead to nontrivial RG constraints, the $\mZ_4$ TY symmetries require the symmetric RG flow to end either in a nontrivial CFT or a spontaneous symmetry broken phase. Below we will see explicitly that for  RG flows out of the orbifold branch preserving $\TY(\mZ_4,\chi_\pm,1)$, the IR phase is described by a TQFT with degenerate ground states, fulfilling the second scenario predicted by the anomalous fusion category symmetry. 

 As discussed in Section~\ref{secZ4parafermion}, the $\mZ_4$ TY symmetries are most transparent at the  $\mZ_4$ parafermion point on the orbifold branch, which is described by the $SU(2)_4/U(1)$ coset CFT. Even though the $\mZ_4$ parafermion point does not have relevant deformations that preserve the $\mZ_4$ duality TDLs, once we have moved to large enough $R$ by the duality invariant marginal deformation, there are plenty of relevant deformations that do commute with these TDLs, namely the momentum operators
 \ie
 V^+_{4n,0}:~h=\bar h={8n^2\over R^2}\,,
 \fe
cf. \eqref{Z4TYonop}.
 Note that $R=2\sqrt{2}$ corresponds to the four-state Potts model in which case $V_{4,0}$ is marginal and related to the marginal operator $\pa X \bar\pa X$ by the enhanced $S_4$ symmetry.
 
 For $R>2\sqrt{2}$, $V^+_{4,0}$ is a relevant deformation that preserves the $\mZ_4$ TY symmetries, and consequently the IR phase must realize the $\mZ_4$ TY symmetries and cannot be trivially gapped! 
  Indeed, in terms of the circle variable $\theta\sim \theta+2\pi$, the $V^+_{4,0}$ deformation takes the form of $\pm \cos{4\theta }$. With $+$ sign, the IR theory is gapped with two vacua located at $\theta=\{\pi/4,3\pi/4\}$, while with $-$ sign, there are three vacua located at $\theta=\{0,{\pi\over 2},\pi\}$.\footnote{Note that we have taken into account the $\mZ_2$ identification of $\theta$ in the orbifold theory.} In either case, the IR theory completes into a TQFT with two or three degenerate ground states.\footnote{Applying the self-duality condition \eqref{eqnpartfnselfdualgauging} directly to the putative TQFT on $\Sigma=T^2$, it is a simple exercise to derive a number of constraints on the TY symmetric TQFT using modular invariance. One of them requires the vacuum degeneracy to be $3d_1+2d_2$ where $d_1,d_2\in \mZ_{\geq 0}$ count the number of ground states with $\mZ_4$ charge 1 and 2 respectively. 
  }
  
  For $\bZ_k$ parafermions of general $k$, the $SU(2)_k/U(1)$ coset CFT contains TY symmetric relevant deformations for $k>5$, for example given by the leading duality-invariant thermal operator $\phi^2_0$ of weight $h=\bar h={6\over k+2}$. Once again the anomalous $\mZ_k$ TY symmetries demands a nontrivial symmetric IR theory that reproduces the same anomaly. In this case, this is achieved by a \textit{direct sum} of $c=1$ CFTs,
 \ie
 SU(2)_k/U(1) \xrightarrow[\text{deformed by}~\phi^2_0]{\TY\text{-symmetric RG}} {\rm KT} \oplus {\rm KT}/\mZ_k\,,
 \fe
 which is suggested by the numerical analysis of the $\mZ_k$ clock model in \cite{Alcaraz:1980sa,Alcaraz:1980bb,Alcaraz:1986hs} and integrability analysis in \cite{Dorey:1996he}.

     	\section{Fusion Categories of the Exceptional Orbifolds}\label{secexceptional}

    In the previous section we have studied TDLs of the circle and $\bZ_2$ orbifold branches of the $c = 1$ moduli space. There are also three isolated points, collectively known as the exceptional orbifolds, which arise from gauging one of three special symmetry subgroups of the $SU(2)_1$ CFT. Recall this theory has a global symmetry
	\ie
	SO(4)={SU(2)_L\times SU(2)_R\over \mZ_2}\,,
	\fe
	of which the diagonal subgroup $SO(3)_{\rm diag}$ is anomaly-free. Finite subgroups of $SO(3)$ have an ADE classification, where the A series are cyclic rotation groups around a fixed axis, whose orbifolds define the series $R = n \sqrt{2}$ on the circle branch; the D series are dihedral groups of rotations around a fixed axis along with a perpendicular $\pi$ rotation, whose orbifolds define the series $R = n \sqrt{2}$ on the $\bZ_2$ orbifold branch; and the exceptional E series, with $E_6$ the symmetry group $T$ of the tetrahedron, $E_7$ the symmetry group $O$ of the octahedron/cube, and $E_8$ the symmetry group $I$ of the icosahedron/dodecahedron, whose orbifolds define the exceptional orbifolds. These three exceptional groups are isomorphic to alternating and symmetric groups
\ie
T=A_4\,,\quad O=S_4\,,\quad I=A_5\,.
\label{TOIgps}
\fe

The cases $G=D_{4n},A_4,S_4,A_5$ all admit nontrivial discrete torsion $H^2(G,U(1))=\mZ_2$. However because of the 't Hooft anomaly for $SO(4)$ symmetry of the $SU(2)_1$ CFT, the discrete torsion can be removed by a chiral $\mZ_2$ rotation, and thus it does not lead to distinct orbifold theories.\footnote{This phenomena is a $D=2$ analog of what happens in $D=4$ nonabelian gauge theories with an Adler-Bell-Jackiw (ABJ) anomaly for certain chiral (axial) $U(1)$ rotation of the charged fermions. The role of the discrete torsion is played by the usual theta angle term ${i\theta\over 8\pi^2} \int \tr F\wedge F$ in the action. The presence of the ABJ anomaly implies that the theta angle $\theta$ can be freely adjusted by a chiral (axial) rotation of the fermions.}

As discussed in Section~\ref{sec:RepG}, the resulting orbifold CFT contain $G$ Wilson lines which make up a fusion category symmetry described by ${\rm Rep}(G)$. For the A series, the corresponding ${\rm Rep}(\bZ_n)$ symmetry is invertible and part of the $U(1) \times U(1)$ symmetry of the circle branch. For the D series, the symmetry ${\rm Rep}(D_{2n})$ contains non-invertible lines and are part of a continuous family of TDLs on the $\bZ_2$ orbifold branch we discussed in Section \ref{seccontinuumZ2orbifold}. Likewise the ${\rm Rep}(A_4),{\rm Rep}(S_4)$ and ${\rm Rep}(S_5)$ fusion category symmetries of the exceptional orbifolds are part of a much larger algebra of TDLs.

Some of the additional TDLs come from Verlinde lines associated to the enhanced chiral algebra $\cV_G$ of the orbifolds. $\cV_G$ can be derived from the $G$-invariant part of the $SU(2)_1$ Kac-Moody current algebra and are given by W-algebras of the following types,
\ie
\cV_{D_{2n}}=\cW(2,4,n^2),~\cV_{A_4}=\cW(2,9,16),~\cV_{S_4}=\cW(2,16),~\cV_{A_5}=\cW(2,36)\,,
\label{orbca}
\fe
where we have listed the spins of the \textit{strong} generators\footnote{All other higher spin currents in the chiral algebra are generated by the normal ordered products of the strong generators and their derivatives.} and the spin $2$ generator is the stress tensor with central charge $c=1$. The characters for the $\cV_G$ representations and the corresponding $S$ matrices were derived in \cite{Dijkgraaf:1989hb,Cappelli:2002wq}. The exceptional orbifold is described by the diagonal modular invariant of the $\cV_G$ chiral algebra. Consequently, as explained in Section~\ref{sec:verlinde}, they host a family of Verlinde lines that are in one-to-one correspondence with the chiral primaries in the orbifold CFT, and generate a fusion category symmetry corresponding to ${\rm Rep}(\cV_G)$, extending the subcategory ${\rm Rep}(G)$ of $G$-Wilson lines.

Beyond these TDLs we also have a whole continuum obtained from the $SO(4)$ TDLs of the $SU(2)_1$ theory by the method of Section \ref{subsecshadownoether}. The corresponding Noether currents lie at the end of $G$ Wilson lines associated to the three dimensional representation of each $G \subset SO(3)$. It is plausible that all the Verlinde lines appear in the fusion products of this continuum of lines. We leave this interesting conjecture to future work.

We summarize the situation below, giving ${\rm Rep}(G)$ fusion rules in these theories. The relevant $S$ matrices are listed in Appendix~\ref{app:Smatrix} and we attach a \texttt{mathematica} notebook that tabulates the complete fusion rules for Verlinde lines in ${\rm Rep}(\cV_G)$.

	\subsection{$A_4$ orbifold} \label{sec:A4orb}
The orbifold ${SU(2)_1/A_4}$ has ${\rm Rep}(A_4)$ symmetry generated by Wilson lines associated to the four irreducible representations of $A_4$: the trivial representation $1$ and two nontrivial one-dimensional representations $R_1,R_2$ as well as one three-dimensional representation $V$. The fusion rules are
\ie
R_1 R_2=1\,,~~R_1^2=R_2\,,~~R_2^2=R_1\,,~~R_iV=V\,,~~V^2=1+R_1+R_2+2V\,.
\label{A4fusion}
\fe
Note that $\{1,R_1,R_2\}$ generates a $\mZ_3$ symmetry of the orbifold theory. This corresponds to the magnetic symmetry of the theory regarded as an $\mZ_3$ orbifold of the $SU(2)_1/D_4$ CFT (four-state Potts), with $A_4 = D_4 \rtimes \bZ_3$. 

As an RCFT with respect to the $\cV_{A_4}$ chiral algebra \eqref{orbca}, the theory contains $21$ chiral primaries which lead to $21$ Verlinde lines defined by the modular $S$-matrix in Table~\ref{tab:SA4} as in \eqref{verlindeonops}. They include the 4 Wilson lines in ${\rm Rep}(A_4)$ and generate the ${\rm Rep}(\cV_{A_4})$ fusion category symmetry.

	\subsection{$S_4$ orbifold} \label{sec:S4orb}
The orbifold ${SU(2)_1/S_4}$ has ${\rm Rep}(S_4)$ symmetry generated by Wilson lines associated to the five irreducible representations of $S_4$. They are the trivial representation, the sign representation $S$, a two-dimensional representation $U$, and two three-dimensional representations $V,W$. The fusion rules are
\ie
S^2=1\,,~~ SU=U\,,~~  S V=W\,,~~ U^2=1+S+U\,,~~ UV=V+W\,,~~ V^2=1+U+V+W\,.
\fe
Note that the ${SU(2)_1/S_4}$ CFT is also realized by an $\mZ_2$ orbifold of the ${SU(2)_1/A_4}$ theory and $S$ generates the corresponding magnetic $\mZ_2$ symmetry. 

The theory contains 28 chiral primaries with respect to the $\cV_{S_4}$ chiral algebra \eqref{orbca} which give rise to 28 Verlinde lines  defined by the $S$-matrix in   Table~\ref{tab:SS4}. They generate  the fusion category symmetry ${\rm Rep}(\cV_{S_4})$ which contains ${\rm Rep}(S_4)$. 

	\subsection{$A_5$ orbifold} \label{sec:A5orb}
The ${\rm Rep}(A_5)$ symmetry of the ${SU(2)_1/A_5}$ orbifold is generated by Wilson lines associated to the five irreducible representations of $A_5$. They are the trivial representation, two three-dimensional representations $V,W$, one four-dimensional representation $T$, and one five-dimensional representation $U$. The fusion rules are
\ie
&V^2=W^2=1+V+U\,,~~VW=T+U\,,~~VT=W+T+U\,,~~WT=V+T+U\,,
\\
&VU=WU=V+W+U+T\,,~~T^2=1+V+W+T+U\,,~~TU=V+W+T+2U\,,
\\
&
U^2=1+V+W+2T+2U\,.
\fe
 The theory contains 37 chiral primaries with respect to the $\cV_{A_5}$ chiral algebra \eqref{orbca} which give rise to 37 Verlinde lines defined by the S-matrix  in Table~\ref{tab:SA5}. They generate  the fusion category symmetry ${\rm Rep}(\cV_{A_5})$ which contains ${\rm Rep}(S_5)$.

 \section{Conclusion and discussion}

 In this paper, we developed general methods to study non-invertible symmetries in QFTs. In particular, we applied these tools to investigate the fusion category symmetries of $c = 1$ CFTs, and have found a surprisingly rich structure, richer than can be accounted for by just studying Verlinde lines at the rational points. We have not fully classified these symmetries, but we have shown that many are within the grasp of methods which generalize broadly and which we believe demonstrate the ubiquity of fusion category symmetry. It is clear that in higher dimensions there will be interesting fusion category (higher) symmetries to explore. One can take any theory with a group-like symmetry and gauge a non-normal finite subgroup, obtaining a theory with non-invertible TDLs. It will be very interesting therefore to investigate the consequences of these symmetries in settings beyond 1+1D CFT. This is just beginning to be explored, for instance, see \cite{Nguyen:2021naa,Heidenreich:2021tna}.
 
 In light of the continua of topological defects, it is interesting to consider how one might extend the usual mathematical definition of a fusion category to these sets of lines. Intuitively the objects are parametrized by topological spaces, and all of the fusion data, such as the fusion products, duals, and F-symbols should be a suitable generalization of continuous maps. Indeed continuity is too strong as the F-symbols for symmetry lines of continuous symmetry groups are already not continuous, but only Borel measurable \cite{Baker1977DifferentialCA}. This is related to the question ``what kind of category is the category of QFTs?" if we demand that TDLs form the morphisms between 1+1D QFTs, with the sort of fusion category we want to study being the endomorphisms of a given object (such as a $c = 1$ CFT). The parametrized fusion and F-symbol data appear to insist that this category is enriched over measurable spaces. We leave this interesting question to future work.

 \appendix
 
 \section{Global Symmetries Of Orbifolds}
 
 \subsection{The $SU(2)_1$ Point}\label{appSU2symmetry}
 
 The $SU(2)_1$ point where the compact boson is self-T-dual is the key to unlocking the symmetries of the $c = 1$ CFTs. It is convenient to write its global symmetry group as $(SU(2) \times SU(2))/\bZ_2 = SO(4)$. The $SO(4)$ symmetry is generated by six conserved currents in the CFT which can be taken to be the chiral and anti-chiral $SU(2)$ currents
  \ie
  \{\pa X_L, \cos \sqrt{2} X_L, \sin  \sqrt{2} X_L \}~~{\rm and}~~\{\bar\pa X_R, \cos \sqrt{2} X_R, \sin  \sqrt{2} X_R \}\,.
  \fe
The usual anomalous rotations generated by the (anti)holomorphic currents $\partial X_L$ and $\bar \partial  X_R$ are represented by
 \[
 e^{i{\A\over \sqrt{2}} \oint dz\pa X_L }=\begin{bmatrix} R_{\alpha/2} && 0 \\ 0 && R_{\alpha/2} \end{bmatrix} \quad {\rm and} \quad 
 e^{i{\A\over \sqrt{2}} \oint d\bar z\pa X_R }=\begin{bmatrix} R_{\alpha/2} && 0 \\ 0 && R_{-\alpha/2} \end{bmatrix} \quad \in SO(4)\,,\]
 respectively, where $R_\alpha$ is a $2\times2$ rotation matrix parametrized by $\A\in [0,2\pi]$. Note that a $2\pi$ rotation generated by either $\pa X_L$ or $\bar\pa X_R$ is the central element $-1 \in SO(4)$. 
 
The non-holomorphic currents $(\partial \theta,\bar \pa \theta)$ and $(\partial \phi,\bar \pa \phi)$ generate rotations by
 \[
 e^{i \A \oint (dz\pa \theta-d\bar z\bar\pa\theta) }=
 \begin{bmatrix} R_\alpha && 0 \\ 0 && 1 \end{bmatrix} \quad {\rm and} \quad e^{i \A \oint (dz\pa \phi-d\bar z\bar\pa\phi) }=\begin{bmatrix} 1 && 0 \\ 0 && R_{\alpha} \end{bmatrix} 
 \quad \in SO(4)
 \,,\]
 matching the identification $X_L = {\theta + \phi\over \sqrt{2}}$, $X_R = {\theta - \phi\over \sqrt{2}}$. We denote the first group $U(1)^\theta$ and the second $U(1)^\phi$. We take charge conjugation to act by
 \[C = \begin{bmatrix} -1 && 0 && 0 && 0 \\ 0 && 1 && 0 && 0 \\ 0 && 0 && -1 && 0 \\ 0 && 0 && 0 && 1  \end{bmatrix}\,,\]
 which we see negates $\alpha$ in all the above formulas when we conjugate by it. It combines with the $U(1)$'s above to form the symmetry $(U(1)^\theta \times U(1)^\phi) \rtimes \bZ_2^C$ of the whole circle branch.

There are various symmetries of the $SU(2)_1$ point that might be called T-duality. An $SO(4)$ matrix exchanging $U(1)^\phi$ and $U(1)^\theta$ takes the form
\[\begin{bmatrix} 0 && A_{2\times 2} \\ B_{2\times 2} && 0 \end{bmatrix}\,.\]
For instance, one which appears in \cite{Harvey:2017rko} is a $\pi$-rotation generated by the current $\cos X_R$:
\[\begin{bmatrix} 0 && 0 && 1 && 0 \\ 0 && 0 && 0 && 1 \\ -1 && 0 && 0 && 0 \\ 0 && -1 && 0 && 0  \end{bmatrix}\,,\]
which is anomalous and squares to the central element, while in \eqref{eqnTdualitybos} we use
\[T = \begin{bmatrix} 0 && 0 && 1 && 0 \\ 0 && 0 && 0 && 1 \\ 1 && 0 && 0 && 0 \\ 0 && 1 && 0 && 0  \end{bmatrix}\,.\]

The decomposition of the Hilbert space into $SO(4)$ representations may be found in \cite{Gaberdiel:2001xm},
\[\bigoplus_{\substack{s, \bar s \in \bZ \\ s + \bar s \in 2\bZ}} \left(\frac{s}{2},\frac{\bar s}{2}\right) \otimes \left(\frac{s^2}{4},\frac{\bar s^2}{4}\right)\,,\]
where the first part denotes the $SU(2) \times SU(2)$ representation, and the second part the Virasoro representation, ie. $h = {s^2\over 4}$, $\bar h = {\bar s^2\over 4}$. Recall an operator of momentum $n$ and winding $w$ has $s = n + m$, $\bar s = n - m$.
 
 \subsection{Chiral Anomaly and $D_8$ symmetry}\label{apporbifoldsymmetry}
 
  When we vary the radius away from the $SU(2)_1$ point, the $SO(4)$ symmetry is broken down to the subgroup
 \[G_{\rm bos} = (U(1)^\theta \times U(1)^\phi) \rtimes \bZ_2^C\,,\]
 where $U(1)^\theta$ (resp. $U(1)^\phi$) is embedded as the top left (resp. bottom right) $2\times2$ block of $SO(4)$, and $C$ is charge conjugation defined above.
 
 The anomaly is captured by the Euler class of the fundamental representation $V_4$,
 \[e(V_4) \in H^4(BSO(4),\bZ)\,.\]
 To restrict the above anomaly to $G_{\rm bos}$, we need to equivalently study the restriction of the representation $V_4$ to $G_{\rm bos}$. It is seen to split into a sum of the two $O(2)$ fundamentals of the two $2\times 2$ blocks $V_4 = V_2^\theta \oplus V_2^\phi$,
 \[e(V_4) = e(V_2^\theta)e(V_2^\phi)\,.\]
 
 Let us consider gauging $C$ at a generic radius. The commutant of $C$ in $G_{\rm bos}$ is generated by the $\pi$-rotations
 \[(-1)^n = \begin{bmatrix} -1 && 0 && 0 && 0 \\ 0 && -1 && 0 && 0 \\ 0 && 0 && 1 && 0 \\ 0 && 0 && 0 && 1  \end{bmatrix} \in SO(4)\,,\]
 and
 \[(-1)^w = \begin{bmatrix} 1 && 0 && 0 && 0 \\ 0 && 1 && 0 && 0 \\ 0 && 0 && -1 && 0 \\ 0 && 0 && 0 && -1  \end{bmatrix} \in SO(4)\,,\]
 as well as $C$ itself. Thus our fundamental $SO(4)$ bundle splits over the classifying space of the corresponding $\bZ_2$ gauge fields $A_n$, $A_w$, and $C$, respectively as $A_n + C$, $A_n$, $A_w + C$, and $A_w$, so the Euler class is
 \ie e(V_4) =& (A_n + C)A_n(A_w+C)A_w = A_n^2 A_w^2 + C^2 A_n A_w + C A_n^2 A_w + C A_n A_w^2  \\
 =& \frac{1}{2}d(A_n A_w^2 + C A_n A_w)\,.
 \fe
 The last equality allows us to identify the anomaly with the element
 \[\frac{1}{2} A_n A_w^2 + \frac{1}{2} C A_n A_w \in H^3(B \bZ_2^3,U(1))\,.\]
 The first piece is the familiar $\bZ_2$ shadow of the chiral anomaly, while the second piece causes symmetry transmutation when we gauge $C$ to go to the orbifold. Indeed, in that case we find that the $\bZ_2^n \times \bZ_2^w$ commutant becomes $D_8$ because of the extension class $A_n A_w$ which appears in the cubic piece. This is the symmetry at a generic orbifold point
 \[G_{\rm orb} = D_8\,.\]
 In terms of the presentation
 \[D_8 = \langle r,s\ |\ r^4 = s^2 = 1,~ srs = r^{-1}\rangle\,,\]
 we can write the two lifts of $\bZ_2^n$ and $\bZ_2^w$ as $s$ and $sr$. Meanwhile the magnetic symmetry is represented
 as $r^2$.
 
 In fact, whenever a symmetry such as $C$ only appears in a linear term in the anomaly polynomial, the result is a kind of symmetry transmutation \cite{Thorngren:2015gtw,Tachikawa:2017gyf}. To be precise, we introduce the $\bZ_2$ gauge field $M$ of the magnetic symmetry. It satisfies
 \[dM = A_n A_w\,,\]
 which leads to the $D_8$ group as claimed. Furthermore, and key for us is that this allows us to write the chiral anomaly as a counterterm:
 \[\frac{1}{2} A_n A_w^2 = d\left(\frac{1}{2} M A_w \right)\,.\]
 There are two choices here, which are evidently exchanged by $T$ duality, since we may also take
 \[\frac{1}{2} A_n A_w^2 = d\left(\frac{1}{2} M A_n + \frac{1}{4} A_n A_w \right)\,.\]
 These also differ by the unique choice of $D_8$ discrete torsion, which may be written as
 \[\frac{1}{2} M A_n + \frac{1}{2} M A_w + \frac{1}{4} A_n A_w\,.\]

 \subsection{The KT Point}\label{appKTsymmetry}
 
The  T-duality on the circle branch exchanges $U(1)^\theta$ and $U(1)^\phi$ but commutes with $C$. It thus descends to a duality of the orbifold which switches the lifts of the $\bZ_2$ subgroups of each of these $U(1)$'s, hence it acts by the outer automorphism of $D_8$ which exchanges $s$ and $sr$. This duality fixes the Kosterlitz-Thouless (KT) point, which is the $\mZ_2^C$ orbifold of the $SU(2)_1$ point, and we expect at this point the $D_8$ symmetry is enhanced. In fact the KT point is a junction between the circle and $\bZ_2$ orbifold branches and we will find the symmetry is enhanced to $G_{\rm bos} = (U(1)^{\tilde \theta} \times U(1)^{\tilde \phi}) \rtimes  \bZ_2^{\tilde C}$, where $\tilde \theta$ and $\tilde \phi$ are the circle branch fields describing the KT point and $\tilde C$ is the charge conjugation $\tilde \theta \mapsto - \tilde \theta$, $\tilde \phi \mapsto - \tilde \phi$. We reserve the untilded fields $\phi,\theta$ to describe the circle branch fields at the $SU(2)_1$ point.
 
  The $D_8$ symmetry is embedded into $G_{\rm bos}$ with $sr$ acting as $\tilde C$ and $r$ acting as a ${\pi\over 2}$ rotation in $U(1)^{\tilde \theta}$. The T-duality of the orbifold branch is realized by a $\pi\over 4$ rotation in $U(1)^{\tilde \theta}$ (which swaps $s$ and $sr$). We derive this as follows.
 
 At the $SU(2)_1$ point, we represent the $SO(4)$ group elements in the basis \eqref{SO4basis}. The normalizer of $\bZ_2^C$ is enlarged to $(U(1) \times U(1)) \rtimes \bZ_2$.  In particular, we see it includes a full rotation group $SO(2)^a$ in the $1-3$ coordinates and another $SO(2)^b$ in the $2-4$ coordinates. The elements $(-1)^n, (-1)^w$ we studied above act as simultaneous reflections in these $2 \times 2$ blocks, while their product, as well as $C$, is in the center. It is convenient to exchange the 2nd and 3rd coordinates so we have:
  \[SO(2)^a = \begin{bmatrix} \star && \star && 0 && 0 \\ \star && \star && 0 && 0 \\ 0 && 0 && 1 && 0 \\ 0 && 0 && 0 && 1  \end{bmatrix} \subset SO(4)\,,\]
  \[SO(2)^b = \begin{bmatrix} 1 && 0 && 0 && 0 \\ 0 && 1 && 0 && 0 \\ 0 && 0 && \star && \star \\ 0 && 0 && \star && \star  \end{bmatrix} \subset SO(4)\,,\]
 \[(-1)^n = \begin{bmatrix} -1 && 0 && 0 && 0 \\ 0 && 1 && 0 && 0 \\ 0 && 0 && -1 && 0 \\ 0 && 0 && 0 && 1  \end{bmatrix} \in SO(4)\,,\]
 \[(-1)^w = \begin{bmatrix} 1 && 0 && 0 && 0 \\ 0 && -1 && 0 && 0 \\ 0 && 0 && 1 && 0 \\ 0 && 0 && 0 && -1  \end{bmatrix} \in SO(4)\,,\]
 \[C = \begin{bmatrix} -1 && 0 && 0 && 0 \\ 0 && -1 && 0 && 0 \\ 0 && 0 && 1 && 0 \\ 0 && 0 && 0 && 1  \end{bmatrix} \in SO(4)\,,\]
  \[T = \begin{bmatrix} 0 && 1 && 0 && 0 \\ 1 && 0 && 0 && 0 \\ 0 && 0 && 0 && 1 \\ 0 && 0 && 1 && 0  \end{bmatrix} \in SO(4)\,.\]
Note that $C \in SO(2)^a$. The full normalizer group is $(SO(2)^a \times SO(2)^b) \rtimes \bZ_2^n$.
 
 In fact, we can re-express the $SU(2)_1$ point in terms of fields $\phi',\theta'$ related to $\phi,\theta$ by an $SO(4)$ rotation which produces the change of basis used above. In these variables, $SO(2)^a$ and $SO(2)^b$ act as shifts of $\phi'$ and $\theta'$, respectively. $C$ is a $\pi$ rotation in $SO(2)^a$, and when we gauge it the effect is to double the radius, such that the new circle branch variables at the KT point may be expressed as $\tilde \phi = \phi'/2$, $\tilde \theta = 2\theta'$ (compare with Section \ref{subsubsecZkselfdualgauging}). The resulting symmetry group is $G_{\rm bos} = (U(1)^{\tilde \theta} \times U(1)^{\tilde \phi}) \rtimes  \bZ_2^{\tilde C}$ as claimed.

  \section{Parafermions and self-dualities}
  \label{app:cosetduality}

In this appendix, we discuss in detail the symmetries in the $\mZ_k$ parafermion theory of \cite{Fateev:1985mm} and the relation to TDLs in the $SU(2)_k/U(1)$ coset CFT upon bosonization. 

\subsection{The $\mZ_k$ parafermions}
\label{app:Zkpf}
As briefly reviewed in Section~\ref{sec:TDLfromPF}, the fundamental fields in the $\mZ_k$ parafermion theory consist of the chiral parafermions $\psi_n(z)$ and its antichiral partners $\bar \psi_n(\bar z)$ which obey fractional mutual statistics as in Figure~\ref{fig:PFstat}.
The parafermions $\psi_n$ generate a chiral algebra that extends the Virasoro algebra at central charge $c$ (and similarly for the antichiral fields $\bar\psi_n)$,
\ie
\psi_n(z)\psi_m(0) =&{c_{n,m} \over z^{\Delta_n+\Delta_m-\Delta_{m+n}}}\times\begin{cases}
  \psi_{m+n} (0) +\cO(z) & {\rm for}~n+m<k\,,
 \\
 \psi_{n-m} (0)+\cO(z) & {\rm for}~m<n~{\rm and}~m+n>k\,,
  \\
 1+{2\Delta_n\over c}z^2T(z)+\cO(z^3) & {\rm for}~m<n~{\rm and}~m+n=k\,.
\end{cases}
\label{PFope}
\fe
In particular $\psi_n$ are Virasoro primaries with respect to the stress tensor $T(z)$
and their scaling dimensions $\Delta_n$ satisfy
\ie
\Delta_n+\Delta_m-\Delta_{m+n}={2mn \over k}~{\rm mod\,} \mZ\,,
\fe
as required for the mutual statistics of the parafermions. The parafermion algebra is completely specified by $(c,\Delta_k)$ and the OPE coefficients $c_{m,n}$ in \eqref{PFope}, which are subjected to constraints from associativity. There are infinitely many solutions to these constraints \cite{Fateev:1985mm}. Here we focus on the simplest parafermion theories given by
\ie
c={2(k-1)\over k+2}\,,\quad \Delta_k={n(k-n)\over k}\,,
\fe 
and the explicit form of $c_{n,m}$ can be found in \cite{Fateev:1985mm} which is not important for our discussion here. The bosonization of this parafermion theory corresponds to the $SU(2)_k/U(1)$ coset CFT which we review in the following.

\subsection{The $SU(2)_k/U(1)$ coset CFT}
  \label{app:coset}

The $SU(2)_k/U(1)$ coset CFT is constructed by taking a quotient of the $SU(2)_k$ WZW CFT by the current subalgebra $U(1)_{2k}$. The resulting theory has central charge
  \ie
  c={2(k-1)\over k+2}\,,
  \fe
and follows an ADE classification \cite{Gepner:1986hr} which is inherited from a choice of modular invariant in the parent $SU(2)_k$ WZW CFT. Here we will focus on the diagonal modular invariant for the $SU(2)_k/U(1)$ coset which is often referred to as the $A_{k+1}$ type.\footnote{The coset CFTs with non-diagonal modular invariants can be obtained from gauging a symmetric Frobenius algebra in the diagonal CFT at special $k$ \cite{Bhardwaj:2017xup}. For $SU(2)_{2k-4}/U(1)$ CFT of the $D_{k}$ type, this amounts to gauging the $\mZ_2$ center of the symmetry \eqref{cosetsym} in the $A_{2k-3}$ type coset CFT, and the resulting theory has $D_{2k-4}\times \mZ_2^M$ global symmetry with a mixed anomaly between the two factors \cite{Tachikawa:2017gyf}.} 

Physically the coset construction amounts to gauging an anomaly-free $U(1)$ subgroup of the $SO(4)={(SU(2)_L\times SU(2)_R)/\mZ_2}$ global symmetry in the WZW model. Such a $U(1)$ subgroup is unique up to conjugation by elements in $SO(4)$ and commonly referred to as the $U(1)_V$ vector subgroup, whose commutant in $SO(4)$ is a $U(1)_A$ axial subgroup.

In general, potential global symmetries of a gauge theory with continuous gauge group $G$  arise from the quotient $N_G(\hat G)/G$ of the normalizer of $G$ in the symmetry $\hat G$ of the ungauged theory. Here the normalizer of $U(1)_V$ in $SO(4)$ is 
\ie
N_{U(1)_V}(SO(4))=(U(1)_V\times U(1)_A)\rtimes \mZ_2\,,
\fe
where the $\mZ_2$ acts on both $U(1)$ factors by complex conjugation. However the 't Hooft anomaly of the WZW model governed by the anomaly four-form ${\cal I}_4=ke_4(SO(4))$
induces an Adler-Bell-Jackiw (ABJ) anomaly for $N_{U(1)_V}(SO(4))/U(1)_V$ upon gauging $U(1)_V$. Consequently, the global symmetry of the coset theory is\footnote{For $k=2$ which corresponds to the Ising CFT, the second $\mZ_2$ factor in \eqref{cosetsym} acts trivially and the faithful global symmetry is just $\mZ_{k=2}$.}
\ie
G_{SU(2)_k/U(1)}=\mZ_k \rtimes \mZ_2 =D_{2k}\,.
\label{cosetsym}
\fe 

Coset CFTs are in general RCFTs with respect to certain chiral algebras that descend from the parent current algebras by taking the commutant of the subalgebras that we gauge. Here the $SU(2)_k/U(1)$ coset CFT bosonizes the $\mZ_k$ parafermions, and consequently the relevant chiral algebra $\cV_k\equiv \cV(SU(2)_k/U(1)) $ is also an integer-spin subalgebra of the chiral parafermion algebra generated by $\psi_n$ \cite{Zamolodchikov:1985wn,Blumenhagen:1990jv,Hornfeck:1992tm,deBoer:1993gd,Blumenhagen:1994wg,Dong:2009xd},
\ie
\cV_k=\begin{cases}
 W(2)\cong {\rm Vir}_{c={1\over2}} & k=2\,,
 \\
 W(2,3) & k=3\,,
 \\
 W(2,3,4) & k=4\,,
 \\
 W(2,3,4,5) & k\geq 5\,,
\end{cases}
\label{cosetchiralalg}
\fe
where $W(2,3,\dots)$ denotes a W-algebra with strong generators of spins $2,3$ and so on.\footnote{The chiral algebra $\cV_k$ is invariant under the $\mZ_k$ symmetry of the coset CFT. Under the $\mZ_2$ conjugation in \eqref{cosetsym}, each generator of $\cV_k$ picks up a sign $(-1)^s$ depending on its spin $s$, similarly for the antichiral algebra.} 
The primary operators in the coset CFT with respect to $\cV_k$ (and its anti-chiral partner) are obtained from the $SU(2)_k$ current primaries $\Phi^{j,\bar j}_{m,\bar m}$ by factoring out $U(1)_{2k}$ charged states which can be parametrized by a compact boson $(X_L,X_R)$ at radius $R=\sqrt{2k}$ \cite{Gepner:1986hr},\footnote{Here our convention is that $j,\bar j=0,{1\over 2},\dots,{k\over 2}$ label the $SU(2)$ spins and $m,\bar m\in \mZ/2$ satisfy $-j\leq m\leq j,-\bar j\leq \bar m\leq \bar j$.} 
  \ie
  \Phi^{j,\bar j}_{m,\bar m}=\phi^{j,\bar j}_{m,\bar m} (z) e^{{2i\over \sqrt{2k}}
  (m X_L(z)+\bar m X_R(\bar z))}\,.
  \fe
The coset primary is denoted by $\phi^{j,\bar j}_{m,\bar m}$ and has scaling dimensions,
  \ie
  (h,\bar h)=\left({j(j+1)\over (k+2)}-{m^2\over k},{\bar j(\bar j+1)\over (k+2)}-{\bar m^2\over k}\right),\quad {\rm for~} |m|<j~{\rm and}~|\bar m|<\bar j\,.
  \fe
With respect to the global symmetry \eqref{cosetsym} of the coset CFT, the operator $\phi^{j,\bar j}_{m,\bar m}$ carries $\mZ_k$ charge $m+\bar m$ and transforms to $\phi^{j,\bar j}_{-m,-\bar m}$ under the $\mZ_2$ conjugation.\footnote{Note that $j-\bar j \in \mZ$ irrespective of the choice of the ADE modular invariant \cite{Gepner:1986hr} and thus $m+\bar m \in \mZ$.} It is convenient to extend the range of the $m,\bar m$ indices with the identifications
  \ie
\phi^{j,\bar j}_{m,\bar m}=\phi^{k/2-j,k/2-\bar j}_{k/2+m,k/2+\bar m}=\phi^{j,\bar j}_{m+k,\bar m}=\phi^{j,\bar j}_{m,\bar m+k}\,.
\label{cosetopid}
\fe
Consequently we use the following convenient set of indices $(j,m)$ with $m+j\in \mZ$ to parameterize the coset primaries,
 \ie
  {\rm Even}~k~\{(j,m)\}=&\{0\leq j \leq  {k-2\over 4},-j\leq m\leq k-j-1\} \cup  \{ j={k\over 4}, -j\leq m\leq {k\over 2}-j-1\}   \,,	   	\\
{\rm Odd}~k~\{(j,m)\}=&\{0\leq j \leq  {k-1\over 4}, -j\leq m\leq k-j-1\}\,,
\label{cosetjmlist}
  	   	   \fe
 and similarly for the antichiral indices $(\bar j,\bar m)$. There are ${k(k+1)\over 2}$ independent primaries on each side. The chiral fusion rule of these primary operators is
  \ie
  \phi^{j_1}_{m_1} \otimes \phi^{j_2}_{m_2} =\bigoplus_{j=|j_1-j_2|}^{\min (j_1+j_2,k-j_1-j_2)} \phi^j_{m_1+m_2}
  \label{cosetfusion}
 \fe  
and similarly for the antichiral side. 
  
The character for the coset chiral primary $\phi^j_m$ is often denoted by $\eta(\tau)c^{2j}_{2m}(\tau)$, which is obtained from factorizing the $SU(2)_k$ characters $\chi_j^{k}(\tau)$ in terms of the $U(1)_{2k}$ characters $K_n^k(\tau)$,
  \ie
 \chi_j^{k}(\tau,z)=\sum_{n=-k+1}^{k} \eta(\tau)c^{2j}_{n} (\tau) K^k_{n}(\tau)\,.
  \fe
The so-called string functions $c^\ell_n(\tau)$ for $\ell-n\in 2\mZ$ satisfy the identities
  \ie
  c^\ell_n=c^{\ell}_{n+2k}=c^{k-\ell}_{k+n}=c^{\ell}_{-n}\,
  \fe
  where the first three equalities follow from \eqref{cosetopid} and the last from the $\mZ_2$ symmetry in \eqref{cosetsym}. 
 
  The local operator content of the $SU(2)_k/U(1)$ CFT of the $A_{k+1}$ type consists of primaries $\phi^{j,j}_{m,m}$ for $(j,m)$ given in \eqref{cosetjmlist} and their $\cV_k$ descendants.\footnote{In the main text, we denote the local operators $\phi^{j,j}_{m,m}$ as $\phi^{j}_{m}$ for convenience.} The torus partition function reads
  \ie
  Z(\tau,\bar \tau)=|\eta(\tau)|^2\sum_{(j,m)}|c^{2j}_{2m}(\tau)|^2\,.
  \fe
  
As explained in Section~\ref{sec:verlinde}, the $SU(2)_k/U(1)$ CFT contains obvious TDLs in the form of Verlinde lines $\cL_{(j,m)}$ associated to each of the coset primaries above labelled by $(j,m)$. These TDLs preserve the extended $\cV_k$ algebra (and its antichiral version) and follow the same fusion rules as in \eqref{cosetfusion}, therefore furnishing a $\Rep(\cV_k)$ fusion category symmetry in the coset CFT. In particular $\Rep(\cV_k)$ contains the $\mZ_k$ symmetry TDLs which correspond to $\cL_{(0,m)}$ with $m=0,1,\dots,k-1$. The full category symmetry of the coset CFT is much richer and come from TDLs that only commute with a subalgebra of $\cV_k$ that contains the Virasoro subalgebra. This includes as independent generators, the $\mZ_2$ conjugation symmetry in \eqref{cosetsym} and also a duality defect $\cN$ which generalizes the Kramers-Wannier duality of the Ising CFT (see Table~\ref{tab:PFsym}). 
 
While only the diagonal coset primaries appear as local operators in the $A_{k+1}$ type $SU(2)_k/U(1)$ CFT, more general coset operators of the type $\phi^{j,j}_{m,\bar m}$ reside in the sector twisted by an element $a_1=m-\bar m\in \mZ_k$. They contribute to the twisted partition function with $\mZ_k$ holonomies $a_1,a_2$ around the spatial and temporal cycles on $T^2$ respectively \cite{Gepner:1986hr},
\ie
Z(\tau,\bar \tau;a_1,a_2)={1\over 2}|\eta(\tau)|^2 \sum_{n=-k+1}^k \sum_{\ell=0}^{k} \omega^{(n+a_1)a_2}c^\ell_{n+2a_1}(\tau) c^{  \ell}_{n}(\bar \tau)\,.
\label{cosettPF}
\fe
The twisted partition function satisfies
\ie
Z(\tau,\bar \tau;a_1,a_2)={1\over k}\sum_{b_1,b_2\in \mZ_k} Z(\tau,\bar \tau;b_1,b_2)\omega^{b_1 a_2-b_2 a_1}={1\over k}\sum_{b_1,b_2\in \mZ_k}  Z(\tau,\bar \tau;b_1,b_2)\bar\omega^{b_1 a_2-b_2 a_1}
\label{T2Zksd}
\fe
which confirms the two self-dualities of the theory under $\mZ_k$ gauging with the  bicharacter in \eqref{eqnpartfnselfdualgauging} given by $\chi_\pm(a,b) \equiv \pm {2\pi ab\over k}$ .

\subsection{Operators and symmetries in the parafermion theory}

Similar to the case of Majorana fermion, a general operator $\cO$ in the parafermion theory leads to a $\mZ_k$ valued monodromy for the parafermion field $\psi_1$,
\ie
\psi_1(e^{2\pi i}z) \cO(0)=\omega^{s}\psi_1(z)\cO(0)
\fe
and corresponds to a state in the Hilbert space $\cH^{PF}_s$ on $S^1$ with the corresponding twisted periodicity for $\psi_1$. For $k=2$, the relations to the usual fermionic Hilbert spaces are $\cH^{PF}_0=H^F_{NS}$ and $\cH^{PF}_1=H^F_{R}$.\footnote{Because of the nontrivial mutual statistics among $\psi_n$ (see Figure~\ref{fig:PFstat}), the chiral parafermion $\psi_1$ is not local with respect to itself unless $k=2$ and thus $\psi_1$ (and general $\psi_n$) are not contained in $\cH_0^{PF}$.} 

Thanks to the bosonization/parafermionization map \eqref{PFandB}, we can describe the parafermion operators using the coset primaries $\phi^{j,\bar j}_{m,\bar m}$ reviewed in the last section.   
We also impose an additional constraint $j=\bar j$ which is consistent with the OPE and write for simplicity
\ie
\phi^j_{m,\bar m}\equiv \phi^{j,j}_{m,\bar m}\,.
\fe
In particular, the parafermion fields  in Section~\ref{app:Zkpf} are realized by
\ie
\psi_n \equiv \phi^{0}_{n,0}\,,\quad \bar \psi_n \equiv \phi^{0}_{0,n}\,.
\fe
The mutual statistics for two such operators (generalizing Figure~\ref{fig:PFstat})  is 
\ie
 \phi^{j}_{m,\bar m}(z,\bar z)\phi^{j'}_{m',\bar m'}(w,\bar w)= \omega^{{\bar m\bar m'-mm'}}\phi^{j'}_{m',\bar m'}(w,\bar w)\phi^{j}_{m,\bar m}(z,\bar z)\,.
\fe
Consequently, the operators corresponding to the states in the twisted parafermion Hilbert space $\cH^{PF}_s$ are $\phi^{j}_{s/2,\bar m}$. There are two distinguished operators in each $\cH^{PF}_s$ which are a pair of parafermion primaries
\ie
\sigma_s = \phi^{s/2}_{s/2,s/2}\,,\quad \mu_s = \phi^{s/2}_{s/2,-s/2}\,,
\fe
that generalize the order and disorder spin operators for the Majorana fermion and its bosonization, the Ising CFT at $k=2$. All other operators arise from consecutive actions of the modes of the basic parafermions $\psi_1,\bar \psi_1$ on these (generalized) spin operators (including the identity operator at $s=0$) \cite{Fateev:1985mm}. In particular, they include energy (thermal) operators $\ep_r$ for $r=1,2,\dots,\left \lfloor{k\over 2}\right \rfloor$,
\ie
\ep_r=\phi^{r}_{0,0}\,.
\fe

\begin{table}[!htb]
    \centering
    \begin{tabular}{c|c|c|c}
         & $\mZ_k\times \bar\mZ_k$ & $\mZ_2^C$ & $\mZ_2^S$ \\\hline 
     $\phi^{j}_{m,\bar m}$ & $({m+\bar m},m-\bar m)$ & $\phi^{j}_{-m,-\bar m}$ & $(-1)^{m+j}\phi^{j}_{m,-\bar m}$
     \\\hline 
     $\psi_n$ &   $(n,n)$  &  $\psi_{k-n}$ & $(-1)^n\psi_n$\\
      $\bar\psi_n$    & $(n,-n)$ &  $\bar\psi_{k-n}$ & $\bar \psi_{k-n}$\\
       $\sigma_n$    & $(n,0)$  & $\sigma_{k-n}$ & $\mu_n$ \\
              $\mu_n$    & $(0,n)$  & $\mu_{k-n}$ & $\sigma_n$\\
       $\epsilon_r$    & $(0,0)$  & $\epsilon_r$ & $(-1)^r\epsilon_{r}$\\ \hline        \end{tabular}
    \caption{Global symmetries of the $\mZ_k$ parafermion theory.}
    \label{tab:PFsym}
\end{table}

The full global symmetry of the parafermion theory is,
\ie
G_{PF}=(\mZ_k\times \bar\mZ_k)\rtimes (\mZ_2^C\times \mZ_2^{S})\,,
\label{PFsym}
\fe
where $\mZ_2^{C}$ and $\mZ_2^{S}$ act on the $\mZ_k\times \bar\mZ_k$ by conjugating and swapping the two factors respectively. The operators of the parafermion theory transform under \eqref{PFsym} as summarized in Table~\ref{tab:PFsym}. Here $\bar \mZ_k$ is the generalization of the $(-1)^F$ symmetry at $k=2$, while $\mZ_k$ is identified with the $\mZ_k^B$ symmetry of the bosonized theory.\footnote{For notational simplicity, $\mZ_k^B$ and $\mZ_2^C$ are denoted as $\mZ_k$ and $\mZ_2$ in the last section.} The bosonization of the parafermion theory amounts to summing over $\bar \mZ_k$ twisted sectors and projecting to the $\bar \mZ_k$ invariant states. Consequently, the bosonized theory only contain operators of the form $\phi^{j}_{m,m}$, which constituents the diagonal modular invariant (of the $A_{k+1}$ type) for the $SU(2)_k/U(1)$ coset CFT. The other more general parafermion operators carry nonzero $\bar\mZ_k$ charges reside in the $\mZ_k^B$ twisted sectors, as reviewed in the last section.

The full global symmetry of the bosonic $SU(2)_k/U(1)$ coset CFT is given in \eqref{cosetsym} and we see the $\mZ_2$ conjugation is also inherited from the $\mZ_2^C$ symmetry in  the parafermion theory. The $\mZ_2^S$ symmetry in \eqref{PFsym} however does not lead to an ordinary symmetry in the bosonized theory. For $k=2$, this is simply the left fermion parity $(-1)^{F_L}$ which is known to generate the Kramers-Wannier duality under bosonization. Here we will see that for general $k$, the $\mZ_2^S$ symmetry of the $\mZ_k$ parafermions corresponds to a generalized Kramers-Wannier duality for $SU(2)_k/U(1)$ coset CFTs, namely self-duality under gauging the $\mZ_k^B$ symmetry with the bicharacter $\chi_+(a,b)={2\pi a b\over k}$, as evident from the first equality in \eqref{T2Zksd} satisfied by the torus partition function in general $\mZ_k^B$ background.

The parafermionic partition function with the paraspin structure $(s_1,s_2)$ on $T^2$ is obtained from \eqref{PFandB} applied to \eqref{cosettPF} \cite{Yao:2020dqx},
\ie
Z_{PF}(\tau,\bar\tau;s_1,s_2)={1\over 2}|\eta(\tau)|^2 \sum_{a=0}^{k-1} \sum_{\ell=0}^{k} \omega^{-(a+s_1)s_2}
\left(
c^\ell_{s_1}(\tau) c^{  \ell}_{s_1+2a}(\bar \tau)+
c^\ell_{s_1+k}(\tau) c^{  \ell}_{s_1+k+2a}(\bar \tau)
\right)\,,
\fe
which counts the states in the twisted sector $\cH_{s_1}^{PF}$ with an insertion of the $\mZ_k$ group element corresponding to $s_2$ along the spatial cycle of the $T^2$.
The condition \eqref{PFrel1} for self-duality in the parafermion theory, after projecting to sectors with definite $\bar \mZ_k$ charges, simply requires a symmetry between states of charge $q$ and $s-q$ in the twisted sector $\cH_s^{PF}$, which follows from the $\mZ_2^S$ symmetry in Table~\ref{tab:PFsym}. Furthermore, the $\mZ_2^S$ symmetry exchanges the order spin operators $\sigma_n$ and disorder spin operators $\mu_n$, which is expected for the $\mZ_k$ duality since the latter live in the $\mZ_k^B$ twisted sector of the bosonic coset CFT. Finally the $\mZ_k^B$ symmetric operators (i.e. the energy operators $\ep_r$) also carry definite parity under the duality  \cite{Fateev:1990bf}.
The duality defect is not a Verlinde line in the $SU(2)_k/U(1)$ coset CFT since $\mZ_2^S$ acts nontrivially on the  chiral algebra $\cV_k$ \eqref{cosetchiralalg}. Under the duality, all left-moving generators are invariant, whereas on the right-moving side, the generators of spin $\ell$ transform by a sign $(-1)^\ell$. This needs to be taken into account when writing down the duality twisted partition functions of the coset CFT. A similar analysis as above applies to the diagonal  $\mZ_2\subset \mZ_2^C\times \mZ_2^S$ symmetry which leads to the condition \eqref{PFrel2}, corresponds to the second self-duality in the coset CFT, and explains the second equality in \eqref{T2Zksd}. For this latter self-duality, it is the left-moving chiral algebra generators that transform with a sign $(-1)^\ell$. 

\subsection{The duality defects and $\mZ_k$ Tambara-Yamagami Symmetries}

As explained in the last section, the $\mZ_2^S$ symmetry generator $S$ of the $\mZ_k$ parafermions, upon bosonization, gives rise to a $\mZ_k$ duality defect $\cN$. For $k>2$ the same procedure also works for the other chiral  symmetry $CS$ where $C$ generates the $\mZ_2^C$ in \eqref{PFsym}. Therefore we have in fact  two such duality defects $\cN$ and $C\cN$ in the $SU(2)_k/U(1)$ coset CFT, which lead to the two equalities between the $\mZ_k$ twisted partition functions in \eqref{T2Zksd} following the general discussion in Section~\ref{subsecoverviewselfdual}.
Since $C$ acts as an automorphism on $\mZ_k$, the two duality defects have identical fusion rules and generate two $\mZ_k$ TY symmetries. 

Such fusion categories are classified in \cite{tambarayam}. 
For $k>2$, they are parametrized by the bicharacter $e^{i\chi(a,b)}=\omega^{ab}$ where $\omega$ is a primitive $k$-root of unity and the Frobenius-Schur indicator $\ep=\pm$. Comparing the $\mZ_k$ and $\bar \mZ_k$ charges of the order and disorder spin operators given in Table~\ref{tab:PFsym} that are related by the parafermionic symmetries $S$ and $CS$, we see that the corresponding bicharacters for the duality defects $\cN$ and $C\cN$ are $\chi_+$ and $\chi_-$ respectively (see Section~\ref{subsecoverviewselfdual}).
Below we provide further evidence that the TY symmetries in the coset CFT correspond to the two categories $\TY(\mZ_k,\chi_\pm,1)$  (see~Table~\ref{table:cosetTDLs}).

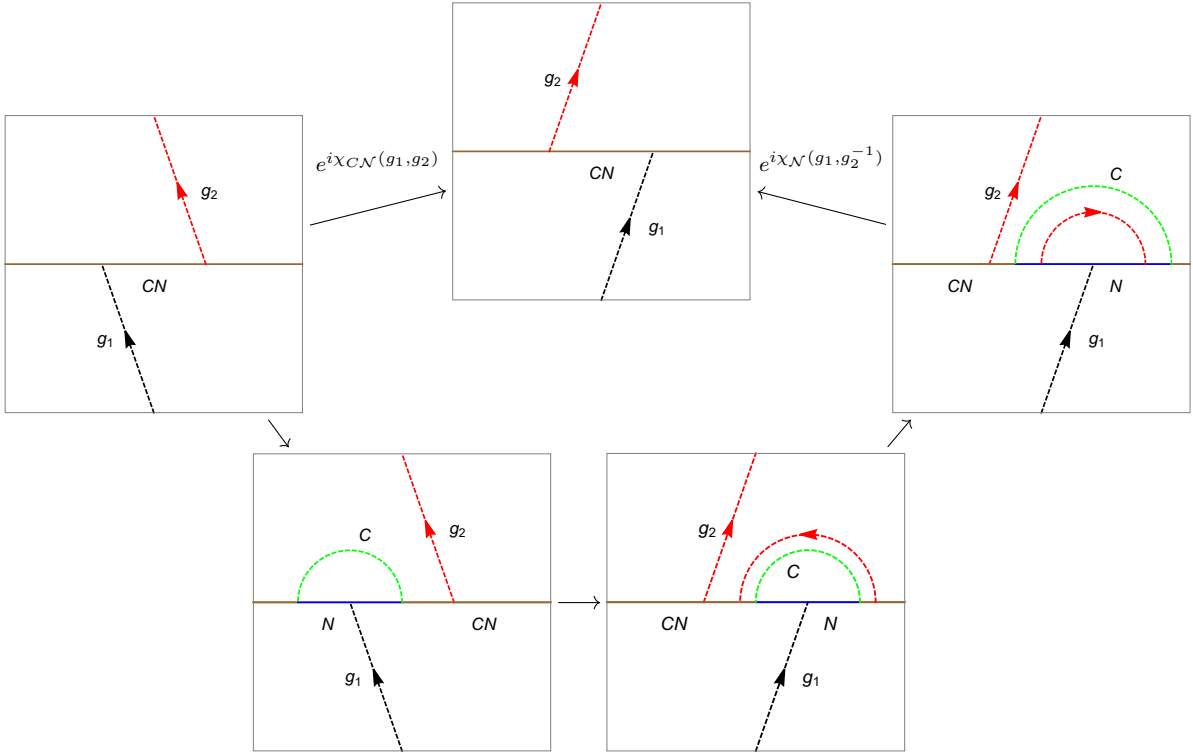
\begin{figure}[htb]
	\centering
	\begin{tikzpicture}
	\node[inner sep=0pt] (A) at (-.5,1.5)
	{\includegraphics[width=.25\textwidth]{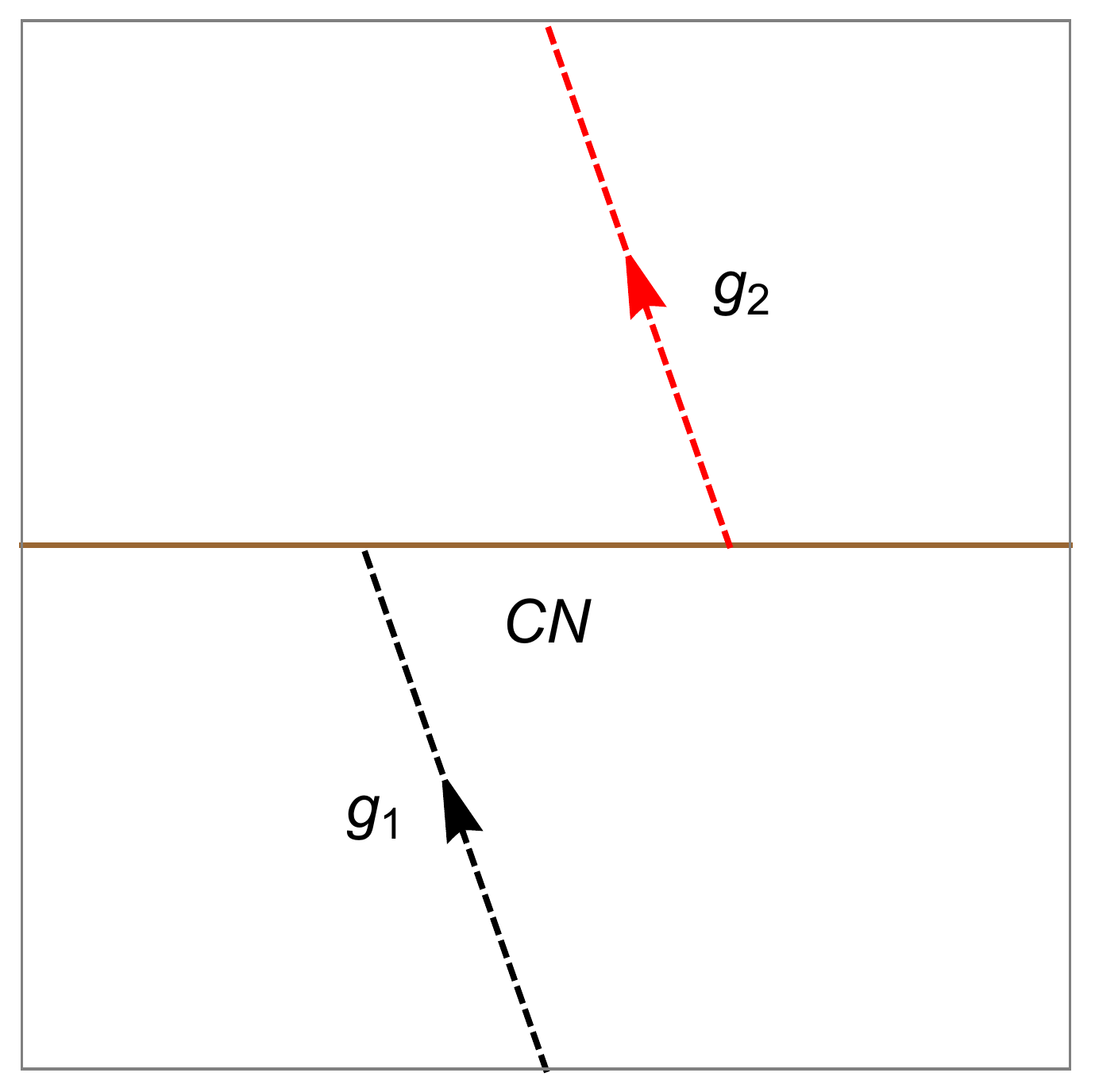}};
	\node[inner sep=0pt] (B) at (2.8,-3)
	{\includegraphics[width=.25\textwidth]{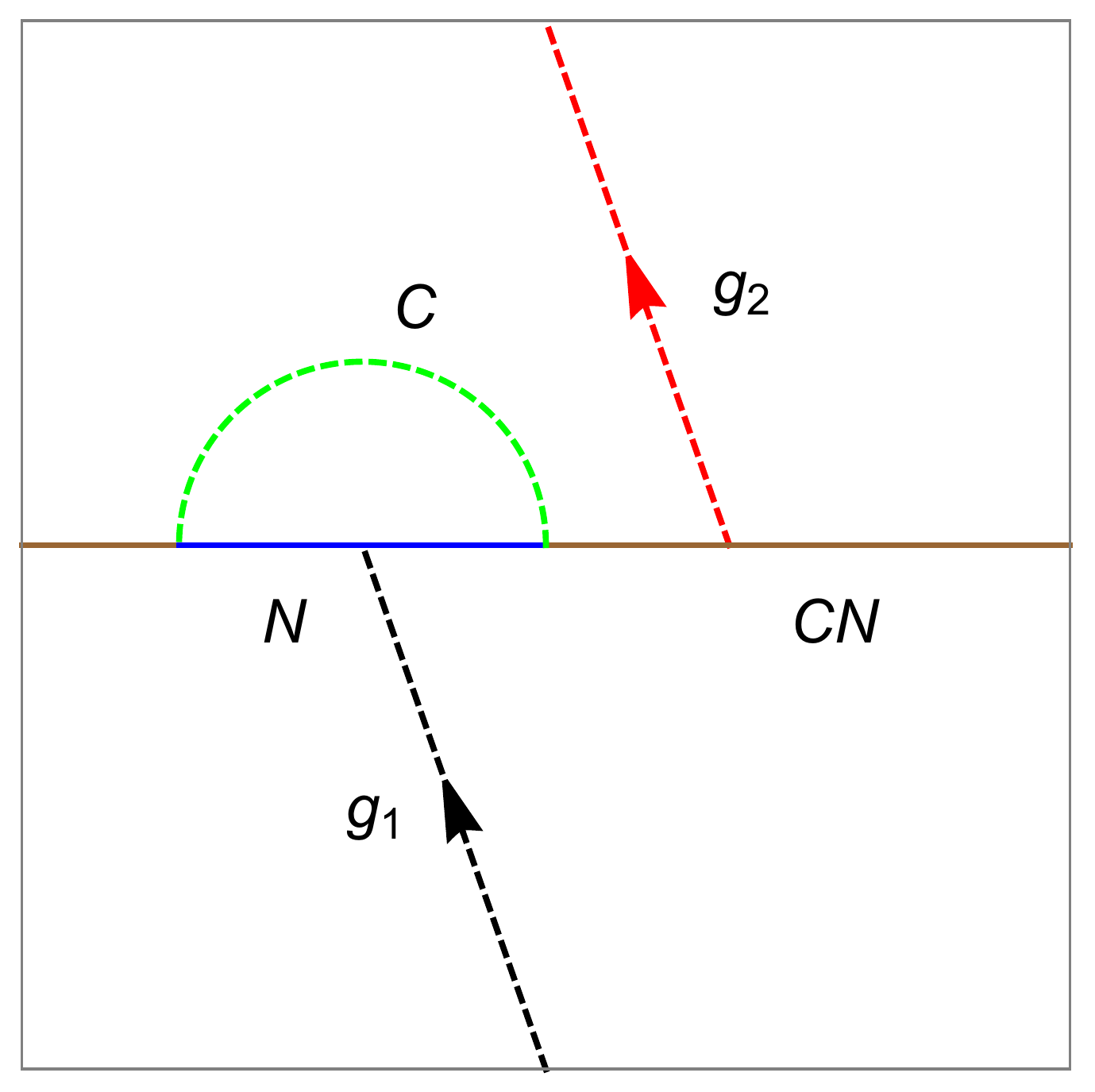}};
	\node[inner sep=0pt] (C) at (7.5,-3)
	{\includegraphics[width=.25\textwidth]{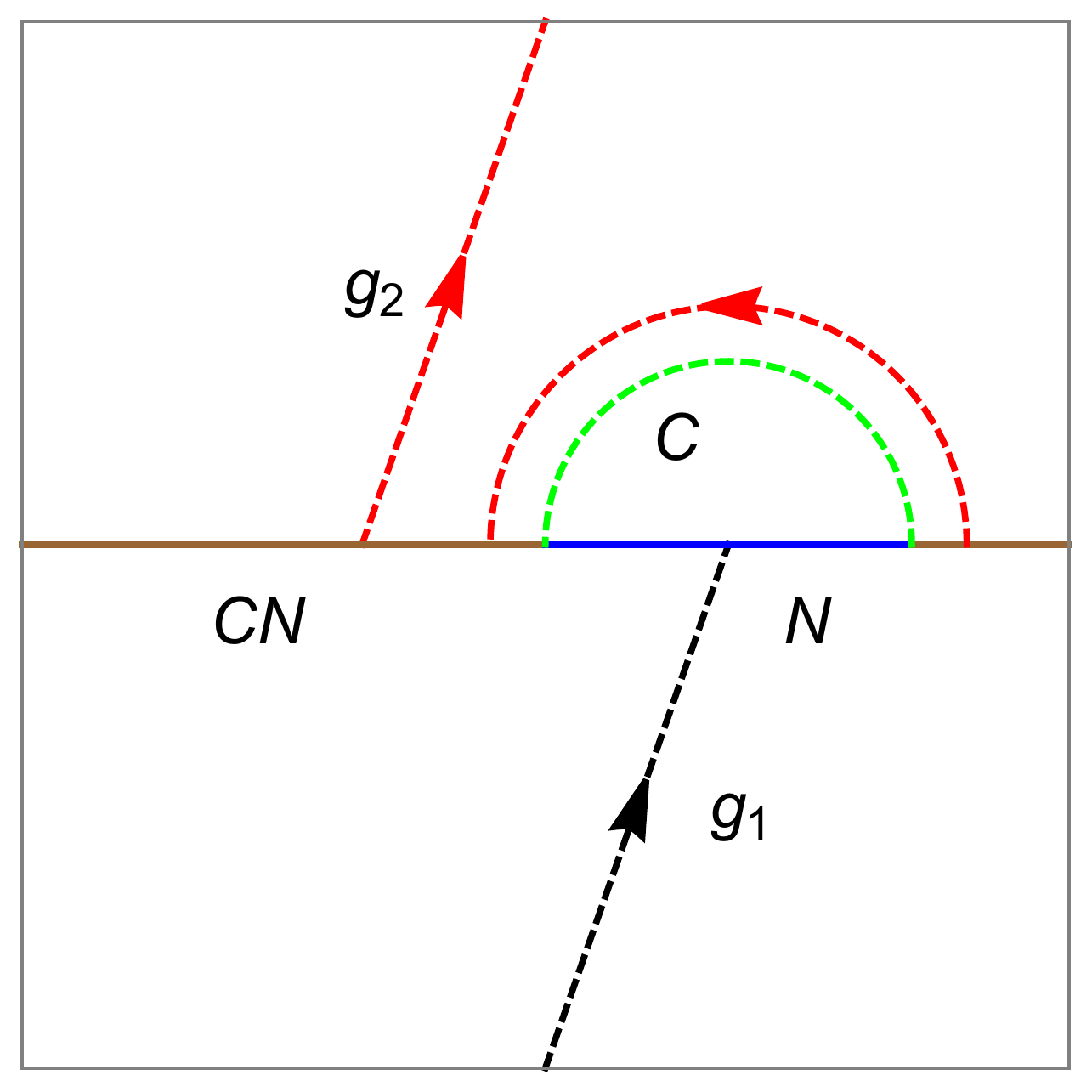}};
	\node[inner sep=0pt] (D) at (11.3,1.5)
	{\includegraphics[width=.25\textwidth]{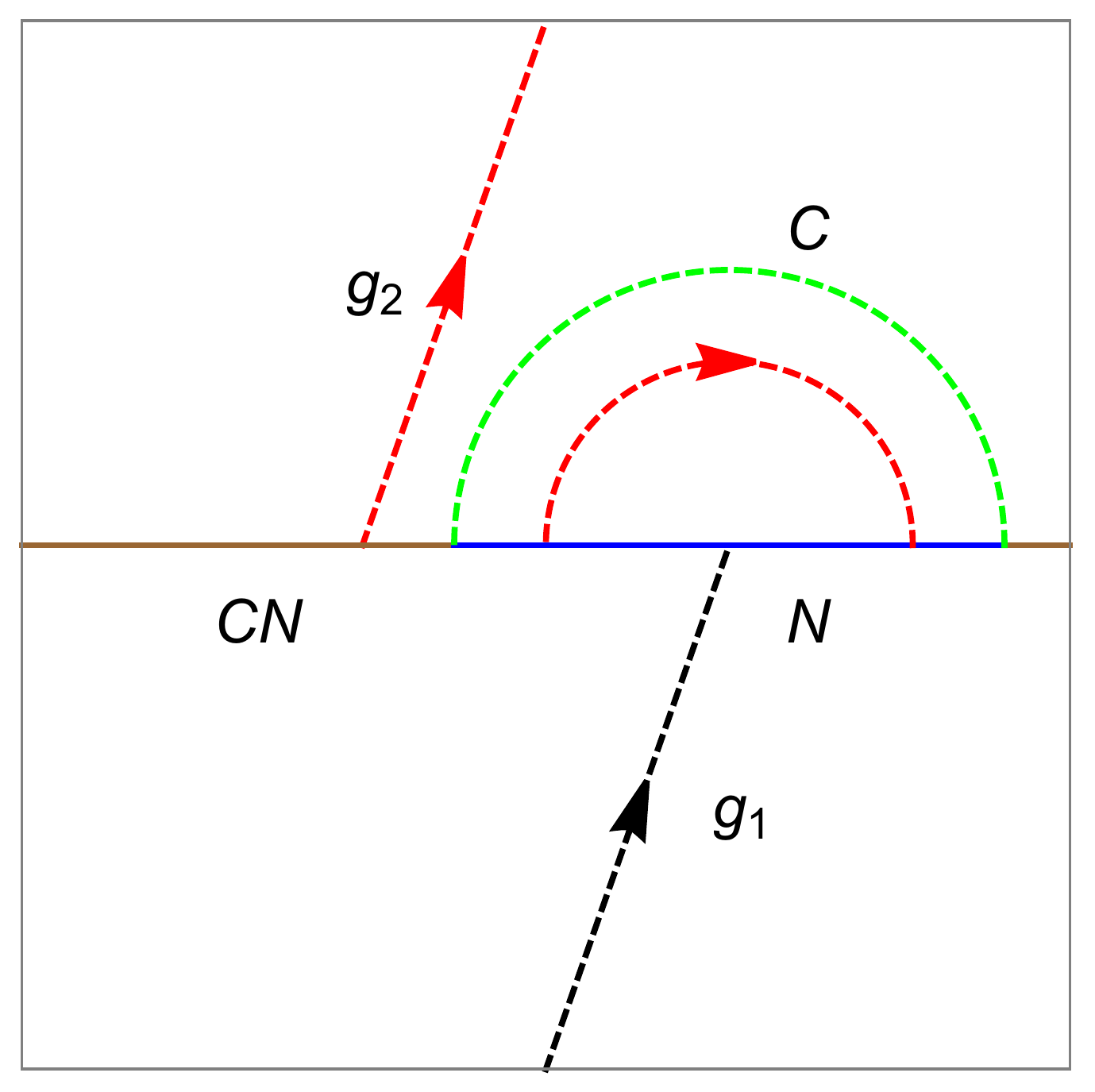}};
	\node[inner sep=0pt] (E) at (5.45,3)
	{\includegraphics[width=.25\textwidth]{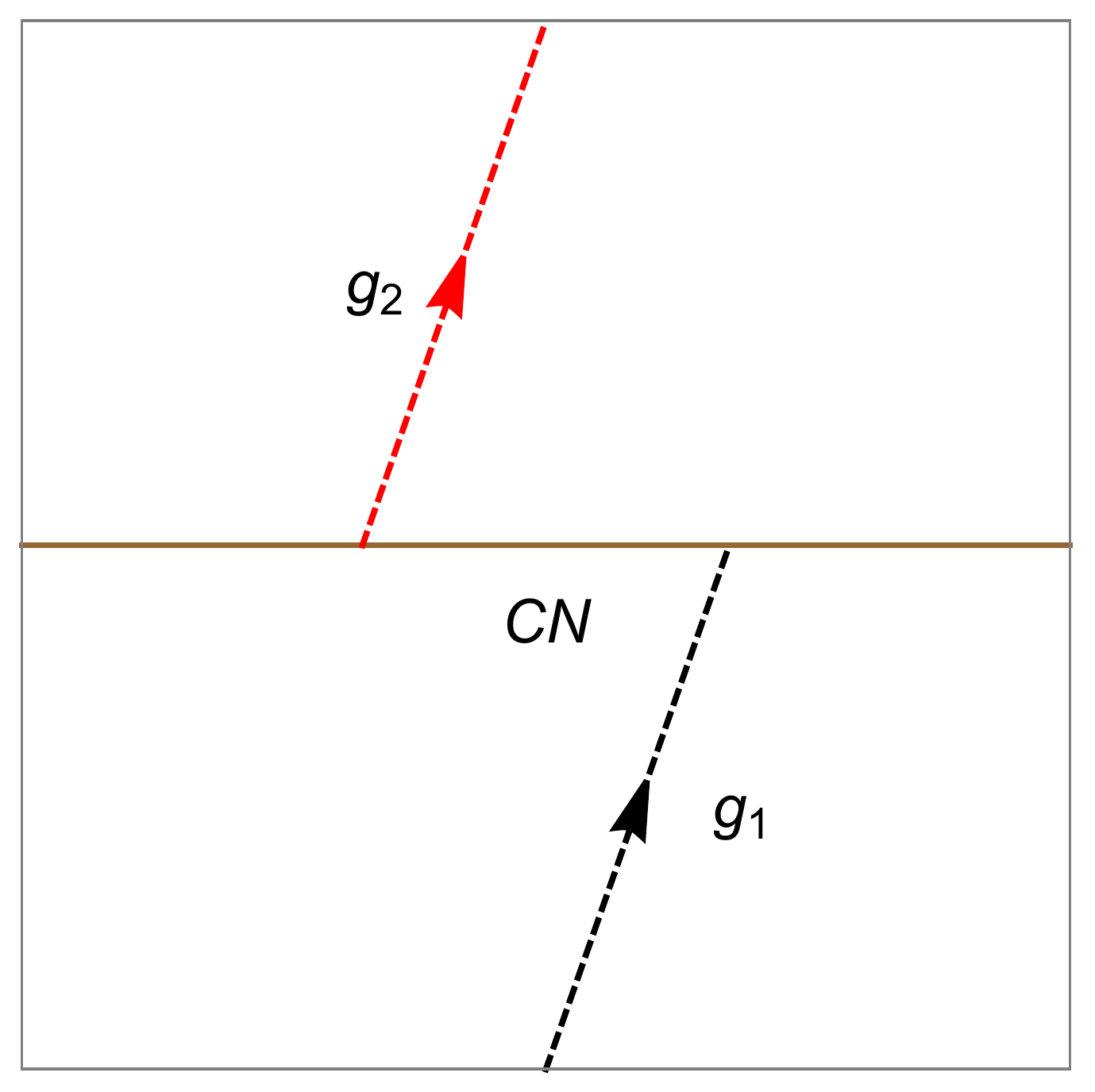}};
	\draw[->] (A) to [anchor=west] node [above right=-2] {} (B);
	\draw[->] (B) to [anchor=west] node [above=-2] {} (C);
	\draw[->] (C) to [anchor=west] node [above left=-2] {} (D);
	\draw[->] (D) to [anchor=west] node [above=10] {\scriptsize $e^{i\chi_{\cN}(g_1,g_2^{-1})}$} (E);
	\draw[<-] (E) to [anchor=west] node [above=10] {\scriptsize $e^{i\chi_{C\cN}(g_1,g_2)}$} (A);
	\end{tikzpicture}
	\caption{Two sequences of F-moves that go from the TDL configuration in the leftmost figure to the top figure. The equivalence between them relates the bicharacters associated to the two duality defects $\cN$ and $C\cN$ by $e^{i\chi_{C\cN}(g_1,g_2)}=e^{i\chi_{\cN}(g_1,g_2^{-1})}=e^{-i\chi_{\cN}(g_1,g_2)}$.
		}
	\label{fig:stackf}
\end{figure}

Firstly, the two duality TDLs $\cN$ and $C\cN$ must have the identical FS indicator but conjugate bicharacters. The former follows since $C$ has trivial F-symbols (no 't Hooft anomaly) whereas the latter can be derived from F-moves as in \cite{Thorngren:2019iar} (see Figure~\ref{fig:stackf}).\footnote{We note that these relations between the F-symbols of the underlying fusion categories do not depend on specifics of the underlying CFT and it would be interesting to understand more general relations of this type.} Next, the spin spectrum in the defect Hilbert space $\cH_\cN$ (or $\cH_{C\cN}$) of the duality TDL is constrained by both $\chi$ and $\ep$. For $\chi=\chi_\pm$, the spin selection rules are given in \eqref{Zkssr},
\ie
e^{4\pi i s}=\begin{cases}
 \ep e^{\pm{ \pi i \over 4k }(2n+1)^2} e^{\mp \pi i /4} & k \in 2\mZ \,,
 \\
 \ep e^{\pm{ \pi i \over k }n^2} e^{\mp \pi i /4} & k \in 2\mZ+1 \,.
\end{cases}
\fe
Therefore knowledge of the partition function for the defect Hilbert space $\cH_\cN$ can be used to fix $\ep$. For $k=2,3$, by inspecting the spin spectrum of $\cH_\cN$ in the Ising and 3-state Potts CFT, one finds that $\ep=1$ \cite{Chang:2018iay}. For $k=4$, we deduce that $\ep=1$ from the bootstrap analysis in Section~\ref{sec:bootstrapZk}. We conjecture that the duality defects for general $k$ in the $SU(2)_k/U(1)$ coset CFT have $\ep=1$. This can be proven for example by working out the duality twisted partition functions $Z_{1\cN}$ and we leave that to the interested readers.

 \section{$c=1$ CFT along the orbifold branch and $D_8$ discrete gauging}\label{appD8gauging}
At a generic point along the orbifold branch of the $c=1$ moduli space, the CFT has $D_8$ global symmetry. Here we consider discrete gauging by subgroups of $D_8$.  

 We adopt the standard representation $D_8=\la r,s|r^4=s^2=(sr)^2=1\ra$, whose conjugacy classes and corresponding stabilizers are
\ie
\{1\}:&~D_8\,,
\\
\{r^2\} :&~D_8\,,
\\
\{s, r^2 s\} :&~\mZ_2\times \mZ_2=\{1,r^2,s,  r^2s\}\,,
\\
\{ rs, r^3 s\} :&~\mZ_2\times \mZ_2=\{1,r^2, rs,  r^3s\}\,,
\\
\{r, r^3\} :&~\mZ_4=\{1,r,r^2,r^3\}\,.
\fe
The distinct gaugings of the CFT will be labeled by subgroups $G\subset D_8$ up to conjugation, and possible discrete torsion classes from $H^2(G,U(1))$ which we will refer to as $\iota(G) \in \mZ_2$. In particular, we have 3 conjugacy classes of $\mZ_2$ subgroups, 2 conjugacy classes of $\mZ_2\times \mZ_2$ subgroups (which we will label as $A$ and $B$ types) and 1 conjugacy class of $\mZ_4$ subgroup. As we will see, depending on these data, the discrete gauging may map one point on the orbifold branch with partition function $Z_{\rm orb}(R_1)$ to a point on the unorbifolded branch $Z_{\rm circ}(R_2)$ or  another point on the orbifold branch $Z_{\rm orb}(R_2)$. Let us briefly summarize our findings below and the details will be giving in the subsequent sections.
\ie
&Z_{\rm orb}(R) \xrightarrow{\mZ_2^{r^2}}Z_{\rm circ}(R)\,,\\ 
&Z_{\rm orb}(R) \xrightarrow{\mZ_2^{s}}Z_{\rm orb}(2R)\,,\quad
Z_{\rm orb}(R) \xrightarrow{\mZ_2^{sr }}Z_{\rm orb}(R/2)\,,
\\
&Z_{\rm orb}(R) \xrightarrow[\iota=0]{(\mZ_2\times \mZ_2)_A}Z_{\rm circ}(2R)\,,\quad Z_{\rm orb}(R) \xrightarrow[\iota=0]{(\mZ_2\times \mZ_2)_B}Z_{\rm orb}(R)\,,
\\
&
Z_{\rm orb}(R) \xrightarrow[\iota=1]{(\mZ_2\times \mZ_2)_A}Z_{\rm orb}(R)\,,\quad Z_{\rm orb}(R) \xrightarrow[\iota=1]{(\mZ_2\times \mZ_2)_B}Z_{\rm circ}(2R)\,,
\\
&Z_{\rm orb}(R)\xrightarrow{\mZ_4}Z_{\rm orb}(R)\,,\quad 
\\
&
Z_{\rm orb}(R) \xrightarrow[\iota=0]{D_8}Z_{\rm orb}(2R)\,,\quad Z_{\rm orb}(R) \xrightarrow[\iota=1]{D_8}Z_{\rm orb}(R/2)\,.
\fe
Note that T-duality acts on this set of transformation rules by swapping $A$ and $B$, $s$ and $sr$, $\iota = 0$ and $\iota = 1$.
 
 \subsection{The $D_8$ twisted torus partition functions}

We label the  torus partition function in the $r^a$ twisted sector with a $sr^b$ symmetry line inserted along the spatial cycle as $Z_{a\bar b}$. Similarly we define the twisted partition functions
$Z_{ab},Z_{\bar a\bar b},Z_{\bar a b}$.

Let $\cZ^{\A\B}_{\C\D}$ denote twisted partition function of the (unorbifolded) $c=1$  compact boson at radius $R$:
\ie
\cZ^{\A\B}_{\C\D}(\tau,R,\omega)
={1\over |\eta|^2} \sum_{n,m\in \mZ} \omega^{(2n+\C)(2m+\D)} 
q^{{1\over 2}\left({n+\A/2\over R}+ {(m+\B/2)R\over 2}\right)^2}
\bar q^{{1\over 2}\left({n+\A/2\over R}-{(m+\B/2)R\over 2}\right)^2}\,,
\fe
with $\omega=e^{\pi i\over 2}$. Under $T$ and $S$  transformations, they transform as
\ie
\cZ^{\A\B}_{\C\D}(\tau+1,R,\omega)=&\omega^{-\A\D-\B\C}\cZ^{\A+\C,\B+\D}_{\C\D}(\tau,R,\omega)\,,
\\
\cZ^{\A\B}_{\C\D}(-1/\tau,R,\omega)=&
 \omega^{(\A-\C)(\B-\D)+2\A\B}\cZ^{\C\D}_{\A\B}(\tau,R,\omega)\,.
\fe
We also recall \cite{Yang:1987wk}
\ie
     	W\equiv {1\over \eta} \sum_{r\in \mZ}(-1)^r q^{r^2}\,,\quad W_\pm \equiv {1\over \eta} \sum_{r\in \mZ}(\pm 1)^r q^{\left(r+{1\over 4}\right)^2}\,.
     	\fe
which transform as
\ie
{}&T: \begin{pmatrix} \sqrt{1\over 2} W(\tau+1) \\ W_+(\tau+1) \\W_-(\tau+1)
\end{pmatrix}
=\begin{pmatrix} e^{-2\pi i/24} & 0 & 0 \\  0 & e^{\pi i/24} & 0\\  0 & 0 & e^{\pi i/24} 
\end{pmatrix}
\begin{pmatrix} \sqrt{1\over 2} W(\tau) \\ W_-(\tau) \\W_+(\tau)
\end{pmatrix}\,,
\\
{}&S: \begin{pmatrix} \sqrt{1\over 2} W(-1/\tau) \\ W_+(-1/\tau) \\W_-(-1/\tau)
\end{pmatrix}
=\begin{pmatrix}  0 & 1 & 0 \\  1 & 0 & 0\\  0 & 0 & 1
\end{pmatrix}
\begin{pmatrix} \sqrt{1\over 2} W(\tau) \\ W_+(\tau) \\W_-(\tau)
\end{pmatrix}\,.
\fe
The explicit forms of these twisted torus partition functions of the $\bZ_2$ orbifold are
\ie
Z_{0\bar 0}=&{1\over 2}\cZ_{00}^{10}+{1\over 2}|W|^2
\,,~Z_{\bar 0 0}={1\over 2}\cZ_{10}^{00}+ |W_+|^2
\,,~Z_{\bar 0\bar 0}={1\over 2}\cZ_{10}^{10}+|W_-|^2\,,
\\
Z_{2\bar 0}=&{1\over 2}\cZ_{00}^{10}-{1\over 2}|W|^2
\,,~Z_{\bar 0 2}={1\over 2}\cZ_{10}^{00}- |W_+|^2
\,,~Z_{\bar 0\bar 2}={1\over 2}\cZ_{10}^{10}-|W_-|^2\,,
\\
Z_{ 0\bar 1}=&{1\over 2}\cZ_{00}^{01}+{1\over 2}|W|^2\,,~
Z_{\bar 1 0}={1\over 2}\cZ_{01}^{00}+|W_+|^2
\,,~Z_{2\bar  1}=-{1\over 2}\cZ_{00}^{01}+{1\over 2} |W|^2
\,,~Z_{\bar 1 2}=Z_{\bar 3 2}=-{1\over 2}\cZ_{01}^{00}+|W_+|^2\,,
\\
Z_{ \bar 1\bar 1}=&{1\over 2}\cZ_{01}^{01}+|W_-|^2\,,~
Z_{ \bar 3\bar 1}=-{1\over 2}\cZ_{01}^{01}+|W_-|^2\,,~
\\
Z_{00}=&{1\over 2}\cZ_{00}^{00}+{1\over 2}|W|^2+|W_+|^2+|W_-|^2\,,~
Z_{20}={1\over 2}\cZ_{00}^{00}-{1\over 2}|W|^2+|W_+|^2-|W_-|^2\,,~
\\
Z_{02}=&{1\over 2}\cZ_{00}^{00}+{1\over 2}|W|^2-|W_+|^2-|W_-|^2\,,~
Z_{22}={1\over 2}\cZ_{00}^{00}-{1\over 2}|W|^2-|W_+|^2+|W_-|^2\,,~
\\
{Z_{10}}=&{ Z_{30}={1\over 2}\cZ_{11}^{00}+ |W_+|^2}\,,~
Z_{01}=Z_{03}={1\over 2}\omega^{-1} \cZ_{00}^{11}+ {1\over 2}|W|^2\,,~
\\
Z_{12}=& -{1\over 2} \cZ_{11}^{00}+ |W_+|^2\,,~
{Z_{21}}={1\over 2}\omega \cZ_{00}^{11}+ {1\over 2}|W|^2\,,~
\\
Z_{13}=& -{1\over 2} \cZ_{11}^{11}+ |W_-|^2\,,~
Z_{23}={1\over 2}\omega \cZ_{00}^{11}+ {1\over 2}|W|^2\,,~
\\
{Z_{31}}=& {-{1\over 2} \cZ_{11}^{11}+ |W_-|^2}\,,~
Z_{32}=-{1\over 2}  \cZ_{11}^{00}+  |W_+|^2\,,~
\\
{ Z_{11}=}&{ {1\over 2}\cZ_{11}^{11}+ |W_-|^2}\,,~
Z_{33} ={1\over 2} \cZ_{11}^{11}+ |W_-|^2\,.
\fe
Note the following identities since $Z(g,h)=Z(c g c^{-1},chc^{-1})$,
\ie
&Z_{10}=Z_{30}\,,~Z_{\bar 10}=Z_{\bar 30}\,,~Z_{\bar 00}=Z_{\bar 20}\,,~Z_{\bar 02}=Z_{\bar 22}\,,~Z_{\bar \bar 12 }=Z_{\bar 32}\,,~
\\
&Z_{\bar 2\bar 2}=Z_{\bar 0 \bar 0}\,,~Z_{\bar 0\bar 2}=Z_{\bar 2 \bar 0}
\fe
and that under S-transform
\ie
Z_{11}(-1/\tau,-1/\bar\tau)=Z_{31}(\tau,\bar\tau)\,,
\fe
since the TDLs are oriented.

\subsection{$\mZ_2$ gauging}
The $\mZ_2$ subgroups of $D_8$ fall into 3 conjugate classes whose representatives are generated by $r^2$ (center $\mZ_2$), $s$ or $rs$.

The center $\mZ_2$ orbifold returns the original partition function between $\mZ_2$ orbifold
\ie
Z^{r^2}_{\mZ_2}(R)={1\over 2}(Z_{00}+Z_{02}+Z_{20}+Z_{22})=Z^{00}_{00}(R)=Z_{\rm circ}(R)\,.
\fe
The reflection $\mZ_2$ orbifold  preserves the orbifold branch but increases or decreases $R$ by 2 respectively
\ie
Z^{s}_{\mZ_2}(R)=&{1\over 2}(Z_{00}+Z_{0\bar 0}+Z_{\bar 0 0}+Z_{\bar 0\bar 0})
\\
=&{1\over 4}(\cZ_{00}^{00}+Z^{10}_{00}+Z^{00}_{10}+Z^{10}_{10})+{1\over 2}|W|^2+|W_+|^2+|W_-|^2
=Z(2R)\,,
\\
Z^{rs}_{\mZ_2}(R)=&{1\over 2}(Z_{00}+Z_{0\bar 1}+Z_{\bar 1 0}+Z_{\bar 1\bar 1})
\\
=&{1\over 4}(\cZ_{00}^{00}+Z^{01}_{00}+Z^{00}_{01}+Z^{01}_{01})+{1\over 2}|W|^2+|W_+|^2+|W_-|^2
=Z(R/2)\,.
\fe
Here we have used
\ie
&\cZ_{00}^{00}+\cZ^{01}_{00}+\cZ^{00}_{01}+\cZ^{01}_{01}
\\
=&
{1\over |\eta|^2} \sum_{n,m\in \mZ}  (1+(-)^n)
q^{{1\over 2}\left({n\over R}+ {mR\over 2}\right)^2}
\bar q^{{1\over 2}\left({n\over R}-{mR\over 2}\right)^2}
+
(1+(-)^n) q^{{1\over 2}\left({n\over R}+ {(m+1/2)R\over 2}\right)^2}
\bar q^{{1\over 2}\left({n\over R}-{(m+1/2)R\over 2}\right)^2}\,,
\\
=&
{2\over |\eta|^2} \sum_{n,m\in \mZ}  
q^{{1\over 2}\left({2n\over R}+ {mR\over 4}\right)^2}
\bar q^{{1\over 2}\left({2n\over R}-{mR\over 4}\right)^2}\,.
\fe

\subsection{$\mZ_4$ gauging}
The orbifold branch is point-wise invariant under $\mZ_4$ gauging
\ie
Z_{\mZ_4}(R)={1\over 4}\sum_{a,b=0}^3 Z_{ab}
={1\over 4}
(2Z^{00}_{00} +2|W|^2+4|W_+|^2+4|W_+|^2)
=Z_{00}(R)=Z(R)
\fe

\subsection{$\mZ_2\times \mZ_2$ gauging}
There are  two $\mZ_2\times \mZ_2$ subgroups of $D_8$ up to conjugation. We will refer to them as
\ie
(\mZ_2\times \mZ_2)^A=&\{1,r^2,s,r^2s\}\,,
\\
(\mZ_2\times \mZ_2)^B=&\{1,r^2,rs,r^3s\}\,.
\fe
Furthermore $H^2(\mZ_2\times \mZ_2,U(1))=\mZ_2$, so we may couple a nontrivial SPT when gauging.

\subsubsection*{Trivial SPT}
In one case the orbifold branch is mapped to the unorbifolded branch with $R$ doubled
 \ie
 Z^A_{\mZ_2 \times \mZ_2}=&
 {1\over 2}(Z_{0\bar 0}+Z_{2\bar 0}+Z_{\bar 0 0}+Z_{\bar 0 2}+Z_{\bar 0\bar 0}+Z_{\bar 0\bar 2})+{1\over 4}(Z_{00}+Z_{02}+Z_{20}+Z_{22})\,,
 \\
 =&{1\over 2}(\cZ_{00}^{00}+\cZ_{10}^{00}+\cZ_{00}^{10}+\cZ_{10}^{10})\,,
 \\
 =& Z^{00}_{00}(2R) =Z_{\rm circ}(2R)\,,
 \fe
while in the other case the orbifold branch is invariant
 \ie
 Z^B_{\mZ_2 \times \mZ_2}(R)=&
 {1\over 2}(Z_{0\bar 1}+Z_{2\bar 1}+Z_{\bar 1 0}+Z_{\bar 1 2}+Z_{\bar 1\bar 1}+Z_{\bar 1\bar 3})+{1\over 4}(Z_{00}+Z_{02}+Z_{20}+Z_{22})\,,
 \\
 =&{1\over 2}(\cZ_{00}^{00}+\cZ_{10}^{00}+\cZ_{00}^{10}+\cZ_{10}^{10})\,,
 \\
 =&{1\over 2} Z^{00}_{00}+{1\over 2}|W|^2+|W_+|^2+|W_-|^2=Z(R)\,.
 \fe

\subsubsection*{Nontrivial SPT}
In one case, the orbifold branch is point-wise invariant
\ie
 Z^{A\star}_{\mZ_2 \times \mZ_2}(R)=&
{1\over 2}(Z_{0\bar 0}-Z_{2\bar 0}+Z_{\bar 0 0}-Z_{\bar 0 2}+Z_{\bar 0\bar 0}-Z_{\bar 0\bar 2})+{1\over 4}(Z_{00}+Z_{02}+Z_{20}+Z_{22})\,,
\\
=&{1\over 2}(\cZ_{00}^{00}+\cZ_{10}^{00}+\cZ_{00}^{10}+\cZ_{10}^{10})-Z_{2\bar 0}-Z_{\bar 0 2}-Z_{\bar 0\bar 2}\,,
\\
=& Z^{00}_{00}(2R)-Z_{2\bar 0}-Z_{\bar 0 2}-Z_{\bar 0\bar 2}\,,
\\
=&{1\over 2}(\cZ_{00}^{00}  +|W|^2+|W_+|^2+|W_-|^2)=Z(R)\,,
\fe
and  in the other case, the orbifold branch is mapped to the unorbifolded branch with $R$ doubled
\ie
  Z^{B\star}_{\mZ_2 \times \mZ_2}(R)=&
{1\over 2}(Z_{0\bar 1}-Z_{2\bar 1}+Z_{\bar 1 0}-Z_{\bar 1 2}+Z_{\bar 1\bar 1}-Z_{\bar 1\bar 3})+{1\over 4}(Z_{00}+Z_{02}+Z_{20}+Z_{22})\,, \\
=&{1\over 2}(\cZ_{00}^{00}+\cZ_{10}^{00}+\cZ_{00}^{10}+\cZ_{10}^{10})=Z^{00}_{00}(2R)=Z_{\rm circ}(2R)\,.
\fe

\subsection{$D_8$ gauging}
 Recall that the nonabelian $G$ orbifold partition function involves sum over the conjugacy classes of $G$ and in each $g$ twisted sector a insertion of projector consisting of stablizers of $g$. Since $H^2(D_8,U(1))=\mZ_2$, we may consider coupling to a nontrivial SPT (discrete torsion) when gauging.
 
\subsubsection*{Trivial SPT} 
\ie
Z_{D_8}=&
{1\over 8} (\sum_{a=0}^4 Z_{0a}+\sum_{\bar a=0}^4 Z_{0\bar a}+\sum_{a=0}^4 Z_{2a}+\sum_{\bar a=0}^4 Z_{2\bar a})
+
{1\over 4} (  Z_{\bar 0 0}+ Z_{\bar 0 2}+ Z_{\bar 0 \bar 0}+ Z_{\bar 0 \bar 2})
\\
+&
{1\over 4} (  Z_{\bar 1 0}+ Z_{\bar 1 2}+ Z_{\bar 1 \bar 1}+ Z_{\bar 1 \bar 3})
+
{1\over 4}   \sum_{a=0}^4 Z_{1a} \,,
\\
=&{1\over 2} Z_{\mZ_4}+
{1\over 8} ( \sum_{\bar a=0}^4 Z_{0\bar a} +\sum_{\bar a=0}^4 Z_{2\bar a})
+
{1\over 4} (  Z_{\bar 0 0}+ Z_{\bar 0 2}+ Z_{\bar 0 \bar 0}+ Z_{\bar 0 \bar 2})
+
{1\over 4} (  Z_{\bar 1 0}+ Z_{\bar 1 2}+ Z_{\bar 1 \bar 1}+ Z_{\bar 1 \bar 3})\,.
\fe
We can rewrite the above as a combination of partition functions from gauging subgroups of $D_8$,
\ie
Z_{D_8}
=&{1\over 2} Z_{\mZ_4}+{1\over 2} Z^A_{\mZ_2\times \mZ_2}+{1\over 2} Z^B_{\mZ_2\times \mZ_2}-{1\over 2} Z^{r^2}_{\mZ_2}\,.
\fe
Hence
\ie
Z_{D_8}(R)=Z(R)+{1\over 2}(\cZ_{00}^{00}({2R})-\cZ_{00}^{00}({R}))
=Z(2R) \,.
\fe
Namely, the $D_8$ gauging preserves the orbifold branch but doubles $R$.

\subsubsection*{Nontrivial SPT} 
The nontrivial SPT associated with the $D_8$ discrete torsion has nontrivial torus partition functions
\ie
Z^{\rm SPT}_{\bar a b}= i^{b}\,,~~
Z^{\rm SPT}_{  a \bar  b} =i^{-a}\,,~~
Z^{\rm SPT}_{\bar a \bar b}= i^{b-a}\,.
\fe
Consequently
\ie
 Z^\star_{D_8}
=&{1\over 2} Z_{\mZ_4}+
{1\over 4} (  Z_{0\bar 0}   -Z_{2\bar 0}  +Z_{0\bar 1}   -Z_{2\bar 1}   )
+
{1\over 4} (  Z_{\bar 0 0}- Z_{\bar 0 2}+ Z_{\bar 0 \bar 0}- Z_{\bar 0 \bar 2})
+
{1\over 4} (  Z_{\bar 1 0}- Z_{\bar 1 2}+ Z_{\bar 1 \bar 1}-Z_{\bar 1 \bar 3})\\
=&Z_{D_8}-{1\over 2}(Z_{2\bar 0} +Z_{2\bar 1}+Z_{\bar 0 2}+Z_{\bar 1 2}+Z_{\bar 1 \bar 3}+Z_{\bar 0 \bar 2})\,,
\\
=&{1\over 2} |W|^2+|W_+|^2+|W_-|^2
+ Z^{00}_{00}(R/2)
=Z(R/2)\,.
\fe
Thus the $D_8$ gauging with discrete torsion preserves the orbifold branch but halves $R$.

\section{Some Lattice Defects}\label{appendixlattice}

\subsection{Ising Duality Defect}

Let us use our recipe from Section \ref{subsecoverviewselfdual} to define the Ising duality defect. We start with the Hamiltonian
\[H = - \sum_j Z_j Z_{j+1} + X_j\,,\]
($X, Z$ etc are Pauli matrices) and apply the $R_x = \prod_j X_j$ gauging procedure
\[X_j \mapsto Z_j Z_{j+1}\,,\]
and
\[Z_j \mapsto \prod_{0 \le k\le j} X_k\,,\]
to all $j \ge 0$. The Hamiltonian is invariant away from the resulting defect at $j = 0$, but near it we have
\[X_{-1} \mapsto X_{-1}\,,\]
\[Z_{-1} Z_0 \mapsto Z_{-1} X_0\,,\]
\[X_0 \mapsto Z_0 Z_1\,,\]
\[Z_0 Z_1 \mapsto X_1\,,\]
so we obtain
\[H' = - \left(\sum_{j \le -1} Z_{j-1} Z_j + X_j \right) - Z_{-1} X_0 - \left(\sum_{j \ge 1} Z_{j-1} Z_j + X_j \right)\,,\]
in agreement with \cite{Oshikawa:1996dj,Aasen:2016dop}.

We see that $H'$ enjoys two anti-commuting $\bZ_2$ symmetries,
\[U_L = \left(\prod_{j \le -1} X_j\right) Z_0\,,\]
and
\[U_R = \left(\prod_{j \ge 0} X_j\right)\,,\]
which we recognize as the left and right $\bZ_2$ action in the duality twisted sector, with the anti-commutation reflecting the nontrivial TY F-symbol.

\subsection{Continuum of topological defects in Ising$^2$}
 
We begin with the XY chain
\[H = - \sum_j X_j X_{j+1} + Y_j Y_{j+1}\,.\]
This can be related to the Ising$^2$ chain by gauging $R_x = \prod_j X_j$. In terms of the operator algebra, this amounts to the transformation
\[X_j \mapsto Z_j Z_{j+1}\,,\]
and
\[Z_j \mapsto \prod_{k\le j} X_k\,,\]
from which we obtain
\[H' = - \sum_j Z_j Z_{j+2} + Z_j X_{j+1} Z_{j+2}\,.\]
This is related to a more familiar Hamiltonian $H''$ after applying the cluster entangler transformation
\[X_j \mapsto Z_{j-1} X_j Z_{j+1}\,,\]
\[H'' = - \sum_j Z_j Z_{j+2} + X_j\,,\]
for which the even and odd sites define two decoupled critical Ising chains.

To construct the symmetry defect for a rotation $U(\theta) = \prod_j e^{i \theta Z_j/2}$, we apply this symmetry to sites $j > 0$. The bond terms are unchanged except for the 01 bond, which becomes
\[-\cos \theta (X_0 X_1 + Y_0 Y_1)-\sin \theta (X_0 Y_1 - Y_0 X_1)\,.\]
For $\theta \neq 0, \pi$, this defect is not invariant under the symmetry $R_x$, indeed $R_x U(\theta) R_x = U(-\theta)$. As in Section \ref{subsecshadownoether}, we want to replace it with a sum of defects labelled by $\theta$ and $-\theta$. We do this by introducing an extra qubit at $j = 1/2$, and writing the modified bond
\[-\cos \theta (X_0 X_1 + Y_0 Y_1)-\sin \theta (X_0 Y_{1/2} Y_1 - Y_0 Y_{1/2} X_1)\,,\]
with the new symmetry $R_x = X_{1/2} \prod_j X_j$.

We gauge this symmetry as above and obtain
\[-\cos \theta (Z_0 Z_{1/2} Z_1 Z_2 + Z_0 Y_{1/2} Z_1 Y_2) - \sin \theta (Z_0 X_1 Z_2 + Z_0 X_{1/2} Z_2)\,.\]
 If we then apply the cluster entangler, it becomes
\[-\cos \theta (Z_0 Z_{1/2} Z_1 Z_2 + Y_{1/2} Z_1 Y_2 Z_3) - \sin \theta (Z_0 Z_{1/2} X_1 + X_{1/2} Z_1 Z_2)\,.\]

\section{Modular $S$ matrices for nonabelian orbifolds}
\label{app:Smatrix}
The modular S-matrices for nonabelian orbifolds of the $SU(2)_1$ CFT were derived for $G=D_{2n}$ in \cite{Dijkgraaf:1989hb} and for $G=A_4, S_4,A_5$ in \cite{Cappelli:2002wq}. Here for completeness we record their results after correcting a number of typos. As explained in Section~\ref{sec:verlinde}, they determine the Verlinde lines in these orbifold theories. 

The $SU(2)_1/D_{2n}$ CFT contains $n^2+7$ chiral primaries with respect to the $\cV_{D_{2n}}$ chiral algebra denoted as
\ie
\{
u_\pm,~\phi_\pm,~\chi_a,~\sigma_r,~\tau_r
\}
\fe
with $1\leq a\leq n^2-1$ and $r=1,2$. The vacuum representation corresponds to $u_+$. The modular $S$ matrix takes slightly different forms for $n$ even versus for $n$ odd and are given in Table~\ref{tab:SDeven} and Table~\ref{tab:SDodd} respectively. Here $\sigma_{rs}=2\D_{rs}-1$.

\begin{table}[!htb]
    \centering
    \begin{tabular}{c|ccccc}
         & $u_\pm $ & $\phi_\pm $ & $\chi_b$ & $\sigma_s$ & $\tau_s$  \\\hline
        $u_\pm $ & 1 & 1 & 2 & $\pm n$ & $\pm n$ \\
        $\phi_\pm $ & 1 & 1 & $2(-1)^b$ & $ \sigma_{rs} n$ & $\sigma_{rs} n$  \\
        $\chi_a$ &  2 & $2(-1)^a$ & $4 \cos{\pi ab\over n}$ & 0 & 0\\
        $\sigma_r$ & $\pm n$ & $\sigma_{rs} n $ & 0 & $\D_{rs} n\sqrt{2}$ & $-\D_{rs} n \sqrt{2}$\\
        $\tau_r$  & $\pm n$ & $\sigma_{rs} n $ & 0 & $-\D_{rs} n\sqrt{2}$ & $\D_{rs} n \sqrt{2}$\\
    \end{tabular}
    \caption{The modular $S$ matrix for $\cV_{D_{2n}}$ multiplied by $2\sqrt{2n}$ for $n\in 2\mZ$.}
    \label{tab:SDeven}
\end{table}

\begin{table}[!htb]
    \centering
   \begin{tabular}{c|ccccc}
         & $u_\pm $ & $\phi_j $ & $\chi_b$ & $\sigma_s$ & $\tau_s$  \\\hline
        $u_\pm $ & 1 & 1 & 2 & $\pm n$ & $\pm n$ \\
        $\phi_\pm $ & 1 & 1 & $2(-1)^b$ & $ i\sigma_{rs} n$ & $i\sigma_{rs} n$  \\
        $\chi_a$ &  2 & $2(-1)^a$ & $4 \cos{\pi ab\over n}$ & 0 & 0\\
        $\sigma_r$ & $\pm n$ & $i\sigma_{rs} n $ & 0 & $e^{{\pi i\over4} \sigma_{rs}} n$ & $-e^{{\pi i\over4} \sigma_{rs}} n $\\
        $\tau_r$  & $\pm n$ & $i\sigma_{rs} n $ & 0 & $-e^{{\pi i\over4} \sigma_{rs}} n$ & $e^{{\pi i\over4} \sigma_{rs}} n$\\
    \end{tabular}
    \caption{The modular $S$ matrix for $\cV_{D_{2n}}$ multiplied by $2\sqrt{2n}$ for $n\in 2\mZ+1$.}
    \label{tab:SDodd}
\end{table}

The $SU(2)_1/A_4$ CFT contains $21$ chiral primaries with respect to the $\cV_{A_4}$ chiral algebra which are denoted as
\ie
\{
u_i\,,~j\,,~\phi_i\,,~\sigma\,,~\tau\,,~\omega_i^\pm\,,~\theta_i^\pm 
\}
\fe
with $i=0,1,2$. The vacuum representation corresponds to $u_0$.  The modular $S$ matrix is given in Table~\ref{tab:SA4}. For convenience, we introduce the following parameters given by various roots of unity,
\ie
\omega=e^{2\pi i\over 3}\,,~\A=e^{\pi i \over 9}\,,~\B=e^{\pi i \over 16}\,,~\zeta=e^{2\pi i \over 5}\,.
\label{rou}
\fe

\begin{table}[!htb]
    \centering
    \begin{tabular}{c|ccccccccc}
         & $u_j$ & $j$ & $\phi_j$ & $\sigma$ & $\tau$ & $\omega_j^+$  & $\omega_j^-$ & $\theta_j^+$  & $\theta_j^-$    \\\hline
     $u_j$    & 1 & 3 & 2 & 6 & 6& $4\omega^{ i}$ & $4\bar\omega^i$ & $4\omega^{i}$ & $4\bar\omega^{ i}$  \\
     $j$  & 3 & 9 & 6 & $-6$ & $-6$ & 0 & 0& 0 & 0\\
      $\phi_j$ & 2 & 6 & $-4$ & 0 & 0& $-4\omega^{ i}$ & $-4\bar\omega^{ i}$ & $4\omega^{ i}$   & $4\bar\omega^{  i}$    \\
      $\sigma$ & 6 & $-6$ & 0 & $6\sqrt{2}$ & $-6\sqrt{2}$ & 0 &  0  & 0 &  0 \\
       $\tau$ &   6 & $-6$ & 0 & $-6\sqrt{2}$ & $ 6\sqrt{2}$ & 0 &  0  & 0 &  0 \\
       $\omega_i^+$ & $4\omega^{ j}$ & 0 &  $-4\omega^{ j}$ & 0 & 0 & $4\bar \A^4\bar \omega^{i+j}$ &$4  \A^4  \omega^{i+j}$ & $4\A^{2}\omega^{2i+j}$ & $4\bar\A^2\bar\omega^{2i+j}$   \\
         $\omega_i^-$ & $4\bar \omega^{ j}$ & 0 &  $-4\bar\omega^{ j}$ & 0 & 0 & $4 \A^4 \omega^{i+j}$ &$4 \bar \A^4  \bar \omega^{i+j}$ & $4\bar \A^{2}\bar \omega^{2i+j}$ & $4\A^2\omega^{2i+j}$   \\
       $\theta_i^+$  & $4\omega^{  j}$ & 0 & $4\omega^{j}$ & 0 & 0 & $4\A^2\omega^{i+2j}$   & $4\bar\A^{ 2}\bar\omega^{i+2j}$  & $4\bar\A\omega^{i+j}$    & $4\A\bar \omega^{i+j}$   \\
        $\theta_i^-$  & $4\bar\omega^{  j}$ & 0 & $4\bar\omega^{j}$ & 0 & 0 & $4\bar\A^2\bar\omega^{i+2j}$   & $4\A^{ 2}\omega^{i+2j}$  & $4\A\bar \omega^{i+j}$    & $4\bar\A \omega^{i+j}$   \\
    \end{tabular}
    \caption{The modular $S$ matrix for $\cV_{A_4}$ multiplied by $12\sqrt{2}$.}
    \label{tab:SA4}
\end{table}

The $SU(2)_1/S_4$ CFT contains $28$ chiral primaries with respect to the $\cV_{S_4}$ chiral algebra which are denoted as
\ie
\{
u_\pm,~u_f\,,~j_\pm\,,~\phi_\pm\,,~\phi_f\,,~\mu_s\,,~\sigma_\pm \,,~\tau_\pm\,,~\omega_i\,,~\theta_i\,,~\A_k\,,~\B_k 
\}
\fe
with $s=1,2$, $i=0,1,2$ and $k=0,1,2,3$. The vacuum representation corresponds to $u_+$. The modular $S$ matrix is given in Table~\ref{tab:SS4} with the parameters $\omega,\A,\B$ as in \eqref{rou}, and in addition 
\ie
&c_k=(-1)^k 12 {\rm Re\,} \B^2 \,,~
s_k=(-1)^k 12   {\rm Im\,} \B^2\,,~
\\
&a_{ij}=16 {\rm Re\,}  \bar\A^{4}\bar\omega^{i+j}\,,~
b_{ij}=16 {\rm Re\,}   \A^2\omega^{2i+j}\,,~
d_{ij}=16 {\rm Re\,}  \bar\A  \omega^{i+j}\,,~
\\
&q_{kl}=12 {\rm Re\,}  \bar\B i^{k+l}\,,~
r_{kl}=12 {\rm Re\,}  \bar\B^3 i^{l-k}\,,~
s_{kl}=12 {\rm Re\,}  \bar\B^9 i^{-l-k}\,.
\fe

 \begin{table}[!htb]
    \centering
  \scalebox{1}{
    \begin{tabular}{c|cccccccccccc}
         & $u_\pm$ & $u_f$ & $j_\pm$ & $\phi_\pm$ & $\phi_f$ & $\mu_s$ & $\sigma_\pm$ & $\tau_\pm$ & $\omega_j$ & $\theta_j$ & $\A_l$ & $\B_l$     \\\hline
     $u_\pm$ & 1 & 2 & 3 & 2 &4 & $\pm 12$ & 6 & 6 & 8 & 8 & $\pm 6$ & $\pm 6$\\
     $u_f$ & 2 & 4 & 6 & 4 &8 & $0$ & 12 & 12 & $-8$ & $-8$ & 0 & 0\\
     $j_\pm$ &  3 & 6 & 9 & 6 & 12 & $\pm 12$ & $-6$ & $-6$ & 0 & 0 & $\mp 6$ & $\mp 6$\\
     $\phi_\pm$ & 2 & 4 & 6 & $-4$ & $-8$ & $0$ & 0 & 0 & $-8$ & 8 & $\pm 6\sqrt{2}$ & $\mp 6\sqrt{2}$\\
     $\phi_f$ & 4 & 8 & 12 & $-8$ & $-16$ & $0$ & 0 & 0 & 8 & $-8$ & 0 & 0\\
     $\mu_r$ & $\pm 12$ & 0 & $\pm 12$ & 0 & 0 & $12\sqrt{2}(-1)^{r+s}$ & 0 & 0 & 0 &0 & 0 & 0\\
     $\sigma_\pm$ & 6 & 12 & $-6$ & 0 & 0 & $0$ & $6\sqrt{2}$ & $-6\sqrt{2}$ & 0 & 0 & $\pm c_l$ &  $\pm s_l$ \\
     $\tau_\pm$ & 6 & 12 & $-6$ & 0 & 0 & $0$ & $-6\sqrt{2}$ & $6\sqrt{2}$ & 0 & 0 & $\pm s_l$ &  $\mp c_l$ \\
     $\omega_i$ &  8 & $-8$ & 0 & $-8$ & $8$ & 0 & 0 & 0 & $a_{ij}$ & $b_{ij}$  & 0 & 0
     \\
     $\theta_i$  &  8 & $-8$ & 0 & $8$ & $-8$ & 0 & 0 & 0 & $b_{ji}$ & $d_{ij}$  & 0 & 0
     \\
     $\A_k$  &  $\pm 6$ & 0 & $\mp 6$ & $\pm 6 \sqrt{2}$  & $0$ & 0 & $\pm c_k$ & $\pm s_k$  & 0 & 0 & $q_{kl}$ & $r_{kl}$  
     \\
     $\B_k$  & $\pm 6$ & 0 & $\mp 6$ & $\mp 6 \sqrt{2}$  & $0$ & 0 & $\pm s_k$ & $\mp c_k$  & 0 & 0 & $r_{lk}$ & $s_{kl}$  
    \end{tabular}}
    \caption{The modular $S$ matrix for $\cV_{S_4}$ multiplied by $24\sqrt{2}$.}
    \label{tab:SS4}
\end{table}

The $SU(2)_1/A_5$ CFT contains $37$ chiral primaries with respect to the $\cV_{A_5}$ chiral algebra which are denoted as
\ie
\{
u_m\,,~\phi_k\,,~\sigma \,,~\tau\,,~\omega_i\,,~\theta_i\,,~\pi_m\,,~\rho_m\,,~\lambda_m\,,~\xi_m 
\}
\fe
with $m=0,1,2,3,4$, $i=0,1,2$ and $k=0,1,2,3$. The vacuum representation corresponds to $u_0$. The modular $S$ matrix is given in Table~\ref{tab:SS4} with $g_\pm=6(1\pm \sqrt{5})$ and
\ie
&P_{mn}^1={\rm Re\,} e^{-{4\pi i \over 25}} \zeta^{2(n+m)}\,,~
P_{mn}^2={\rm Re\,} e^{-{2\pi i \over 25}} \zeta^{ 2n+m}\,,~
P_{mn}^3={\rm Re\,} e^{-{8\pi i \over 25}} \zeta^{ 2n-m}\,,~
P_{mn}^4={\rm Re\,} e^{-{6\pi i \over 25}} \zeta^{2(n-m)}\,,~
\\
&R_{mn}^1={\rm Re\,} e^{-{\pi i \over 25}} \zeta^{n+m}\,,~
R_{mn}^2={\rm Re\,} e^{-{4\pi i \over 25}} \zeta^{n-m}\,,~
R_{mn}^3={\rm Re\,} e^{-{3\pi i \over 25}} \zeta^{n-2m}\,,~
L_{mn}^1={\rm Re\,} e^{ {16\pi i \over 25}} \zeta^{n+m}\,,~
\\
&
L_{mn}^2={\rm Re\,} e^{ {12\pi i \over 25}} \zeta^{2m+n}\,,~
X_{mn}={\rm Re\,} e^{ {9\pi i \over 25}} \zeta^{2(n+m)}\,.
\fe

 \begin{table}[!htb]
    \centering
  \scalebox{.8}{
    \begin{tabular}{c|ccccccccccccccccc}
         & $u_0$ & $u_1$ & $u_2$ & $u_3$ & $u_4$ & $\phi_1$  & $\phi_2$  & $\phi_3$ & $\phi_4$ & $\sigma$ & $\tau$
         & $\omega_j$ & $\theta_j$ & $\pi_n$ & $\rho_n$ & $\lambda_n$ & $\xi_n$ 
         \\\hline
   $u_0$ & 1 & 3 & 3& 4 & 5 & 2 & 2 & 4 & 6 & 30 & 30 & 20 & 20 & 12 & 12 & 12 & 12\\
   $u_1$ & 3 & 9 & 9 & 12 & 15 & 6 & 6 & 12 & 18 & $-30$ &  $-30$ & 0 & 0 & $g_-$ & $g_+$ & $g_+$ & $g_-$  \\
   $u_2$  & 3 & 9 & 9 & 12 & 15 & 6 & 6 & 12 & 18 & $-30$ &  $-30$ & 0 & 0 & $g_+$ & $g_-$ & $g_-$ & $g_+$  \\
   $u_3$ & 4 & 12 & 12 & 16 & 20 & 8   & 8 & 16 & 24 & 0 & 0 & $20$ &  $20$ &  $-12$ &    $-12$ &  $-12$ &  $-12$   \\
   $u_4$ & 5& 15 & 15 & 20 & 25 & 10   & 10 & 20 & 30 & 30 & 30 & $-20$ &  $-20$ &  0 & 0 &  0 &  0   \\
   $\phi_1$  & 2 & 6 & 6 & 8 & 10 & $-4$ & $-4$ & $-8$ & $-12$ & 0 & 0 & $-20$ & $20$ & $-g_+$ & $g_-$ & $-g_-$ & $g_+$\\
   $\phi_2$  & 2 & 6 & 6 & 8 & 10 & $-4$ & $-4$ & $-8$ & $-12$ & 0 & 0 & $-20$ & $20$ & $-g_-$ & $g_+$ & $-g_+$ & $g_-$\\
   $\phi_3$ & 4 & 12 & 12 & 16 & 20 & $-8$ & $-8$ & $-16$ & $-24$ & 0 & 0 & 20 & $-20$ & $-12$ & $12 $ & $-12$ & $12$ \\
   $\phi_4$ & 6 & 18 & 18 & 24 & 30 & $-12$ & $-12$ & $-24$ & $-36$ & 0 & 0 & 0 & $0$ & $12$ & $-12 $ & $12$ & $-12$ \\
   $\sigma$ & 30 & $-30$ & $-30$ & 0 & 30 & 0 & 0 &0 &0 & $30\sqrt{2}$ & $-30\sqrt{2}$ & 0 & 0 & 0 & 0 &0 & 0 \\
   $\tau$
        & 30 & $-30$ & $-30$ & 0 & 30 & 0 & 0 &0 &0 & $-30\sqrt{2}$ & $30\sqrt{2}$ & 0 & 0 & 0 & 0 &0 & 0 \\
 $\omega_j$ & 20 & $0$ & $0$ & 20 & $-20$ & $-20$ & $-20$ & $20$ & 0 & 0 &0 & ${5\over 2}a_{ij}$ & ${5\over 2}b_{ij}$ &  0 & 0 &0 & 0 \\
 $\theta_j$  & 20 & $0$ & $0$ & 20 & $-20$ & $ 20$ & $ 20$ & $-20$ & 0 & 0 &0 & ${5\over 2}b_{ji}$ & ${5\over 2}d_{ij}$ &  0 & 0 &0 & 0 \\
 $\pi_m$  & 12 & $g_-$ & $g_+$ & $-12$ & 0 & $-g_+$ & $-g_-$ & $-12$ & $12$ & 0 & 0 & 0 &0 & $P_{mn}^1$ & $P_{mn}^2$ & $P_{mn}^3$ & $P_{mn}^4$ \\
 $\rho_m$ & 12 & $g_+$ & $g_-$ & $-12$ & 0 & $g_-$ & $g_+$ & $ 12$ & $-12$ & 0 & 0 & 0 &0 & $P_{nm}^2$ & $R_{mn}^1$ & $R_{mn}^2$ & $R_{mn}^3$ \\
 $\lambda_m$ & 12 & $g_+$ & $g_-$ & $-12$ & 0 & $-g_-$ & $-g_+$ & $-12$ & $12$ & 0 & 0 & 0 &0 & $P_{nm}^3$ & $R_{nm}^2$ & $L_{mn}^1$ & $L_{mn}^2$ \\
 $\xi_m$ & 12 & $g_-$ & $g_+$ & $-12$ & 0 & $g_+$ & $g_-$ & $12$ & $-12$ & 0 & 0 & 0 &0 & $P_{nm}^4$ & $R_{nm}^3$ & $L_{nm}^2$ & $X_{mn}$  
    \end{tabular}}
    \caption{The modular $S$ matrix for $\cV_{A_5}$ multiplied by $60\sqrt{2}$.}
    \label{tab:SA5}
\end{table}

\bibliography{refs}

\providecommand{\href}[2]{#2}\begingroup\raggedright\begin{thebibliography}{10}

\bibitem{Gukov:2013zka}
S.~Gukov and A.~Kapustin, \emph{{Topological Quantum Field Theory, Nonlocal
  Operators, and Gapped Phases of Gauge Theories}},
  \href{https://arxiv.org/abs/1307.4793}{{\ttfamily 1307.4793}}.

\bibitem{Kapustin:2013uxa}
A.~Kapustin and R.~Thorngren, \emph{{Higher symmetry and gapped phases of gauge
  theories}},  \href{https://arxiv.org/abs/1309.4721}{{\ttfamily 1309.4721}}.

\bibitem{Gaiotto:2014kfa}
D.~Gaiotto, A.~Kapustin, N.~Seiberg and B.~Willett, \emph{{Generalized Global
  Symmetries}}, \href{https://doi.org/10.1007/JHEP02(2015)172}{\emph{JHEP}
  {\bfseries 02} (2015) 172} [\href{https://arxiv.org/abs/1412.5148}{{\ttfamily
  1412.5148}}].

\bibitem{Kapustin_2017}
A.~Kapustin and R.~Thorngren, \emph{Higher symmetry and gapped phases of gauge
  theories}, \href{https://doi.org/10.1007/978-3-319-59939-7_5}{\emph{Progress
  in Mathematics} (2017) 177–202}.

\bibitem{Gaiotto:2017yup}
D.~Gaiotto, A.~Kapustin, Z.~Komargodski and N.~Seiberg, \emph{{Theta, Time
  Reversal, and Temperature}},
  \href{https://doi.org/10.1007/JHEP05(2017)091}{\emph{JHEP} {\bfseries 05}
  (2017) 091} [\href{https://arxiv.org/abs/1703.00501}{{\ttfamily
  1703.00501}}].

\bibitem{Kapustin:2009av}
A.~Kapustin and M.~Tikhonov, \emph{{Abelian duality, walls and boundary
  conditions in diverse dimensions}},
  \href{https://doi.org/10.1088/1126-6708/2009/11/006}{\emph{JHEP} {\bfseries
  11} (2009) 006} [\href{https://arxiv.org/abs/0904.0840}{{\ttfamily
  0904.0840}}].

\bibitem{Kapustin:2010if}
A.~Kapustin and N.~Saulina, \emph{{Surface operators in 3d Topological Field
  Theory and 2d Rational Conformal Field Theory}},
  \href{https://arxiv.org/abs/1012.0911}{{\ttfamily 1012.0911}}.

\bibitem{Fuchs:2012dt}
J.~Fuchs, C.~Schweigert and A.~Valentino, \emph{{Bicategories for boundary
  conditions and for surface defects in 3-d TFT}},
  \href{https://doi.org/10.1007/s00220-013-1723-0}{\emph{Commun. Math. Phys.}
  {\bfseries 321} (2013) 543}
  [\href{https://arxiv.org/abs/1203.4568}{{\ttfamily 1203.4568}}].

\bibitem{Aasen:2016dop}
D.~Aasen, R.~S.~K. Mong and P.~Fendley, \emph{{Topological Defects on the
  Lattice I: The Ising model}},
  \href{https://doi.org/10.1088/1751-8113/49/35/354001}{\emph{J. Phys. A}
  {\bfseries 49} (2016) 354001}
  [\href{https://arxiv.org/abs/1601.07185}{{\ttfamily 1601.07185}}].

\bibitem{Bhardwaj:2017xup}
L.~Bhardwaj and Y.~Tachikawa, \emph{{On finite symmetries and their gauging in
  two dimensions}}, \href{https://doi.org/10.1007/JHEP03(2018)189}{\emph{JHEP}
  {\bfseries 03} (2018) 189}
  [\href{https://arxiv.org/abs/1704.02330}{{\ttfamily 1704.02330}}].

\bibitem{Bal:2018wbw}
R.~Vanhove, M.~Bal, D.~J. Williamson, N.~Bultinck, J.~Haegeman and
  F.~Verstraete, \emph{{Mapping topological to conformal field theories through
  strange correlators}},
  \href{https://doi.org/10.1103/PhysRevLett.121.177203}{\emph{Phys. Rev. Lett.}
  {\bfseries 121} (2018) 177203}
  [\href{https://arxiv.org/abs/1801.05959}{{\ttfamily 1801.05959}}].

\bibitem{Chang:2018iay}
C.-M. Chang, Y.-H. Lin, S.-H. Shao, Y.~Wang and X.~Yin, \emph{{Topological
  Defect Lines and Renormalization Group Flows in Two Dimensions}},
  \href{https://doi.org/10.1007/JHEP01(2019)026}{\emph{JHEP} {\bfseries 01}
  (2019) 026} [\href{https://arxiv.org/abs/1802.04445}{{\ttfamily
  1802.04445}}].

\bibitem{Lin:2019hks}
Y.-H. Lin and S.-H. Shao, \emph{{Duality Defect of the Monster CFT}},
  \href{https://arxiv.org/abs/1911.00042}{{\ttfamily 1911.00042}}.

\bibitem{Thorngren:2019iar}
R.~Thorngren and Y.~Wang, \emph{{Fusion Category Symmetry I: Anomaly In-Flow
  and Gapped Phases}},  \href{https://arxiv.org/abs/1912.02817}{{\ttfamily
  1912.02817}}.

\bibitem{Lichtman:2020nuw}
T.~Lichtman, R.~Thorngren, N.~H. Lindner, A.~Stern and E.~Berg, \emph{{Bulk
  Anyons as Edge Symmetries: Boundary Phase Diagrams of Topologically Ordered
  States}},  \href{https://arxiv.org/abs/2003.04328}{{\ttfamily 2003.04328}}.

\bibitem{Aasen:2020jwb}
D.~Aasen, P.~Fendley and R.~S.~K. Mong, \emph{{Topological Defects on the
  Lattice: Dualities and Degeneracies}},
  \href{https://arxiv.org/abs/2008.08598}{{\ttfamily 2008.08598}}.

\bibitem{Komargodski:2020mxz}
Z.~Komargodski, K.~Ohmori, K.~Roumpedakis and S.~Seifnashri, \emph{{Symmetries
  and Strings of Adjoint QCD${}_2$}},
  \href{https://arxiv.org/abs/2008.07567}{{\ttfamily 2008.07567}}.

\bibitem{Chang:2020imq}
C.-M. Chang and Y.-H. Lin, \emph{{Lorentzian Dynamics and Factorization Beyond
  Rationality}},  \href{https://arxiv.org/abs/2012.01429}{{\ttfamily
  2012.01429}}.

\bibitem{Huang:2021ytb}
T.-C. Huang and Y.-H. Lin, \emph{{Topological Field Theory with Haagerup
  Symmetry}},  \href{https://arxiv.org/abs/2102.05664}{{\ttfamily 2102.05664}}.

\bibitem{Ji:2019jhk}
W.~Ji and X.-G. Wen, \emph{{Categorical symmetry and non-invertible anomaly in
  symmetry-breaking and topological phase transitions}},
  \href{https://arxiv.org/abs/1912.13492}{{\ttfamily 1912.13492}}.

\bibitem{Kong:2020cie}
L.~Kong, T.~Lan, X.-G. Wen, Z.-H. Zhang and H.~Zheng, \emph{{Algebraic higher
  symmetry and categorical symmetry -- a holographic and entanglement view of
  symmetry}},
  \href{https://doi.org/10.1103/PhysRevResearch.2.043086}{\emph{Phys. Rev.
  Res.} {\bfseries 2} (2020) 043086}
  [\href{https://arxiv.org/abs/2005.14178}{{\ttfamily 2005.14178}}].

\bibitem{Kapustin:2010ta}
A.~Kapustin, \emph{{Topological Field Theory, Higher Categories, and Their
  Applications}},  in \emph{{International Congress of Mathematicians}}, 4,
  2010, \href{https://arxiv.org/abs/1004.2307}{{\ttfamily 1004.2307}}.

\bibitem{Johnson-Freyd:2020usu}
T.~Johnson-Freyd, \emph{{On the classification of topological orders}},
  \href{https://arxiv.org/abs/2003.06663}{{\ttfamily 2003.06663}}.

\bibitem{etingof2015tensor}
P.~Etingof, S.~Gelaki, D.~Nikshych and V.~Ostrik, \emph{Tensor Categories},
  Mathematical Surveys and Monographs. American Mathematical Society, 2015.

\bibitem{Verlinde:1988sn}
E.~P. Verlinde, \emph{{Fusion Rules and Modular Transformations in 2D Conformal
  Field Theory}},
  \href{https://doi.org/10.1016/0550-3213(88)90603-7}{\emph{Nucl. Phys. B}
  {\bfseries 300} (1988) 360}.

\bibitem{Petkova:2000ip}
V.~Petkova and J.~Zuber, \emph{{Generalized twisted partition functions}},
  \href{https://doi.org/10.1016/S0370-2693(01)00276-3}{\emph{Phys. Lett. B}
  {\bfseries 504} (2001) 157}
  [\href{https://arxiv.org/abs/hep-th/0011021}{{\ttfamily hep-th/0011021}}].

\bibitem{Ginsparg:1987eb}
P.~H. Ginsparg, \emph{{Curiosities at c = 1}},
  \href{https://doi.org/10.1016/0550-3213(88)90249-0}{\emph{Nucl. Phys. B}
  {\bfseries 295} (1988) 153}.

\bibitem{Moore:1988qv}
G.~W. Moore and N.~Seiberg, \emph{{Classical and Quantum Conformal Field
  Theory}}, \href{https://doi.org/10.1007/BF01238857}{\emph{Commun. Math.
  Phys.} {\bfseries 123} (1989) 177}.

\bibitem{Moore:1989vd}
G.~W. Moore and N.~Seiberg, \emph{{LECTURES ON RCFT}},  in \emph{{1989 Banff
  NATO ASI: Physics, Geometry and Topology}}, 9, 1989.

\bibitem{Fuchs:2002cm}
J.~Fuchs, I.~Runkel and C.~Schweigert, \emph{{TFT construction of RCFT
  correlators 1. Partition functions}},
  \href{https://doi.org/10.1016/S0550-3213(02)00744-7}{\emph{Nucl. Phys. B}
  {\bfseries 646} (2002) 353}
  [\href{https://arxiv.org/abs/hep-th/0204148}{{\ttfamily hep-th/0204148}}].

\bibitem{Fuchs:2003id}
J.~Fuchs, I.~Runkel and C.~Schweigert, \emph{{TFT construction of RCFT
  correlators. 2. Unoriented world sheets}},
  \href{https://doi.org/10.1016/j.nuclphysb.2003.11.026}{\emph{Nucl. Phys. B}
  {\bfseries 678} (2004) 511}
  [\href{https://arxiv.org/abs/hep-th/0306164}{{\ttfamily hep-th/0306164}}].

\bibitem{Fuchs:2004dz}
J.~Fuchs, I.~Runkel and C.~Schweigert, \emph{{TFT construction of RCFT
  correlators. 3. Simple currents}},
  \href{https://doi.org/10.1016/j.nuclphysb.2004.05.014}{\emph{Nucl. Phys. B}
  {\bfseries 694} (2004) 277}
  [\href{https://arxiv.org/abs/hep-th/0403157}{{\ttfamily hep-th/0403157}}].

\bibitem{Fuchs:2004xi}
J.~Fuchs, I.~Runkel and C.~Schweigert, \emph{{TFT construction of RCFT
  correlators IV: Structure constants and correlation functions}},
  \href{https://doi.org/10.1016/j.nuclphysb.2005.03.018}{\emph{Nucl. Phys. B}
  {\bfseries 715} (2005) 539}
  [\href{https://arxiv.org/abs/hep-th/0412290}{{\ttfamily hep-th/0412290}}].

\bibitem{Fjelstad:2005ua}
J.~Fjelstad, J.~Fuchs, I.~Runkel and C.~Schweigert, \emph{{TFT construction of
  RCFT correlators. V. Proof of modular invariance and factorisation}},
  {\emph{Theor. Appl. Categor.} {\bfseries 16} (2006) 342}
  [\href{https://arxiv.org/abs/hep-th/0503194}{{\ttfamily hep-th/0503194}}].

\bibitem{Bachas:2009mc}
C.~Bachas and S.~Monnier, \emph{{Defect loops in gauged Wess-Zumino-Witten
  models}}, \href{https://doi.org/10.1007/JHEP02(2010)003}{\emph{JHEP}
  {\bfseries 02} (2010) 003} [\href{https://arxiv.org/abs/0911.1562}{{\ttfamily
  0911.1562}}].

\bibitem{tambarayam}
D.~Tambara and S.~Yamagami, \emph{Tensor categories with fusion rules of
  self-duality for finite abelian groups},
  \href{https://doi.org/https://doi.org/10.1006/jabr.1998.7558}{\emph{Journal
  of Algebra} {\bfseries 209} (1998) 692 }.

\bibitem{Recknagel:2013uja}
A.~Recknagel and V.~Schomerus, \emph{{Boundary Conformal Field Theory and the
  Worldsheet Approach to D-Branes}}, Cambridge Monographs on Mathematical
  Physics. Cambridge University Press, 11, 2013,
  \href{https://doi.org/10.1017/CBO9780511806476}{10.1017/CBO9780511806476}.

\bibitem{Oshikawa:1996dj}
M.~Oshikawa and I.~Affleck, \emph{{Boundary conformal field theory approach to
  the critical two-dimensional Ising model with a defect line}},
  \href{https://doi.org/10.1016/S0550-3213(97)00219-8}{\emph{Nucl. Phys. B}
  {\bfseries 495} (1997) 533}
  [\href{https://arxiv.org/abs/cond-mat/9612187}{{\ttfamily
  cond-mat/9612187}}].

\bibitem{Tachikawa:2017gyf}
Y.~Tachikawa, \emph{{On gauging finite subgroups}},
  \href{https://doi.org/10.21468/SciPostPhys.8.1.015}{\emph{SciPost Phys.}
  {\bfseries 8} (2020) 015} [\href{https://arxiv.org/abs/1712.09542}{{\ttfamily
  1712.09542}}].

\bibitem{Lin:2019kpn}
Y.-H. Lin and S.-H. Shao, \emph{{Anomalies and Bounds on Charged Operators}},
  \href{https://doi.org/10.1103/PhysRevD.100.025013}{\emph{Phys. Rev. D}
  {\bfseries 100} (2019) 025013}
  [\href{https://arxiv.org/abs/1904.04833}{{\ttfamily 1904.04833}}].

\bibitem{Heidenreich:2021tna}
B.~Heidenreich, J.~Mcnamara, M.~Montero, M.~Reece, T.~Rudelius and
  I.~Valenzuela, \emph{{Non-Invertible Global Symmetries and Completeness of
  the Spectrum}},  \href{https://arxiv.org/abs/2104.07036}{{\ttfamily
  2104.07036}}.

\bibitem{Thorngren:2018bhj}
R.~Thorngren, \emph{{Anomalies and Bosonization}},
  \href{https://doi.org/10.1007/s00220-020-03830-0}{\emph{Commun. Math. Phys.}
  {\bfseries 378} (2020) 1775}
  [\href{https://arxiv.org/abs/1810.04414}{{\ttfamily 1810.04414}}].

\bibitem{Karch:2019lnn}
A.~Karch, D.~Tong and C.~Turner, \emph{{A Web of 2d Dualities: ${\bf Z}_2$
  Gauge Fields and Arf Invariants}},
  \href{https://doi.org/10.21468/SciPostPhys.7.1.007}{\emph{SciPost Phys.}
  {\bfseries 7} (2019) 007} [\href{https://arxiv.org/abs/1902.05550}{{\ttfamily
  1902.05550}}].

\bibitem{Ji:2019ugf}
W.~Ji, S.-H. Shao and X.-G. Wen, \emph{{Topological Transition on the Conformal
  Manifold}},
  \href{https://doi.org/10.1103/PhysRevResearch.2.033317}{\emph{Phys. Rev.
  Res.} {\bfseries 2} (2020) 033317}
  [\href{https://arxiv.org/abs/1909.01425}{{\ttfamily 1909.01425}}].

\bibitem{Thorngren:2015gtw}
R.~Thorngren and C.~von Keyserlingk, \emph{{Higher SPT's and a generalization
  of anomaly in-flow}},  \href{https://arxiv.org/abs/1511.02929}{{\ttfamily
  1511.02929}}.

\bibitem{Fidkowski:2009dba}
L.~Fidkowski and A.~Kitaev, \emph{{The effects of interactions on the
  topological classification of free fermion systems}},
  \href{https://doi.org/10.1103/PhysRevB.81.134509}{\emph{Phys. Rev. B}
  {\bfseries 81} (2010) 134509}
  [\href{https://arxiv.org/abs/0904.2197}{{\ttfamily 0904.2197}}].

\bibitem{Ryu:2012he}
S.~Ryu and S.-C. Zhang, \emph{{Interacting topological phases and modular
  invariance}}, \href{https://doi.org/10.1103/PhysRevB.85.245132}{\emph{Phys.
  Rev. B} {\bfseries 85} (2012) 245132}
  [\href{https://arxiv.org/abs/1202.4484}{{\ttfamily 1202.4484}}].

\bibitem{Yao:2012dhg}
H.~Yao and S.~Ryu, \emph{{Interaction effect on topological classification of
  superconductors in two dimensions}},
  \href{https://doi.org/10.1103/PhysRevB.88.064507}{\emph{Phys. Rev. B}
  {\bfseries 88} (2013) 064507}
  [\href{https://arxiv.org/abs/1202.5805}{{\ttfamily 1202.5805}}].

\bibitem{Qi:2013dsa}
X.-L. Qi, \emph{{A new class of (2 + 1)-dimensional topological superconductors
  with $\mathbb{Z}$ topological classification}},
  \href{https://doi.org/10.1088/1367-2630/15/6/065002}{\emph{New J. Phys.}
  {\bfseries 15} (2013) 065002}
  [\href{https://arxiv.org/abs/1202.3983}{{\ttfamily 1202.3983}}].

\bibitem{Kapustin:2014dxa}
A.~Kapustin, R.~Thorngren, A.~Turzillo and Z.~Wang, \emph{{Fermionic Symmetry
  Protected Topological Phases and Cobordisms}},
  \href{https://doi.org/10.1007/JHEP12(2015)052}{\emph{JHEP} {\bfseries 12}
  (2015) 052} [\href{https://arxiv.org/abs/1406.7329}{{\ttfamily 1406.7329}}].

\bibitem{Numasawa:2017crf}
T.~Numasawa and S.~Yamaguchi, \emph{{Mixed global anomalies and boundary
  conformal field theories}},
  \href{https://doi.org/10.1007/JHEP11(2018)202}{\emph{JHEP} {\bfseries 11}
  (2018) 202} [\href{https://arxiv.org/abs/1712.09361}{{\ttfamily
  1712.09361}}].

\bibitem{Runkel:2018feb}
I.~Runkel and L.~Szegedy, \emph{{Topological field theory on $r$-spin surfaces
  and the Arf invariant}},  \href{https://arxiv.org/abs/1802.09978}{{\ttfamily
  1802.09978}}.

\bibitem{Radicevic:2018okd}
D.~Radicevic, \emph{{Spin Structures and Exact Dualities in Low Dimensions}},
  \href{https://arxiv.org/abs/1809.07757}{{\ttfamily 1809.07757}}.

\bibitem{Yao:2020dqx}
Y.~Yao and A.~Furusaki, \emph{{Parafermionization, bosonization, and critical
  parafermionic theories}},  \href{https://arxiv.org/abs/2012.07529}{{\ttfamily
  2012.07529}}.

\bibitem{Fradkin:1980th}
E.~H. Fradkin and L.~P. Kadanoff, \emph{{DISORDER VARIABLES AND PARAFERMIONS IN
  TWO-DIMENSIONAL STATISTICAL MECHANICS}},
  \href{https://doi.org/10.1016/0550-3213(80)90472-1}{\emph{Nucl. Phys. B}
  {\bfseries 170} (1980) 1}.

\bibitem{Fateev:1985mm}
V.~A. Fateev and A.~B. Zamolodchikov, \emph{{Parafermionic Currents in the
  Two-Dimensional Conformal Quantum Field Theory and Selfdual Critical Points
  in Z(n) Invariant Statistical Systems}}, {\emph{Sov. Phys. JETP} {\bfseries
  62} (1985) 215}.

\bibitem{Harvey:2017rko}
J.~A. Harvey and G.~W. Moore, \emph{{An Uplifting Discussion of T-Duality}},
  \href{https://doi.org/10.1007/JHEP05(2018)145}{\emph{JHEP} {\bfseries 05}
  (2018) 145} [\href{https://arxiv.org/abs/1707.08888}{{\ttfamily
  1707.08888}}].

\bibitem{Carqueville:2017ono}
N.~Carqueville, I.~Runkel and G.~Schaumann, \emph{{Line and surface defects in
  Reshetikhin-Turaev TQFT}},
  \href{https://arxiv.org/abs/1710.10214}{{\ttfamily 1710.10214}}.

\bibitem{Longo:1994xe}
R.~Longo and K.-H. Rehren, \emph{{Nets of subfactors}},
  \href{https://doi.org/10.1142/S0129055X95000232}{\emph{Rev. Math. Phys.}
  {\bfseries 7} (1995) 567}
  [\href{https://arxiv.org/abs/hep-th/9411077}{{\ttfamily hep-th/9411077}}].

\bibitem{Bockenhauer:1998ca}
J.~Bockenhauer and D.~E. Evans, \emph{{Modular invariants, graphs and alpha
  induction for nets of subfactors. 1.}},
  \href{https://doi.org/10.1007/s002200050455}{\emph{Commun. Math. Phys.}
  {\bfseries 197} (1998) 361}
  [\href{https://arxiv.org/abs/hep-th/9801171}{{\ttfamily hep-th/9801171}}].

\bibitem{Bockenhauer:1998in}
J.~Bockenhauer and D.~E. Evans, \emph{{Modular invariants, graphs and alpha
  induction for nets of subfactors. 2.}},
  \href{https://doi.org/10.1007/s002200050523}{\emph{Commun. Math. Phys.}
  {\bfseries 200} (1999) 57}
  [\href{https://arxiv.org/abs/hep-th/9805023}{{\ttfamily hep-th/9805023}}].

\bibitem{Bockenhauer:1998ef}
J.~Bockenhauer and D.~E. Evans, \emph{{Modular invariants, graphs and alpha
  induction for nets of subfactors. 3.}},
  \href{https://doi.org/10.1007/s002200050673}{\emph{Commun. Math. Phys.}
  {\bfseries 205} (1999) 183}
  [\href{https://arxiv.org/abs/hep-th/9812110}{{\ttfamily hep-th/9812110}}].

\bibitem{Bockenhauer:1999wt}
J.~Bockenhauer, D.~E. Evans and Y.~Kawahigashi, \emph{{Chiral structure of
  modular invariants for subfactors}},
  \href{https://doi.org/10.1007/s002200050798}{\emph{Commun. Math. Phys.}
  {\bfseries 210} (2000) 733}
  [\href{https://arxiv.org/abs/math/9907149}{{\ttfamily math/9907149}}].

\bibitem{2001math.....11139O}
V.~{Ostrik}, \emph{{Module categories, weak Hopf algebras and modular
  invariants}}, {\emph{arXiv Mathematics e-prints} (2001) math/0111139}
  [\href{https://arxiv.org/abs/math/0111139}{{\ttfamily math/0111139}}].

\bibitem{milnor1973symmetric}
J.~Milnor and D.~Husem{\"o}ller, \emph{Symmetric Bilinear Forms}, Ergebnisse
  der Mathematik und ihrer Grenzgebiete. Springer-Verlag, 1973.

\bibitem{WALL1963281}
C.~Wall, \emph{Quadratic forms on finite groups, and related topics},
  \href{https://doi.org/https://doi.org/10.1016/0040-9383(63)90012-0}{\emph{Topology}
  {\bfseries 2} (1963) 281 }.

\bibitem{wang2020abelian}
L.~Wang and Z.~Wang, \emph{In and around abelian anyon models},  2020.

\bibitem{turaev_1998}
V.~TURAEV, \emph{Reciprocity for gauss sums on finite abelian groups},
  \href{https://doi.org/10.1017/S0305004198002655}{\emph{Mathematical
  Proceedings of the Cambridge Philosophical Society} {\bfseries 124} (1998)
  205–214}.

\bibitem{Tambara2000}
D.~Tambara, \emph{Representations of tensor categories with fusion rules of
  self-duality for abelian groups},
  \href{https://doi.org/10.1007/BF02803515}{\emph{Israel Journal of
  Mathematics} {\bfseries 118} (2000) 29}.

\bibitem{Dijkgraaf:1987vp}
R.~Dijkgraaf, E.~P. Verlinde and H.~L. Verlinde, \emph{{C = 1 Conformal Field
  Theories on Riemann Surfaces}},
  \href{https://doi.org/10.1007/BF01224132}{\emph{Commun. Math. Phys.}
  {\bfseries 115} (1988) 649}.

\bibitem{Gepner:1986hr}
D.~Gepner and Z.-a. Qiu, \emph{{Modular Invariant Partition Functions for
  Parafermionic Field Theories}},
  \href{https://doi.org/10.1016/0550-3213(87)90348-8}{\emph{Nucl. Phys. B}
  {\bfseries 285} (1987) 423}.

\bibitem{Fateev:1991bv}
V.~A. Fateev and A.~B. Zamolodchikov, \emph{{Integrable perturbations of Z(N)
  parafermion models and O(3) sigma model}},
  \href{https://doi.org/10.1016/0370-2693(91)91283-2}{\emph{Phys. Lett. B}
  {\bfseries 271} (1991) 91}.

\bibitem{Fuchs:2007tx}
J.~Fuchs, M.~R. Gaberdiel, I.~Runkel and C.~Schweigert, \emph{{Topological
  defects for the free boson CFT}},
  \href{https://doi.org/10.1088/1751-8113/40/37/016}{\emph{J. Phys. A}
  {\bfseries 40} (2007) 11403}
  [\href{https://arxiv.org/abs/0705.3129}{{\ttfamily 0705.3129}}].

\bibitem{Dijkgraaf:1989hb}
R.~Dijkgraaf, C.~Vafa, E.~P. Verlinde and H.~L. Verlinde, \emph{{The Operator
  Algebra of Orbifold Models}},
  \href{https://doi.org/10.1007/BF01238812}{\emph{Commun. Math. Phys.}
  {\bfseries 123} (1989) 485}.

\bibitem{Yang:1987wk}
S.-K. Yang, \emph{{$Z$(4) X $Z$(4) Symmetry and Parafermion Operators in the
  Selfdual Critical Ashkin-teller Model}},
  \href{https://doi.org/10.1016/0550-3213(87)90359-2}{\emph{Nucl. Phys. B}
  {\bfseries 285} (1987) 639}.

\bibitem{Meir_2012}
E.~Meir and E.~Musicantov, \emph{Module categories over graded fusion
  categories}, \href{https://doi.org/10.1016/j.jpaa.2012.03.014}{\emph{Journal
  of Pure and Applied Algebra} {\bfseries 216} (2012) 2449–2466}.

\bibitem{deepak}
D.~{Naidu}, \emph{{Categorical Morita equivalence for group-theoretical
  categories}}, {\emph{arXiv Mathematics e-prints} (2006) math/0605530}
  [\href{https://arxiv.org/abs/math/0605530}{{\ttfamily math/0605530}}].

\bibitem{Alcaraz:1980sa}
F.~C. Alcaraz and R.~Koberle, \emph{{Duality and the Phases of $Z$(n) Spin
  Systems}}, \href{https://doi.org/10.1088/0305-4470/13/5/008}{\emph{J. Phys.
  A} {\bfseries 13} (1980) L153}.

\bibitem{Alcaraz:1980bb}
F.~C. Alcaraz and R.~Koberle, \emph{{The Phases of Two-dimensional Spin and
  Four-dimensional Gauge Systems With $Z(N)$ Symmetry}},
  \href{https://doi.org/10.1088/0305-4470/14/5/036}{\emph{J. Phys. A}
  {\bfseries 14} (1981) 1169}.

\bibitem{Alcaraz:1986hs}
F.~C. Alcaraz, \emph{{The Critical Behavior of Selfdual $Z(N$) Spin Systems:
  Finite Size Scaling and Conformal Invariance}},
  \href{https://doi.org/10.1088/0305-4470/20/9/035}{\emph{J. Phys. A}
  {\bfseries 20} (1987) 2511}.

\bibitem{Dorey:1996he}
P.~Dorey, R.~Tateo and K.~E. Thompson, \emph{{Massive and massless phases in
  selfdual Z(N) spin models: Some exact results from the thermodynamic Bethe
  ansatz}}, \href{https://doi.org/10.1016/0550-3213(96)00183-6}{\emph{Nucl.
  Phys. B} {\bfseries 470} (1996) 317}
  [\href{https://arxiv.org/abs/hep-th/9601123}{{\ttfamily hep-th/9601123}}].

\bibitem{Cappelli:2002wq}
A.~Cappelli and G.~D'Appollonio, \emph{{Boundary states of c = 1 and 3/2
  rational conformal field theories}},
  \href{https://doi.org/10.1088/1126-6708/2002/02/039}{\emph{JHEP} {\bfseries
  02} (2002) 039} [\href{https://arxiv.org/abs/hep-th/0201173}{{\ttfamily
  hep-th/0201173}}].

\bibitem{Nguyen:2021naa}
M.~Nguyen, Y.~Tanizaki and M.~\"Unsal, \emph{{Non-invertible 1-form symmetry
  and Casimir scaling in 2d Yang-Mills theory}},
  \href{https://arxiv.org/abs/2104.01824}{{\ttfamily 2104.01824}}.

\bibitem{Baker1977DifferentialCA}
D.~Baker, \emph{Differential characters and borel cohomology}, {\emph{Topology}
  {\bfseries 16} (1977) 441}.

\bibitem{Gaberdiel:2001xm}
M.~R. Gaberdiel, A.~Recknagel and G.~M.~T. Watts, \emph{{The Conformal boundary
  states for SU(2) at level 1}},
  \href{https://doi.org/10.1016/S0550-3213(02)00033-0}{\emph{Nucl. Phys. B}
  {\bfseries 626} (2002) 344}
  [\href{https://arxiv.org/abs/hep-th/0108102}{{\ttfamily hep-th/0108102}}].

\bibitem{Zamolodchikov:1985wn}
A.~B. Zamolodchikov, \emph{{Infinite Additional Symmetries in Two-Dimensional
  Conformal Quantum Field Theory}},
  \href{https://doi.org/10.1007/BF01036128}{\emph{Theor. Math. Phys.}
  {\bfseries 65} (1985) 1205}.

\bibitem{Blumenhagen:1990jv}
R.~Blumenhagen, M.~Flohr, A.~Kliem, W.~Nahm, A.~Recknagel and R.~Varnhagen,
  \emph{{W algebras with two and three generators}},
  \href{https://doi.org/10.1016/0550-3213(91)90624-7}{\emph{Nucl. Phys. B}
  {\bfseries 361} (1991) 255}.

\bibitem{Hornfeck:1992tm}
K.~Hornfeck, \emph{{W algebras with set of primary fields of dimensions (3, 4,
  5) and (3, 4, 5, 6)}},
  \href{https://doi.org/10.1016/0550-3213(93)90281-S}{\emph{Nucl. Phys. B}
  {\bfseries 407} (1993) 237}
  [\href{https://arxiv.org/abs/hep-th/9212104}{{\ttfamily hep-th/9212104}}].

\bibitem{deBoer:1993gd}
J.~de~Boer, L.~Feher and A.~Honecker, \emph{{A Class of W algebras with
  infinitely generated classical limit}},
  \href{https://doi.org/10.1016/0550-3213(94)90388-3}{\emph{Nucl. Phys. B}
  {\bfseries 420} (1994) 409}
  [\href{https://arxiv.org/abs/hep-th/9312049}{{\ttfamily hep-th/9312049}}].

\bibitem{Blumenhagen:1994wg}
R.~Blumenhagen, W.~Eholzer, A.~Honecker, K.~Hornfeck and R.~Hubel, \emph{{Coset
  realization of unifying W algebras}},
  \href{https://doi.org/10.1142/S0217751X95001157}{\emph{Int. J. Mod. Phys. A}
  {\bfseries 10} (1995) 2367}
  [\href{https://arxiv.org/abs/hep-th/9406203}{{\ttfamily hep-th/9406203}}].

\bibitem{Dong:2009xd}
C.~Dong, C.~H. Lam, Q.~Wang and H.~Yamada, \emph{{The Structure of parafermion
  vertex operator algebras}},
  \href{https://doi.org/10.1007/s00220-010-1114-8}{\emph{Commun. Math. Phys.}
  {\bfseries 299} (2010) 783}
  [\href{https://arxiv.org/abs/0904.2758}{{\ttfamily 0904.2758}}].

\bibitem{Fateev:1990bf}
V.~A. Fateev, \emph{{Integrable deformations in Z(N) symmetrical models of
  conformal quantum field theory}},
  \href{https://doi.org/10.1142/S0217751X91001052}{\emph{Int. J. Mod. Phys. A}
  {\bfseries 6} (1991) 2109}.

\end{thebibliography}\endgroup
\bibliographystyle{JHEP}

\end{document}